\documentclass[twoside,fleqn]{ActaPS}
\usepackage[colorlinks=true, pdfstartview=FitV, linkcolor=black,
            citecolor=black, urlcolor=black]{hyperref}

\usepackage{latexsym}
\usepackage{amsthm}
\usepackage{calc}
\usepackage{epsfig}
\usepackage{subfig}
\usepackage{amsmath, amssymb, amsthm, euscript} 
\usepackage{psfrag}
\usepackage{caption}
\usepackage{txfonts}
\usepackage{textcomp}
\usepackage{pdfpages}

\usepackage{dsfont}

\usepackage{breakcites}
\usepackage{graphicx}
\usepackage{times,cite}
\usepackage{dsfont}
\usepackage{fancyhdr}

\renewcommand{\subsectionmark}[1]{}
\renewcommand{\thesection}{\arabic{section}}
\renewcommand{\thesubsection}{\thesection.\arabic{subsection}}

\renewcommand{\thefigure}{\thesection.\arabic{figure}}
\renewcommand{\thetable}{\thesection.\arabic{table}}

\newcommand{\bra}[1]{\langle #1|\,}
\newcommand{\ket}[1]{\,|#1 \rangle}

\def\shorttitle{Tensor Networks}

\begin{document}

\pagerange{1}{121}

\vspace{-0.8cm}
\title{TENSOR NETWORKS: PHASE TRANSITION PHENOMENA\\ON HYPERBOLIC AND FRACTAL GEOMETRIES}

\vspace{-0.3cm}
\author{Jozef~Genzor$^{\;a}$, Tomotoshi~Nishino$^{\;a}$, Andrej~Gendiar$^{\;b,}$\email{andrej.gendiar@savba.sk}}
{$^{a}$Department of Physics, Graduate School of Science, Kobe University\\ Kobe 657-8501, Japan\\
$^{b}$Institute of Physics, Slovak Academy of Sciences, D\'{u}bravsk\'{a} cesta 9\\ SK-845~11 Bratislava, Slovakia}

\abstract{
One of the challenging problems in the condensed matter physics is to understand the
quantum many-body systems, especially, their physical mechanisms behind. Since there are
only a few complete analytical solutions of these systems, several numerical simulation
methods have been proposed in recent years. Amongst all of them, the {\it Tensor Network}
algorithms have become increasingly popular in recent years, especially for their
adaptability to simulate strongly correlated systems. The current work focuses on the
generalization of such Tensor-Network-based algorithms, which are sufficiently robust
to describe critical phenomena and phase transitions of multistate spin Hamiltonians
in the thermodynamic limit. Therefore, one has to deal with systems of infinitely many
interacting spin particles. For this purpose, we have chosen two algorithms: the Corner
Transfer Matrix Renormalization Group and the Higher-Order Tensor Renormalization Group.
The ground state of those multistate spin systems in the thermodynamic equilibrium is
constructed in terms of a tensor product state Ansatz in both of the algorithms. The main
aim of this work is to generalize the idea behind these two algorithms in order to be able
to calculate the thermodynamic properties of non-Euclidean geometries. In particular,
the tensor product state algorithms of hyperbolic geometries with negative Gaussian
curvatures as well as fractal geometries will be theoretically analyzed followed
by extensive numerical simulations of the multistate spin models. These spin systems
were chosen for their applicability to mimic intrinsic properties of more complex
systems, such as social behavior, neural network, the holographic principle, including
the correspondence between the anti-de Sitter and conformal field theory of quantum
gravity. This work is based on tensor-network analysis and opens doors for
the understanding of phase transition and entanglement of the interacting systems
on the non-Euclidean geometries. We focus on three main topics: A new thermodynamic
model of social influence, free energy is analyzed to classify the phase transitions
on an infinite set of the negatively curved geometries where a relation between the
free energy and the Gaussian radius of the curvature is conjectured, a unique
tensor-based algorithm is proposed to study the phase transition on fractal structures.}


\begin{minipage}{2.5cm}
\quad{\small {\sf KEYWORDS:}}
\end{minipage}
\begin{minipage}{10.0cm}
Classical Statistical Mechanics, Phase Transitions and Criticality, Spin Systems,
Tensor Networks, Density Matrix Renormalization Group, Hyperbolic Geometry,
Fractal Lattices
\end{minipage}

\tableofcontents

\newpage

\section{Introduction}

The mathematical treatment of the collective behavior of many-body systems is a highly nontrivial task. 
Even knowing the underlying laws of microscopic interactions does not guarantee that we can say 
anything specific about the large-scale behavior of the studied system. 
The application of the laws might lead to equations which are too complex to be solved.  
Even worse, the difficulty is usually one level deeper, the Hilbert space is far too large. 
Imagine having $N$ particles with spin one-half. 
To describe a state of such a system would require knowing $2^N$ complex amplitudes. 
For realistic systems (like a piece of a magnet), the number of particles is $N \sim 10^{23}$, 
which makes the number of basis states larger than the number of all particles in the observable universe. 

The nature usually prefers systems with local interactions (i.e., the nearest and/or the second-nearest neighbors are assumed to interact preferably). Moreover, not all states can be considered equal. As a consequence, Hilbert space of realistic systems can become significantly reduced. Low-energy states of such systems, which have gapped Hamiltonian spectra, constitute only a tiny fraction of all the possible states of the entire Hilbert space. Those states satisfy an \textit{area law}, which applies to entanglement entropy. It means that the entanglement entropy obeys a specific rule, in which the entanglement entropy scales with respect to the surface of a subsystem if embedded in the entire system, provided that the gapped systems are considered. (We remark here that the entanglement entropy is not proportional to the volume of the subsystem.)

The area law is useful in an efficient quantifying of the entanglement of various quantum systems. This is the reason why tensor networks have been successful in description of the quantum systems. The tensor networks can be also applied to the systems, for which the area law is not satisfied, for instance, at phase transitions, topological phases, etc. It is worth mentioning that the gapless systems studied by the tensor networks cannot reach as high accuracy as the gapped ones. The tensor-network formalism follows interaction geometry among particles, which is recognized as the lattice structure. This formalism, in connection with the real-space renormalization-group methods, allows us to perform numerical calculations efficiently in the thermodynamic limit, i.e., $N\rightarrow\infty$.  

The underlying interaction topology of a system under study plays a crucial role in determining its thermodynamic properties. This is related to the lattice dimensionality. For example, there is no phase transition at nonzero temperature in the classical one-dimensional Ising model, whereas, there exists a finite critical temperature at higher dimensions. We intend to focus on studies of the phase transitions on non-Euclidean lattices. In particular, we plan to investigate and classify hyperbolic surface geometries, which have the infinite effective spatial dimension ($d\rightarrow\infty$) with negative Gaussian curvatures measured on curved lattice surfaces. Fractal geometries with fractional dimensions $1 < d < 2$ are of our interest in this review, too. The main purpose for researching the phase transition phenomena of spin systems on the non-Euclidean lattice geometries is the fact that these systems are neither exactly solvable nor numerically feasible by standard methods such as Monte Carlo simulations, exact diagonalization, Density Matrix Renormalization Group, etc. We therefore propose a few generalized numerical algorithms based on Tensor Network ideas, which enable us to solve the spin systems on hyperbolic and fractal lattices in the thermodynamic limit. The algorithms reach a sufficiently high numerical accuracy, which allow us to classify their phase transitions and evaluate the associated critical exponents. We show that we have successfully achieved novel results, which have been missing in the theory of the solid state physics, statistical mechanics, quantum information, as well as in the anti-de Sitter space, which is useful in the general theory of relativity. The main results of this review have been published in Refs.~\cite{Serina, Genzor, Fractal}. 

This work is structured in the following. Section~\ref{chap1} contains basic definitions and notations of the phase transition theory, including the Suzuki-Trotter mapping. The tensor-network theory is explained in Section~\ref{chap2}. This Section is meant as a tutorial, where we tried to include many practical details and comments related to the numerical calculations with the aim to explain missing information for those who are interested in this area of research. The three conceptually different numerical methods are explained, i.e., infinite Time-Evolving Block Decimation (iTEBD), Corner Transfer Matrix Renormalization Group (CTMRG), and Higher-Order Tensor Renormalization Group (HOTRG). Exclusively for demonstrative reasons, this Section also contains supporting numerical results. These explanations are also complemented by the source codes which can be found in the online repositories~\cite{CTMRG_source, HoTRG_source, iTEBD_source}.

We encourage the readers who are experienced in the statistical physics and tensor networks to skip Section~\ref{chap2} and proceed with the next Section, where novel results are presented. Section~\ref{chap3} generalizes the CTMRG method to multiple hyperbolic geometries and investigates the relationship between the lattice curvature and the free energy. Section~\ref{chap4} is concerned with the models of social behavior. We propose a unique thermodynamic model of social influence, being inspired by the well-known Axelrod model. We adapt the CTMRG method in this study. The phase transitions on fractal geometries are studied in Section~\ref{chap5}, where a simple fractal lattice is proposed. The HOTRG algorithm, which has been developed for the two-dimensional (square) and the three-dimensional (cubic) lattices, is modified to be applied to the the fractal lattice. In addition, we propose two infinite series of fractal lattices converging either to one-dimensional or to two-dimensional regular lattices (this research is subject to our ongoing research and is to be published elsewhere \cite{GGT}). The Sections \ref{chap3}, \ref{chap4}, and \ref{chap5} contain the research results which have been published as a part of the PhD thesis of the first author.

\newpage\setcounter{equation}{0} \setcounter{figure}{0} \setcounter{table}{0}
\section{General introduction and concepts} \label{chap1}

\subsection{Theory of phase transitions}\label{PT_Section}

\begin{figure}[b!]
 \centering
 \includegraphics[width=3in]{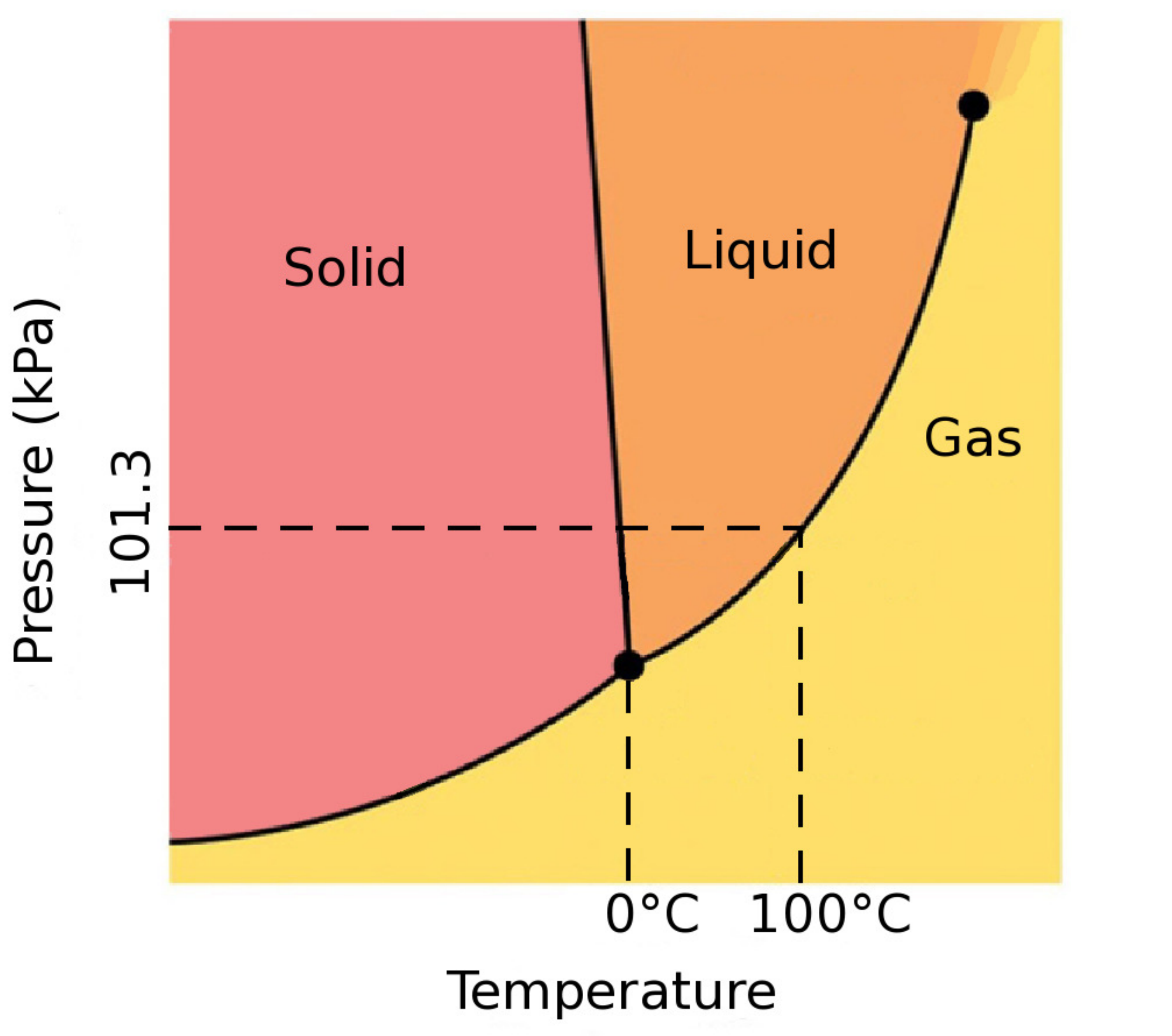}
   \caption{The temperature-pressure phase diagram for water.
}\label{phasediagram}
 \end{figure}

The \textit{phase transition} phenomena have a long history of the study. 
The term `phase transition' refers to an abrupt change in properties of a system 
induced by changes in external parameters like temperature or pressure.
The exhibited abrupt change can be described in terms of a certain non-analyticity of the thermodynamic functions derived from the free energy, i.e., a discontinuity observed in these functions. A few types of various phase transitions can be distinguished. 
The phase transition exhibiting a discontinuity in the first derivative of the free energy is classified as a \textit{first-order phase transition}, according to the Ehrenfest classification. One of the examples is solid/liquid/gas phase transitions, see Fig.~\ref{phasediagram}. 
 A sudden discontinuity (jump) in the first derivative of the free energy (being the internal energy) at a phase transition is associated with the latent heat, which needs to be exchanged to the transition to occur. The type of the phase transition is known as the discontinuous one.
 
\begin{figure}[tb]
 \centering
 \includegraphics[width=3.5in]{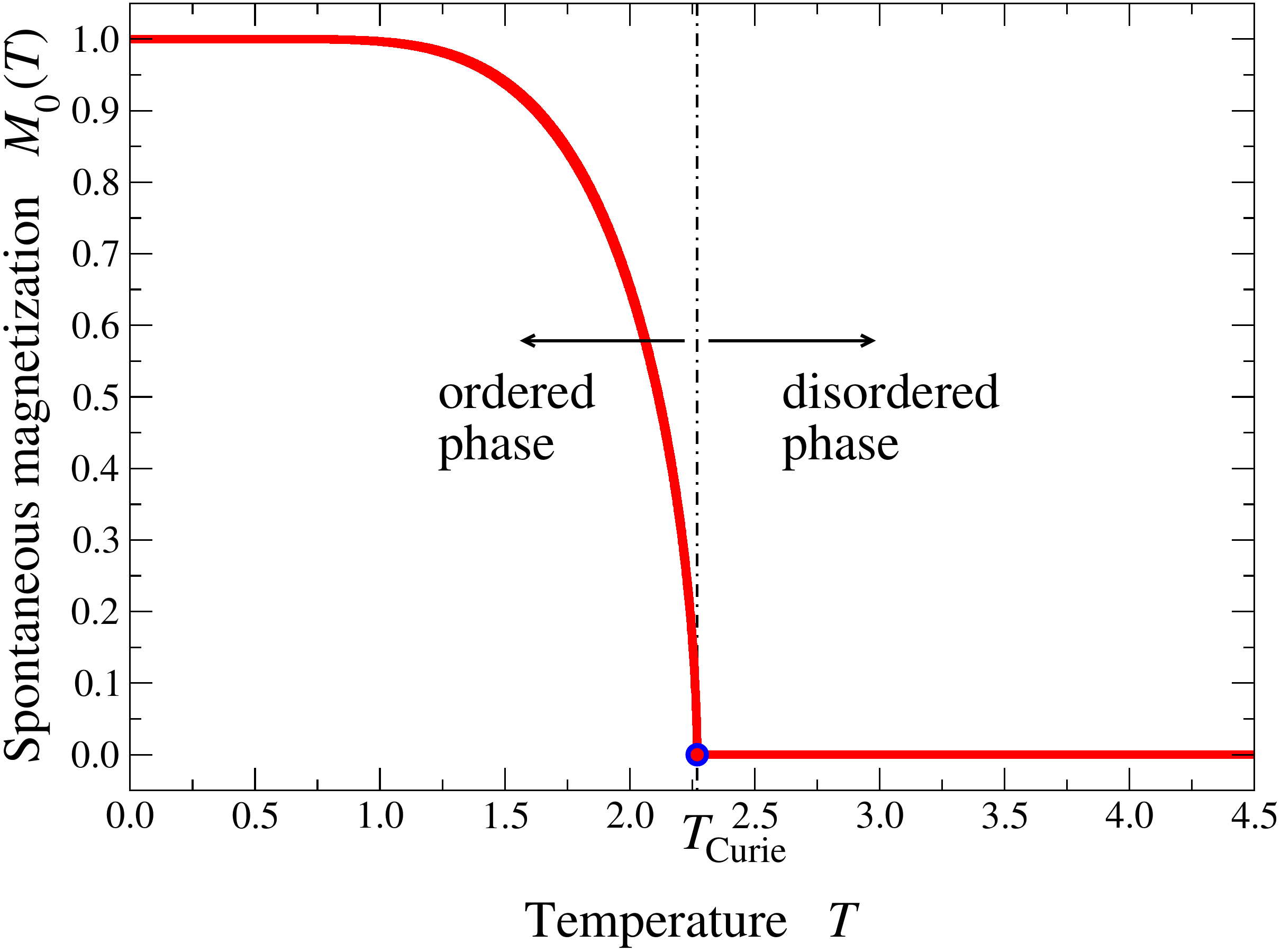}
	 \caption{The phase diagram of a magnetic material. Below the Curie temperature $0 \leq T < T_{\text{Curie}}$, the ordered magnetic phase emerges, which is typical for the nonzero spontaneous magnetization $M_0(T)$. The spontaneous magnetization of a ferromagnetic material is used as the typical order parameter. Above the Curie temperature $T > T_{\text{Curie}}$, the disordered phase is located, where the order parameter $M_0(T)=0$; this phase is often referred to as the paramagnetic phase. The sign of $M_0(T)$ below $T_{\text{Curie}}$ is determined by the spontaneous symmetry-breaking mechanism in the thermodynamic limit. In this particular case, we plot $M_0(T)>0$, which occurs if $h \to 0^+$.}
\label{magnetization}
 \end{figure}

Another type of the phase transition is a \textit{second-order phase transition}, when the first derivative of the free energy remains continuous; however, the second derivative is discontinuous. This phase transition is known as the continuous one. An important example of the second-order phase transition is a magnetic material, which exhibits nonzero macroscopic spontaneous magnetization, $M_0(T)$, emerging below a specific temperature (the so-called Curie temperature $T_{\text{Curie}}$), cf. Fig.~\ref{magnetization}. If $T < T_{\text{Curie}}$, the spontaneous magnetization is nonzero (can be either positive or negative). The sign of $M_0(T)$ can be determined by a symmetry-breaking mechanism. It can be initialized by a small external magnetic field $h\neq0$. Hence, the final sign of $M_0(T)$ at zero field obeys the rule
\begin{equation}
\lim_{h \to 0^{\pm}} M(h, T) = \pm M_0(T) \, , 
\end{equation}
where $M_0(T)$ is called the \textit{spontaneous magnetization}, see Fig.~\ref{magnetization}. 
However, for temperatures above the Curie temperature $T > T_{\text{Curie}}$, the spontaneous magnetization is strictly zero. This spontaneous magnetization is a typical example of the order parameter, which is nonzero below and zero above a phase transition.

\subsubsection{Ising model}\label{Ising_model_subs}

A simple example of magnetic materials is modeled by the spin Ising model. In this model, the spins are placed on the sites of the lattice. The interaction between spins is limited to the nearest neighbors only in this model. A spin variable $\sigma_i$ at a lattice position $i$ can assume only two values, either $+1$ or $-1$. In this simplified model of a magnetic material, the spin Hamiltonian is defined as
\begin{equation} \label{J}
{\cal H}(\sigma) = - \displaystyle \sum_{i \neq j} J_{ij}\sigma_i \sigma_j 
 - \sum_{i} h_{i} \sigma_i \, ,
\end{equation}
where
\begin{equation} 
  J_{ij}^{~}\begin{cases}
    >0 & \text{$i,j$ are the nearest neighbors} \, ,\\
    =0 & \text{otherwise} \, .
  \end{cases}
\end{equation}
For tutorial purposes, we consider the simplest case of a constant spin interaction $J_{ij}=J$ and constant magnetic field $h_i = h$. The ferromagnetism or antiferromagnetism, respectively, is represented by setting $J>0$ or $J<0$.

At $T = 0$, the system tends to the minimum energy that is achieved when all the spins are aligned (either in the $+1$ state or in the $-1$ state). The order parameter is given by averaging the spontaneous magnetization.

\begin{figure}[tb] 
\centering
 \includegraphics[width=4.022cm]{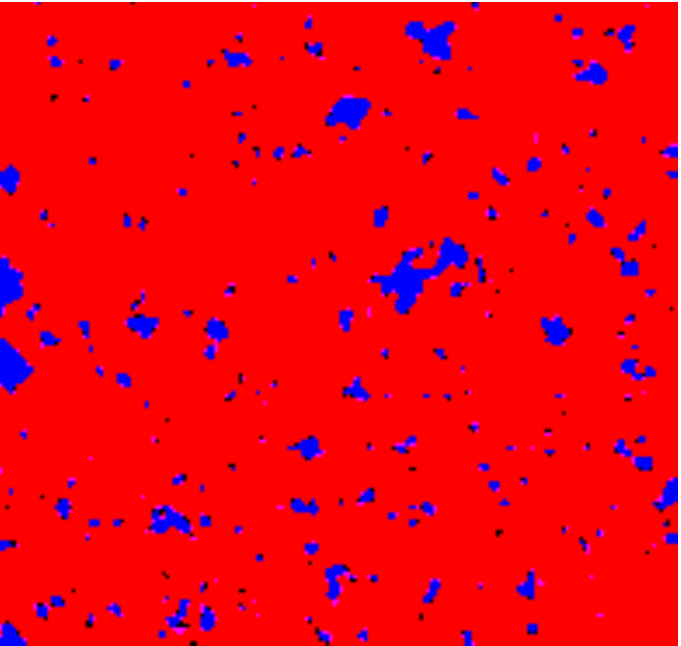}
 \includegraphics[width=4.000cm]{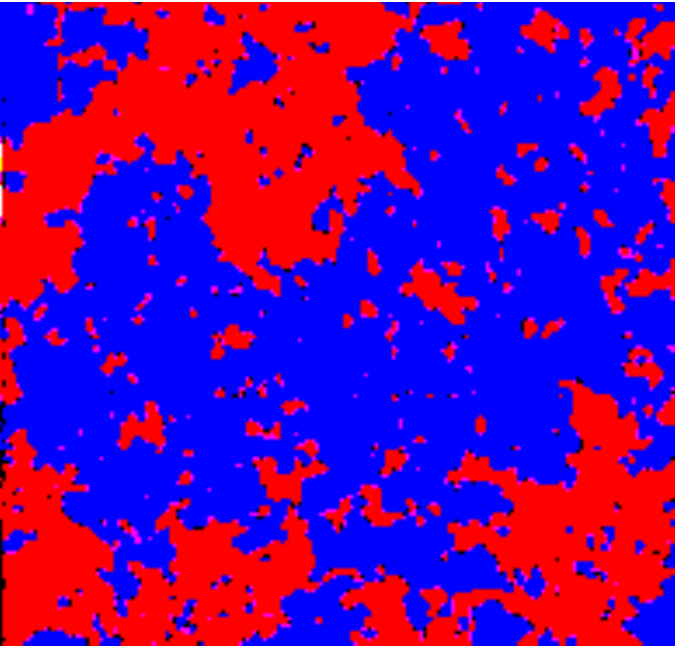}
 \includegraphics[width=4.063cm]{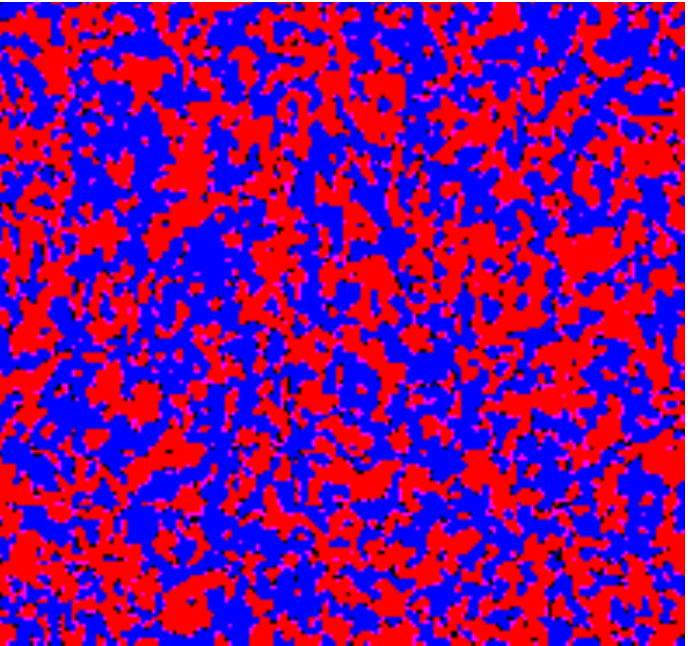}\\
\hfill $T<T_c$ \hfill \qquad $T=T_c$ \hfill \qquad $T>T_c$ \hfill \phantom{.}
 \caption{Typical snapshots of the Ising system on the square lattice below, at, and above the critical temperature $T_c$ at zero magnetic field $h=0$.}\label{Ising}
 \end{figure}

There is no phase transition in the one-dimensional Ising model; the ordered configuration
is actually present at $T = h = 0$ only. The situation is radically different in higher
dimensions, however. From the analytical point of view, the phase transitions are rigorously determined in the thermodynamic limit ($N \rightarrow \infty$). At finite but large enough $N$, however, a qualitative change in behavior can be observed as the temperature is lowered, see Fig.~\ref{Ising}. The red and the blue regions on the square lattice system represent, respectively, the spin up and down of a particular spin configuration (the so-called snapshot) calculated by the Monte Carlo simulations at three different temperatures $T$.

At high temperatures, the system is in a disordered phase and the spin configurations with a globally vanishing total magnetization ($M=0$ as the number of up and down spins is equal and form a random pattern with small spin-sized domains of spins with the same orientation). At low temperatures, a symmetry breaking between two spin states $+1$ and $-1$ takes place. In the particular case, large-sized domains of up spins can be formed (yielding the spontaneous magnetization $M>0$). In the thermodynamic limit, a phase transition occurs at a certain critical temperature $T_c$.  Below the critical temperature $T < T_c$, an ordered ferromagnetic phase with the spontaneous magnetization $M \neq 0$ is present, whereas above the critical temperature $T > T_c$, a disordered paramagnetic phase with $M = 0$ occurs. Analytic solutions for the Ising model exist only for the one-dimensional spin chain and for the two-dimensional lattice at zero magnetic field $h=0$~\cite{Baxter}.

The Ising model was originally defined in the physical context of the magnetism. However, the
concept of the phase transitions is much wider -- the Ising model itself has various
applications. Amongst all of them, we later present its usefulness when applying the Ising model to the thermodynamic systems of social behavior and opinion dynamics~\cite{BBV}.
 
\subsubsection{Equilibrial statistical physics and critical phenomena}

Let us begin with definition of the canonical partition function
\begin{equation} \label{partition function}
{\cal Z} = \sum_{\{\sigma\}} \exp\left( - \frac{{\cal H}(\sigma)}{k_{\rm B} T} \right) \equiv \exp \left(- \frac{F}{k_{\rm B} T} \right)\, ,
\end{equation}
where ${\cal H}(\sigma)$ can be an arbitrary spin Hamiltonian, and the summation runs over
entire range of all spin configurations $\{\sigma\}$. This is often expressed
in terms of the free energy
\begin{equation}
F = - k_{\rm B}T\ln{\cal Z} \, .
\label{freng}
\end{equation}
Our analysis is concentrated on the so-called {\em Helmholtz free energy} $F=U-TS$, which
is the function of the internal energy $U$, temperature $T$, and the entropy $S$. The free
energy contains sufficient information on the spin system and can be used to determine other
thermodynamic functions. Out of the non-analytic behavior at phase transitions, one can
analyze the thermodynamic functions and classify them, provided that the system is in the thermodynamic limit. It means that the total number of all the spin variables
$\sigma$ has to be infinite.

Let us list the most commonly used thermodynamic functions. The thermodynamic entropy
\begin{equation}
 \label{th_entr}
S = -\dfrac{\partial F}{\partial T}
\end{equation}
is usually an increasing analytic function, such that $S=0$ at zero temperature and gets
saturated at large temperatures. In particular, $S\to\ln\,2$ at $T\to\infty$ for the classical two-state spin models, like the Ising model. The first derivative with
respect to temperature $T$ results in the
{\em internal energy}
\begin{equation}
 \label{int_eng}
U = - T^2 \dfrac{\partial \left(F/T\right)}{\partial T}\, .
\end{equation}

The consequent temperature derivative of the internal energy yields the {\em specific heat}
\begin{equation}
 \label{spec_heat}
C = \dfrac{\partial U}{\partial T}=-T\dfrac{\partial^2 F}{\partial T^2}\, ,
\end{equation}
which has a non-analytic (divergent) behavior at a phase transition.
Analogously, the first derivative of the free energy with respect to an external field $h$ results in the magnetization
\begin{equation}
M = -\left. \frac{\partial F(T, h)}{\partial h} \right|_{h \to 0} \, , 
\end{equation}
and the second derivatives of the free energy
specifies the \textit{magnetic susceptibility}
\begin{equation}
\chi = \left. \frac{\partial M}{\partial h} \right|_{h \to 0} \, . 
\end{equation}

These thermodynamic functions can be equivalently derived through the probability of the
system to be in a given spin microstate
\begin{equation}
{\cal P}(\sigma) = \frac{1}{{\cal Z}}\exp\left[- \frac{{\cal H}(\sigma)}{k_{\rm B} T} \right] \, .
\end{equation}
An observable (i.~e. an averaged thermodynamic function) $O$ at temperature $T$ is given by the formula
\begin{equation}
\left<O\right> = \sum_{\{\sigma\}} O(\sigma)\ {\cal P}(\sigma) = \frac{1}{\cal Z} \sum_{\{\sigma\}} O(\sigma)\exp\left[- \frac{{\cal H}(\sigma)}{k_{\rm B}T}\right] \, .
\end{equation}
Then, equivalently, the internal energy (the energy per interacting bond) can by expressed as
\begin{equation}
U = \left<{\cal H}\right> = \frac{1}{\cal Z} \sum_{\{\sigma\}} {\cal H}(\sigma) \exp \left( - \frac{\cal H(\sigma)}{k_{\rm B} T} \right) \, , 
\end{equation}
and, for an example, the magnetization $M$ per spin site as
\begin{equation}
M = \left<{\sigma}\right> = \frac{1}{\cal Z} \sum_{\{\sigma\}} \frac{1}{N} \sum\limits_{i=1}^{N} \sigma_i \exp \left( - \frac{\cal H(\sigma)}{k_{\rm B} T} \right) \, ,
\label{sp_mag}
\end{equation}
where $N$ denotes the total number of spins $\sigma_i$ placed at lattice sites $i$.

\subsubsection{Correlation function}

The correlation function $G_{i,j}$ is a symmetric function $G_{i,j}=G_{j,i}$ acting between two spins $\sigma_i$ and $\sigma_j$ on the lattice 
\begin{equation}
G_{i,j} = \left< \sigma_i^{~} \sigma_j^{~} \right> - \left< \sigma_i^{~} \right>\left< \sigma_j^{~} \right> \, .
\end{equation}
If $r$ is a distance, $r=|i-j|=\sqrt{|\vec{r}_i|^2 + |\vec{r}_j|^2}$, between two spins $\sigma_i$ and $\sigma_j$ placed at vector positions $\vec{r}_i$ and $\vec{r}_j$ on the square lattice, the correlation function $G_{i,j}$, as the function of the two-spin position, decreases to zero if $r \rightarrow \infty$. In general, the correlation function
\begin{equation}
|G_{i,j}| \equiv g(r) \propto \frac{1}{r^{\,\tau}} \exp(-r/\xi) \, , 
\end{equation}
where $\xi$ is known as the \textit{correlation length}, and $\tau$ is an exponent,
which becomes dominant at the critical phase transition temperature $T_c$. Then, the
correlation length $\xi$ diverges (according to Eq.~\eqref{corr_len} and the correlation 
function behaves according to Eq.~\eqref{corr_func}. 

\subsubsection{Critical exponents}

Here, we briefly recall the definition of the critical exponents as introduced in~\cite{Baxter}, as they hold for the following thermodynamic functions
\begin{align}
C(h=0, T) &\propto |T-T_c|^{-\alpha} & \text{as\quad}  &T \to T_c^{~} , \\ \label{beta_exponent}
M_0(T) &\propto \left(T_c - T\right)_{~}^{\beta} & \text{as\quad}  &T \to T_c^- , \\
\chi(h=0, T) &\propto \left(T-T_c\right)^{-\gamma} & \text{as\quad}  &T \to T_c^{~} , \\
M(h, T=T_c) &\propto (h)^{1/\delta} & \text{as\quad}  &h \to 0^{~} , \\ 
\xi(h=0, T) &\propto \left(T-T_c\right)^{-\nu} & \text{as\quad} &T \to T_c^{~} \label{corr_len} , \\
g(\emph{r}) &\propto r^{-d+2-\eta} & \text{as\quad} &T \to T_c^{~} \label{corr_func} , \\
s(h=0, T) &\propto \left(T-T_c\right)^{\mu} & \text{as\quad} &T \to T_c^- . \label{inter_tension}
\end{align}
The above relations can be understood as the definitions of the critical exponents. 
In Eq.~\eqref{corr_func}, the power-law decay of the correlation function also depends on $d$, which denotes the lattice dimension of the system. 
The last quantity we have not yet defined is the interfacial tension per unit area $s$ in Eq.~\eqref{inter_tension}. 
It is defined only for $h=0$ and $T < T_c$ and represents the surface free energy due to the interface between the domains. 

The critical exponents are not entirely independent on each other. The relations between them are given by various scaling assumptions, for instance, by assuming the scaling near the critical temperature $T_c$
\begin{equation}
\frac{h}{k_{\rm B}T_c} = M |M|^{\delta-1} \omega \left[ \left(T-T_c\right) |M|^{-1/\beta} \right].
\end{equation}
We have assumed a dimensionless positive monotonic increasing function $\omega(x)$
in the interval $-x_0 < x < \infty$, whereas $\omega(x)=0$ if $-\infty < x \leq -x_0$.
The critical exponents defined in Eqs.~\eqref{beta_exponent}--\eqref{inter_tension}
satisfy the following rules
\begin{align}
\gamma &= \beta \left( \delta - 1 \right) \label{scaling} \, , \\
\alpha + 2\beta + \gamma &= 2 \, , \\
\left(2 - \eta \right) \nu &= \gamma \, , \\
\mu + \nu &= 2 - \alpha \, , \\
d \nu &= 2 - \alpha \label{hyperscaling} \, . 
\end{align}
The last equation, which involves the system (lattice) dimension $d$, can be derived by
making further assumptions, known as the \textit{hyperscaling hypothesis}. Moreover,
if just two independent critical exponents are known, the remaining exponents
can be derived form Eqs.~\eqref{scaling}--\eqref{hyperscaling}. 

The critical exponents are determined by the lattice dimensionality of the system $d$ and the
symmetry of the order parameter, e.g., $M_0$. However, they do not depend on the detailed
form of the microscopic interactions. This concept, known as the \textit{universality}, allows
replacing a complicated system by a much simpler one of the identical dimensionality and
symmetry in order to obtain the correct behavior at the critical point. It means that behavior
of the thermodynamic functions at a critical point of, for instance, a fluid system is identical to a certain ferromagnetic material. The collection of the models with the identical critical exponents is said to constitute the so-called \textit{universality class}. One often encounters a mean-field approximation of spin models. Here, the mean-field models and the Ising models
exhibit two different sets of the critical exponents, which attribute them to two typical
universality classes studied in this work.
 
\subsubsection{Mean-field theory of phase transitions}

First we recall the main features of the mean-field approximation (theory). Consider a system
consisting of $N$ interacting spins $\sigma_i$, where $i=1,2,\dots,N$, and each spin has $q$
neighbors. The number $q$ is known as \textit{coordination number}. Within the mean-field
approximation, each spin $\sigma_i$ interacts with the averaged spin polarization $M$ of
all the remaining spins $\frac{q}{N-1}\sum_{j\neq i}\sigma_j\equiv qM$.

In equilibrium, we obtain the self-consistent equation (cf. Eq.~\eqref{sp_mag})
\begin{eqnarray} \nonumber \label{M}
M = \left< \sigma \right> &=& \frac{1}{{\cal Z}}\sum_{\sigma_i = \pm 1} \sigma_i \exp\left(\frac{q J M}{k_{\rm B}T}\sigma_i \right)\\
                                            &=&  \frac{\exp\left(\frac{q J}{k_{\rm B}T}M\right) -\exp\left(-\frac{q J}{k_{\rm B}T}M\right) }{\exp\left(\frac{q J}{k_{\rm B}T}M\right) +\exp\left(-\frac{q J}{k_{\rm B}T}M\right) }   \nonumber \\
                                           &=&  \tanh\left(\frac{q J}{k_{\rm B}T}M\right)  \, . 
\end{eqnarray}
\begin{figure}[tb]
\centering
 \includegraphics[width=2.4in]{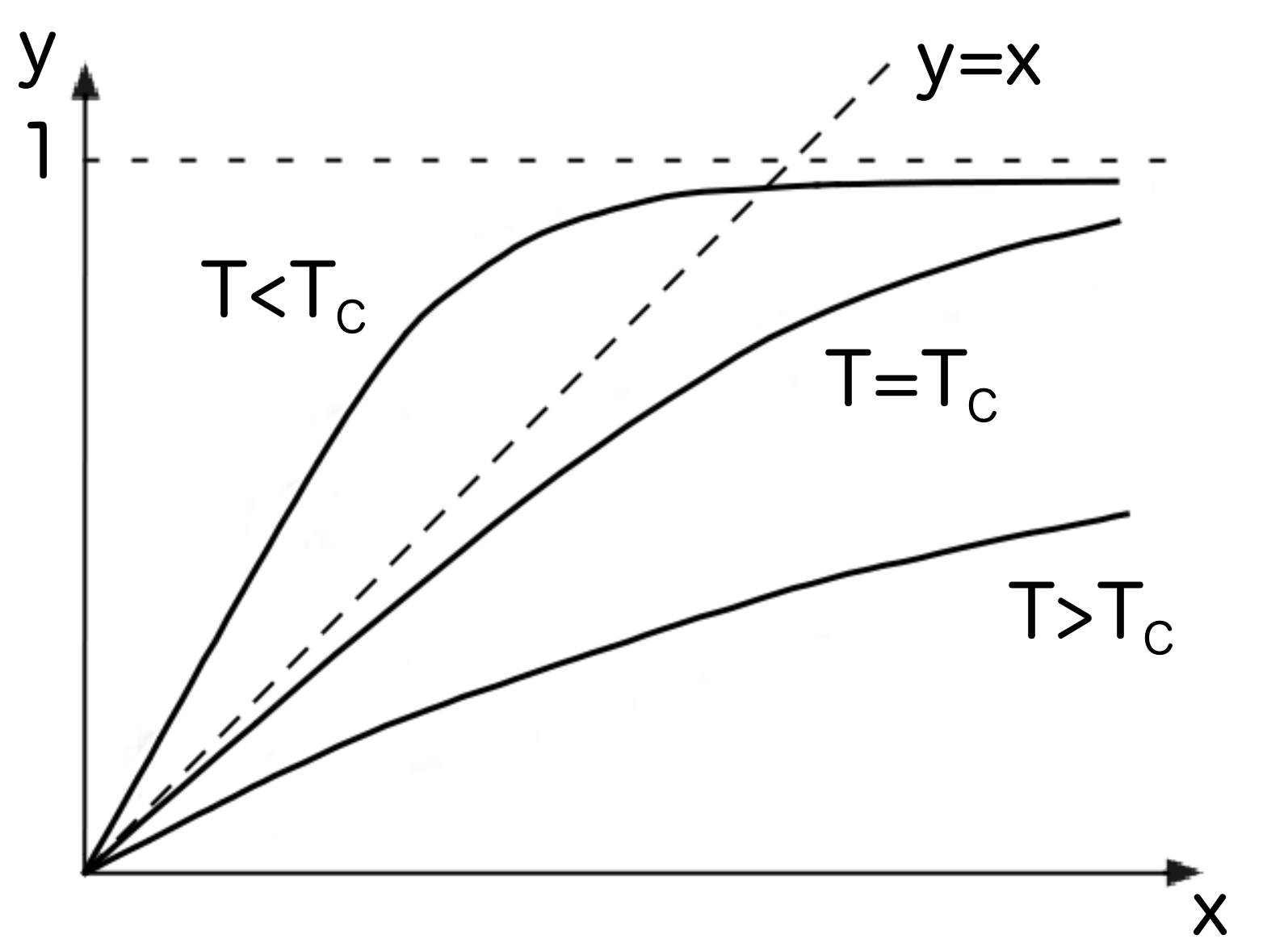}
  \caption{Schematic representation of Eq.~\eqref{M}, where we simplified the notations
$M\equiv x$ and $\tanh\left(\frac{q J}{k_{\rm B}T}M\right)\equiv y$ for brevity.}
\label{lines}
 \end{figure}
The solution of Eq.~\eqref{M} is graphically shown in Fig.\ref{lines} and it exists only if
\begin{equation}
\frac{\text{d}}{\text{d}M} \left[ \tanh\left(\frac{q J}{k_{\rm B}T}M\right) \right]_{M=0} > 1 \, . 
\end{equation}
This condition is satisfied if and only if ${q J}/{k_{\rm B} T} > 1$. Since we are interested
in temperature dependence of the spin model, whereas $q$, $J$, and $k_{\rm B}$ are fixed,
the nonzero solution of the order parameter $M$ is found for $T < T_\text{c}$. Thus, the
phase transition of the mean-field approximation happens at temperature
\begin{equation}
T_{\text{c}} = \frac{q J}{k_{\rm B}} \, . 
\end{equation}
The resulted phase transition temperature $T_{\rm c}$ does not depend on the lattice
dimension $d$. Instead, it is a function of the coordination number $q$, which may partially
reflect the dimension $d$. In particular, one-dimensional chain ($d=1$), two-dimensional
square lattice ($d=2$), three-dimensional cubic lattice ($d=3$), etc. correspond to the
coordination numbers $q=2,4,6,\dots$, respectively. On the other hand, one cannot
distinguish between the two-dimensional triangular lattice and the three-dimensional
cubic lattice if studied by the mean-field approximation, since both of them yield the
same $q=6$, hence, the identical $T_{\rm c}$. For completeness of the solution, we list the
mean-field (classical) exponents, which can be easily derived~\cite{Baxter}, resulting
in $\alpha = 0$, $\beta = \frac{1}{2}$,
$\gamma = 1$, and $\delta = 3$. 

At this stage we make a comment to be discussed in detail later. There is a critical
lattice dimension $d_{\rm c}=4$, at and above which classical spin models with short
range interactions always belong to the mean-field universality class, thus yielding the
identical set of the critical exponents, as if they are treated by the mean-field
approximation~\cite{Yeomans}. Later, we will refer to the Bethe and hyperbolic lattices
which exhibit the lattice (Hausdorff) dimension $d=\infty$, which means that their
critical exponents have to belong to the mean-field universality class despite the
fact they are not solved by the mean-field approximation at all.

\subsection{Transfer-matrix formalism for classical systems}

\subsubsection{The transfer matrix}

Let us briefly introduce an important concept known as the \textit{transfer matrix}
formalism which is, as an analytical method, used to solve spin models exactly (i.e.,
without the simplified mean-field approximation). It also serves as a prerequisite for
better understanding of the \textit{corner transfer matrix} formalism, which is to be
discussed in the following. And, last but not least, the corner transfer matrix formalism
has become a source for further improvements resulting in another numerical algorithm,
which is employed extensively throughout our study. 

The power and elegance of the transfer matrix approach is demonstrated on the simple spin
Ising model. First, we consider an analytic solution of the Ising model on one-dimensional
spin chain. Later, we generalize the transfer matrix approach to the Ising model on the
two-dimensional square lattice.

\paragraph{One-dimensional case:}

The Hamiltonian of the one-dimensional Ising model with nearest-neighbor coupling $J$ and magnetic field $h$ acting on $N$ spins reads
\begin{equation}
{\cal H}_N = - J \sum_{i=1}^{N} \sigma_i \sigma_{i+1} - h \sum_{i=1}^{N} \sigma_i \, .
\end{equation}
We consider the ferromagnetic case (i.~e., $J>0$) with an external magnetic field $h$ on
a ring with $N$ spins (sites). For this reason, we assume the periodic boundary conditions
i.~e., $\sigma_{N+1} \equiv \sigma_1$. This assumption enables that the solution on the ring
becomes translationally invariant. The statistical partition sum (according to
Eq.~\eqref{partition function}) has a simple form
\begin{equation}
{\cal Z}_N = \sum_{\{\sigma\}} \exp \left( K \sum_{j=1}^{N} \sigma_j \sigma_{j+1} + G \sum_{j=1}^{N} \sigma_j  \right) \, ,
\end{equation}
where we introduced the notation $K=J/{k_{\rm B} T}$ and $G = h/{k_{\rm B} T}$. 

The statistical sum ${\cal Z}_N$ can be factorized into the product of the identical
symmetric matrices $V_{\sigma_i,\sigma_{i+1}}^{~}$ acting on two nearest spins
\begin{equation}
{\cal Z}_N = \sum_{\{\sigma\}}^{} V_{\sigma_1,\sigma_2}^{~} V_{\sigma_2,\sigma_3}^{~} \dots
V_{\sigma_{N-1},\sigma_N}^{~} V_{\sigma_N,\sigma_1}^{~}
=\sum\limits_{\{\sigma\}}^{} \prod\limits_{i=1}^{N} V_{\sigma_i,\sigma_{i+1}}^{~}\, ,
\label{V_mul}
\end{equation}
where
\begin{equation}
V_{\sigma^{~}_i,\sigma^{~}_{i+1}} = \exp \left[ K \sigma^{~}_i \sigma^{~}_{i+1} + \frac{G}{2}
(\sigma^{~}_i + \sigma^{~}_{i+1}) \right] \, .
\end{equation}
For convenience, each $V_{\sigma^{~}_{i}, \sigma^{~}_{i+1}}$ represents a $2\times2$ matrix
with row and column indices to be $\sigma^{~}_{i}$ and $\sigma^{~}_{i+1}$, respectively, 
\begin{equation}
V_{\sigma^{~}_i, \sigma^{~}_{i+1}} =\left(\begin{array}{cc} 
V_{+1, +1} & V_{+1, -1}  \\
V_{-1, +1} & V_{-1, -1} \end{array} \right) = 
 \left(\begin{array}{cc} 
e^{K+G} & e^{-K}  \\
e^{-K} & e^{K-G}\end{array} \right) \, . 
\end{equation}
Hence, $V$ is the above-mentioned transfer matrix.
The complete matrix multiplication in Eq.~\eqref{V_mul} is equivalent to taking trace
\begin{equation}
{\cal Z}_N = \text{Tr} \left( V ^{N} \right) \, .
\end{equation}
In order to solve this problem, it is convenient to diagonalize the $2\times2$ matrix $V$
\begin{equation}
V = P\Lambda P^{-1}=
P \left(\begin{array}{cc} 
\lambda_+ & 0 \\
0 & \lambda_- \end{array} \right)  P^{-1} \, , 
\end{equation}
where $P$ is a unitary matrix containing the two eigenvectors in its columns which are
associated with the two eigenvalues $\lambda_+$ and $\lambda_-$ of the diagonal matrix
$\Lambda$. Using the fact that the trace is invariant under cyclic permutations, we can
rewrite the statistical sum
\begin{equation}
{\cal Z}_N = \text{Tr } \left( \Lambda^N \right)
=\text{Tr } \left(\begin{array}{cc} 
\lambda_+^N & 0 \\
0 & \lambda_-^N \end{array} \right) = 
\lambda_+^{N} + \lambda_-^{N} \, ,
\end{equation}
where
\begin{equation}
\lambda_{\pm} = e^{K} \cosh G \pm \sqrt{ e^{2K} \sinh^2 G + e^{-2K} } \, ,
\end{equation}
In the thermodynamic limit ($N \to \infty$), the free energy per spin site (cf. Eq.~\eqref{freng}) has the form
\begin{equation}
f = \lim_{N \rightarrow \infty} - \frac{k_{\rm B}T}{N} \ln {\cal Z}_N
=\lim_{N \rightarrow \infty} - \frac{k_{\rm B}T}{N}
\left\{
   N \ln\lambda_+ + \ln \left[ 1 + \left(\frac{\lambda_-}{\lambda_+}\right)^N \right]
\right\}
=- k_{\rm B} T \ln \lambda_1 \, , 
\end{equation}
where we assumed that $\lambda_+ \geq \lambda_-$, which is justified for any $T \geq 0$. 
Note that the free energy per site is an analytic function for $T>0$ and arbitrary $h$. 
The critical point referring to a phase transition is defined at such temperature, which
leads to the divergence of the correlation length, i.e., $\xi = 1/\ln \left(
\lambda_+/\lambda_- \right)\to\infty$. This is satisfied if $\lambda_+ = \lambda_-$,
which is true only at $T=0$ and $h=0$. Since $T_{\rm c}=0$, there is no ordered phase
below the critical point, the classical Ising model on one-dimensional chain exhibits no
phase transition.

\paragraph{Two-dimensional case:}

\begin{figure}[tb]
 \centering
 \includegraphics[width=2.5in]{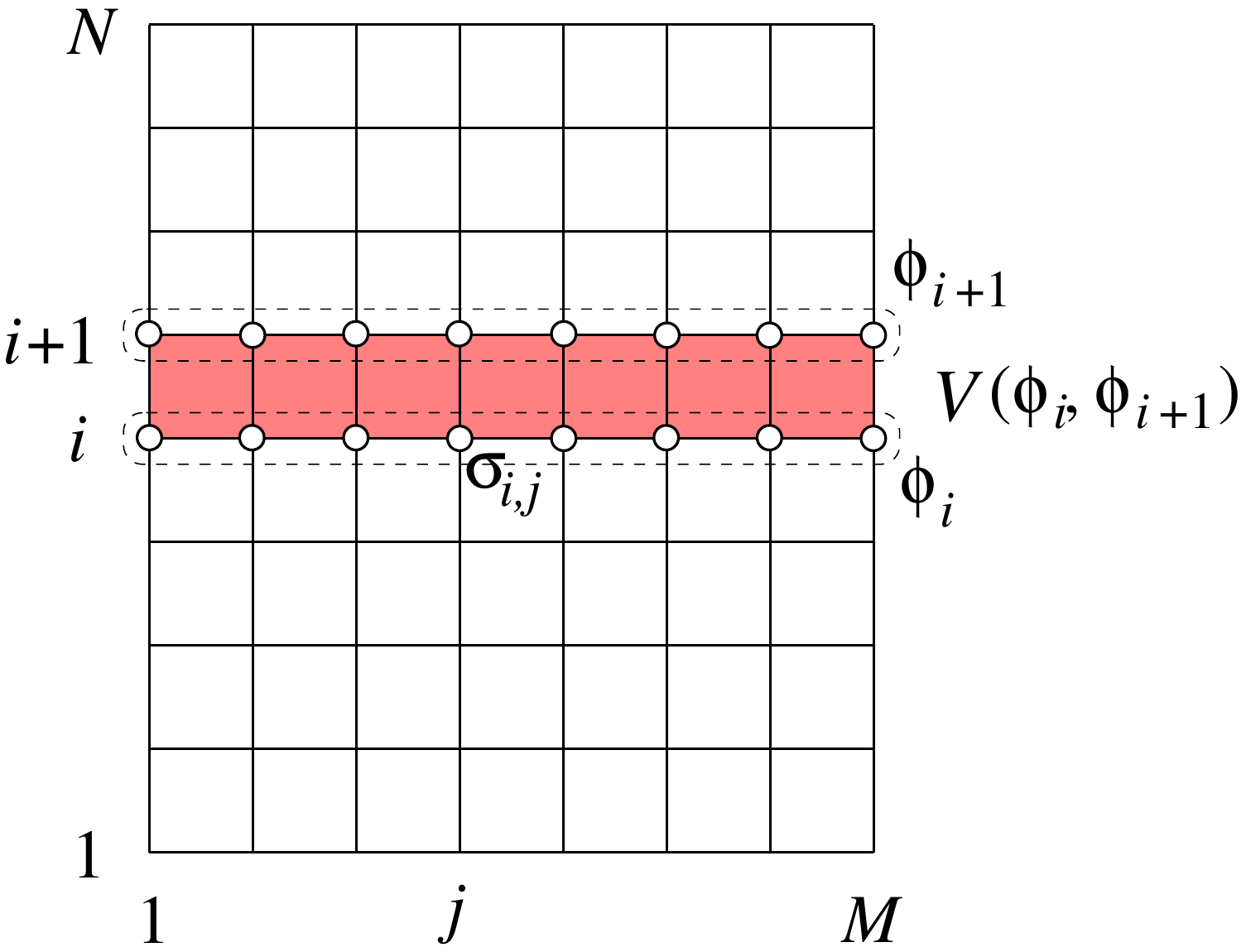}
   \caption{Graphical representation of the transfer matrix $V(\phi_i, \phi_{i+1})$ on the square lattice $M \times N$.}\label{TM_2D}
 \end{figure}

We again consider the Ising model on the regular $N\times M$ (rectangular-shaped) lattice with
the Hamiltonian
\begin{equation}
{\cal H}_{N\times M} = - J \sum_{i=1}^{N} \sum_{j=1}^{M} \left( \sigma_{i, j} \sigma_{i+1, j} + \sigma_{i, j} \sigma_{i, j+1} \right) - h \sum_{i=1}^{N}\sum_{j=1}^{M} \sigma_{i,j} \, , 
\end{equation}
where a spin $\sigma_{i, j}$ is located at row $i$ and column $j$ of the lattice, see
Fig.~\ref{TM_2D}. Likewise in the one-dimensional case, we suppose the periodic boundary
conditions in both directions. 
Again, the statistical sum can be factorized in terms of the transfer matrices $V$. This time,
however, it is a large $2^M \times 2^M$ matrix. 

Using the similar procedure as in the one-dimensional case, we express the statistical
sum by taking the trace of the matrix product\footnote{For later convenience, we introduce
a 4-spin square-shaped lattice object (being the Boltzmann weight) defined as the $4\times4$
matrix
$
{\cal W}(\sigma_{i, j}^{~}, \sigma_{i, j+1}^{~}, \sigma_{i+1, j}^{~}, \sigma_{i+1, j+1}^{~})=$\vspace{-0.15cm}
$$
 = \exp \left\{\frac{1}{k_{\rm B}T}\left[ \frac{J}{2} \left( \sigma_{i, j} \sigma_{i, j+1} + \sigma_{i, j+1} \sigma_{i+1, j+1} +
 \sigma_{i+1, j+1}\sigma_{i+1, j} + \sigma_{i+1, j}\sigma_{i, j} \right)  
+ \frac{h}{4} \left( \sigma_{i, j} + \sigma_{i, j+1} + \sigma_{i+1, j} + \sigma_{i+1, j+1} \right) \right] \right\}.
\vspace{-0.07cm}$$
Since the 4-spin square-shaped Boltzmann weights are used to build up the entire $N\times M$ lattice, the prefactors $\frac{J}{2}$ and $\frac{h}{4}$, respectively, adjust the interactions on the bonds and the sites (excluding the boundary ones, as discussed later in Section~\ref{CTMRG_Section}). Therefore, one can express the transfer matrix $V$ as the product form of the above-defined Boltzmann weights

$\vspace{-0.3cm}\hspace{0.6cm}
V(\phi_i, \phi_{i+1}) = \prod\limits_{j=1}^{M-1} {\cal W} \left( \sigma_{i, j}^{~}, \sigma_{i, j+1}^{~}, \sigma_{i+1, j}^{~}, \sigma_{i+1, j+1}^{~} \right) \, . 
$}
\begin{equation}
{\cal Z}_{N\times M} = \sum_{\{ \phi \}}^{} V\left( \phi_1, \phi_2 \right) V\left( \phi_2, \phi_3 \right) \dots V\left( \phi_{N-1}, \phi_N \right) V\left( \phi_N, \phi_1 \right) 
= \text{Tr} \left( V^{N} \right) \, ,
\end{equation}
where the spin configurations on the entire row $i$ is grouped into
\begin{equation}
\phi_i = \left\{ \sigma_{i,1} \sigma_{i,2} \dots \sigma_{i,M} \right\} \, ,
\end{equation}
which becomes a compound of the multi-state variable with $2^M$ spin degrees of freedom.
The free energy per site, $f_{N\times M}=-k_{\rm B}T(NM)^{-1}\ln{\cal Z}_{N\times M}$,
on a finite lattice is
\begin{equation}
f_{N\times M} = - \frac{k_{\rm B} T}{N M} \ln \left( V^{N} \right) = - \frac{k_{\rm B} T}{N M} \ln \sum_{i=1}^{2^M} \lambda_i^{N} =
- \frac{k_{\rm B} T}{M} \left\{ \ln \lambda_1 + \frac{1}{N} \ln \left[ 1 +  \sum_{i=2}^{2^M} \left(\frac{\lambda_i}{\lambda_1}\right)^{N}\right] \right\} \, ,
\label{fN}
\end{equation}
We suppose a decreasing ordering of the eigenvalues of the transfer matrix $V$ defined
between two adjacent lattice rows, and each row contains $M$ spin sites. In particular,
$\lambda_1 \geq \lambda_2 \geq \dots \geq \lambda_{\left(2^M\right)}$, provided that
homogeneous spin models with short-range interactions never exhibit a complete degeneracy
of the eigenvalue spectra. Then, if approaching the thermodynamic limit $N \rightarrow \infty$,
the second term on the right-hand side of Eq.~\eqref{fN} vanishes leading to a simple
formula for the free energy per row with $M$ sites 
\begin{equation}
f_M = \lim\limits_{N\to\infty} f_{N\times M} = - \frac{k_{\rm B} T}{M} \ln \lambda_1 \, .
\label{fe1}
\end{equation}
Hence, the free-energy calculation requires to find out an appropriate way of obtaining
the largest eigenvalue $\lambda_1$ of the transfer matrix $V$. This task is far easier
if only $\lambda_1$ is calculated than if one has to obtain the complete set of eigenvalues
$\lambda_i$ for all $i=1,2,\dots,2^M$. 

\subsubsection{The corner transfer matrix}\label{CTM_subsec}

 \begin{figure}[tb]
 \centering
 \includegraphics[width=4.6in]{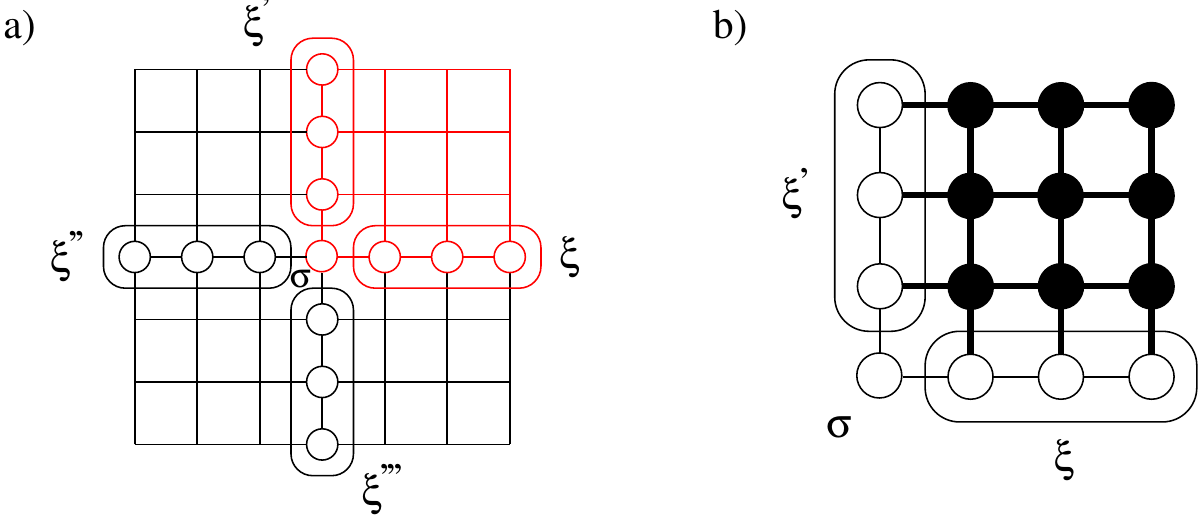}
   \caption{Graphical representation of the corner transfer matrix on the square lattice: a) The division of the lattice into the four identical quadrants. 
   The spin $\sigma$ positioned in the middle of the system is shared 
   by the four quadrants, whereas $\xi$, $\xi'$, $\xi''$, $\xi'''$ denote the multi-spin variables on the boundaries shared by the two 
    adjacent quadrants. In red color, one of the quadrants (upper-right) is highlighted. 
    b) The upper-right corner transfer matrix ${\cal C} \left(\left. \sigma\xi \right| \sigma\xi^{\prime}  \right)$ 
    (or alternatively, the upper-right corner transfer tensor ${\cal C}_{\sigma\xi\xi^{\prime}}$, see Eq.~\eqref{C_as_tensor}). 
    The thick lines represent the terms from the first product in Eq.~\eqref{corner_matrix}, 
    whereas the thin lines represent the terms from the second product (with the factor $\frac{1}{2}$). 
    The sum is taken over the spin configurations of the spins inside the quadrant, which are indicated by the filled circles.}\label{corner_matrix_fig}
 \end{figure}   
The concept of the corner transfer matrix ${\cal C}$ is best suitable for two-dimensional
lattice structures. As a typical example, we again consider the Ising model on the square
lattice $N\times N$ for open or fixed boundary conditions.
If dividing the lattice into four identical quadrants, see Fig.~\ref{corner_matrix_fig}a), the four corner transfer matrices are used to describe the quadrants (i.e., the corners).
The aim of this approach is to express the statistical sum in a product form of the four
corner transfer matrices, in particular
\begin{equation} \label{c^4}
{\cal Z}_{N\times N} = \sum\limits_{\{\sigma\}} {\cal C}^{\,4} \, .
\end{equation}
The corner transfer matrix is expressed as a matrix
\begin{equation} \label{corner_matrix}
{\cal C} \left(\left. \sigma \xi \right|  \sigma \xi^{\prime}  \right) = 
{\sum_{\{ \sigma \}}}^{\prime}
\prod_{\left< ij \right>} \exp \left( \frac{J}{k_{\rm B} T} \sigma_i \sigma_j + 
\frac{h}{k_{\rm B} T} \sigma_i \right) \, ,
\end{equation}
where the sum $\sum_{\{ \sigma \}}^{\prime}$ is taken over those spin configurations in the
quadrant, which are indicated by the filled (black) circles in Fig.~\ref{corner_matrix_fig}b)
leaving the spin variables $\sigma$, $\xi$, and $\xi^\prime$ not summed up. The corner spin
$\sigma$ is the spin in the square lattice center (the spin is shared by all four quadrants).
The multi-spin variables $\xi$ and $\xi'$ represent grouped spins with the analogous
definition as the variable $\phi$ mentioned earlier. Each grouped variable $\xi$ is
thus shared by the same spin grouped spins on the adjacent quadrants\footnote{Specifically,
if $N$ and $M$ are even, the central spin $\sigma$ is positioned at $\sigma_{N/2, M/2}$ on
the lattice, and the multi-spin variables $\xi$ and $\xi'$ of the upper-right corner transfer
matrix group the following spins
\begin{eqnarray}
\xi = \left\{\sigma_{N/2,\ M/2+1} \sigma_{N/2,\ M/2+2} \dots \sigma_{N/2, M}\right\} \, ,\\
\xi' = \left\{\sigma_{N/2+1,\ M/2} \sigma_{N/2+2,\ M/2} \dots \sigma_{N, M/2}\right\} \, . 
\end{eqnarray}}.

In accord with Eq.~\eqref{c^4}, the statistical sum can be rewritten in terms of the
relevant spin variables
\begin{equation}
{\cal Z}_{N\times N} = \sum\limits_{\sigma \xi \xi' \xi'' \xi'''} {\cal C} \left(\left.\sigma\xi\right|\sigma\xi' \right) {\cal C} \left(\left.\sigma\xi'\right|\sigma\xi'' \right) 
{\cal C} \left(\left.\sigma\xi''\right|\sigma\xi''' \right) {\cal C} \left(\left.\sigma\xi'''\right|\sigma\xi \right).
\end{equation}
We notice that the central spin $\sigma$ is a duplicated index used in the notation of
each corner transfer matrix ${\cal C}$, which makes this square matrix object defined
inefficiently, as the index $\sigma$ appears twice. In the following, we will regard the
corner transfer matrix ${\cal C}$ as a tensor and will refer to it as the corner transfer
tensor instead\footnote{
In the tensor language, the statistical sum is expressed as
\begin{equation}
{\cal Z}_{N\times N}= 
\sum_{\sigma \xi \xi' \xi'' \xi'''} 
 {\cal C}_{\sigma\xi\xi'} {\cal C}_{\sigma\xi'\xi''}
{\cal C}_{\sigma\xi''\xi'''} {\cal C}_{\sigma\xi'''\xi} \, .  
\end{equation}
}
\begin{equation} \label{C_as_tensor}
{\cal C}_{\sigma \xi \xi^{\prime}} \equiv {\cal C} \left(\left. \sigma \xi \right| \sigma \xi^{\prime}  \right) \, . 
\end{equation}

The advantage of the corner transfer matrix formalism rests in its simplicity and
suitability to adapt to non-Euclidean lattices, where the transfer-matrix formalism
cannot be applied. This formalism was introduced by Baxter~\cite{Baxter} and later
implemented into a numerical algorithm Corner Transfer Matrix Renormalization Group,
which is described in Section~\ref{CTMRG_Section}. 

\subsection{Suzuki-Trotter mapping}
\label{STm}

One can find a correspondence between $d$-dimensional quantum spin models and
$d+1$-dimensional classical spin models. It means that both spin models belong
to the same universality class. This
quantum-classical correspondence is also known as the Suzuki-Trotter mapping.
We shortly derive this correspondence on a simple example, where the one-dimensional
quantum Ising model in the transverse magnetic field $h$ is mapped on the
two-dimensional classical Ising model on the rectangular lattice.

The Hamiltonian of the one-dimensional transverse-field Ising model reads
\begin{equation}
{\cal H}_N = -J\sum_{j=1}^{N} {\cal S}^{\,z}_{j} {\cal S}^{\,z}_{j+1} -h\sum_{j=1}^N
{\cal S}^{\,x}_j \equiv {\cal H}_J + {\cal H}_h \, ,
\end{equation}
where the indices in the square brackets label spin positions on an $N$-site chain.
If imposing the periodic boundary conditions to the spin chain, i.e., 
${\cal S}^{\,z}_{N+1} \equiv {\cal S}^{\,z}_1$, the chain becomes the ring.
The standard Pauli matrix operators
\begin{equation} \label{pauli_x}
{\cal S}^{\,x} = \left( \begin{matrix}
  0 & 1 \\
  1 & 0 
 \end{matrix} \right) \, , 
 \end{equation}
 \begin{equation}
{\cal S}^{\,y} = \left( \begin{matrix}
  0 & -i \\
  i & 0 
 \end{matrix} \right) \, , 
 \end{equation}
 
 \begin{equation} \label{pauli_z}
{\cal S}^{\,z} = \left( \begin{matrix}
  1 & 0 \\
  0 & -1 
 \end{matrix} \right) \,
 \end{equation}
enter the spin model including the identity operator
  \begin{equation} \label{pauli_id}
I = \left( \begin{matrix}
  1 & 0 \\
  0 & 1 
 \end{matrix} \right) \, . 
 \end{equation}

Now, we express the partition function of the spin system

\begin{equation}
{\cal Z}_N = \text{Tr} \,\, \exp \left( - \frac{\cal H_N}{k_{\rm B} T} \right)
=\text{Tr} \left[ \exp \left( - \Delta \tau\, {\cal H_N} \right) \right]^{M_{\tau}} \, ,
\label{STexp}
\end{equation}
where $1/k_{\rm B}T = M_{\tau} \Delta\tau$ is fixed to a constant to satisfy
$0 \leq \Delta \tau \ll 0$ and $1 \ll M_{\tau} < \infty $. 
Thus rewritten partition function represents a two-dimensional spin system
$N \times M_{\tau}$ (recall that $M_{\tau}$ will denote the second perpendicular
direction of the lattice system being known as Trotter direction or, equivalently,
imaginary-time evolution).

The Suzuki-Trotter mapping is an expansion, which approximates the partition function in
Eq.~\eqref{STexp} by the product of two exponential operators ${\cal V}_J$ and
${\cal V}_h$,
\begin{equation}
\exp \left( -\Delta\tau\, {\cal H_N} \right)
= \exp \left( -\Delta\tau\, {\cal H}_J \right)
  \exp \left( -\Delta\tau\, {\cal H}_h \right)
+ {\cal O}\left( \Delta \tau^2 \right)
\equiv {\cal V}_J {\cal V}_h + {\cal O}\left( \Delta \tau^2 \right)
\, , 
\end{equation}
which can be identified as the following transfer matrices 
\begin{equation}
{\cal V}_J = \exp\left(\Delta\tau J \sum_{i=1}^{N} {\cal S}^{\,z}_i {\cal S}^{\,z}_{i+1} \right) \, . 
\end{equation}

\begin{equation}
{\cal V}_h = \exp \left(\Delta\tau h \sum_{i=1}^{N} {\cal S}^{\,x}_i \right) \, , 
\end{equation}

Let $\ket{S_j}$ be an eigenstate of the Pauli operator ${\cal S}^{\,z}_j$ acting
on the $j{\rm-th}$ spin site. Its corresponding eigenvalue is represented by
a scalar variable $\sigma_j = \pm 1$. One can immediately find out that ${\cal V}_J$ is
 a diagonal matrix operator with respect to that basis for $j=1,2,\dots,N$, i.e.,
\begin{equation} \label{T2_diagonal}
{\cal V}_J \ket{ S_j } = e^{\Delta\tau J \sum\limits_{i=1}^{N} S_i S_{i+1}} \ket{ S_j } \, . 
\end{equation}
Inserting the complete basis set $M_{\tau}$ times into the partition function, we obtain
\begin{equation}
{\cal Z_{N\times M_{\tau}}} = \sum_{S_1=\pm1}\sum_{S_2=\pm1}\cdots\sum_{S_{M_{\tau}}=\pm1}
\prod_{j=1}^{M_{\tau}} \bra{ S_j} {\cal V}_J {\cal V}_h \ket{ S_{j+1}} \, , 
\end{equation}
where the integer $j$ labels the imaginary-time step. 
Using Eq.~\eqref{T2_diagonal}, we can write
\begin{equation}
\bra{S_j} {\cal V}_J {\cal V}_h \ket{S_{j+1}} = e^{\Delta\tau J \sum_{i=1}^{N}
\sigma_{i,j} \sigma_{i+1,j}} \bra{S_j} {\cal V}_h \ket{S_{j+1}} \, . 
\end{equation}
To find the elements of the transfer matrix ${\cal V}_h$, we use the formula
\begin{equation} \label{taylor_e}
e^{\Delta\tau h {\cal S}^{\,x}} = I \cosh (\Delta \tau h) + {\cal S}^{\,x} \sinh (\Delta\tau h) \, . 
\end{equation}
If assuming the following form of the matrix elements 
\begin{equation}
\bra{S_j} e^{\Delta \tau {\cal S}^{\,x}} \ket{S_k} \eqqcolon \Lambda e^{\gamma \sigma_j \sigma_k} \, , 
\end{equation}
where $S_j$, $S_k = \pm 1$, and using Eq.~\eqref{taylor_e}, we obtain 

\begin{equation}
\bra{S_j} e^{\Delta \tau {\cal S}^{\,x}} \ket{S_j} = \cosh (\Delta \tau h) = \Lambda e^{\gamma} \, , 
\end{equation}

\begin{equation}
\bra{-S_j} e^{\Delta \tau {\cal S}^{\,x}} \ket{S_j} = \sinh (\Delta \tau h) = \Lambda e^{-\gamma} \, , 
\end{equation}
and thus
\begin{equation}
\Lambda = \sqrt { \sinh (\Delta \tau h) \cosh (\Delta \tau h) } \qquad
\gamma = \ln \sqrt { 1 / \tanh (\Delta\tau h) } \, . 
\end{equation}
Now, we can rewrite the statistical sum in the form
\begin{equation}
{\cal Z} = \Lambda^{N M_{\tau}} \sum_{\{ S_{i,j} = \pm 1 \}}
\exp \left(\Delta\tau J \sum_{i=1}^{N}\sum_{j=1}^{M_{\tau}} \sigma_{i,j} \sigma_{i+1,j} 
+ \gamma \sum_{i=1}^{N} \sum_{j=1}^{M_{\tau}} \sigma_{i,j} \sigma_{i,j+1}\right) \, . 
\end{equation}
This statistical sum is identical to the two-dimensional classical Ising model with the Hamiltonian (defined on a rectangular $N$ by $M_{\tau}$ lattice)
\begin{equation}
{\cal H} = - J_1\sum_{i=1}^{N}\sum_{j=1}^{M_{\tau}} \sigma_{i,j} \sigma_{i+1,j} 
- J_2 \sum_{i=1}^{N} \sum_{j=1}^{M_{\tau}} \sigma_{i,j} \sigma_{i,j+1} \, , 
\end{equation}
where $J_1 = \Delta\tau J\,k_{\rm B}T$ and $J_2 = \gamma \,k_{\rm B}T$ with $T$ being the
thermodynamic temperature.

\newpage
\newpage\setcounter{equation}{0} \setcounter{figure}{0} \setcounter{table}{0}
\section{Tensor networks} \label{chap2}

\subsection{Matrix Product State}

This Section provides a description of the infinite Time-Evolving Block Decimation (iTEBD) algorithm introduced in~\cite{Vidal}.  
The algorithm is used for efficient simulations of one-dimensional quantum lattice systems when representing a quantum state (typically the ground state) as the product of matrices. This representation is called the Matrix Product State (MPS). 
We focus on the computation of the ground state only which is carried out as the evolution of an initial state in imaginary time. It is noteworthy to add that considering time evolution follows the identical rules as used in the imaginary time~\cite{Vidal}. In the following, we demonstrate the application of the MPS algorithm on the transverse field quantum Ising and Heisenberg models. 

Let us consider an infinite one-dimensional (1D) spin chain, where each spin site carries $d$ degrees of the freedom (spin-$\frac{d-1}{2}$ model), i.~e., the physical dimension of each site is $d$. Hence, the typical spin-$\frac{1}{2}$ models have $d=2$, as it is for the Ising and Heisenberg models. For simplicity, let us assume that only the nearest-neighbor spins are allowed to interact. Thus, the interactions in 1D are given by the translational invariant Hamiltonian ${\cal H}$, which is given by sum of identical local Hamiltonians ${\cal H}_{\rm loc}$

\begin{equation}
{\cal H} = \sum_{r=-\infty}^{\infty} {\cal H}_{\text{loc}}^{[r, r+1]} \, ,
\end{equation}
where $r$ locates a spin site on the infinite chain with all the sites positioned equidistantly. Since the chain size is infinite, it is natural to assume that local observables measured on an arbitrary spin site remain positionally independent; hence the translational invariance of the state applies. If $\tau=-\frac{i}{\hbar}t$ denotes the imaginary time, a normalized initial pure state $\ket{\Psi_0}$ evolves as
\begin{equation} \label{img_evolution}
\ket{\Psi_{\tau}} = \frac{\exp{\left(-{\cal H}\tau\right)} \ket{\Psi_0} }{ || \exp\left( - {\cal H}\tau\right) \ket{\Psi_0} || } \, . 
\end{equation}

\paragraph{Expansion:} The Matrix Product State can be applied to any pure state $\ket{\Psi}$ by a series of Schmidt decompositions. First, let us divide the chain into two parts $\{ -\infty, ..., r \}$ and $\{r+1, ..., \infty \}$ to obtain 
\begin{equation} \label{psitot}
\ket{\Psi} = \sum_{\alpha=1}^{\chi} \lambda_{\alpha}^{[r]} \ket{\Phi_{\alpha}^{[\vartriangleleft r]}} \ket{\Phi_{\alpha}^{[r+1 \vartriangleright ]}} \, .
\end{equation}
Here, $\lambda_{\alpha}$ are known as the Schmidt coefficients with the properties $\lambda_1 \geq \lambda_2 \geq \cdots \geq \lambda_{\chi} > 0$ and $\sum_{\alpha=1}^{\chi} \lambda_{\alpha}^{2} = 1$ provided that $\langle\Psi\,\vert\,\Psi\rangle=1$, and the integer $\chi$ is the Schmidt rank. The states $\ket{\Phi_{\alpha}^{[\vartriangleleft r]}}$ and $\ket{\Phi_{\alpha}^{[r+1 \vartriangleright ]}}$ form the orthonormal basis vectors of the left and right sublattice, respectively. To express the state $\ket{\Psi}$ in the local basis $\ket{i^{[r]}}$ and $\ket{i^{[r+1]}}$ of sites $r$ and $(r+1)$, respectively, we use the following decompositions 
\begin{equation} \label{psi1}
\ket{\Psi_{\alpha}^{[\vartriangleleft r]}} = \sum_{\beta=1}^{\chi} \sum_{i=1}^{d} \lambda_{\beta}^{[r-1]} \Gamma_{i \beta \alpha}^{[r]} \ket{\Phi_{\beta}^{[\vartriangleleft r - 1]}} \ket{i^{[r]}} \, , 
\end{equation}
\begin{equation}\label{psi2}
\ket{\Psi_{\alpha}^{[r+1 \vartriangleright ]}} = \sum_{\beta=1}^{\chi} \sum_{i=1}^{d} \Gamma_{i \beta \alpha}^{[r+1]} \lambda_{\beta}^{[r+1]} \ket{i^{[r+1]}} \ket{\Phi_{\beta}^{[r + 2 \vartriangleright ]}} \, , 
\end{equation}
where $\Gamma_{i\beta\alpha}^{[\ast]}$ is a three-index $\{i\beta\alpha\}$ tensor of the respective dimensions $d$, $\chi$, and $\chi$ (the symbol $[\ast]$ is used to denote position on the chain). Here, we reserve the Roman indices (e.g. $i,j=1,\dots,d$) to represent the physical (spin) degrees of the freedom, whereas the Greek indices (e.g. $\alpha,\beta=1,2,\dots,\chi$) are ancillary (non-physical) degrees of the freedom, which are often referred to as the bond dimensions. Thus, in general, the bond dimension $\chi$ can be infinite. However, in real numerical calculations, a finite $\chi$ (cut-off) is inevitable so that the maximal allowed Schmidt rank $\chi_{\rm max}$ is denoted by the variable $D$. Inserting Eqs.~\eqref{psi1} and \eqref{psi2} into Eq.~\eqref{psitot} yields the complete expansion of $\ket{\Psi}$ for the sites $\{r, r+1\}$ 
\begin{equation}
\ket{\Psi} = \sum_{\alpha , \beta, \gamma = 1}^{\chi} \sum_{i, j = 1}^{d} \lambda^{[r-1]}_{\alpha} \Gamma^{[r]}_{i \alpha \beta} \lambda^{[r]}_{\beta} \Gamma^{[r+1]}_{j \beta \gamma} \lambda^{[r+1]}_{\gamma} \ket{\Phi_{\alpha}^{[\vartriangleleft r-1]}} \ket{i^{[r]}} \ket{j^{[r+1]}} \ket{\Phi^{[r+2 \vartriangleright]}_{\gamma}} \, . 
\end{equation}
 \begin{figure}[tb]
 \centering
 \includegraphics[width=2.1in]{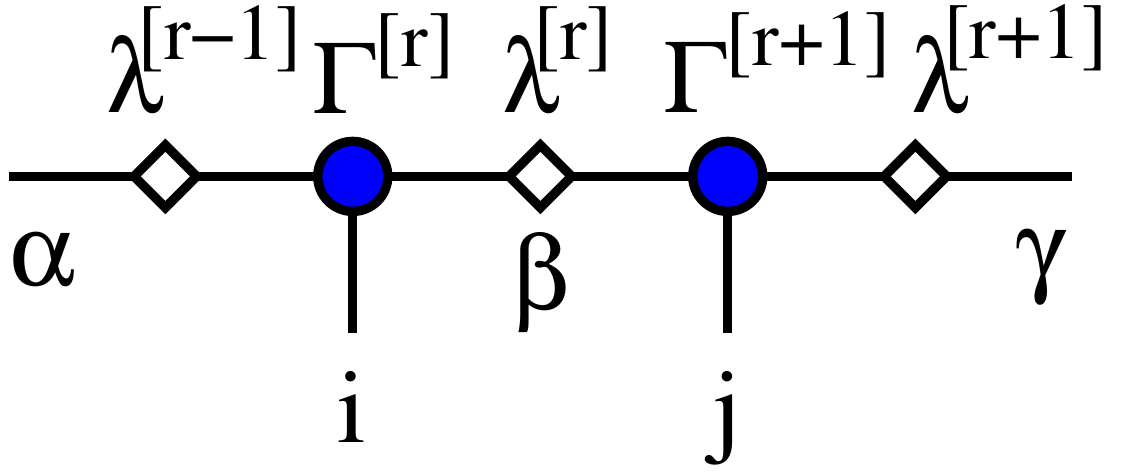}
   \caption{Graphical representation of the expansion of a pure state $\ket{\Psi}$ with respect to the neighboring sites $r$ and $r+1$, which is expressed as the tensor $\Psi_{\alpha i j \gamma}$ of the rank four. 
   }\label{MPS_Psi_fig}
 \end{figure}   
The pure state $\ket{\Psi}$ can be represented as a single tensor given by the coefficients of the above expansion (see Fig.~\ref{MPS_Psi_fig})
\begin{equation}
\Psi_{\alpha i j \gamma} = \sum_{\beta = 1}^{\chi} \lambda^{[r-1]}_{\alpha} \Gamma^{[r]}_{i \alpha \beta} \lambda^{[r]}_{\beta} \Gamma^{[r+1]}_{j \beta \gamma} \lambda^{[r+1]}_{\gamma} \, . 
\end{equation} 

A non-unitary evolution operator $V$ acting in Eq.~\eqref{img_evolution} can be expanded by Suzuki-Trotter decomposition into a sequence of two-site gates
\begin{equation} \label{non_unitary_2s_gate}
V^{[r,r+1]} \equiv \exp \left( - {\cal H}_{\text{loc}}^{[r, r+1]} \delta \tau \right) \, , \quad 0 < \delta \tau \ll 1 \, ,
\end{equation}
where $V^{[r,r+1]}$ is a square $d^2 \times d^2$ matrix acting on the two neighboring sites $[r,r+1]$; if expressed by the physical indices $i,i^\prime=1,\dots,d^2$, the matrix has the components $V_{i,i^\prime}$.
The two-site gates are arranged into the gates $V^{AB}$ and $V^{BA}$, which act on the alternating pair of the sites $\{2r, 2r+1\}$ and $\{2r+1, 2r+2\}$, respectively, hence
\begin{equation}
V^{AB} \equiv \bigotimes_{r=-\infty}^{+\infty} V^{[2r,2r+1]} \qquad {\rm and} \qquad V^{BA} \equiv \bigotimes_{r=-\infty}^{+\infty} V^{[2r+1,2r+2]} \, . 
\end{equation}
This construction breaks the assumed translational symmetry (i.~e., independence on $r$), 
and is taken into account by the following Ansatz of the MPS (the left site, $[2r]$, being always even)
\begin{equation}
\Gamma^{[2r]} = \Gamma^A \, , \qquad \lambda^{[2r]} = \lambda^{A}
\end{equation}
and  (the left site, $[2r+1]$, being odd)
\begin{equation}
\Gamma^{[2r+1]} = \Gamma^B \, , \qquad \lambda^{[2r+1]} = \lambda^{B} \, . 
\end{equation}

\paragraph{MPS update:} 
For the even-odd pair of sites $\{2r, 2r+1\}$, we rewrite the state expansion
\begin{equation} \label{MPS_update_eq1}
\Psi_{\alpha i j \gamma}^{AB} = \sum_{\beta = 1}^{\chi} \lambda^{B}_{\alpha} \Gamma^{A}_{i \alpha \beta} \lambda^{A}_{\beta} \Gamma^{B}_{j \beta \gamma}  \lambda^{B}_{\gamma} \, .
\end{equation}
The application of the non-unitary operator $V$ on the state yields (a new tensor)
\begin{equation}
\Theta_{\{\alpha i\} \{j \gamma\}}^{AB} =  V_{\{i j\} \{k l\}} \Psi_{\alpha k l \gamma}^{AB} \, , 
\end{equation}
where we regrouped the indices to form a matrix $\Theta_{\{\alpha i\} \{j \gamma\}}^{AB}$ (for later applicability of singular value decomposition), instead of keeping the fourth-order tensor form $\Theta_{\alpha i  j \gamma}^{AB}$. We used a special notation for the matrix $V_{\{i j\} \{k l\}}$, in which the grouped indices $\{ij\}=1,\dots,d^2$ and $\{kl\}=1,\dots,d^2$ are understood in the following enumeration $\{ij\} \equiv \{d (i-1) + j\}$ and $\{kl\} \equiv \{d (k-1) + l\}$, respectively, where $i,j,k,l=1,\dots,d$.
 
Applying the Singular Value Decomposition (SVD) to the matrix $\Theta^{AB}$,
\begin{equation}
\Theta_{\{\alpha i\} \{j \gamma\}}^{AB} = X_{\{\alpha i\} \beta}^{~} \tilde{\lambda}_{\beta}^{A} Y_{\beta \{j \gamma\}}^{~} \, ,
\label{ThSVD}
\end{equation}
results in the three-matrix factorization $\Theta=X \Lambda Y^{\rm T}$ so that the unitary matrices satisfy $XX^{\rm T}=I$, $YY^{\rm T}=I$, and $\Lambda$ is a diagonal matrix with non-negative real (singular) values $\tilde{\lambda}_{\beta}^{A}$. The index $\beta$ takes the identical number of the degrees of the freedom as the two grouped indices $\{\alpha i\}\equiv\{d(\alpha-1)+i\}$ and $\{j\gamma\}\equiv\{\chi(j-1)+\gamma\}$, i.e., $\beta = 1,\dots,d\chi$.

To keep all calculations numerically feasible, a truncation process has to be performed in order to decrease the exponential increase of the degrees of freedom in the variable $\chi\propto d^n$, where $n$ is an iteration cycle. The truncation proceeds in one of the two consecutive steps. First, if $d \chi > D$, the matrix dimension is truncated down $D$, otherwise no truncation is performed. Second, truncate the dimension up to the largest $\beta$ index for which $\lambda_{\beta} > \epsilon$ such that $0<\epsilon\lll 1$. Subsequently, both the $\Gamma$ tensors need to be updated
\begin{equation}
\tilde{\Gamma}^{A}_{i \alpha \beta} = X_{\{\alpha i\} \beta}^{~} / \lambda_{\alpha}^{B} \, , \quad \tilde{\Gamma}^{B}_{j \beta \gamma} = Y_{\{j \beta\} \gamma}^{~} / \lambda_{\gamma}^{B} \, .
\end{equation}
The reason of the division by $\lambda^{B}$ in the last pair of equations lies in the re-introduction of $\lambda^{B}$ back into the network. 
Next, we normalize the updated $\tilde{\lambda}_{\beta}^{A}$ coefficients by dividing each coefficient by the norm 
$$\sqrt{\sum_{\beta=1}^{min(D,d\chi)} \left(\tilde{\lambda}_{\beta}^{A} \right) ^2} \, . $$ 
There is no need for any normalization of the $\Gamma$ tensors.

\paragraph{Expectation values:} We demonstrate a way of obtaining expectation values for a state $\ket{\Psi}$ represented as the MPS. It is sufficient to mention two specific examples only, in particular, the mean value of the local energy and the spontaneous magnetization measured on a spin site. The energy per one site corresponding to the state can be calculated by simply taking the sum
\begin{equation}
\langle{\cal H}_{\rm loc}\rangle = \bra{\Psi} {\cal H}_{\rm loc} \ket{\Psi} = \sum_{\alpha, \gamma = 1}^{\chi} \sum_{i, j, k, l = 1}^{d} \Psi_{\alpha k l \gamma} \left({\cal H}_\text{loc}\right)_{kl, ij}^{~} \Psi_{\alpha i j \gamma} \, , 
\end{equation}
where we used the two-site MPS expansion $\Psi_{\alpha i j \gamma} $ defined in Eq.~\eqref{MPS_update_eq1}. The magnetization is just an expectation value of a chosen Pauli matrix ${\cal S}^{\,\Omega}$, where $\Omega=x$, $y$ , or $z$. For such a one-site observable, we use the expansion of the state $\Psi_{\alpha i \beta} = \lambda^{B}_{\alpha} \Gamma^{A}_{i \alpha \beta} \lambda^{A}_{\beta}$. Hence, the spontaneous magnetization is
\begin{equation}
\langle{\cal S}^{\,\Omega}\rangle = \bra{\Psi} {\cal S}^{\,\Omega} \ket{\Psi} = \sum_{\alpha, \beta = 1}^{\chi} \sum_{i, k = 1}^{d} \lambda^{B}_{\alpha} \Gamma^{A}_{k \alpha \beta} \lambda^{A}_{\beta} 
{\cal S}^{\,\Omega}_{ki} \lambda^{B}_{\alpha} \Gamma^{A}_{i \alpha \beta} \lambda^{A}_{\beta} \, .
\end{equation}

\subsection{Ising model}

 The Hamiltonian of the one-dimensional quantum Ising model in the transverse magnetic field $h$ is defined as 
 \begin{equation}
 {\cal H} = \sum\limits_{r = -\infty}^{+\infty} -J{\cal S}^{\,x}_{r} {\cal S}^{\,x}_{r+1} - h {\cal S}^{\,z}_{r} \, . 
 \end{equation}
In the language of the Pauli matrices defined in Eq.~\eqref{pauli_x}--\eqref{pauli_z} including the identity matrix, the local Hamiltonian is a $4\times4$ matrix
\begin{equation}
{\cal H}_{\text{loc}}^{[r,r+1]} = -J {\cal S}^{\,x}_{r} \otimes {\cal S}^{\,x}_{r+1} - 
\frac{h}{2} \left( {\cal S}^{\,z}_{r} \otimes I_{r+1}^{~} + I_{r}^{~} \otimes {\cal S}^{\,z}_{r+1} \right) \, .
\end{equation}
The imaginary-time evolution is realized by applying the non-unitary operator $\exp{\left( -\tau\, {\cal H}^{~}_{\text{loc}} \right) }$, which has the matrix form
\begin{equation}
 \left( \begin{matrix}
\cosh ( \tau s) - hs^{-1}\sinh ( \tau s ) & 0 & 0 & -s^{-1}\sinh(\tau s) \\
  0 & \phantom{-}\cosh(\tau) & -\sinh(\tau) & 0 \\
  0 & -\sinh(\tau) & \phantom{-}\cosh(\tau) & 0 \\ 
  -s^{-1}\sinh(\tau s) & 0 & 0 & \cosh ( \tau s) + hs^{-1} \sinh ( \tau s ) &  
 \end{matrix} \right) \, ,
\end{equation}
where $s=\sqrt{1+h^2}$ and the coupling constant $J=1$.

\paragraph{MPS initialization:} Typically, the numerical calculations require the following MPS initialization Ansatz proved to be appropriate for the Ising model (with the initial bond dimension $\chi=1$)
\begin{equation} \label{MPS_Ising_Init}
\Gamma^{A}_{~} = [1, 0] \, , \quad \Gamma^{B}_{~} = [1, 0] \, \quad \lambda^{A} = [1] \, , \quad \lambda^{B} = [1] \, . 
\end{equation}

\paragraph{Numerical results:}

\begin{figure}[tb]
 \centering
 \includegraphics[width=3.8in]{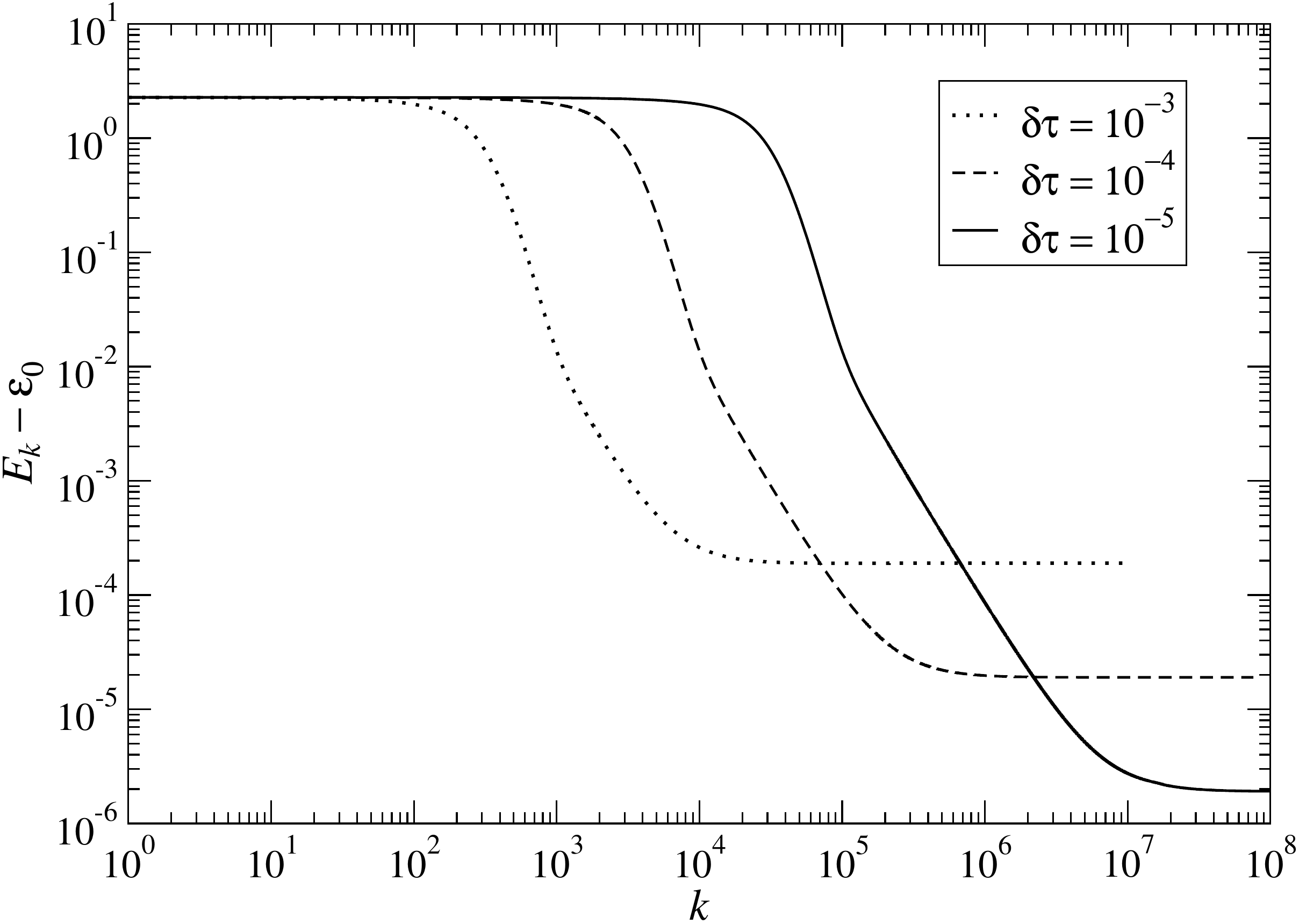}
   \caption{
The imaginary-time evolution of the transverse field Ising model at criticality ($h=1$) for three selected the time steps $\delta \tau = \{10^{-3}, 10^{-4}, 10^{-5}\}$. The referenced analytical ground-state energy is $\varepsilon_0 = -4/\pi$.}
 \label{1d_ising_energy}
 \end{figure}

\begin{figure}[tb]
 \centering
 \includegraphics[width=3.8in]{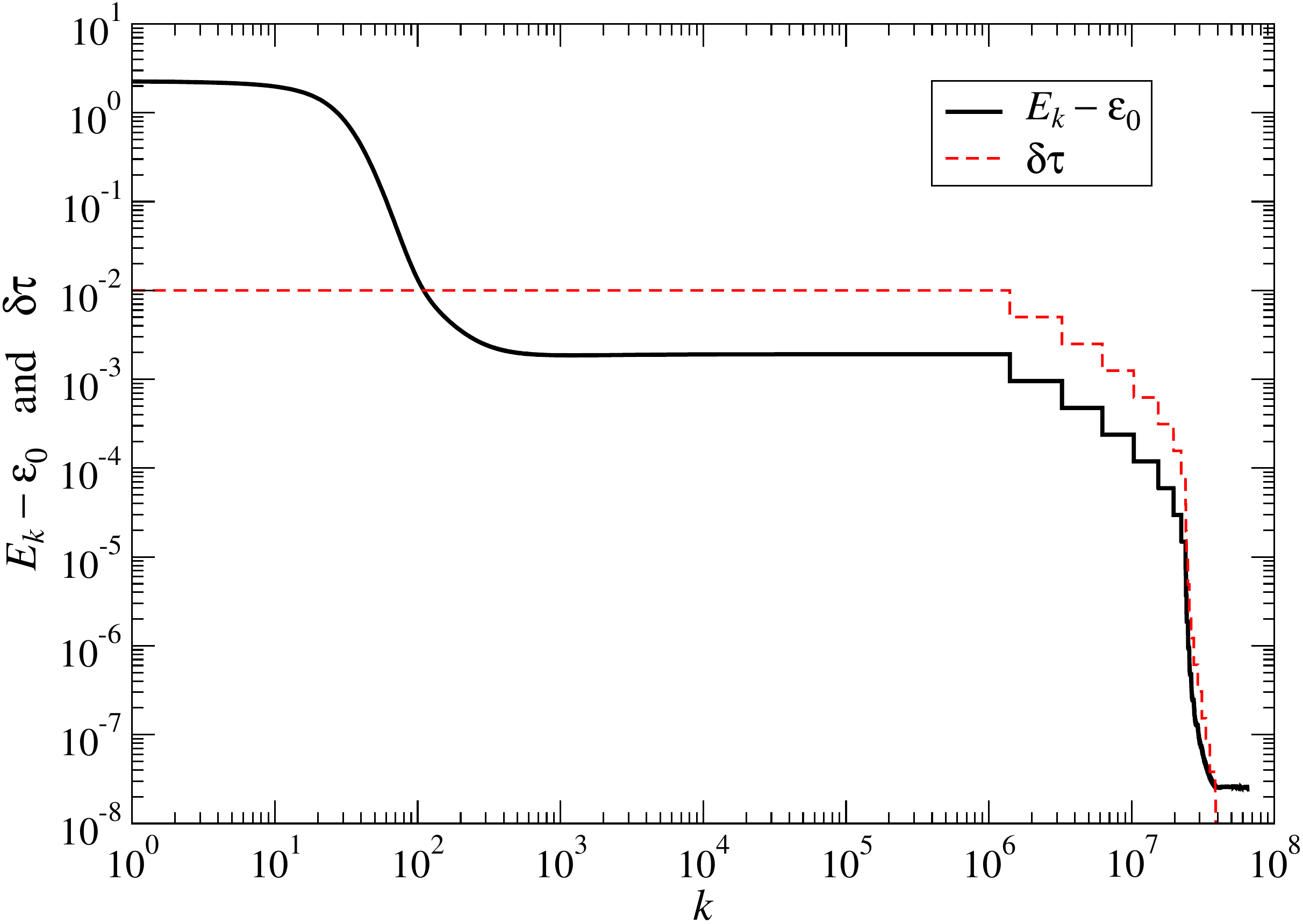}
   \caption{
   The imaginary-time evolution of the Ising model at the phase transition ($h=1$), where 
   the adaptive step is initialized to $\delta\, \tau = 10^{-2}$ for $D=32$. 
   The consequent decrease $\delta\, \tau \to \delta\, \tau/2$ is applied after converging
   $E_k$ for preceding $\delta\, \tau$.}
   \label{1d_ising_adaptive}
 \end{figure}

 We focus on the particular numerical calculation of the ground-state energy $\varepsilon_0$ of the one-dimensional transverse field Ising model at the criticality $h=1$, which is known to be nontrivial with the longest computational time and the lowest numerical accuracy.
 Two strategies are at hand. The first one is to fix the imaginary-time step $\delta\tau$ to a small value and observe convergence of the numerical ground-state energy $E_k$ with respect to the iteration step $k$, as plotted in Fig.~\ref{1d_ising_energy}. The second (adaptive) strategy, uses such $\delta\tau$, which is not constant, but it gradually decreases whenever $E_k$ has converged for a particular $\delta\tau$. For illustration, we applied a simple mechanism $\delta\tau \to \delta\tau/2$, as shown in Fig.~\ref{1d_ising_adaptive}. 
The maximal bond dimension $D=32$ was applied in the numerical calculations (while considering
the truncation threshold as small as $\epsilon=10^{-32}$ in all of the MPS extensions). 
One can achieve a better accuracy (i.e., a smaller absolute error $|E_k-\varepsilon_0|$) whenever $\delta\tau$ get small enough, which requires enormous increase of the iterative steps to
reach the complete convergence. Hence, For fixed $\delta\tau = \{10^{-3}, 10^{-4}, 10^{-5}\}$, respectively, the absolute errors of the numerical lowest energies cannot be lowered anymore resulting in $E_k-\varepsilon_0 \sim \{10^{-4}, 10^{-5}, 10^{-6} \}$, as depicted in Fig.~\ref{1d_ising_energy}. The exact value of the ground-state energy is $\varepsilon_0 = -4/\pi \approx -1.27324$~\cite{1dqising}. If using the adaptive strategy, we have reached a better accuracy $E_k-\varepsilon_0 \lesssim 10^{-7}$ (cf. Fig.~\ref{1d_ising_adaptive}). 

\subsection{Heisenberg model}

The one-dimensional Heisenberg model (including a term with the magnetic field $h$),
\begin{equation}
{\cal H} = \sum_{r = -\infty}^{+\infty} -J \left( {\cal S}^{\,x}_{r} {\cal S}^{\,x}_{r+1} + 
{\cal S}^{\,y}_{r} {\cal S}^{\,y}_{r+1} + {\cal S}^{\,z}_{r} {\cal S}^{\,z}_{r+1} \right)
- h\, {\cal S}^{\,z}_{r} \, ,
\end{equation}
has, in the notation of the local Hamiltonian as the $4\times4$ matrix, form
\begin{equation}
{\cal H}^{[r,r+1]}_{\text{loc}} = -J \left(
  {\cal S}^{\,x}_{r} \otimes {\cal S}^{\,x}_{r+1}
+ {\cal S}^{\,y}_{r} \otimes {\cal S}^{\,y}_{r+1}
+ {\cal S}^{\,z}_{r} \otimes {\cal S}^{\,z}_{r+1} \right)
+ \frac{h}{2} \left( {\cal S}^{\,z}_{r} \otimes I_{r+1}^{~}
+ I_{r}^{~} \otimes {\cal S}^{\,z}_{r+1} \right)  \, .
\end{equation}
By exponentiating the local Hamiltonian, we obtain the non-unitary (the imaginary-time
evolution) operator in the matrix form
\begin{equation}
\exp{\left( -\tau\, {\cal H}^{~}_{\text{loc}} \right) } = \left( \begin{matrix}
e^{- \tau (1+h) }  & 0 & 0 & 0 \\
  0 & \phantom{-}e^{\tau} \cosh(2\tau) & -e^{\tau}\sinh(2\tau) & 0 \\
  0 & -e^{\tau}\sinh(2\tau) & \phantom{-}e^{\tau}\cosh(2\tau) & 0 \\ 
  0 & 0 & 0 & e^{-\tau (1-h)} &  
 \end{matrix} \right) \, ,
\end{equation}
which is used in numerical analyses. We again simplified the form by setting $J=1$.

\paragraph{MPS initialization:} The MPS is initialized (with $\chi=1$) by setting
\begin{equation}
\Gamma^{A}_{~} = [1, 0] \, , \quad \Gamma^{B}_{~} = [0, 1] \, \quad \lambda^{A} = [1] \, , \quad \lambda^{B} = [1] \, ,
\end{equation}
as this particular Ansatz provides a stable solution for the Heisenberg model. Compare the difference for the transverse field Ising model in Eq.~\eqref{MPS_Ising_Init}. Inappropriate use of the Ising-model initialization in the Heisenberg model leads to trivial MPS updates, which fails to obtain the correct $\varepsilon_0$. 

\paragraph{Numerical results:}

\begin{figure}[tb]
 \centering
 \includegraphics[width=3.8in]{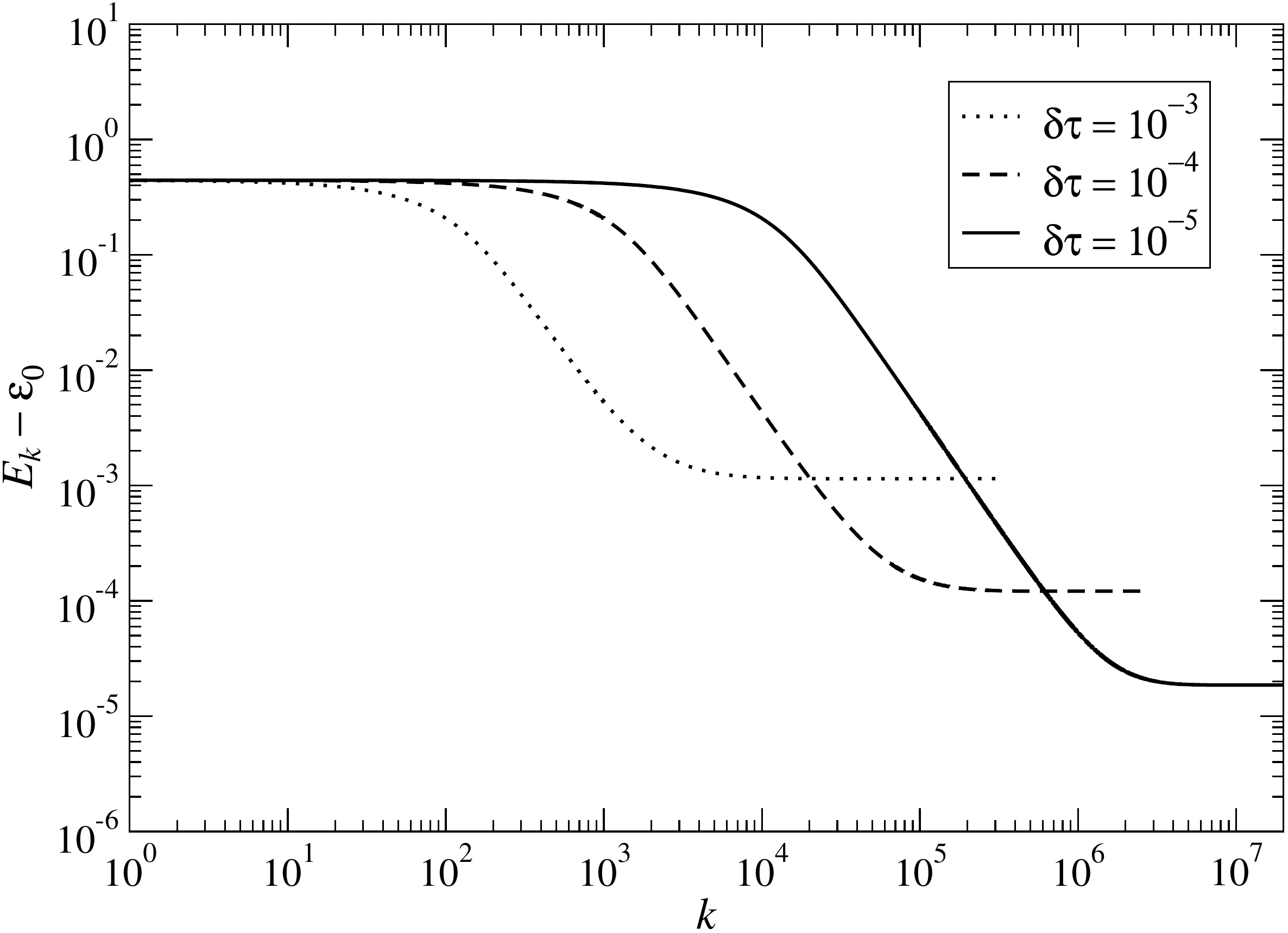}
   \caption{The imaginary-time evolution for the Heisenberg model at $h=0$ for the time steps
   $\delta \tau = \{10^{-3}, 10^{-4}, 10^{-5}\}$. The improving energies $E_k$ towards the
   the analytical ground-state energy $\varepsilon_0 = 1/4 - \ln 2$ depends on $\delta\,\tau$.
   }
   \label{1d_heisenberg_energy}
 \end{figure}

The Heisenberg model is gapless at zero magnetic field $h=0$. If this case is compared with the transverse field Ising model at $h=1$, the computational cost for the Heisenberg model at $h=0$ is even higher. In order to compare efficiency of both the Ising and the Heisenberg models, the identical bond dimension $D=32$ and the parameter $\epsilon=10^{-32}$ are set. For fixed $\delta\tau = \{10^{-3}, 10^{-4}, 10^{-5}\}$, the respective absolute errors of the energies are $E_k-\varepsilon_0 \sim \{10^{-3}, 10^{-4}, 10^{-5} \}$, as shown in Fig.~\ref{1d_heisenberg_energy}). The referenced exact value of the ground-state energy is $\varepsilon_0 = 1/4 - \ln 2 \approx -0.44315$~\cite{1dheisenberg}. 
 \begin{figure}[htb!]
 \centering
 \includegraphics[width=3.8in]{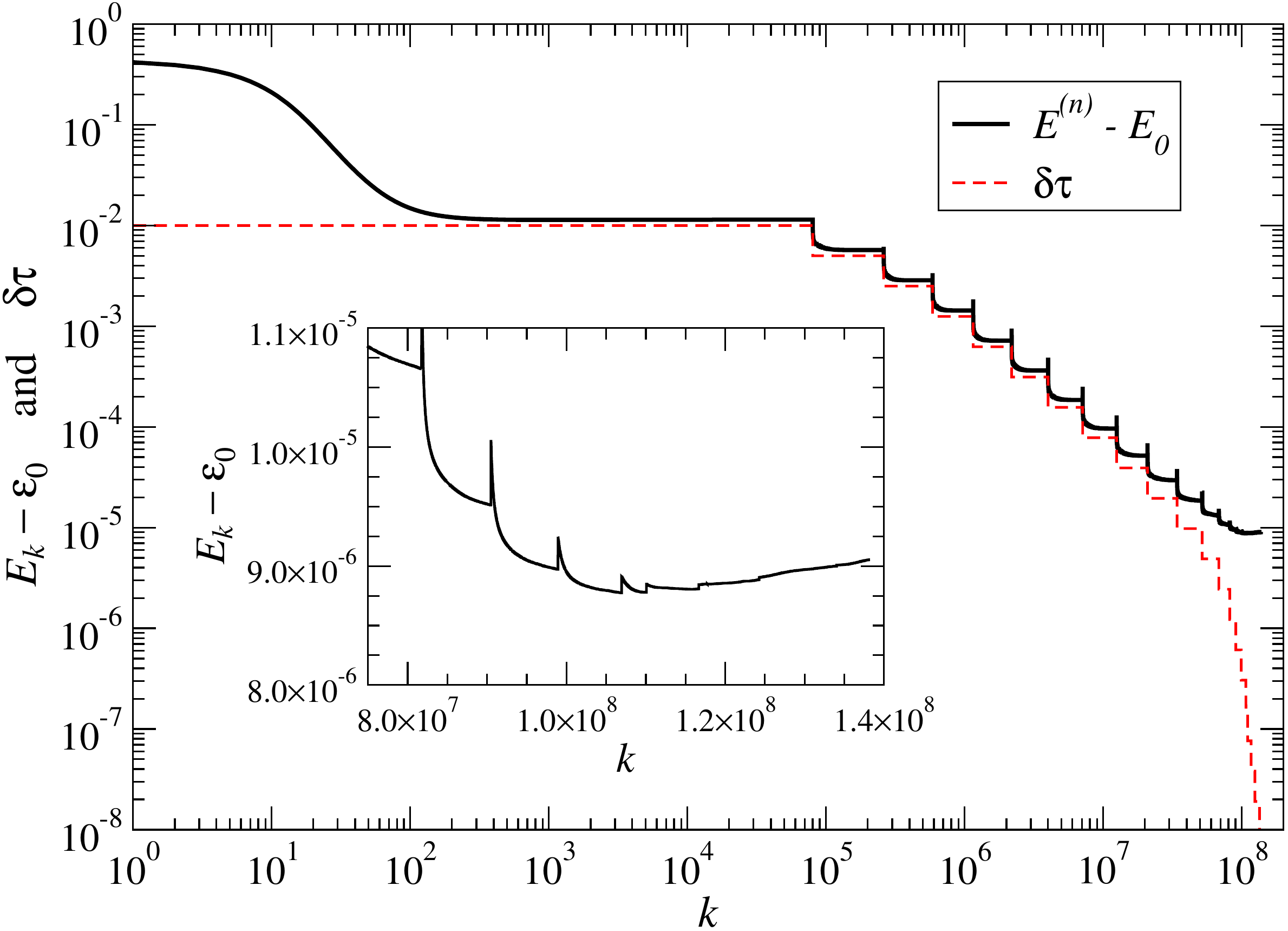}
   \caption{The imaginary-time evolution of the Heisenberg model at $h=0$, where 
   the adaptive time step is applied and initialized to $\delta \tau = 10^{-2}$.
   The time step decreases as $\delta\, \tau \to \delta\, \tau/2$. If $\delta\tau
   \lesssim 10^{-8}$, the energies $E_k$ do not improve anymore to $\varepsilon_0$
   while keeping $D=24$ unchanged.}
   \label{1d_heisenberg_adaptive}
 \end{figure}   

 The absolute errors of the approximated ground-state energy are roughly an order of the magnitude larger than those for the Ising model. The minimal absolute error $E_k-\varepsilon_0 \lesssim 10^{-5}$ (cf. Fig.~\ref{1d_heisenberg_adaptive}) is again reached for the adaptive strategy. However, the numerical accuracy of the lowest energy for the identical $D=24$ is about two orders of the magnitude larger than it is in the Ising model. A certain numerical instability is obvious from the behavior of the energy $E_k$ after performing $k \sim 10^8$ iterative steps, as plotted in the inset of Fig.~\ref{1d_heisenberg_adaptive}. The reason for this behavior lies in accumulating smallest numerical errors, which cause the gradual increase of $E_k$ at large $k$.

\subsection{Corner Transfer Matrix Renormalization Group}\label{CTMRG_Section}

The numerical algorithm Corner Transfer Matrix Renormalization Group (CTMRG) is a composition of two techniques: the analytical method Corner Transfer Matrix proposed by Baxter (see Subsection~\ref{CTM_subsec}) and a powerful numerical method Density Matrix Renormalization Group~\cite{White}. The CTMRG algorithm was developed by Nishino and Okunishi to study two-dimensional 
classical spin systems~\cite{ctmrg1, ctrmg2}. 

It is an iterative numerical procedure, which accurately calculates the partition function of spin systems on two-dimensional discrete lattices. For simplicity, we explain the CTMRG algorithm applied the classical Ising model on the square lattice. If each iterative step of CTMRG is enumerated by integer $k=1, 2, 3, \dots$, the size of the square lattice gradually expands its size as $\left(2k+1\right)\times\left(2k+1\right)$. An additional small modification to the algorithm enables to expands the square lattices $2k\times2k$ as well.

\paragraph{Initialization:}

\begin{figure}[tb]
 \centering
 \includegraphics[width=3.2in]{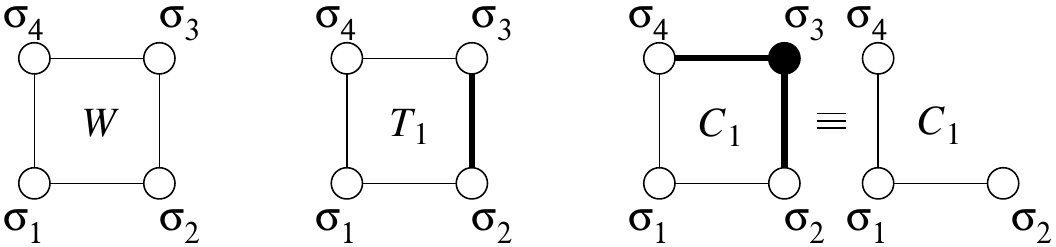}
   \caption{Graphical representation of the Boltzmann weight ${\cal W}$ (left), 
   the initial forms of the half-row transfer tensor ${\cal T}_1$ (middle), 
   as well as the corner transfer tensor ${\cal C}_1$ (right) at step $k=1$. 
   The thin lines correspond to the shared links, which are accounted for by the
   overall factor of $\frac{J}{2}$ in Eq.~\eqref{T_init} and Eq.~\eqref{C_init}
   (and, analogously, for the prefactors $\frac{h}{4}$ and $\frac{g}{4}$ on the spin
   sites denoted by the circles). The thick lines and the spin sites connected to them
   are always positioned on the lattice boundary. 
   The filled circle in the corner transfer tensor ${\cal C}_1$ indicates the spin variable
   $\sigma_3$, in which the summation has to be taken.}
   \label{ctrmg_init}
 \end{figure}

At the first iteration ($k=1$), we create the $3\times3$ lattice by joining four quadrants, i.e., each quadrant is a basic lattice cell as small as $2 \times 2$ and is composed of four spins $\sigma_{1}^{~}$, $\sigma_{2}^{~}$, $\sigma_{3}^{~}$, and $ \sigma_{4}^{~}$. The basic lattice cell can be expressed by the statistical Boltzmann weight ${\cal W}$ (see Fig.~\ref{ctrmg_init} on the left) such that
\begin{equation}
{\cal W}_{\sigma_{1}\sigma_{2}\sigma_{3}\sigma_{4}} =
 \exp \left\{\frac{1}{k_{\rm B}T} \left[ \frac{J}{2} \left( \sigma_{1} \sigma_{2} + \sigma_{2} \sigma_{3} +
 \sigma_{3}\sigma_{4} + \sigma_{4}\sigma_{1} \right) 
+ \frac{h}{4} \left( \sigma_{1} + \sigma_{2} + \sigma_{3} + \sigma_{4} \right) \right] \right\} .
\label{W_B}
\end{equation}
The CTMRG algorithm requires to define two types of tensor operators (expressed in terms of the Boltzmann weights): a {\it half-row transfer tensor} ${\cal T}_k$ and a {\it corner transfer tensor} ${\cal C}_k$.

Before the first iteration step is carried out, the half-row transfer tensor ${\cal T}_{k=1}$ has to be initialized (see the graphical representation of ${\cal T}_{1}$ in Fig.~\ref{ctrmg_init} in the middle) so that
\begin{eqnarray}\label{T_init}
{\cal T}_{1,\,\sigma_{1}\sigma_{2}\sigma_{3}\sigma_{4}}=
 \exp \left\{ \frac{1}{k_{\rm B}T} \left[ \frac{J}{2} \left( \sigma_{1} \sigma_{2} + 2\sigma_{2} \sigma_{3} +
 \sigma_{3}\sigma_{4} + \sigma_{4}\sigma_{1} \right)  + \right. \right. \nonumber \hspace{1.05cm}\\
+ \left. \left. \frac{h}{4} \left( \sigma_{1} + 2\sigma_{2} + 2\sigma_{3} + \sigma_{4} \right) 
+ \frac{g}{4}(2\sigma_2 + 2\sigma_3) \right] \right\} ,
\end{eqnarray}
where $g$ can be thought of as a small external magnetic field imposed on two boundary spins (used to accelerate the spontaneous-symmetry breaking when reaching the thermodynamic limit numerically). The prefactor $2$ in the spin-spin interaction term $J\sigma_2 \sigma_3$ is used to adapt the correct interaction (bond) weight acting on the square lattice boundary, as it will be more clear later on. 
Analogously, the two boundary spins $\sigma_2$ and $\sigma_3$ in ${\cal T}_{1}$ have the adjusting prefactors $2$ to keep correct the energy contributions from both the magnetic fields $h$ and $g$ when sharing the two lattice spins with adjacent Boltzmann-weight blocks. Let us remark here that the ${\cal W}$ blocks never appear on the lattice boundary; therefore, no field term $g$ is present in Eq.~\eqref{W_B}, which is always imposed on the boundary spins only. (If $g$ is set to zero, the numerical convergence of the CTMRG algorithm may become logarithmically slow.)

The initial form of the corner transfer tensor ${\cal C}_1$ (cf. Fig.~\ref{ctrmg_init} on the right) is expressed in term of the Boltzmann weight in the following
\begin{eqnarray}\label{C_init}
{\cal C}_{1,\,\sigma_{1}\sigma_{2}\sigma_{4}}=
 \sum_{\sigma_3} \exp \left\{ \frac{1}{k_{\rm B}T} \left[ \frac{J}{2} \left( \sigma_{1} \sigma_{2} + 2\sigma_{2} \sigma_{3} +
 2\sigma_{3}\sigma_{4} + \sigma_{4}\sigma_{1} \right)  + \right. \right. \nonumber \hspace{2.0cm}\\
+ \left. \left. \frac{h}{4} \left( \sigma_{1} + 2\sigma_{2} + 4\sigma_{3} + 2\sigma_{4} \right) 
+ \frac{g}{4}(2\sigma_2 + 4\sigma_3 + 2\sigma_4) \right]  \right\} .
\end{eqnarray}
An analogous reasoning, we have mentioned for initializing ${\cal T}_1$, also holds for the usage of the prefactors $2$ and $4$ in the definition of ${\cal C}_1$. The initial corner transfer tensor ${\cal C}_1$ is defined on the three spins $\sigma_{1}^{~}$, $\sigma_{2}^{~}$, and $ \sigma_{4}^{~}$ while the configuration sum is taken over the Ising spin $\sigma_{3}^{~}=\pm1$ in accord with the definition in Eq.~\eqref{corner_matrix}. (It is convenient to keep using the term `corner transfer tensor' ${\cal C}_{\sigma1\sigma_2\sigma_4}$ rather than the formerly used term `corner transfer matrix' ${\cal C}(\sigma_1\sigma_2|\sigma_1\sigma_4)$, which had required to duplicate the spin variable $\sigma_1$ in order to keep its matrix formalism.)

\paragraph{Density matrix:} 

\begin{figure}[tb]
 \centering
 \includegraphics[width=1.6in]{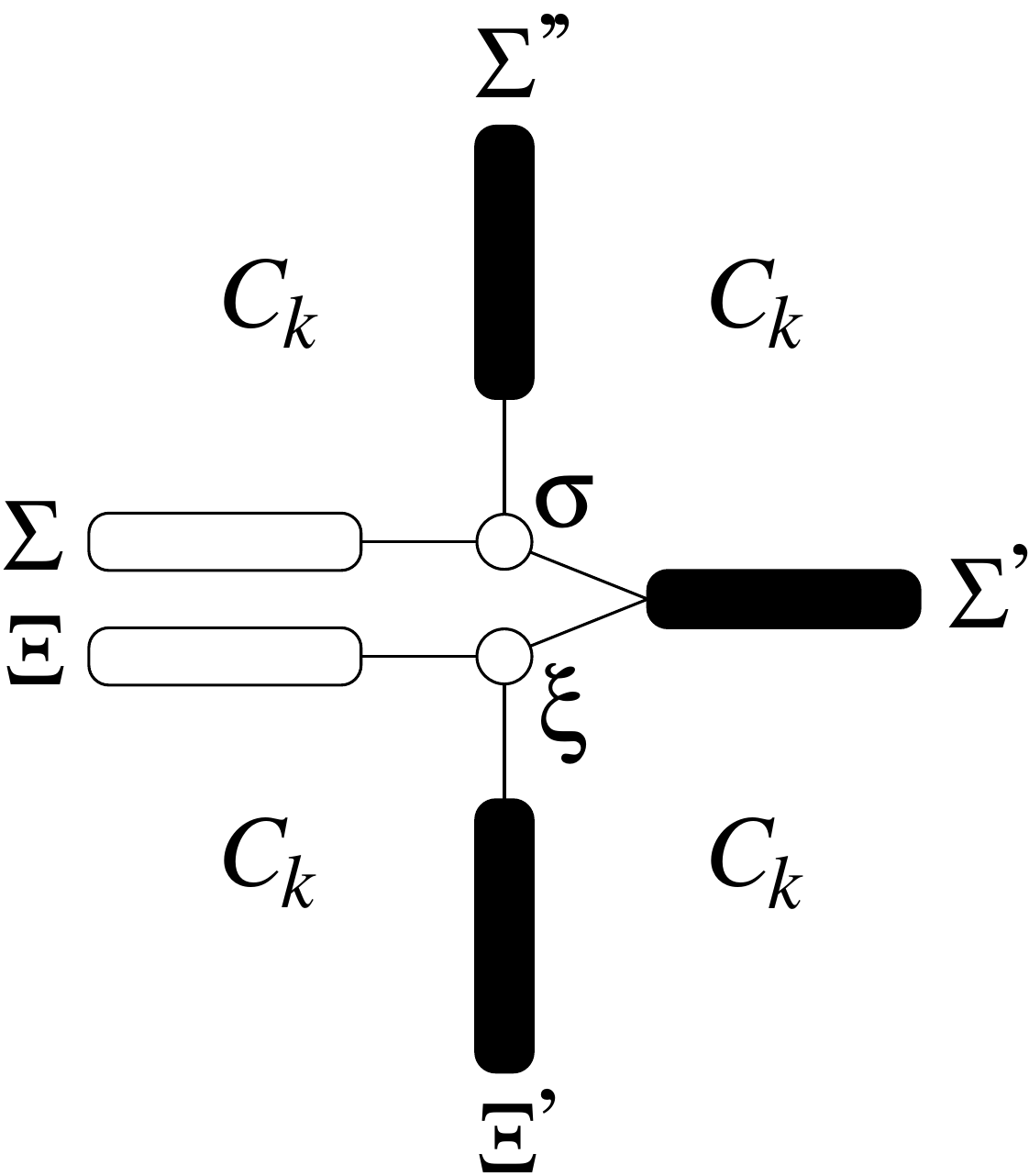}
   \caption{Graphical representation of the density matrix $\rho_k \left( \Sigma\,\sigma
\vert \Xi\,\xi \right)$ constructed in Eq.~\eqref{RDM}. It is formed by the four corner
transfer tensors ${\cal C}_k$. The summation is taken over the variables $\Xi^\prime$, $\Sigma^\prime$, and $\Sigma^{\prime\prime}$ illustrated by the filled ovals.}
   \label{DM_total}
 \end{figure}

A reduced density matrix $\rho_k$ (defined within the quantum mechanics) is a partial
trace over certain group of spin degrees of the freedom (the reservoir). At arbitrary
iteration step $k$ of CTMRG, the four partially traced corner transfer tensors can form
the reduced density matrix expressed as $\rho_k = \text{Tr}\, {\left({\cal C}_k\right)}^4$.
If spin degrees of the freedom are introduced in the corner transfer matrices, the
schematic picture in Fig.~\ref{DM_total} of the reduced density matrix is used as
a helpful illustration the following expression
\begin{equation}
\rho_k \left( \left. \Sigma\, \sigma \right| \Xi\, \xi \right) = 
\sum\limits_{\Xi^{\prime}\Sigma^{\prime}\Sigma^{\prime\prime}}
{\cal C}_{k,\, \sigma \Sigma^{\prime} \Sigma^{\prime\prime}}
{\cal C}_{k,\, \sigma \Sigma^{\prime\prime} \Sigma^{~}}
{\cal C}_{k,\, \xi \Xi^{~} \Xi^{\prime}}
{\cal C}_{k,\, \xi \Xi^{\prime} \Sigma^{\prime\prime}}.
\label{RDM}
\end{equation} 
The Greek lower and uppercase symbols, respectively denote the single-spin and the
multi-spin variables, as has been discussed in Subsec.~\ref{CTM_subsec}; cf.
Eqs.~\eqref{corner_matrix}-\eqref{C_as_tensor}. For the tutorial purposes, let the
single-spin variables have two degrees of the freedom (such as in the Ising model)
and the multi-state spin variables have $m$ degrees of the freedom (such that $m\ll2^k$
for large $k$).

It is obvious that ${\rm Tr}\,\rho_k={\cal Z}_{{(2k+1)}^2}$ by definition. Hence, it is
straightforward to find out the meaning of the reduced density matrix in the classical
statistical mechanics: It represents the statistical probability. Hence, one can evaluate
mean values $\langle X\rangle$ of observables (both single- and multi-site ones) to get
$\langle X\rangle = {\cal Z}^{-1} {\rm Tr}\,\left( X\rho \right)$ in the identical way
as in the quantum mechanics.

\paragraph{Projection operator:}

The purpose of the construction of the reduced density matrix is such that it represents
statistical probabilities of spin configurations. Having diagonalized a reduced density
matrix, one obtains a spectrum of eigenvalues and the corresponding eigenvectors. The
properties of the spectrum can be summarized into the following statement: Each eigenvalue
corresponds to a particular spin configuration so that the larger the eigenvalue, the
higher probability of the related spin configuration. After sorting the eigenvalues together
with the corresponding eigenvectors, one can select, for instance, $m$ eigenvectors, which
correspond to the $m$ largest eigenvalues. The set of such $m$ eigenvectors forms a
projection operator $P$, which can be applied to a huge configurational space, which is
reduced to a subspace with the most probable spin configurations (discarding/truncating the
subspace with the insignificant spin configurations for the particular model parameters).

Formally, if a reduced density matrix $\rho_k$ of a subsystem is described by the two-state
spins $\sigma,\xi=1,2$ and $m$-state spins $\Sigma,\Xi=1,2,\dots,m$, then it represents
a real symmetric $2m\times2m$ matrix. Diagonalization of the reduced density matrix,
$\rho_k^{~}=P_k^{-1}D_k^{~}P_k^{~}$, where $P_k^{-1}P_k^{~}=I$, yields a diagonal matrix
$D$ consisting of non-negative eigenvalues $d_i$. They are sorted decreasingly
$d_1\geq d_2\geq \dots, \geq d_{2m}\geq0$. The eigenvalues satisfy the rule $\sum_{i=1}^{2m}
d_i=1$, provided that the corner transfer tensors ${\cal C}_k$ are appropriately
normalized before the reduced density matrix is constructed in Eq.~\eqref{RDM}.

Having sorted the eigenvalues $d_i$ in the decreasing order, the truncation error
$\eta=\sum_{i=m+1}^{2m}d_i$ provides a useful information to quantify reliability
of the real-space renormalization. The calculations are usually considered very reliable
if $0\leq\eta\lesssim10^{-10}$. The renormalization process is carried out by creating the
projection operators $P_k$, which project the full configurational space to a restricted
subspace. In particular, the projection operators are rectangular $2m\times m$
matrices
\begin{equation}
P_k = \left(
\begin{array}{cccc}
    | & | & \dots & | \\
    \phi_1 & \phi_2 & \dots & \phi_m \\
    | & | & \dots & | \\
\end{array}
\right)\, ,
\end{equation}
in which we keep those eigenvalues $\phi_i$, ($i=1,2,\dots,m$) in $P_k$, which correspond
to the first $m$ largest eigenvalues $d_1 \geq d_2 \geq \dots \geq d_m$. The projection
operators $P_k$ are applied to expanded corner transfer tensors ${\cal C}_{k+1}$ and the
transfer tensors ${\cal T}_{k+1}$ while keeping fixed the dimension of the tensors
(i.e., the degrees of the freedom of the multi-state spin indices $\Sigma$ and $\Xi$
are fixed to $m$ at each iteration step $k$, otherwise they grow exponentially fast,
as we have discussed within the iTEBD algorithm).

 \paragraph{The tensor extension and renormalization:}

\begin{figure}[tb]
 \centering
 \includegraphics[width=2.5in]{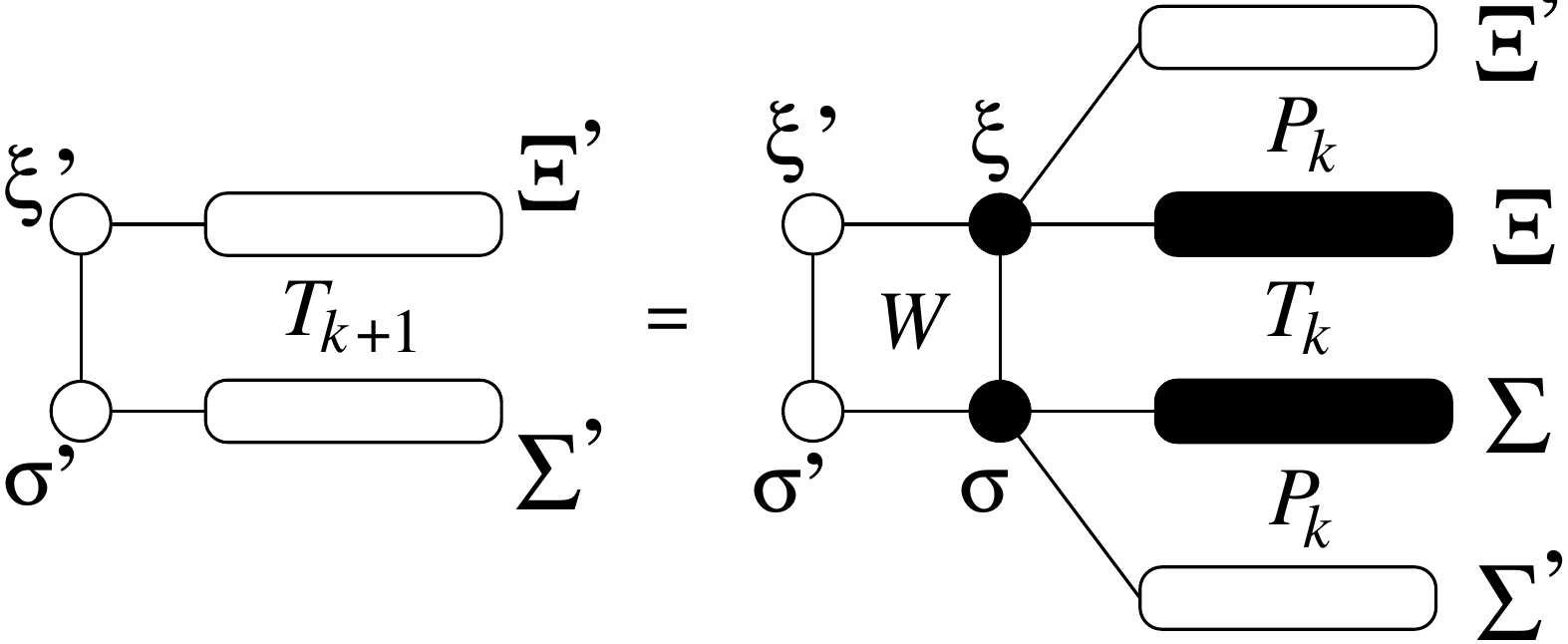}
   \caption{Extension and renormalization process takes the original transfer tensor
${\cal T}_{k,\,\sigma \Sigma \xi \Xi}$ to form the expanded and renormalized transfer
tensor ${\cal T}_{k+1,\, \sigma^{\prime} \Sigma^{\prime}\xi^{\prime} \Xi^{\prime}}$
in accord with Eq.~\eqref{T_exten_eq}. The circles and ovals filled in black denote
those spin variables, which are summed over.}
\label{T_exten_fig}
 \end{figure}
The extended transfer tensor, ${\cal T}_{k+1}={\cal W}{\cal T}_{k}$,
requires to add an extra Boltzmann weight, as depicted in Fig.~\ref{T_exten_fig}. 
The renormalization process (performed by applying the projection operators $P_k$)
can be simultaneously included in the extension step to make this process concise,
formally written as
\begin{equation}\label{T_exten_eq}
{\cal T}_{k+1,\, \sigma^{\prime} \Sigma^{\prime} \xi^{\prime} \Xi^{\prime}}
= \sum_{ \sigma  \Sigma \xi \Xi }  P_{k,\, \Xi \xi \Xi^{\prime}} 
\left({\cal W}_{\sigma^{\prime} \sigma \xi \xi^{\prime}} 
{\cal T}_{k,\, \sigma \Sigma \xi \Xi}\right)
P_{k,\, \Sigma \sigma \Sigma^{\prime}} \, . 
\end{equation}
The extension part is emphasized by inserting the parenthesis in it.
 
\begin{figure}[tb]
 \centering
 \includegraphics[width=2.7in]{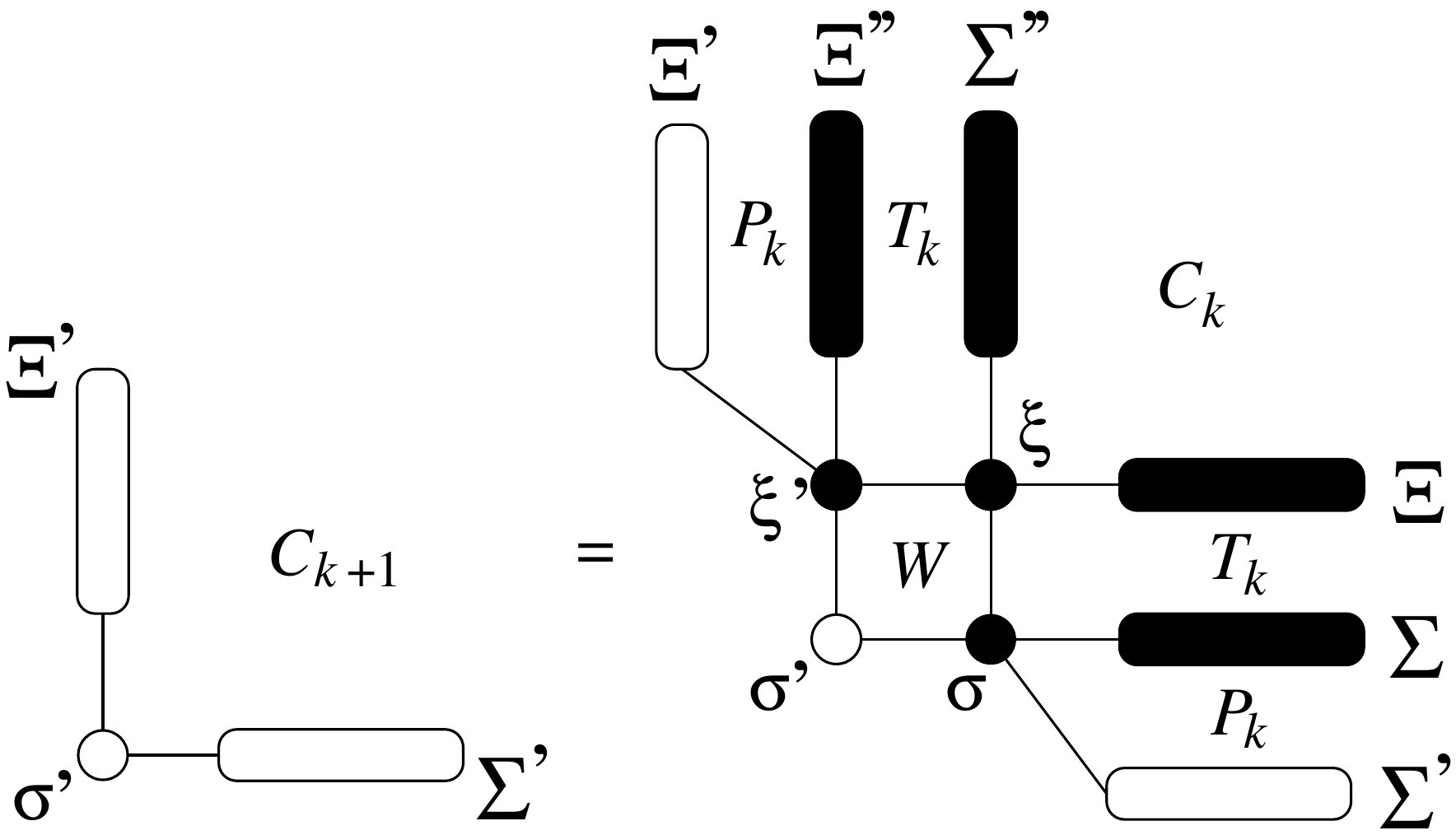}
   \caption{Extension and renormalization of the corner transfer tensor visualizes
   Eq.~\eqref{C_exten_eq}.}
   \label{C_exten_fig}
 \end{figure}

To create the extended corner transfer tensor ${\cal C}_{k+1,\, \sigma^{\prime}
\Sigma^{\prime} \Xi^{\prime}}$ requires a more complex procedure. Again, the parenthesis
enclose the extension part only
\begin{equation}\label{C_exten_eq}
{\cal C}_{k+1,\, \sigma^{\prime} \Sigma^{\prime} \Xi^{\prime}} = 
\sum_{\sigma \xi \xi^{\prime} \Sigma \Xi \Sigma^{\prime\prime} \Xi^{\prime\prime} } 
P_{k,\, \Xi^{\prime\prime} \xi^{\prime} \Xi^{\prime} } \left(
{\cal T}_{k,\, \xi \Sigma^{\prime\prime} \xi^{\prime} \Xi^{\prime\prime}}
{\cal C}_{k,\, \xi \Xi \Sigma^{\prime\prime} }
{\cal W}_{\sigma^{\prime} \sigma \xi \xi^{\prime} }
{\cal T}_{k,\, \sigma \Sigma \xi \Xi } \right)
P_{k,\, \Sigma \sigma \Sigma^{\prime}} \, ,
\end{equation}
where the two additional transfer tensors ${\cal T}_k$ are necessary. The equation is
graphically represented in Fig.~\ref{C_exten_fig}).

Since no tensor normalization has been carried out yet, these extension and renormalization
procedures in Eqs.~\eqref{T_exten_eq} and~\eqref{C_exten_eq} automatically lead to numerical
overflows (a fast divergence of the tensor elements in ${\cal T}_{k+1}$ and ${\cal C}_{k+1}$)
already after a couple of the iterations steps (typically $k\gtrsim10$). For this reason,
an appropriate normalization of the tensors is inevitable to be introduced. We have been
implementing the normalization by the maximum tensor element
\begin{equation}\label{norm_CT}
\begin{split}
& t_k=\max\limits_{\sigma\Sigma\xi\Xi}\,|{\cal T}_{k,\,\sigma\Sigma\xi\Xi}| \, ,\\
& c_k=\max\limits_{\sigma\Sigma\Xi}\,|{\cal C}_{k,\,\sigma\Sigma\Xi}|.
\end{split}
\end{equation}
It means that the absolute value of the respective largest tensor elements $t_k$ and
$c_k$ at each iteration step $k$ are found.
From now on we keep using the notation 'tilde' reserved for the normalized tensors
\begin{equation}\label{norm_CT1}
\begin{split}
{\widetilde{\cal T}}_k & \equiv \frac{{\cal T}_k}{t_k}\, \\
{\widetilde{\cal C}}_k & \equiv \frac{{\cal C}_k}{c_k}\, .
\end{split}
\end{equation}

\paragraph{Free energy calculation:}

Since the square-shaped lattice expands its size as $3 \times 3$, $5 \times 5$, $7 \times 7$,
$\dots$, $(2k+1)\times(2k+1)$, the free energy per spin site $f_k$ (enumerated by the
iteration step $k$) is
\begin{equation} \label{ctmrg_free_energy}
f_k = -\frac{k_{\rm B} T}{N_k}
\ln{\cal Z}_{k} \equiv -\frac{k_{\rm B} T}{N_k}
\ln \left[ {\rm Tr} {\left({\cal C}_k\right)}^4 \right] \, ,
\end{equation}
where
\begin{equation}
N_k = (2k+1)^2
\label{N44}
\end{equation}
is the number of the spin sites for the square lattice at the iteration step $k$.

If (for better tutorial reasons) omitting all the indices including the projectors $P_k$,
the extension procedures in Eqs.~\eqref{T_exten_eq} and~\eqref{C_exten_eq} can be
schematically abbreviated in these two recurrent relations
\begin{eqnarray}
{\cal C}_{k+1}&=&{\cal W} {\cal T}_{k}^{2} {\cal C}_{k}^{~} \, , \label{C_recurrent} \\
{\cal T}_{k+1}&=&{\cal W} {\cal T}_{k}^{~} \, , \label{T_recurrent}
\end{eqnarray}
which are initialized by setting ${\cal C}_1$ and ${\cal T}_1$ (see the definitions in
Eqs.~\eqref{T_init} and \eqref{C_init} for details).

\begin{figure}[tb]
 \centering
 \includegraphics[width=0.8in]{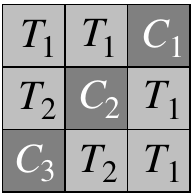}
   \caption{Decomposition of the corner transfer tensor ${\cal C}_3$ following
            Eq.~\eqref{c3_decomp_eqn}.}
   \label{C3_fig}
 \end{figure}

As an example, let us explicitly express the free energy in the third iteration step
($k=3$), i.e., $f_3 = -k_{\rm B}T\ln\left[{\rm Tr}\ ({\cal C}_3)^4\right]/7^2$.
Tracing the tensors back to the initial step $k=1$, the corner tensors ${\cal C}_3$ are
recursively decomposed into the product of the normalized tensors ${\widetilde{\cal C}}_2$
and ${\widetilde{\cal T}}_2$, and they again depend on ${\widetilde{\cal C}}_1$ and
${\widetilde{\cal T}}_1$ according to Eq.~\eqref{C_recurrent} and Eq.~\eqref{T_recurrent},
respectively. These recurrence relations result in the recursive dependence of the
normalization factors $c_k$ and $t_k$, all of them being crucial for the free-energy
calculation. The decomposition of the corner transfer tensor ${\cal C}_3$ yields (for
the square lattice)
\begin{equation} \label{c3_decomp_eqn}
\begin{split}
{\widetilde{\cal C}}_3^{~} & = \frac{{\cal C}_3^{~}}{c_3^{~}}
 = \frac{{\cal W}{\widetilde{\cal T}}_2^2{\widetilde{\cal C}}_2^{~}}
          {c_3^{~}}
 =\frac{{\cal W}{\cal T}_2^2{\cal C}_2^{~}}
             {t_2^2 c_2^{~} c_3^{~}} \\
& = \frac{{\cal W}{\left({\cal W}{\widetilde{\cal T}}_1^{~}\right)}^2
{\left({\cal W}{\widetilde{\cal T}}_1^2{\widetilde{\cal C}}_1^{~}\right)}}
{t_2^2 c_2^{~} c_3^{~}}\\
& = \frac{{\cal W}^4 {\cal T}_1^4{\cal C}_1^{~}}
{t_1^4 t_2^2 c_1^{~} c_2^{~} c_3^{~}}
 = \frac{{\cal W}^9}
{t_1^4 t_2^2 t_3^0 c_1^{1} c_2^{1} c_3^{1}}\, ,
\end{split}
\end{equation}
as conveniently visualized in Fig.~\ref{C3_fig}.
Substituting ${\cal C}_3$ into Eq.~\eqref{ctmrg_free_energy}, the explicit
expression for the free energy per site when $k=3$ leads to
\begin{equation}
f_3 = -\frac{k_{\rm B} T}{7^2}
\left[
\ln {\rm Tr}\ \left(\widetilde{\cal C}_{3}^{~}\right)^4
+ 4 \ln \prod\limits_{j=1}^{3}
c_{j}^{~} t_{j}^{2(3-j)}
\right]\, .
\label{fe44}
\end{equation}

For an arbitrary $k$, the free energy per site can be expressed in terms
of the normalization factors of the four central tensors ${\widetilde{\cal C}_{k}^{~}}$
(at step $k$ only)
\begin{equation}\label{FE44}
f_k = -k_{\rm B} T \frac{\ln{\rm Tr}\ \left( \widetilde{\cal C}_{k}^{~}\right)^4 }{(2k+1)_{~}^{2}}
- 4k_{\rm B} T \frac{ \sum_{j=0}^{k-1} \left( \ln c_{k-j}^{~} + \ln t_{k-j}^{2j} \right) }{(2k+1)_{~}^{2}} \, .
\end{equation}
This expression of the free energy becomes important later when calculating the free energy
per site for arbitrary lattice geometry. The numerical derivatives of $f_k$ specify the
thermodynamic functions, which become non-analytic at phase transitions, provided that
the thermodynamic limit ($k\to\infty$) is taken.

\paragraph{Observables:}

\begin{figure}[tb]
 \centering
 \includegraphics[width=3.5in]{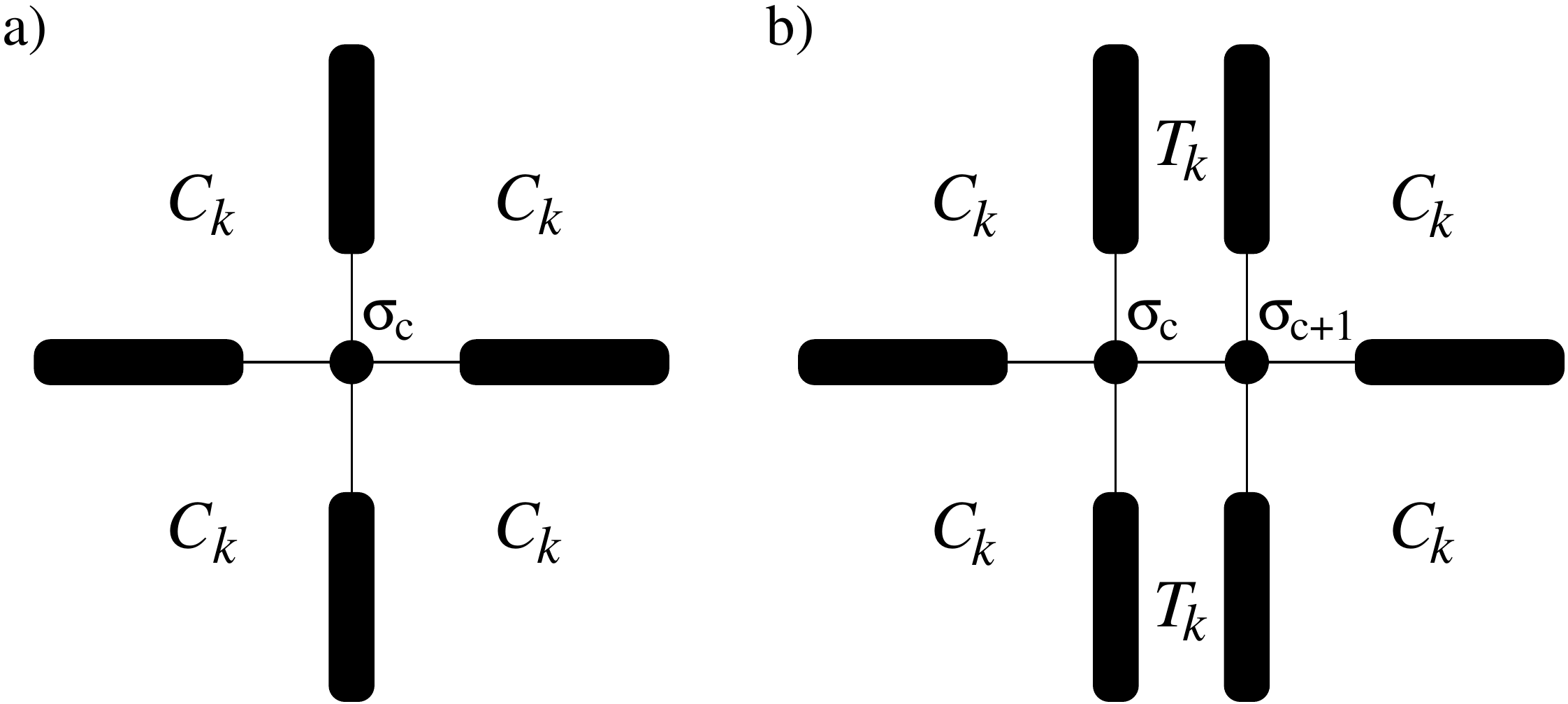}
   \caption{Graphical representation of the calculation of the spontaneous magnetization
$\left< \sigma_{c}\right>$ in the center of the square $(2k+1)\times(2k+1)$ lattice (a)
and the nearest-neighbor correlation function $\left< \sigma_{c}\sigma_{c+1}\right>$
in the two central spin of the rectangular $(2k+2)\times(2k+1)$ lattice (b).}
   \label{U_calculation_fig}
 \end{figure}

If the simplified notation we have employed is considered, the spontaneous magnetization
$M$ measured in the central spin $\sigma_c$ of the lattice can be expressed by the reduced
density matrix in Eq.~\eqref{RDM},
\begin{equation}
M = \left< \sigma_{c} \right>=
\frac{\text{Tr} \left[ \sigma_c ({\cal C}_{k})^4 \right] }
{\text{Tr} \left[ ({\cal C}_{k})^4 \right] }
= \sum_{\Sigma \sigma_c}  \sigma_c\ \rho_k (\Sigma\,\sigma_c|\Sigma\,\sigma_c) \, ,
\end{equation}
where $\sigma_c$ denotes the central spin (shown in Fig.~\ref{U_calculation_fig}a).
If the internal energy is evaluated by means of the nearest-neighbor correlation function,
$U=-2J\langle \sigma_c\sigma_{c+1}\rangle \equiv -T^2\partial_T (f/T)$ (valid
for the Ising model on the square lattice), one can slightly rearrange the geometry
of the lattice by modifying the CTMRG algorithm (cf. Fig.~\ref{U_calculation_fig}b).
The correlation relations is then given by
\begin{equation} \label{U_eqn}
\left< \sigma_{c} \sigma_{c+1} \right> = 
\frac{\text{Tr} \left[ \sigma_c \sigma_{c+1} ({\cal C}_{k})^4 ({\cal T}_{k})^2 \right] }
{\text{Tr} \left[ ({\cal C}_{k})^4 ({\cal T}_{k})^2 \right] } \, . 
\end{equation}
Note, that as the lattice system becomes infinite, the inclusion of the additional transfer
tensors ${\cal T}_k$ does not change the thermodynamic properties. It is so because there
is no difference in the bulk properties of a spin system considered on the square lattice
$N\times N$ and the rectangular one $(N+1) \times N$ when $N\to\infty$.

\newpage
\newpage\setcounter{equation}{0} \setcounter{figure}{0} \setcounter{table}{0}

\section{Higher-Order Tensor Renormalization Group}\label{HOTRG_Section}

\subsection{Tensor networks} \label{TN_representation}

\begin{figure}[b!]
 \centering
 \includegraphics[width=2.0in]{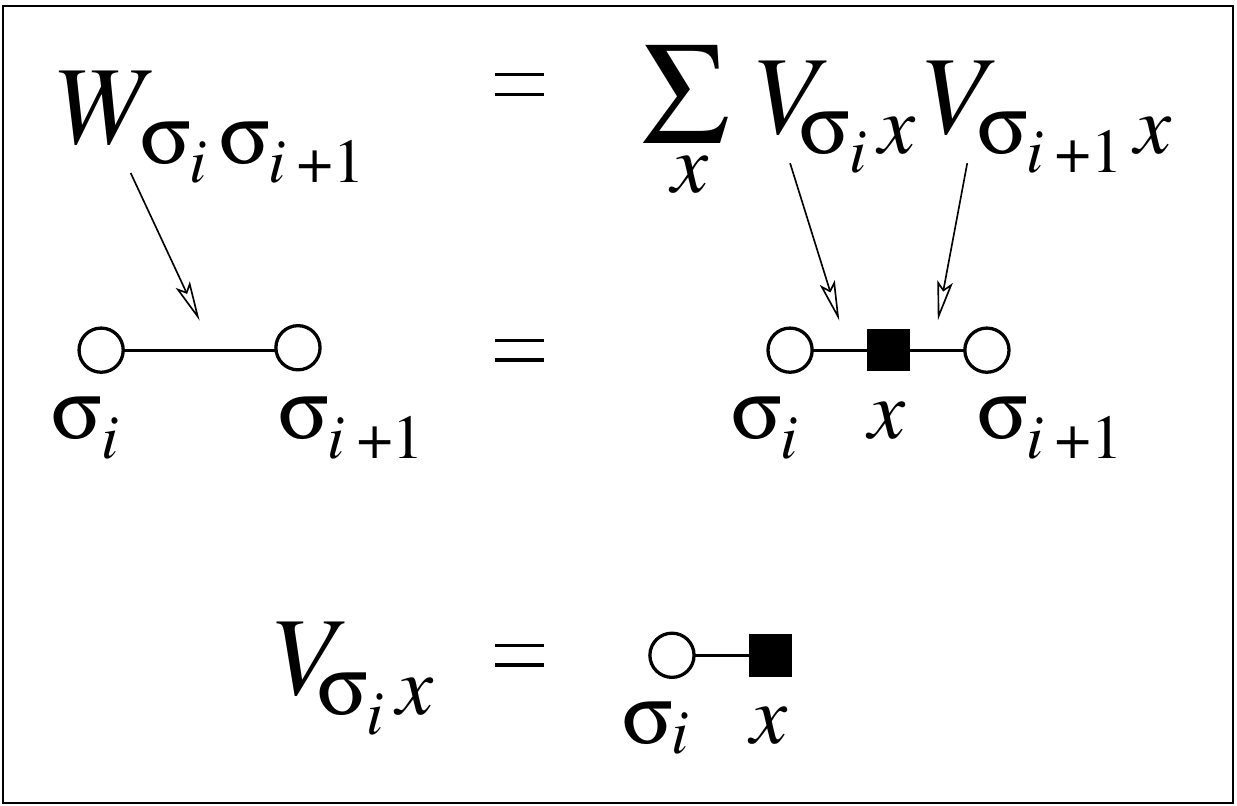}
   \caption{Graphical representation of the decomposition in Eq.~\eqref{eKWW_fact}.}\label{WSVV}
 \end{figure}

The tensor-network representation can be employed to both quantum and classical statistical systems with local (i.e. short-range) interactions. As we have mentioned earlier, such Hamiltonians can be written by means of local Hamiltonians,
\begin{equation}
{\cal H} = \sum_{\left<i, j\right>}^{~} {\cal H}_{\text{loc}}^{[i,j]}\, ,
\end{equation}
where we consider nearest-neighbor interactions only for tutorial purposes. Therefore, let the local Hamiltonian describe a classical spin lattice system with two-spin Boltzmann weight ${\cal W}_{\sigma_i^{~}\sigma_j^{~}}$ defined on the bond. The symbol $\left<i, j\right>$ is used to simplify the summation running over the nearest-neighbor spins $\sigma_i^{~}$ and $\sigma_j^{~}$ on arbitrary lattices. The partition function can be expressed as 
\begin{equation}
{\cal Z} = \sum_{\{\sigma\}} \prod_{\left<i,j\right>} {\cal W}_{\sigma_i^{~} \sigma_j^{~}}
=\sum_{\{\sigma\}} \prod_{\left<i,j\right>}\exp\left[-\frac{{\cal H}_{\text{loc}}^{[i,j]}
\left(\sigma_i^{~}, \sigma_j^{~}\right)}{k_{\rm B} T} \right] \, ,
\end{equation}
and the sum is taken over all spin configurations $\{\sigma\}$. It is convenient to regard
the Boltzmann weight ${\cal W}_{\sigma_i^{~} \sigma_j^{~}}$ as a square matrix whose rows and columns
are indexed by $\sigma_{i}^{~}$ and $\sigma_{j}^{~}$, respectively. The matrix can be decomposed into
two-matrix product,
\begin{equation}
{\cal W}_{\sigma_i^{~} \sigma_j^{~}} = \sum_{x} \, V_{\sigma_{i} x_{~}} V_{\sigma_{j} x_{~}} \, .
\label{eKWW_fact}
\end{equation}
where both of the two square matrices $V_{\sigma_i^{~} x}$ are identical, as shown schematically in Fig.~\ref{WSVV}. The bond between two physical variables $\sigma_{i}^{~}$ and $\sigma_{j}^{~}$ is broken and a non-physical (auxiliary) variable $x$ is added in between.

\begin{figure}[tb]
 \centering
 \includegraphics[width=3.5in]{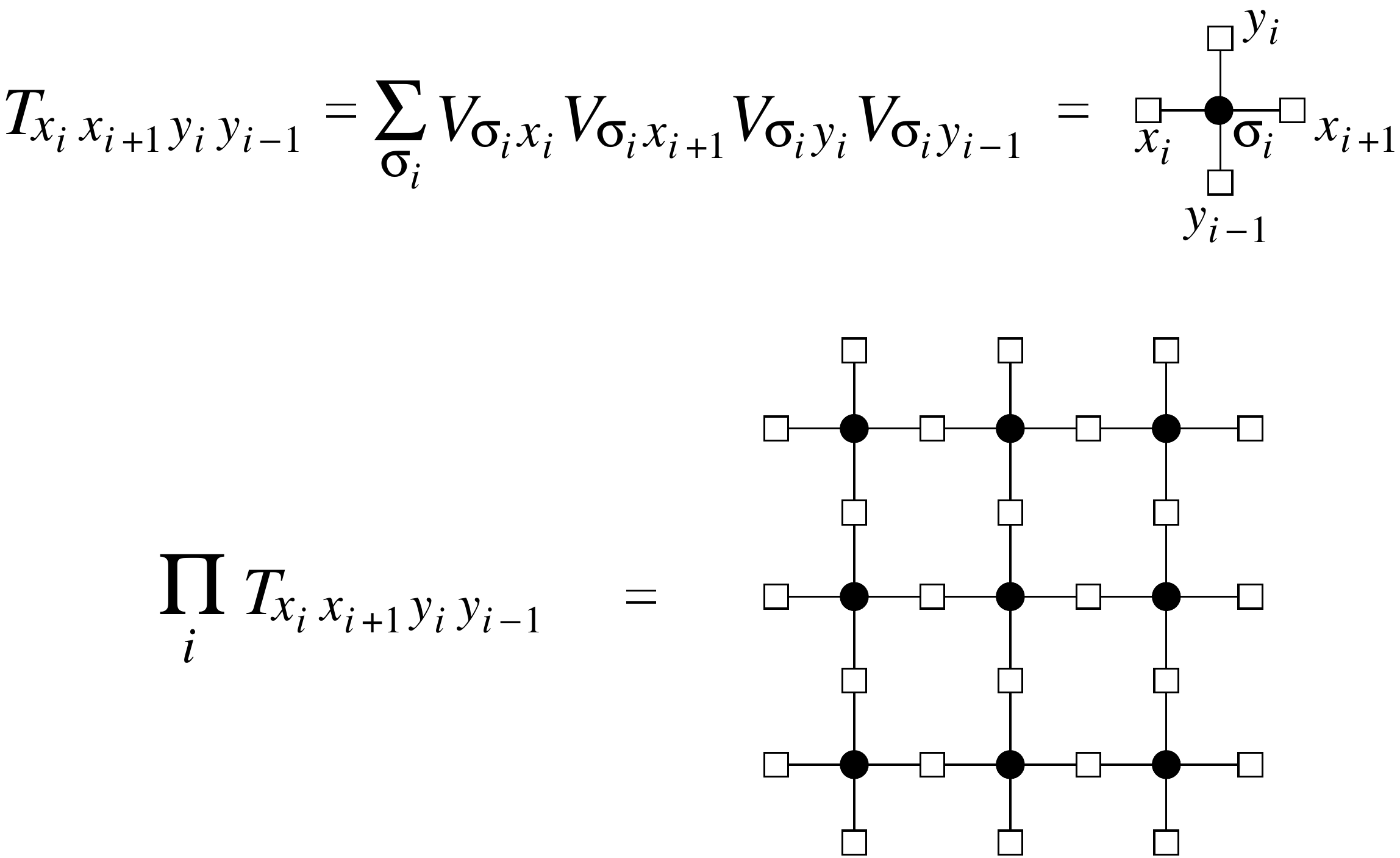}
   \caption{Graphical representation of the tensor decomposition in Eq.~\eqref{TPSe} and the particular square lattice geometry created by the product of the tensors in Eq.~\eqref{ZTPSe}.}
\label{figTPSe}
 \end{figure}

For illustration and tutorial purposes, we consider a classical spin model on the square lattice, where each spin $\sigma_i^{~}$ interacts with four nearest spins, and the partition function of the spin model can be represented as a tensor-network state,
\begin{equation}
{\cal Z} = \sum\limits_{\{x\},\{y\}} \, \prod_{i} \, T_{x_i^{~} x_{i+1}^{~} y_i^{~} y_{i-1}^{~}}
\equiv \text{Tr} \, \prod_{i} \, T_{x_i^{~} x_i^{\prime} y_i^{~} y_i^{\prime}} \, .
\label{ZTPSe}
\end{equation}
Therefore, instead of the typical calculation of the partition function by the single-bond Boltzmann weights ${\cal W}$, let us represent the partition function in terms of local tensors $T$ of the fourth order (due to the square lattice geometry), so that each tensor $T$ is positioned on the lattice spin sites $\sigma_i^{~}$, see Fig.~\ref{figTPSe}. The four identical matrices $V_{\sigma_i^{~} x}$, which we have obtained by the decomposition in Eq.~\eqref{eKWW_fact}, are employed to form the tensor $T$ with four auxiliary variables $x_i^{~}$, $x_i^\prime$, $y_i^{~}$, and $y_i^\prime$ so that
\begin{equation}
T_{x_i^{~} x_i^{\prime} y_i^{~} y_i^{\prime}} = \sum_{\sigma_i^{~}} V_{\sigma_i^{~} x_i^{~}} V_{\sigma_i^{~} x_i^{\prime}} V_{\sigma_i^{~} y_i^{~}} V_{\sigma_i^{~} y_i^{\prime}} \, .
\label{TPSe}
\end{equation}
The tensor network formulated by the tensors $T$ creates a vertex representation of a given model, cf. Fig.~\ref{figTPSe}. The partition functions calculated by both the vertex representation and the original by the Boltzmann weights are identical. The structure and dimensionality of the lattice, on which the model is defined, are specified by the order of the tensor $T$ (or equivalently, number of its components/indices) including the way the tensors are mutually connected.

In the following, the tensor-network representations for the Ising and Potts models on the square lattice are introduced. For both the models, symmetric and asymmetric bond factorizations are explained. In this context, the symmetric factorization leads to the local tensors $T$ that are invariant under arbitrary permutations of the indices. 
The asymmetric factorization is often employed in those tensor network formulations, which do not require any typical symmetry for the local weights, as long as the numerical treatment is concerned. In case of the asymmetric factorization, one has to be careful about the ordering of the tensor indices~\footnote{The symmetry in the local tensors is not always preserved when one performs the renormalization group transformation in the Higher-Order Tensor Renormalization Group method. Thus, in majority of numerical calculations, the symmetry is suppressed.}.

\subsubsection{Ising model}

Let us consider a simple classical Ising model on the square lattice, of which Hamiltonian written in terms of the local Hamiltonians ${\cal H}_{\rm loc}$ is
\begin{equation}
{\cal H} = \sum_{\left<i j\right>}^{~} \left[ - J \sigma_i^{~} \sigma_j^{~} - \frac{h}{4} \left( \sigma_i^{~} + \sigma_j^{~} \right)\right]  \, ,
\end{equation}
where the spin variable $\sigma$ takes either $+1$ or $-1$, the interaction term $J > 0$ represents the ferromagnetic ordering, and $h$ is a constant external magnetic field imposed to each spin. If no magnetic field is present, the local Boltzmann weight between two adjacent spins is
\begin{equation}\label{WB_Ising}
{\cal W}_{\sigma_i^{~} \sigma_j^{~}} = \exp \left( \frac{J\sigma_i^{~} \sigma_j^{~}}{k_{\rm B} T} \right) = e^{K \sigma_i^{~} \sigma_j^{~}} \, ,
\end{equation}
where the parameter $K = J/k_{\rm B} T$. 

\paragraph{Symmetric factorization:}

One can factorize the bond weight $e^{K \sigma_i^{~} \sigma_j^{~}}_{~}$ into two parts, 
by introducing an auxiliary variable $x = \pm 1$, which is often called `ancilla', and 
which is located between $\sigma_i^{~}$ and $\sigma_j^{~}$~\cite{Fisher_M}, as also depicted in Fig.~\ref{WSVV}. The key relation is
\begin{equation}
e^{K \sigma_i^{~} \sigma_j^{~}}_{~} =
\frac{1}{2 \sqrt{\cosh 2 {\overline K} } } \, 
\sum_x e^{{\overline K} x ( \sigma_i^{~} + \sigma_j^{~} )} \, ,
\label{expKsss}
\end{equation}
where the right-hand side is either
  $\sqrt{\cosh 2 {\overline K}}$ if $\sigma_i^{~} =  \sigma_j^{~}$ or
$1/\sqrt{\cosh 2 {\overline K}}$ if $\sigma_i^{~} = -\sigma_j^{~}$. At the same time,
Eq.~\eqref{expKsss} holds under the condition
\begin{equation}
e^{K} = \sqrt{ \cosh 2 {\overline K} } \, .
\end{equation}
The new parameter ${\overline K}$ is then expressed as follows
\begin{equation}
e^{\overline K}_{~} = \sqrt{ e^{2K}_{~} + \sqrt{ e^{4K}_{~} - 1 } } \, .
\end{equation}
As we have introduced the decomposition of the Boltzmann weight in Eq.~\eqref{eKWW_fact}, the $2\times2$ matrix $V$ has the explicit form~\footnote{If the external magnetic field $h$ is nonzero, the explicit form of the $V$ matrix is 
\begin{equation*}
V = \frac{1}{ \sqrt{ 2 \sqrt{ \cosh 2 {\overline K} } } }  \left(\begin{array}{rr} 
e^{\Gamma} e^{{\overline K}} , & e^{\Gamma} e^{-{\overline K}}  \\
e^{-\Gamma} e^{-{\overline K}}  ,& -e^{-\Gamma} e^{{\overline K}} \end{array} \right) \, ,
\end{equation*}
where $\Gamma = h/4k_{\rm B}T$ and we have used the matrix notation for the weight $V_{\sigma x}$.}
\begin{equation} \label{W_sym}
V_{\sigma x}^{~} = \frac{e^{\overline K \sigma x}_{~}}{ \sqrt{ 2 \sqrt{ \cosh 2 {\overline K} }}}
\end{equation}
for each division of a bond, and rewrite the Ising interaction to be of the form of Eq.~\eqref{eKWW_fact}.

\paragraph{Asymmetric factorization:}

The asymmetric decomposition of the Boltzmann weight for $h=0$ results in the asymmetric
matrix~\footnote{If the external magnetic field $h \neq 0$, the matrix becomes 
\begin{equation*}
V =  \left(\begin{array}{rr} 
e^{\Gamma} \sqrt{\cosh K}, & \phantom{-}e^{\Gamma} \sqrt{\sinh K}  \\
e^{-\Gamma} \sqrt{\cosh K} ,& -e^{-\Gamma} \sqrt{\sinh K} \end{array} \right) \, .
\end{equation*}
}
\begin{equation}
V =  \left(\begin{array}{rr} 
\sqrt{\cosh K}, & \phantom{-}\sqrt{\sinh K}   \\
\sqrt{\cosh K} ,& -\sqrt{\sinh K} \end{array} \right) \, .
\end{equation}

\subsubsection{Potts model}

Another spin model we use is a $q$-state Potts model with the Hamiltonian
\begin{equation}
{\cal H} = 
- \sum_{\left<i j\right>}^{~} \left\{  J \delta_{\sigma_i^{~}\sigma_j^{~}} + 
\frac{h}{4} \left[\delta_{\sigma_i^{~}\omega^{~}}+ \delta_{\sigma_j^{~}\omega^{~}}\right] \right\} \, ,
\end{equation}
where $\sigma$ takes values $\sigma = 0, 1, 2, \dots, q-1$ for a given integer $q$ and $\omega$ is fixed to a particular state $\sigma$, usually $\omega=0$ is set. 
Notice also that $\delta$ is the Kronecker delta
\begin{equation}
  \delta_{\sigma_i\sigma_j} =\begin{cases}
    0 & \text{if $\sigma_i \neq \sigma_j$} \, , \\
    1 & \text{if $\sigma_i = \sigma_j$} \, .
  \end{cases}
\end{equation}
Again, we start with the case without the external field, i.e. $h=0$. The Boltzmann weight reads
\begin{equation}\label{W_Potts}
  e^{K\delta_{\sigma_i^{~}\sigma_j^{~}}} =\begin{cases}
    1 & \text{if $\sigma_i \neq \sigma_j$} \, ,\\
    e^{K} & \text{if $\sigma_i = \sigma_j$} \, .
  \end{cases}
\end{equation}

\subparagraph{Symmetric factorization:}

By introducing an auxiliary variable $x = 0, 1, 2, \dots, q-1$, we obtain the key relation
\begin{equation} \label{Potts_primed}
  \sum\limits_{x=0}^{q-1} \exp \left[ {{\overline K} \left\{ \delta_{\sigma_i^{~}x_{~}^{~}} + \delta_{\sigma_i^{~}x_{~}^{~}} \right\}} \right] =\begin{cases}
    q - 2 + 2e^{{\overline K}} & \text{if $\sigma_i \neq \sigma_j$} \, ,\\
    q - 1 + e^{2{\overline K}} & \text{if $\sigma_i = \sigma_j$} \, .
  \end{cases}
\end{equation}
In order to make the expression for the Boltzmann weight in Eq.~\eqref{W_Potts} consistent with Eq.~\eqref{Potts_primed}, the following condition has to be satisfied 
\begin{equation}
e^{-K} = \frac{q-2+2e^{{\overline K}}}{q-1+e^{{\overline K}}} \, .
\end{equation}
By inverting the last equation, one obtains the relation for ${\overline K}$
\begin{equation}
e^{{\overline K}} = e^K + \sqrt{\left(e^K + q - 1\right)\left(e^K-1\right)} \, .
\end{equation}
Thus, the Boltzmann weight can be decomposed in terms of the $q\times q$ matrices~\footnote{
For a nonzero external magnetic field $h \neq 0$, one finds out 
\begin{equation*}
V_{\sigma x} = \frac{e^{{\overline K} \delta_{\sigma x }} 
e^{\Gamma \delta_{\sigma \omega }}}{\sqrt{ q - 2 + 2 e^{\overline K} }} \, .
\end{equation*}
Notice that $\Gamma$ here is not rescaled, whereas the rescaled ${\overline K}$ is used.
}
\begin{equation}
V_{\sigma x} = \frac{e^{{\overline K} \delta_{ \sigma x }}}{ \sqrt{ q - 2 + 2 e^{\overline K} } } \, .
\end{equation}

\subparagraph{Asymmetric (numerical) factorization}

In general, the asymmetric factorization can be always carried out numerically by the simple matrix diagonalization of the Boltzmann weight (being the $q\times q$ real symmetric matrix). Therefore,

\begin{equation}
{\cal W} = P D P_{~}^{\intercal} = \underbrace{\left( P\sqrt{D} \right) }_{V} \underbrace{ \left( P \sqrt{D} \right)^{\intercal}  }_{V_{~}^{\rm T}} \, , 
\end{equation}
in particular, 
\begin{equation}
{\cal W}_{\sigma_i \sigma_j} = \sum_{xy} P_{\sigma_i x}^{~} D_{xy}^{~} P_{y\,\sigma_j}^{\intercal} = 
\sum_x \underbrace{\left( \sum_m P_{\sigma_i\, m}^{~} \sqrt{D_{mx}} \right)}_{V_{\sigma_i \, x}^{~}}
\underbrace{\left( \sum_n \sqrt{D_{xn}} P_{n\, \sigma_j} \right) ^{\intercal}}_{V_{x\, \sigma_j}^{\intercal}} \, . 
\end{equation}

\subsection{Coarse graining}

One of the simple powerful iteration ways of how to evaluate the partition function is a coarse-graining renormalization procedure~\cite{HOTRG}. 
The lattice iteratively contracts either along the horizontal ($x$ axis) or vertical ($y$ axis) directions if a two-dimensional lattice is considered. The two directions alternate while iterating. At each iteration step $k$, a new tensor $T^{(k+1)}$ is created out of two joined tensors $T^{(k)}$ calculated in the previous step. It means the two tensors $T^{(k)}$ are contracted and then renormalized first, say, along the $x$ axis and, subsequently, the resulted tensor is again contracted and renormalized along the other $y$ axis (or vice versa). 
The lattice size effectively expands by a factor of two (the number of rows or columns doubles alternately) within each contraction of the renormalization procedure. And the same time, the size of two joined tensors shrinks into a single tensor after the renormalization step is performed. Since the coarse-graining procedure has an iteration character, the procedure is terminated if demanded observable(s) converged, i.e., reached its fixed point. 
In the following, a step-by-step description of the coarse-graining procedure is explained in detail. 

 \begin{figure}[tb]
 \centering
 \includegraphics[width=3.6in]{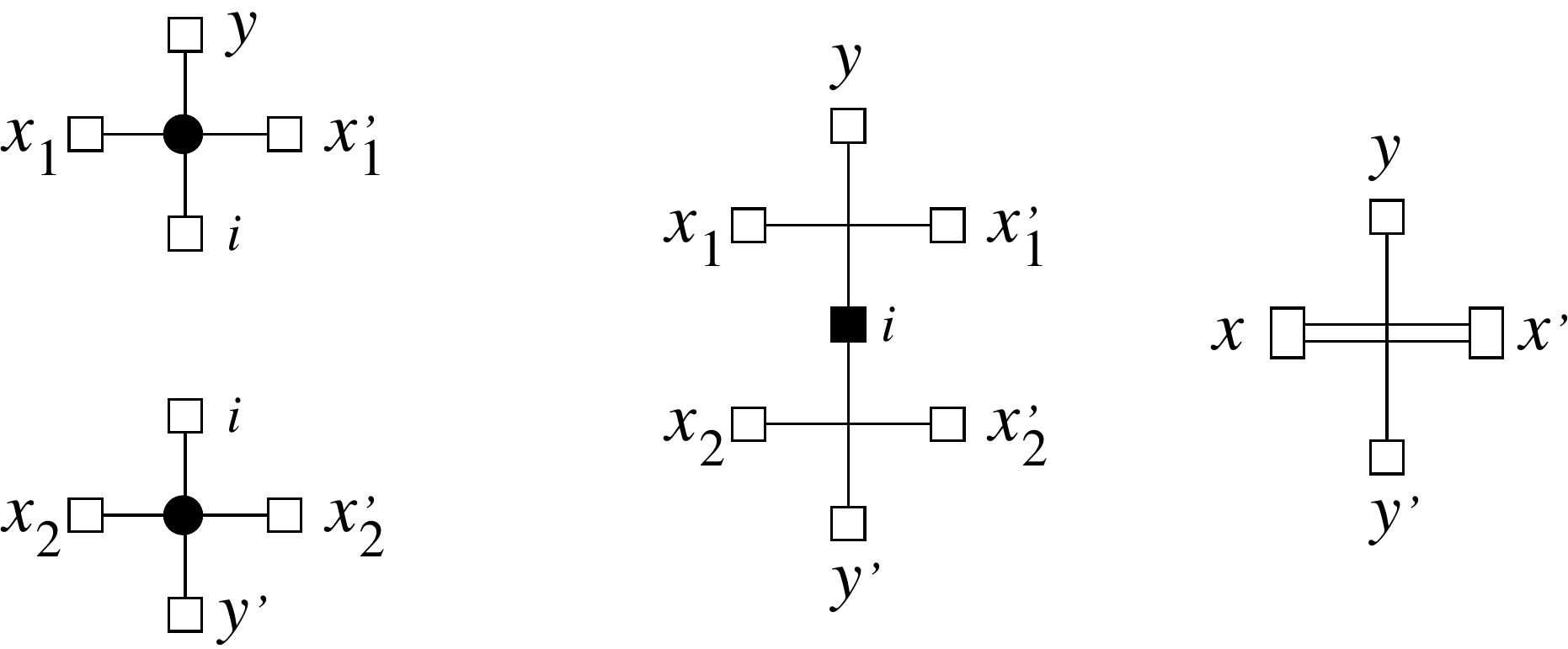}
   \caption{Graphical representation of Eq.~\eqref{eqMTT} demonstrating the detailed description of how the two tensors $T^{(k)}_{x_1^{~} x_1^{\prime} y_{~}^{~} i_{~}^{~}}$ and $T^{(k)}_{x_2^{~} x^{\prime}_2 i_{~}^{~} y_{~}^{\prime}}$ (on the left) are joined and contracted by summing up over their common tensor variable $i$ (in the middle) so as to create the tensor $M^{(k)}_{x x^{\prime} y y^{\prime}}$ (on the right).}
   \label{MTT}
 \end{figure}   

By the term {\it contraction} of two tensors at the $k^{\rm th}$ iteration step along, say, the $y$ axis, we understand the partial trace of the tensor product
\begin{equation}
M^{(k)}_{x x^{\prime} y y^{\prime}} = \sum_i T^{(k)}_{x_1^{~} x_1^{\prime} y_{~}^{~} i_{~}^{~}}  T^{(k)}_{x_2^{~} x^{\prime}_2 i_{~}^{~} y_{~}^{\prime}} \, ,
\label{eqMTT}
\end{equation}
as we have graphically depicted in Fig.~\ref{MTT}. The two grouped indices satisfy the expressions $x = x_1 \otimes x_2$ and $x' =x'_1 \otimes x'_2$. The superscript $k$ corresponds to the $k^{\rm th}$ iteration step. The tensor $M^{(k)}$ is truncated by means of the algorithm Higher-Order Singular Value Decomposition (HOSVD)~\cite{HOSVD}. The HOSVD algorithm requires to employ two matrix unfoldings, which we here distinguish by adding the subscript symbols $(1)$ and $(2)$
\begin{equation}
M _{\left(1\right); \{x\} \{x^{\prime} y y^{\prime}\}} = M_{x x^{\prime} y y^{\prime}} \, , 
\end{equation}
and
\begin{equation}
M_{\left(2\right); \{x^{\prime}\} \{y y^{\prime} x\}} = M_{x x^{\prime} y y^{\prime}} \, .
\end{equation}
The matrix unfolding is a mathematical prescription which enables to re-order matrix indices (by applying a cyclic permutation). In our particular case, we change a tensor of the fourth order $M$ into a rectangular matrix $M$ by reassembling the indices, where the two matrix indices are grouped into the curly parentheses $\{.\}$, as we have used in Eq.~\eqref{ThSVD}.
Then, having applied SVD to the both rectangular matrices, we obtain

\begin{equation}
M_{\left(1\right)}^{~} = U_{\left(1\right)}^{~} \Sigma_{\left(1\right)}^{~} V_{\left(1\right)}^{\dag} \, ,
\end{equation}
\begin{equation}
M_{\left(2\right)}^{~} = U_{\left(2\right)}^{~} \Sigma_{\left(2\right)}^{~} V_{\left(2\right)}^{\dag} \, ,
\end{equation}
where $U_{\left(1\right)}$, $V_{\left(1\right)}$, $U_{\left(2\right)}$, and $V_{\left(2\right)}$ are unitary matrices of the respective dimensions. The diagonal matrices $\Sigma_{\left(1\right)}$ and $\Sigma_{\left(2\right)}$, both of them denoted $\Sigma_{\left(\text{.}\right)}$, contain singular values
\begin{equation}
\Sigma_{\left(\text{.}\right)} = \text{diag}\,\left[\sigma_{(.); 1}, \sigma_{(.); 2}, \sigma_{(.); 3}, \dots\right] \, .
\end{equation}
The non-negative singular values are supposed to be ordered decreasingly whenever SVD has been applied. The optimal approximation of the two tensors $M^{(k)}$ is decided according to the errors
\begin{equation}
\varepsilon_{(1)} = \sum_{i>D} \sigma_{(1); i}^2 \, ,
\end{equation}
and
\begin{equation}
\varepsilon_{(2)} = \sum_{i>D} \sigma_{(2); i}^2 \, ,
\end{equation}
where $D$ represents the dimension threshold of the truncated tensor dimension. If $\varepsilon_{(1)} < \varepsilon_{(2)}$, the grouped matrix index in $U_{\left(1\right)}$, i.e. $\{x^{\prime} y y^{\prime}\}$, has to be truncated down to $D^2$ degrees of the freedom. Otherwise (for $\varepsilon_{(1)} \geq \varepsilon_{(2)}$), the grouped matrix index $\{y y^{\prime} x\}$ of $U_{\left(2\right)}$ is truncated down to $D^2$ degrees of the freedom.

After the unitary matrix $U$ is truncated, it forms a rectangular matrix of the size $D \times D^2$, and an updated tensor in the consequent iteration step $k+1$ is created
\begin{equation}
T^{(k+1)}_{x x' y y'} = \sum_{ij} U^{(k)}_{i x} M^{(k)}_{i j y y'} U^{(k)}_{j x'} \, .
\label{Tnp1}
\end{equation}
For simplicity, we have dropped the subscript index $(.)$ of the unitary projection matrix $U_{(.)}$ in Eq.~\eqref{Tnp1} and we apply such $U$, which correspond to the lower error $\varepsilon_{(.)}$.

The contraction and the renormalization processes along the $x$ axis have to be carried out accordingly. By the contraction of the two tensors $T^{(k)}$ along the $x$ axis, we mean

\begin{equation}
M^{(k)}_{x x' y y'} = \sum_i T^{(k)}_{x i y_1 y'_1}  T^{(k)}_{i x' y_2 y'_2} \, ,
\end{equation}
where $y = y_1 \otimes y_2$ and $y' = y'_1 \otimes y'_2$. Again, applying the matrix unfoldings
\begin{equation}
M_{\left(3\right); y , y' x x' } = M_{x x' y y'} \, ,
\end{equation}
\begin{equation}
M_{\left(4\right); y' , x x' y } = M_{x x' y y'} \, ,
\end{equation}
we reshape the tensors into rectangular matrices to enter SVD. Having evaluated and compared the associated errors $\varepsilon_{(.)}$, distinct unitary projection matrices $U_{(.)}$ are produced. Then, the expression for the contracted tensor for the next iteration step is
\begin{equation}
T^{(k+1)}_{x x' y y'} = \sum_{kl} U^{(k)}_{k y} M^{(k)}_{x x' k l} U^{(k)}_{l y'} \, .
\label{Tnp2}
\end{equation}

A remark for the truncation error $\varepsilon_{(.)}$. In the case of the simple Ising model, the truncation errors are equal~\cite{CDL2}. However, when dealing with models without translational symmetry, for instance, spin-glass models ~\cite{Wang}, the truncation errors differ.

\subsection{Free energy}

In numerical calculations, the tensors are normalized after each coarse-graining step is performed, otherwise the tensor elements may become infeasible large after a couple of steps. The normalization procedure can be done in different ways. For instance, we used the absolute value of the largest element in each tensor as the tensor norm $\lambda$. When the updated renormalized tensors in Eqs.~\eqref{Tnp1} and \eqref{Tnp2} are formed, they are consequently normalized by dividing each tensor element by the tensor norm $\lambda$. Therefore, after a coarse-graining procedure is completed at an iteration step $k$, the corresponding norm
\begin{equation}
\lambda_{k}=\max\limits_{x x' y y'} \left\vert T^{(k)}_{x x' y y'} \right\vert
\end{equation}
is evaluated. Since the HOTRG algorithm is initialized with the tensor $T^{(k=0)}$, the calculation of the norm $\lambda_0$ is not inevitable (and we do not consider it in the following). The free-energy calculation requires to keep track of all the normalization coefficients $\left(\lambda_1, \lambda_2, ..., \lambda_k\right)$, including the final renormalized tensor $T^{(k)}$. Thus, at the $k^{\rm th}$ iteration step, the total number of the sites on the square lattice is exactly $2^k$, so the partition function can be expressed as
\begin{eqnarray}
\nonumber
{\cal Z}_k &=&  \text{Tr} \underbrace{ \left[ T^{(0)} T^{(0)} T^{(0)} \cdots T^{(0)} \right] }_{2^k} \\ \nonumber
            &=&  \lambda_1^{2^{k-1}} \,\text{Tr} \underbrace{ \left[ T^{(1)} T^{(1)} T^{(1)} \cdots T^{(1)} \right] }_{2^{k-1}} \\ \nonumber
            &\vdots&   \\ \nonumber
            &=&  \lambda_1^{2^{k-1}} \lambda_2^{2^{k-2}} \cdots \lambda_k^{2^1} \,\text{Tr} \left[T^{(k)}\right] \\
            &=& \sum_{xy} T^{(k)}_{x x y y} \, \prod\limits_{i=1}^{k} \lambda_i^{2^{k-i}} \, ,
\end{eqnarray}
The trace is equivalent to imposing the periodic boundary conditions, which changes the square lattice shaped into a torus.

\begin{figure}[tb]
 \centering
 \includegraphics[width=3.8in]{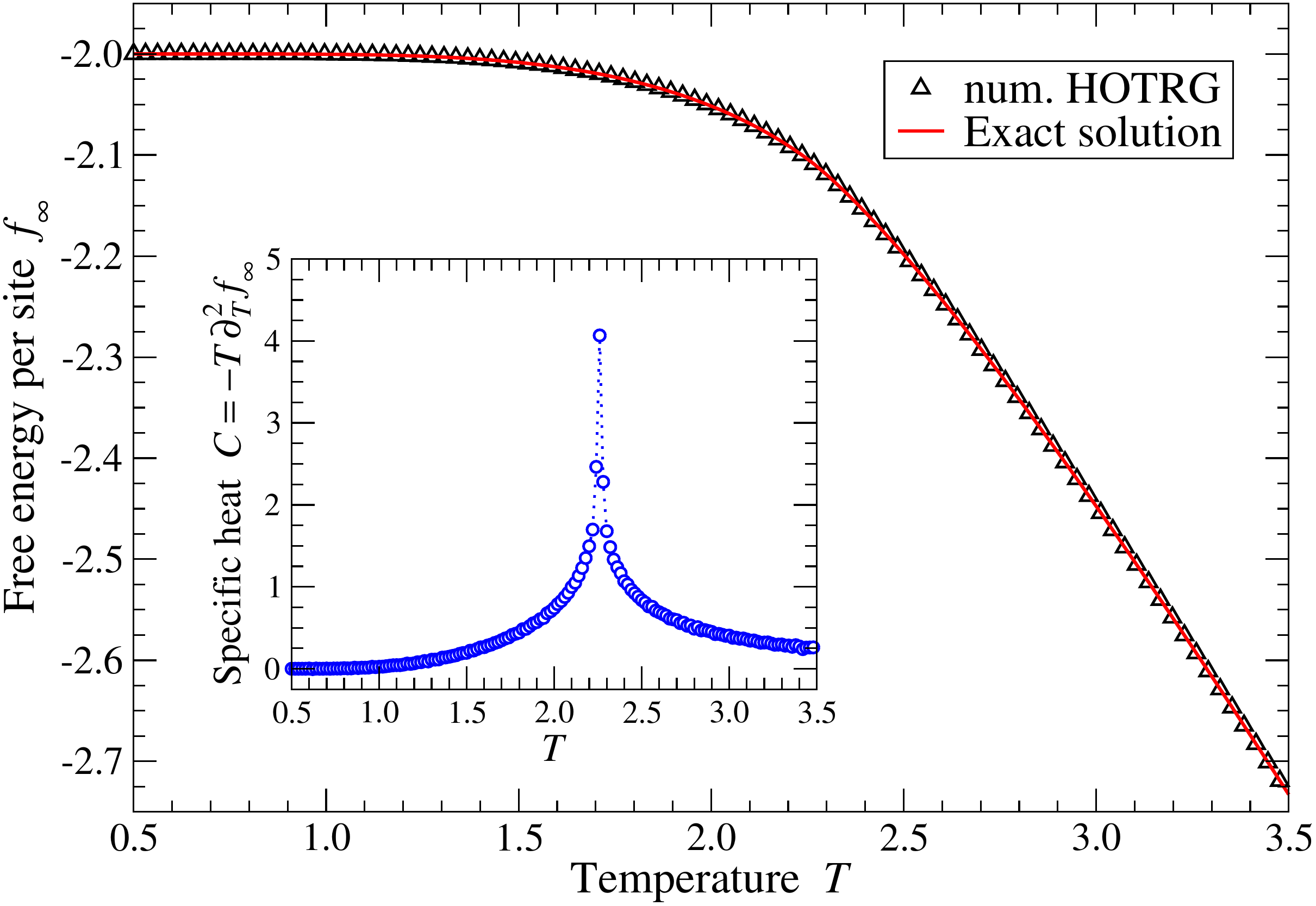}
   \caption{Temperature dependence of the free energy per site calculated by HOTRG in the thermodynamic limit ($k\to\infty$) for $D = 16$. The inset shows the specific heat, which exhibits a sharp peak at the phase transition temperature $T_{\rm c} = 2\,/\ln\,(1+\sqrt{2}) = 2.269185...$}\label{hotrg_fe}
 \end{figure}

\begin{figure}[htb!]
 \centering
 \includegraphics[width=3.8in]{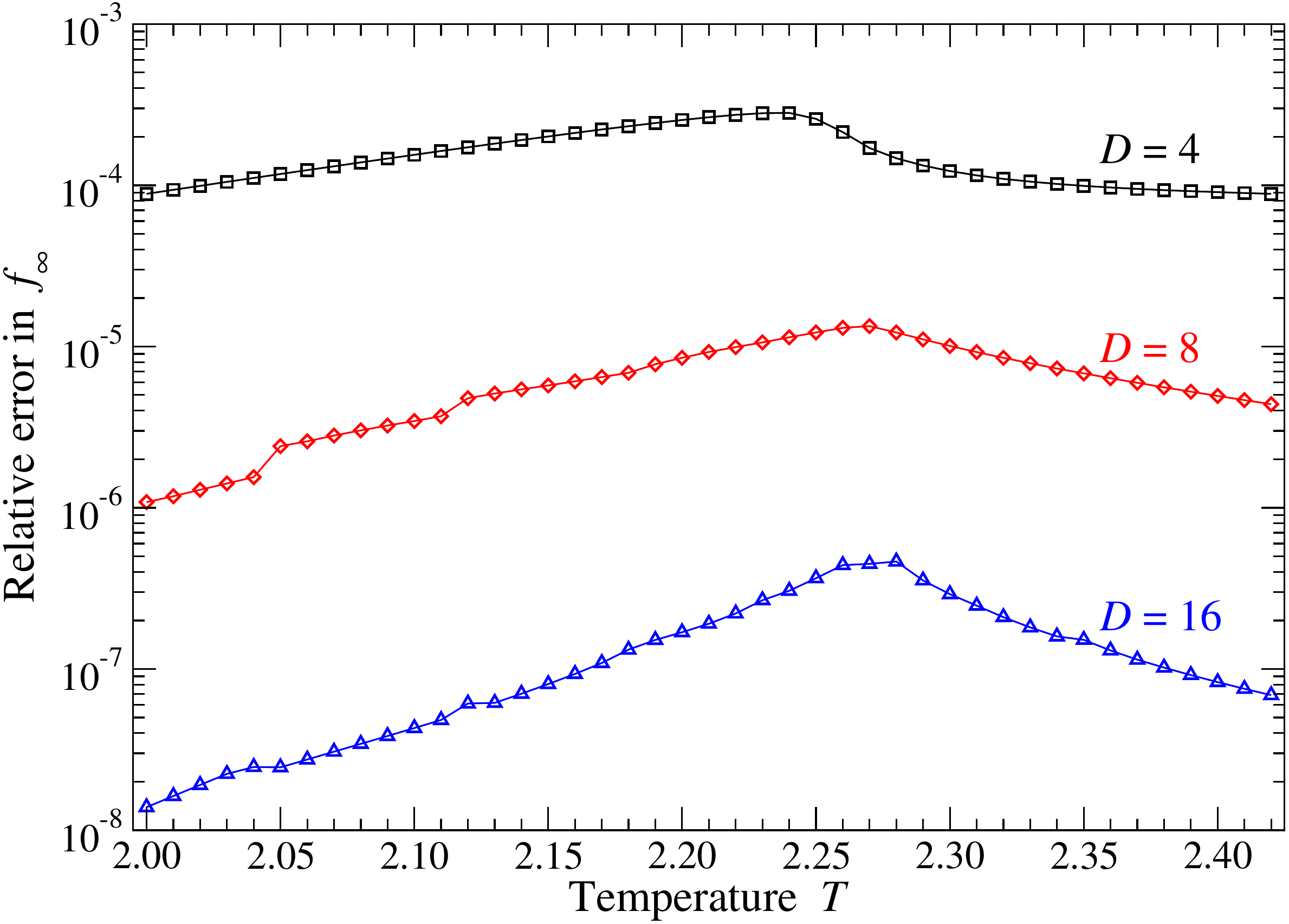}
   \caption{Comparison of the relative errors of the free energy for $D = \left\{4, 8, 16\right\}$. The critical temperature of the square-lattice Ising model corresponds to $T_{\rm c}$.}\label{hotrg_compar}
 \end{figure}

If recalling the expression for the free energy (per lattice site),
\begin{equation}
f_k = - \frac{k_{\rm B} T}{2^k} \ln {\cal Z}_k \, ,
\end{equation}
one immediately finds out that
\begin{equation} \label{f_n_hotrg}
f_k = - k_{\rm B} T \left\{ 2^{-k}\ln \text{Tr}\, \left[T^{(k)}\right] + \sum_{i=1}^{k} \frac{\ln{\lambda_i}}{2^{i}} \right\} \, . 
\end{equation}
Since the trace of the normalized tensor $T^{(k)}$ rapidly converges to a finite number, the first term becomes zero after a few tens of the iteration steps $k$ (usually $k \gtrsim 40$). The free energy calculation is numerically determined by the norms $\lambda_i$ only. We refer the reader to compare the two different expressions for the free energy given in Eqs.~\eqref{fe} and \eqref{f_n_hotrg}. The free energy calculated for the Ising model on the square lattice by HOTRG is depicted in Fig.~\ref{hotrg_fe}. The relative error with respect to the exact solution is shown in Fig.~\ref{hotrg_compar}

\subsection{Impurity tensors}

\paragraph{Magnetization:}
The spontaneous magnetization is a typical order parameter, which is often evaluated in order to determine phase transition. It is nonzero in one phase and zero in the other. For instance, the simple Ising model on the square lattice exhibits an ordered ferromagnetic ($J>0$) phase, for which the spontaneous magnetization normalized per one spin site is a decreasing function of temperature so that $1\leq M < 0$ right below the critical phase transition temperature $T_{\rm c}$, whereas $M=0$ in the disordered paramagnetic phase at $T \geq T_{\rm c}$. The magnetization can be also calculated by HOTRG, although it requires to prepare a special $\sigma$-dependent local tensor, the impurity tensor $\tilde{T}$, which has to be inserted into an appropriate place labeled by integer $i$ on the lattice system. In particular,
\begin{equation} \label{Ising_Imp_init}
\tilde{T}_{x_i^{~} x_i^{\prime} y_i^{~} y_i^{\prime}} = \sum_{\sigma=\pm1} \sigma W_{\sigma x_i^{~}} W_{\sigma x_i^{\prime}} W_{\sigma y_i^{~}} W_{\sigma y_i^{\prime}} \, ,
\end{equation}
for the 2-state Ising spin $\sigma=\pm1$, or,
\begin{equation} \label{Potts_Imp_init}
\tilde{T}_{x_i x_i^{\prime} y_i^{~} y_i^{\prime}} = \sum_{\sigma=0}^{q-1} \delta_{\sigma\omega} W_{\sigma x_i^{~}} W_{\sigma x_i^{\prime}} W_{\sigma y_i^{~}} W_{\sigma y_i^{\prime}} \, ,
\end{equation}
for the $q$-state Potts model. Let us recall that the Potts-model magnetization is measured with respect to a specified spin state $\omega$ (being usually set to zero).  

After the impurity tensor is initialized by either Eqs.~\eqref{Ising_Imp_init} or \eqref{Potts_Imp_init}, it undergoes a contraction with the local tensor at the same coarse-graining step ($k$) in order to form an updated impurity tensor, i.e.,
\begin{eqnarray}
\tilde{T}^{(1)} &=& \tilde{T}^{(0)} * T^{(0)} \, , \\ \nonumber
\tilde{T}^{(2)} &=& \tilde{T}^{(1)} * T^{(1)} \ , \\ \nonumber
& \vdots & \\ \nonumber
\tilde{T}^{(k)} &=& \tilde{T}^{(k-1)} * T^{(k-1)} \, .
\end{eqnarray}
Within the contraction procedures, there is no need to run the HOSVD separately; the unitaries (the projection unitary matrices $U$) taken from the process of creating the local tensors $T^{(k)}$ at each step $k$ suffice for this purpose. The impurity tensors are normalized at each step, i.e., $\tilde{\lambda}_k = \max \vert \tilde{T}^{(k)} \vert$. The impurity tensor is advised to be placed and kept at the center of the lattice system by an appropriate rotation of the local tensors in each coarse-graining iteration step. This is a key point since the spontaneous magnetization can be strongly affected by boundary effects emerging when taking the final trace of the local tensor. 

Then, the spontaneous magnetization for the Ising model is
\begin{equation}
M = \frac{\text{Tr}\left[\tilde{T}^{(k)}\right]}{\text{Tr} \left[{T}^{(k)}\right]} \, , 
\end{equation}
and the $q$-state Potts model requires the additional reformulation
\begin{equation}
M_{\rm Potts} = \frac{ q M - 1}{ q - 1 }
\end{equation}
so as to satisfy the normalization of the paramagnetic ordering, where $M=1$ at $T=0$ in the thermodynamic limit below the phase transition temperature.

\subsection{Numerical results}

 \begin{figure}[tb]
 \centering
 \includegraphics[width=3.8in]{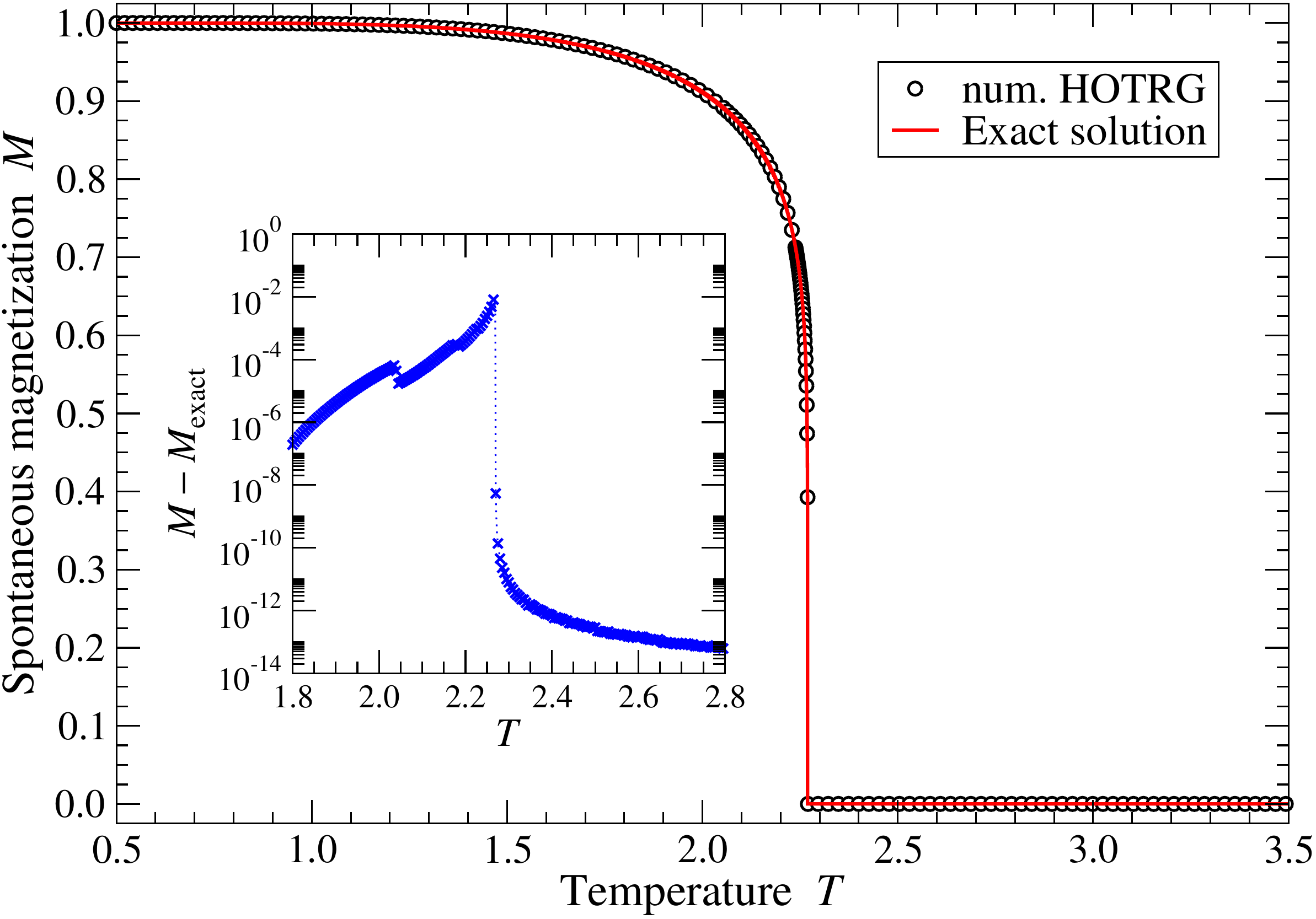}
   \caption{Temperature dependence of the spontaneous magnetization for the square-lattice Ising model ($D=16$). The inset shows the difference between the numerically obtained magnetization $M$ and the analytical result $M_{\rm exact}$.}
\label{hotrg_mag}
 \end{figure}   
If comparing the relative errors of the free energy with the exact solution by Onsager~\cite{Onsager} for the square-lattice Ising model, we can directly see how the relative error improves if $D$ increases, as plotted in Fig.~\ref{hotrg_compar}. We obtained the relative error of the free energy at the critical temperature $T_{\rm c}$ of an order of magnitude about $10^{-6}$ when $D=16$ (stressing the fact that the numerical accuracy is the lowest at the critical phase transition).

The comparison of the spontaneous magnetization obtained numerically (using the impurity tensor) for $D=16$ with the exact solution~\cite{Onsager} is depicted in Fig.~\ref{hotrg_mag}. The spontaneous magnetization right above the critical temperature $T_c$ is $M\approx10^{-9}$ (at $T=2.27$) and it gradually decreases to zero, as the temperature grows.

\newpage\setcounter{equation}{0} \setcounter{figure}{0} \setcounter{table}{0}
\section{Free energy on hyperbolic geometries} \label{chap3}

Having briefly surveyed the tensor-network algorithms used, we now focus our attention
on a particular physical task, in which the tensor network is employed to treat
multi-state spin systems on non-Euclidean (negatively curved) lattice surfaces.
In particular, we intend to relate the free energy of such systems to the radius of the
Gaussian curvature. In addition, the phase-transitions are classified for the multi-state
spin systems when geometric curvature of underlying lattices varies.

These goals are achieved by deriving a set of recurrence relations analytically. The
set is associated with a particular lattice geometry. The structure of recurrence relations
we derive has to reflect the lattice geometry. We also specify such conditions, under which
the CTMRG algorithm enables us to build up the curved lattice geometry. The free-energy is
analyzed numerically (see Introduction to the Euclidean case for the square lattice
Section~\ref{CTMRG_Section}). A particular hyperbolic lattice is constructed by a regular
tessellation of various congruent polygons while keeping the coordination number fixed
at each lattice vertex.

The two specific tasks are to be addressed in this analysis. First, we calculate the
free energy per site for each hyperbolic lattice geometry in order to target a
lattice geometry that gives the lowest free energy per site. Second, we relate
the free energy, the radius of the Gaussian curvature, and the phase-transition
temperature. These two tasks were formulated in order to investigate
the fact whether the physical properties (for instance, the free energy and phase
transition temperature) can intrinsically inherit information on the lattice geometry.
In particular, how they reflect the geometrical structure of the lattice. These tasks
are meant to be confronted when viewed from the condensed matter physics and the new
perspectives in the AdS/CFT correspondence~\cite{AdS-CFT}.

The critical phenomena and the phase transitions observed in various magnetic systems
on two-dimensional non-Euclidean surfaces have been attracting attention both theoretically
and experimentally for a couple of decades. If focusing on the studies with negatively
curved surfaces only (Lobachevski or hyperbolic geometry), the theoretical research is
focused on the open problems in quantum gravity, in which, the anti-de Sitter (AdS)
hyperbolic spatial geometry plays an essential role. The interplay among the solid-state
physics, general theory of relativity, and the conformal field theory (CFT) is of
increasing interest~\cite{AdS-CFT, AdS-CFT1}. Despite there are no directly related
experiments of this interdisciplinary work on large scales, a few experiments have been
carried out on magnetic nanostructures~\cite{exp1,exp2,exp3}, soft materials with a
conical geometry~\cite{conic}. Also, local changes of the Euclidean geometry are
typically known as lattice dislocations of solid-state crystals or complex
networks~\cite{CN1,CN2}, or theories describing the quantum gravity~\cite{QG1,QG2}.

As a simple theoretical example of the hyperbolic surface geometry we have chosen a
two-dimensional (discrete) hyperbolic lattice exhibiting constant negative
Gaussian curvature. We consider an infinite set of regular hyperbolic surfaces
constructed by the tessellation of the congruent polygons. The polygons are mutually
connected without leaving empty spaces or without any lattice irregularities (dislocations)
at the lattice sites (the vertices), and the coordination number of the lattice is
considered constant in this work. Each hyperbolic lattice of a finite size is characteristic
for an enormous number of the boundary sites with respect to the number of the inner (bulk)
sites. The number of the boundary sites is always larger than the remaining number of all
the inner sites on any hyperbolic lattice.

When classifying phase transitions of various multi-spin models on hyperbolic lattices,
the CTMRG algorithm has been used as the only accurate numerical tool, which is capable
of accurate numerical calculations in the thermodynamic limit. Therefore, neither the
standard transfer-matrix diagonalization is numerically feasible (due to a highly
non-trivial way of the transfer-matrix construction) nor the Monte Carlo simulations
(due to the insufficiency of performing the finite-size scaling at the phase transitions).

The CTMRG algorithm was originally developed to treat spin models on the two-dimensional
square lattice~\cite{ctmrg1,ctrmg2}. The first successful application of the CTMRG has
been carried out for the Ising model on hyperbolic lattice, which was made of the pentagonal
polygons with the uniform coordination number equal to four~\cite{hCTMRG54}. Consequently,
the CTMRG algorithm was extended to treat the Ising-like systems on specific types of the
hyperbolic lattices: The lattice coordination number was fixed while varying the
order of the polygons~\cite{hCTMRGp4} and vice versa, the polygon was fixed to be the
triangle while varying the coordination number~\cite{hCTMRG3q}. Our aim is
to expand the two mentioned studies to the multi-state spin Hamiltonians on an infinite
set of the hyperbolic lattices, when both the coordination number $q$ and the order of
the polygons $p$ are allowed to vary.

We derive a set of the generalized recurrence relations analytically, which enables us
to study phase transitions of the $Q$-state\footnote{In order to avoid confusion in
the notation of $q$, we make the difference between the $q$-state spin models we have
used so far and the coordination number $q$. Only in Sec.~5 we denote the multistate
spin variable by the uppercase $Q$.} spin models ($Q\geq2$) 
on arbitrary regular lattice geometries $(p,q)$, where the polygons can have
$p\geq4$ sides and the coordination number can be $q\geq4$.

The numerical evaluation of the free-energy per spin is a well-conditioned thermodynamic
quantity, which does not diverge in the thermodynamic limit. Numerical accuracy of
the free energy will be evident from observing a singular behavior of the specific heat
at a phase transition as a smooth function, even after taking the second derivative of
the free energy with respect to temperature numerically. The analysis of the spin models
in this study serves as an accurate and complementary source of
information for non-integrable spin systems on hyperbolic surfaces. Notice that the
free energy also reflects a complexity of the boundary structure. We remark that the
free-energy analysis has not been considered in any non-Euclidean lattice systems yet.

If considering single-site expectation values, such as the spontaneous magnetization
(measured at the lattice center), the boundary effects become negligibly small; this
state can be reached in the thermodynamic limit only. However, there is a specific
case for the hyperbolic lattices: It is the Bethe lattice, which is equivalent to
taking the limit $p\to\infty$, as discussed in Ref.~\cite{Mosseri}. Since the phase
transition temperature and critical exponents are known exactly (for the Ising model
on the Bethe lattice for any coordination number $q$~\cite{Baxter}), we refer to this
fact, which will serve as a benchmark of the high numerical accuracy of CTMRG.

As discussed later, enormous boundary length of the hyperbolic lattices may affect
the bulk properties so significantly that even the phase transition can be completely
suppressed if analyzed from the free-energy point of view. A way of eliminating the
boundary effects is to be proposed: we redefine the free energy appropriately so that
the the boundary contributions are removed to preserve the properties of the infinite
bulk. The phase transition(s) can be thus restored and observed numerically.

If reversing the order of our considerations, another non-trivial question may arise,
which we also wish to answer in the current study. The question is associated with
the AdS/CFT correspondence, i.e., the gauge duality~\cite{AdS-CFT, AdS-CFT1}. A
complicated boundary structures of a finite anti-de Sitter space (reproduced
by the discrete hyperbolic lattice geometries) is locally viewed as the Minkovski-like
space. Thus, it can be regarded as a spacetime for the conformal field theory, which
is related to the gravitational theory. Our work has been primarily focused
on the thermodynamic features of the complex boundary structures in the thermal
equilibrium, and we point out that no time evolution is considered in thus study.
As a tractable physical model, we have selected a regular hyperbolic network
(the AdS space) with the multistate spin interactions. We keep a sufficient
numerical accuracy, which can be determined by comparing our results with the
integrable spin models~\cite{Baxter}. Since deeper theoretical and numerical studies
are still missing, was focus on the free-energy analysis of the AdS spaces.
A condensed-matter viewpoint on the AdS/CFT correspondence encounters conceptual
difficulties, one of them being the problem of a preferred coordinate system,
i.e. a lattice~\cite{PWA}. We have, therefore, selected an infinite set of
two-dimensional hyperbolic surfaces, where the underlying lattice geometries
are specified by two integers $p$ and $q$, which define the ($p,q$) geometry.

Another question is related to a more concrete physical problem. We consider
multistate spin Hamiltonians specified on infinite lattices ($p,q$). Such a lattice
network allows each multi-state spin to interact with $q$ nearest-neighbors while
the spin interaction coupling $J$ is fixed. The ($p,q$) lattice geometries
correspond to different hyperbolic surfaces, whose properties can be described by
an analytic expression of the Gaussian curvatures $K_{(p,q)}$. The free-energy
study of a classical spin system can be related to the ground-state energy of
a quantum spin system. This relation is given via the so-called quantum-classical
correspondence (see Sec.~\ref{STm}).

The following Chapter is organized as follows. In Section~\ref{hyper_ctmrg_section} we
classify the lattice geometries by the pair ($p,q$) for $Q$-state spin clock and Potts
Hamiltonians. The recurrence relations required by the CTMRG algorithm are derived
gradually, starting from the three simplest cases: ($4,4$), ($5,4$), and ($4,5$),
recalling that the square lattice ($4,4$) has been explained in Sec.~\ref{CTMRG_Section}.
Graphical representations of the recurrence relations are given to simplify the
analytic formulae. Their correctness is checked in Section~\ref{PTA_section} by
calculating phase transition temperatures for the Ising model on sequences of the
selected ($p,q$) lattices. We also support our calculations by studying the asymptotic
regimes ($\infty,7$), ($7,\infty$), and ($\infty,\infty$). We remark here that
the lowest numerical accuracy always occurs at phase transitions, however, the relative
errors of the free energy are still as small as $10^{-5}$. Expressions of the free
energy for any ($p,q$) lattices are given in Section~\ref{free_energy_calc_section},
and the numerical results are summarized in Section~\ref{num_reslts_section}. We propose
a {\it bulk} free energy for the purpose of suppressing strong boundary effects on the
hyperbolic lattices. The free-energy dependence on the two geometry parameters $p$
and $q$ is calculated at the final stage. We conjecture an asymptotic expression
(due to lacking exact solutions). This analytic expression relates the free
energy per site to the radius of the Gaussian curvature of the ($p,q$) lattices.
We show that the numerical free energy per site and the exact Gaussian radius of the
curvature have common features, which are specified by the network of the spin
interactions. In addition, the phase transition temperature (derived from the free
energy at different temperatures, for various spin models, and for the ($p,q$) geometries)
can also reproduce the free-energy structure of the underlying lattice geometries
in asymptotic regions of ($p,q$).

\subsection{Hyperbolic Corner Transfer Matrix Renormalization}\label{hyper_ctmrg_section}

Standard transfer-matrix formulation of the classical spin systems can be modified
by introducing the corner-transfer-matrix formalism, as had been suggested
by Baxter~\cite{Baxter}. Nishino and Okunishi reformulated the original study
of Baxter's into the numerical algorithm Corner Transfer Matrix Renormalization
Group (CTMRG)~\cite{ctmrg1,ctrmg2}. This algorithm also incorporates the ideas of
the well-established Density Matrix Renormalization Group method~\cite{White}.
In 2007 the CTMRG algorithm was applied to the Ising model in order to study phase
transitions on the pentagonal $(5,4)$ lattice~\cite{hCTMRG54}.

The generalized principles of the CTMRG algorithm rely on the precise determination
of recurrence relations, which are required when expanding the corner transfer tensors.
Let us first describe regular lattice geometries made by congruent polygons. It means
that the entire lattice is constructed by the tessellation of regular polygons
with a constant coordination number $q$. The lattice is thus characterized by the
Schl\"{a}fli symbol ($p,q$), where $p$ is the number of the sides (or vertices) of
a regular polygon (often abbreviated as $p$-gon), and the coordination number
$q$ remains unchanged for each lattice site (except for those on the boundary).

Provided that the integers $p>2$ and $q>2$, three types of the curved surfaces are
possible to create a particular ($p,q$) lattice geometry. (1) The condition
$(p-2)(q-2)=4$ refers to the two-dimensional Euclidean (flat) geometry. In this study,
we consider the square lattice ($4,4$) only, which satisfies the condition. The two
remaining cases, the triangular ($3,6$) and the honeycomb ($6,3$) lattices, are to be
studied elsewhere. (2) If $(p-2)(q-2)>4$, an infinite set of the hyperbolic geometries
can satisfy the condition. The infinite-size lattices can be spanned in the
infinite-dimensional space only; this is associated with the infinite Hausdorff
dimension.
None of the infinitely large hyperbolic lattices can be endowed in the three-
(or finite-) dimensional space. (3) The case $(p-2)(q-2)<4$ refers to the five
spherically curved geometries ($3,3$), ($3,4$), ($3,5$), ($4,3$), and ($5,3$).
They form a closed system known as the regular convex Platonic solids (polyhedra)
inside the unit sphere (these five geometries are not considered in the current study).

\subsubsection{The Lattice Model}

Each vertex on the infinitely large ($p,q$) lattice carries a classical multi-spin
variable $\sigma$, which interacts with the $q$ nearest-neighboring spins. The
Hamiltonian ${\cal H}_{(p,q)}$ can be decomposed into sum of identical local
Hamiltonians ${\cal H}_p$. They exclusively act on the local $p$-gons, which
form the basic elements in the construction of the entire ($p,q$) lattice. In
particular, the decomposition of the full Hamiltonian reads
\begin{equation}
{\cal H}_{(p,q)}\{\sigma\} = \sum\limits_{(p,q)} {\cal H}_p[\sigma],
\label{HfHl}
\end{equation}
where the sum is carefully taken over the lattice geometry ($p,q$) accordingly.
The spin notations $[\sigma]$ and $\{\sigma\}$, respectively, are ascribed
to the $p$ spins within each local Hamiltonian ${\cal H}_p[\sigma]\equiv{\cal H}_p
(\sigma_1\sigma_2\cdots\sigma_p)$ and the infinitely many spins $\{\sigma\}$ of the
entire system ${\cal H}_{(p,q)}\{\sigma\}\equiv{\cal H}_{(p,q)}(\sigma_1\sigma_2\cdots
\sigma_{\infty})$. We consider two types of the multi-state spin models: the $Q$-state
clock model with the local Hamiltonian on the $p$-gonal spin lattice
\begin{equation}
{\cal H}_p[\sigma] = -J\sum\limits_{i=1}^{p}
\cos\left[\frac{2\pi}{Q}(\sigma_{i}-\sigma_{i+1})\right]
\label{clock}
\end{equation}
and the $Q$-state Potts model
\begin{equation}
{\cal H}_p[\sigma] = -J\sum\limits_{i=1}^{p} \delta_{\sigma_{i},\sigma_{i+1}}\, ,
\label{Potts}
\end{equation}
where $\sigma_{p+1}^{~}\equiv\sigma_{1}^{~}$ within the $p$-gon, and where each $Q$-state
spin variables $\sigma=0,1,2,\dots,Q-1$. (For the Ising model $Q=2$.)
We consider the ferromagnetic interaction $J>0$ to avoid frustration for odd $p$.
The $q$ dependence does not explicitly enter the local Hamiltonian
${\cal H}_p[\sigma]$ because it is given by the manner of how the local $p$-gons are
connected or, equivalently, how the local Hamiltonians are summed up in Eq.~(\ref{HfHl}).

The basic Boltzmann weight ${\cal W}_{\rm B}[\sigma]=\exp(-{\cal H}_p[\sigma]/k_{\rm B}T)$
is defined on the $p$-gon of the local Hamiltonian, Eq.~(\ref{HfHl}), where $k_{\rm B}$
and $T$ correspond to the Boltzmann constant and temperature, respectively. We use the
dimensionless units throughout this work and set $J=k_{\rm B}=1$. For brevity we drop
the index $p$ in the local Boltzmann weight.

The CTMRG algorithm is an RG-based iterative numerical method which evaluates the
partition function ${\cal Z}_{(p,q)}^{[k]}$ and the other thermodynamic functions in high
accuracy~\cite{Genzor}. Let the iteration step in CTMRG be enumerated by integer
$k=1,2,3,\dots$, which appears as the superscript in the partition function and
has nothing to do with the exponentiation. The CTMRG process is initialized by forming
a lattice as small as the $q$ identical $p$-gons surrounded around one central spin.
This initial step is referred to as the first iteration step, $k=1$. In the consequent
iteration steps, $k>1$, the lattice gradually expands the size. The number of the spin
sites increases either algebraically, which is satisfied in the Euclidean case ($4,4$) only
or they grow exponentially for all the remaining ($p,q$) cases, which obey the condition
$(p-2)(q-2)>4$. The lattice expansion can be regarded with respect to the growing number
of the Boltzmann weights (and the proportionality to the total number of the spins is
straightforward. Since we are interested in the phase-transition analysis, the thermodynamic
limit has to be considered, i.e. $k\to\infty$. This asymptotics is numerically equivalent
to terminating the CTMRG iterations whenever all of the thermodynamic functions normalized
per spin site completely converge.

\begin{figure}[tb]
\begin{center}
\includegraphics[width=4in]{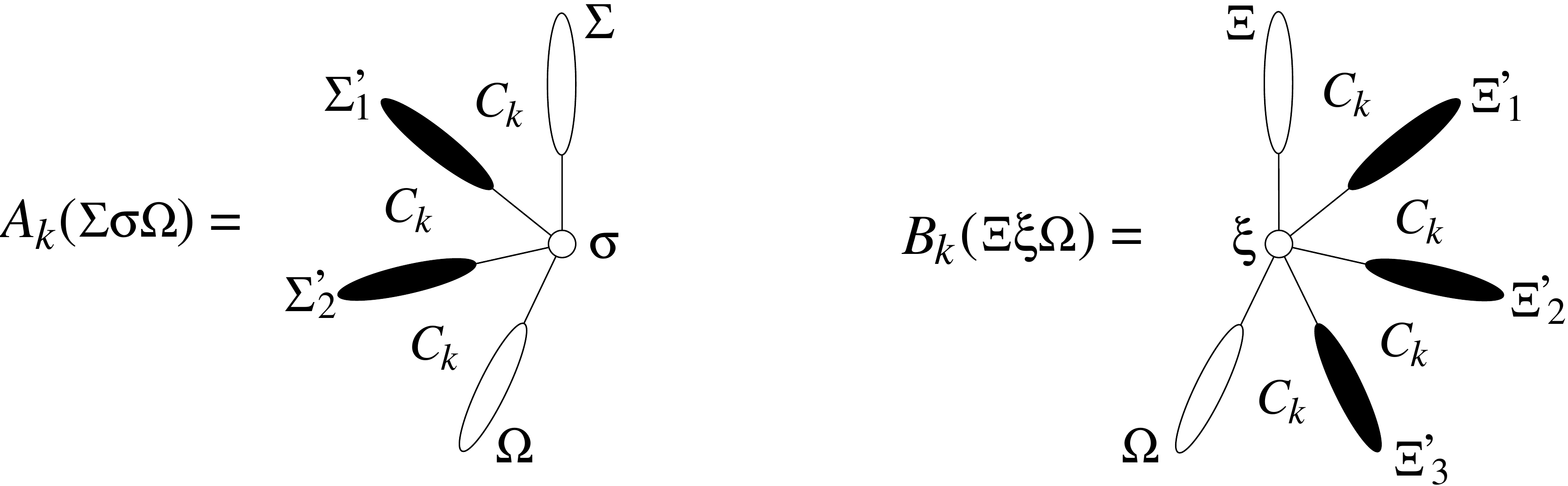}
\end{center}
  \caption{Graphical representation of Eqs.~\eqref{A_k} and \eqref{B_k}.}
\label{fig:51}
\end{figure}

Since CTMRG algorithm was originally applied to the square lattice ($4,4$), as described
in Section~\ref{CTMRG_Section}, this algorithm can be adapted to the hyperbolic lattices
as well. Let us recall that the reduced density matrix was proportional to taking the
partial trace over the four corner transfer tensors, cf. Eq.~\eqref{RDM}. For an arbitrary
coordination number $q>4$, the reduced density matrix evaluated at an iteration step $k$
satisfies the identical rule, i.e., $\rho_k={\rm Tr}^{\prime}\,({\cal C}_k^q)$. Thus
constructed reduced density matrices $\rho_k$ are real and symmetric, only if $q$ is even.
However, if $q$ is an odd integer, $\rho_k$ is not a symmetric matrix anymore, because
of the lattice construction made by the corner transfer tensors. It means that the entire
lattice cannot be divided into two identical halves whenever $q$ is an odd number.
Such an asymmetry can be, however, recovered by symmetrizing the reduced density
matrix~\cite{Uli1, hCTMRG3q}
\begin{equation}
\rho_k \left( \left. \Sigma \sigma \right| \Xi\, \xi \right) = \frac{1}{2} \sum_{\Omega}
\left[
A_k( \Sigma \sigma \Omega) B_k( \Xi\, \xi \Omega ) + 
B_k( \Sigma \sigma \Omega) A_k( \Xi\, \xi \Omega )
\right]\, . 
\end{equation}
As a typical example of the asymmetry, let us consider $q=7$. The two tensors $A_k$ and $B_k$
are formed by joining the corner transfer tensors
\begin{eqnarray}
\label{A_k}
A_k( \Sigma \sigma \Omega) & = & \sum\limits_{\Sigma^\prime_1 \Sigma^\prime_2}
{\cal C}_{k,\,\sigma\Sigma\Sigma^\prime_1}{\cal C}_{k,\,\sigma\Sigma^\prime_1\Sigma^\prime_2}
{\cal C}_{k,\,\sigma\Sigma^\prime_2\Omega}\, ,\\
\label{B_k}
B_k( \Xi \xi \Omega) & = & \sum\limits_{\Xi^\prime_1 \Xi^\prime_2 \Xi^\prime_3}
{\cal C}_{k,\,\xi\Xi^\prime_1\Xi}{\cal C}_{k,\,\xi\Xi^\prime_2\Xi^\prime_1}
{\cal C}_{k,\,\xi\Xi^\prime_3\Xi^\prime_2}{\cal C}_{k,\,\xi\Omega\Xi^\prime_3}\, ,
\end{eqnarray}
as graphically represented in Fig.~\ref{fig:51}. In general, the tensor
$A_k = {\cal C}_k^{\lfloor q/2 \rfloor}$ represents a statistical weight of the lattice system
consisting of $\lfloor q/2 \rfloor$ corner transfer tensors and the $B_k={\cal C}_k^{\lfloor q/2 \rfloor +r}$
is the other part of the lattice made of $\lfloor q/2 \rfloor + r$ corner transfer tensors.
The notation $\lfloor q/2 \rfloor$ denotes the floor function, being the lower integer part,
i.e., $\lfloor q/2 \rfloor \equiv \max\{i \in {\mathbb{Z}} \,\, \vert\,\, i \leq q/2 \}$.
The boolean variable $r$ is either zero or one if $q$ is even or odd, respectively. It is
equivalent to a reminder (modulo) so that $r=(q\mod 2)$; hence, if $q$ is even, $A_k \equiv B_k$.

\subsubsection{Recurrence Relations}

The complete iteration process is given by a set of recurrence relations as we
specify below. For deeper instructive understanding, the final derivation of the
recurrence relations is gradually structured into the following three steps,
which are grouped by the increasing complexity of the lattice geometries
\begin{itemize}
\item[(i)] $(4,4)$, $(5,4)$, and $(4,5)$,
\item[(ii)] $(4,4)\to(5,4)\to(6,4)\to\cdots\to(\infty,4)$,
\item[(iii)] $(p,q)$.
\end{itemize}

\begin{figure}[tb]
\begin{center}
\begin{tabular}{cccc}
& $(4,4)$ & $(5,4)$ & $(4, 5)$ \\[1.85cm]
{\vspace{-0.0cm} $k=1$} & & & \\[-1.9cm]
&
\includegraphics[width=1.24in]{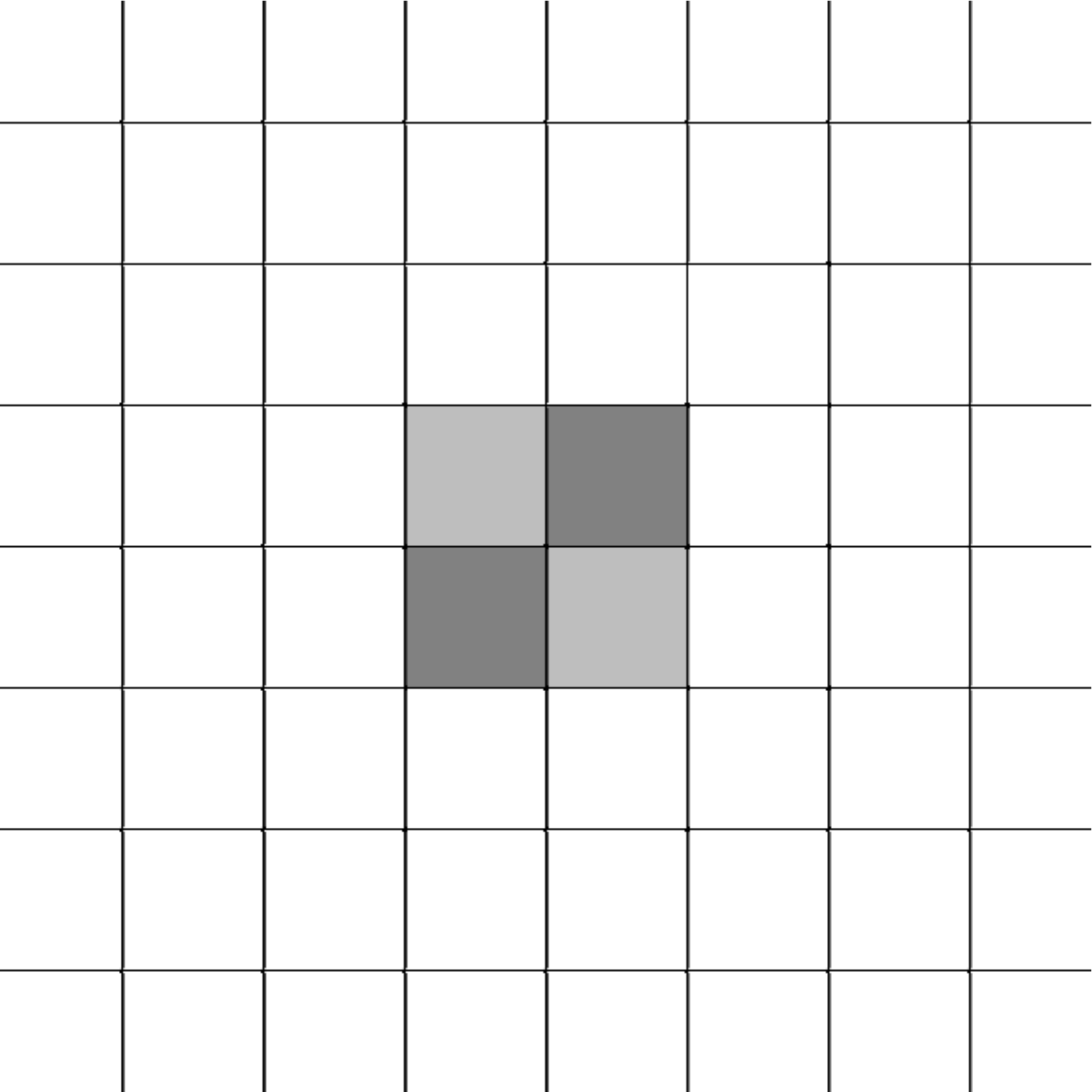} &
\includegraphics[width=1.24in]{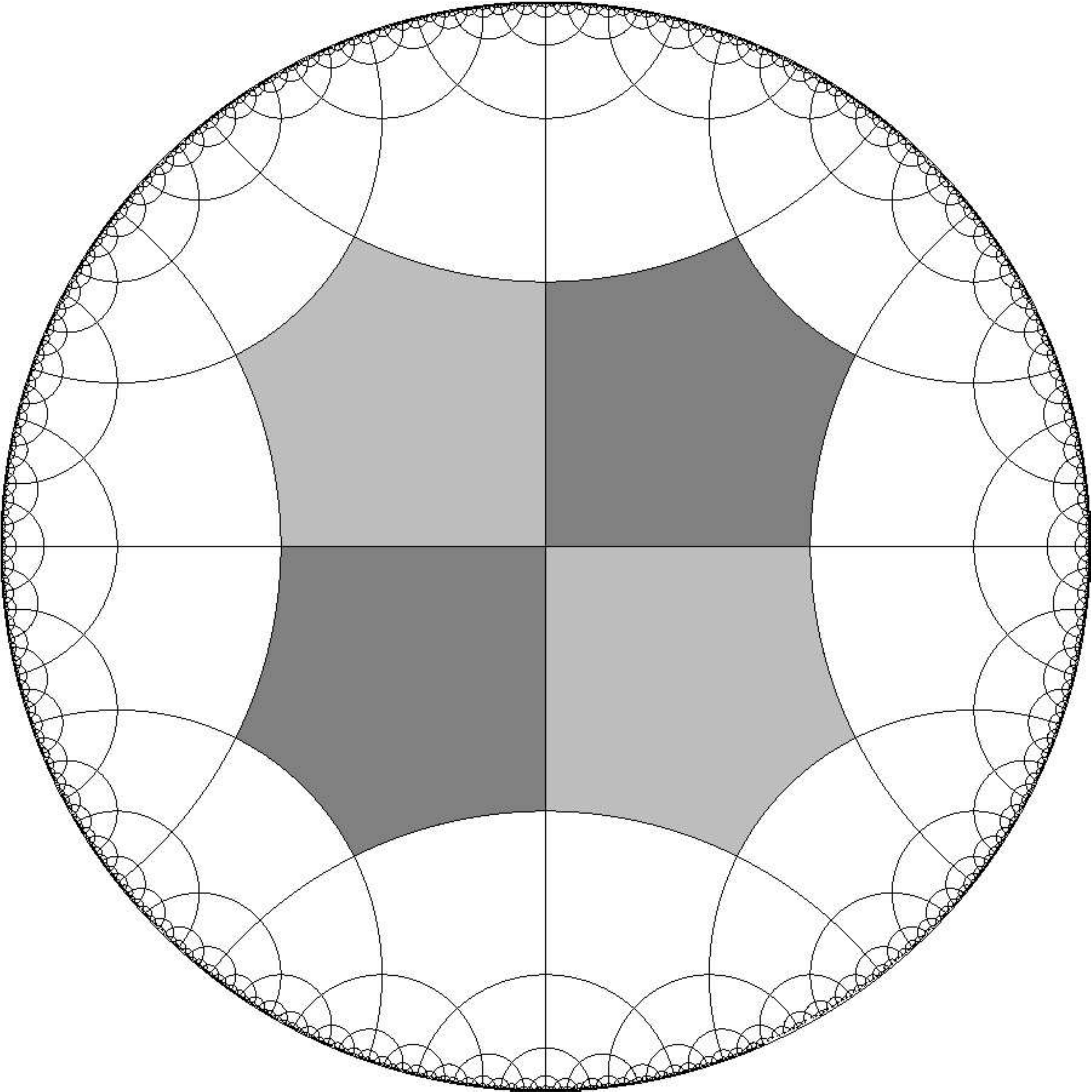} &
\includegraphics[width=1.24in]{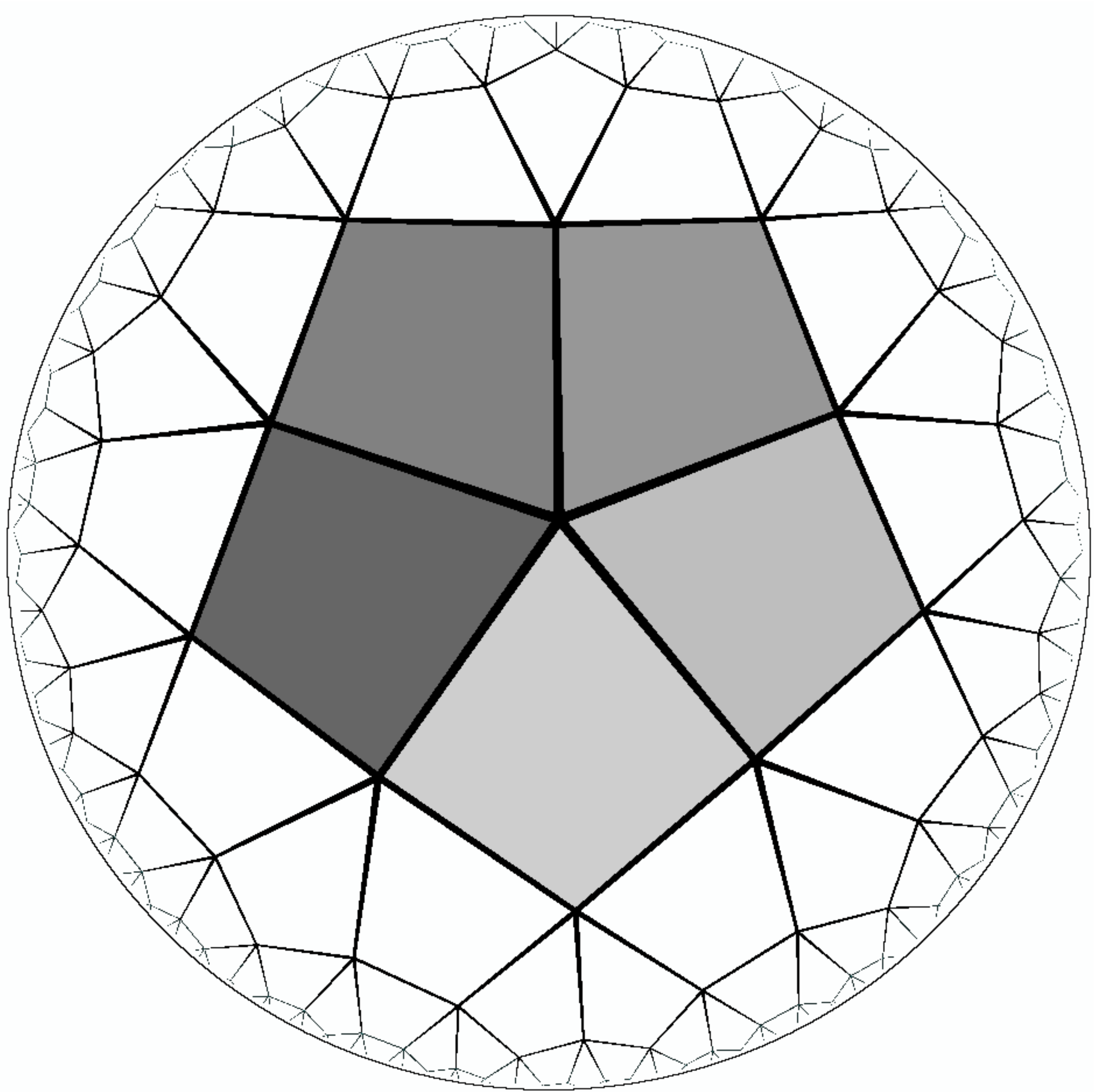} \\[1.85cm]
{\vspace{-0.0cm} $k=2$} & & & \\[-1.9cm]
&
\includegraphics[width=1.24in]{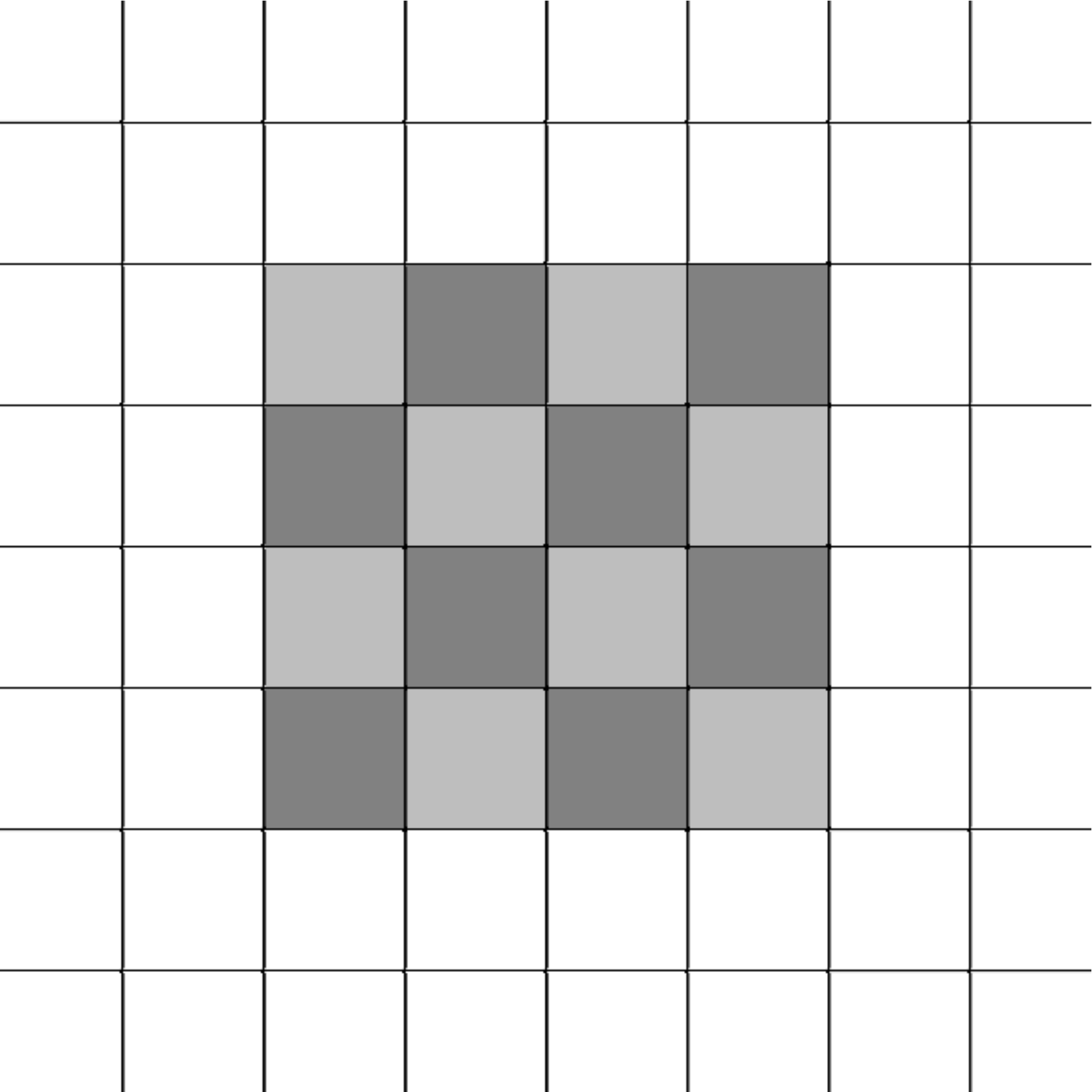} &
\includegraphics[width=1.24in]{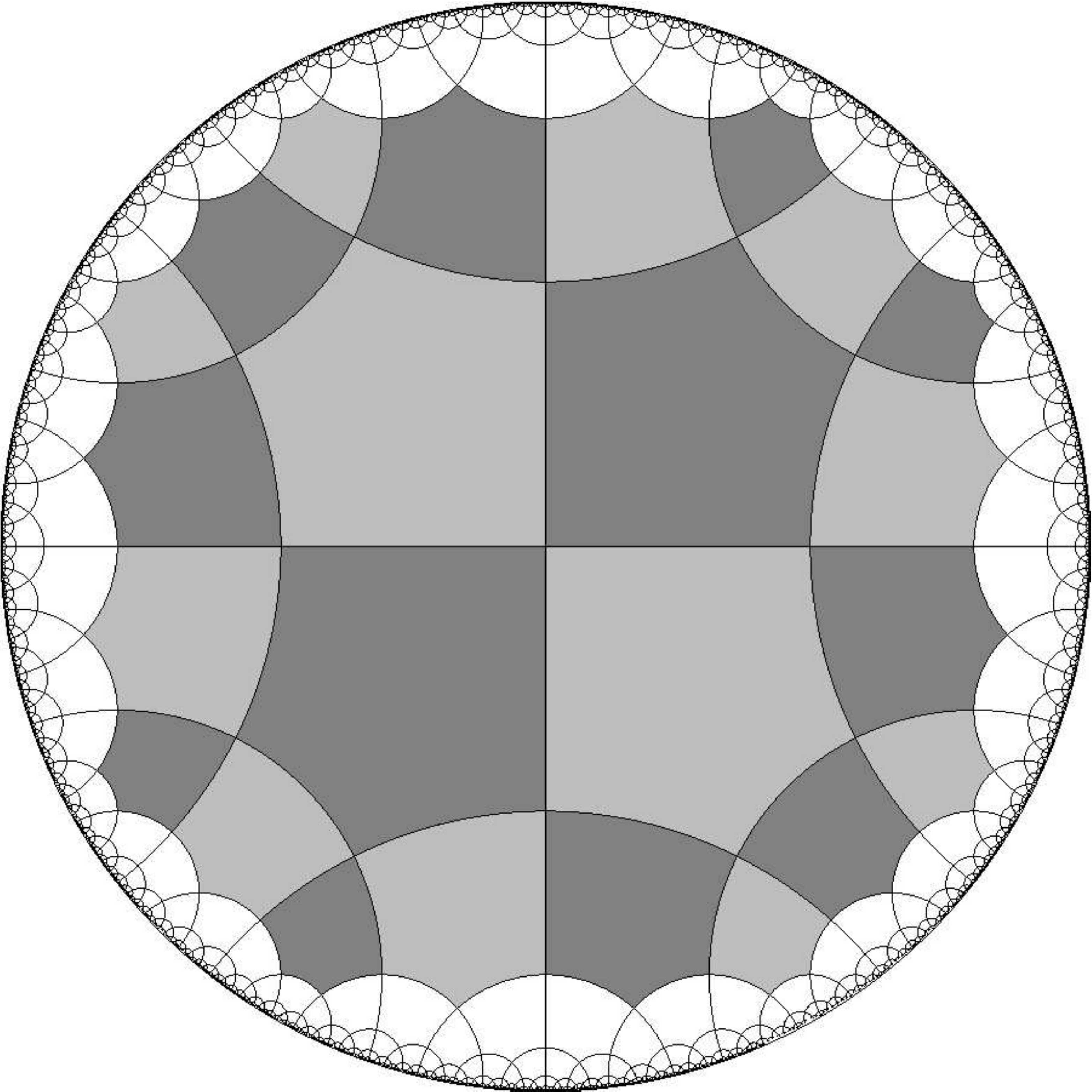} &
\includegraphics[width=1.24in]{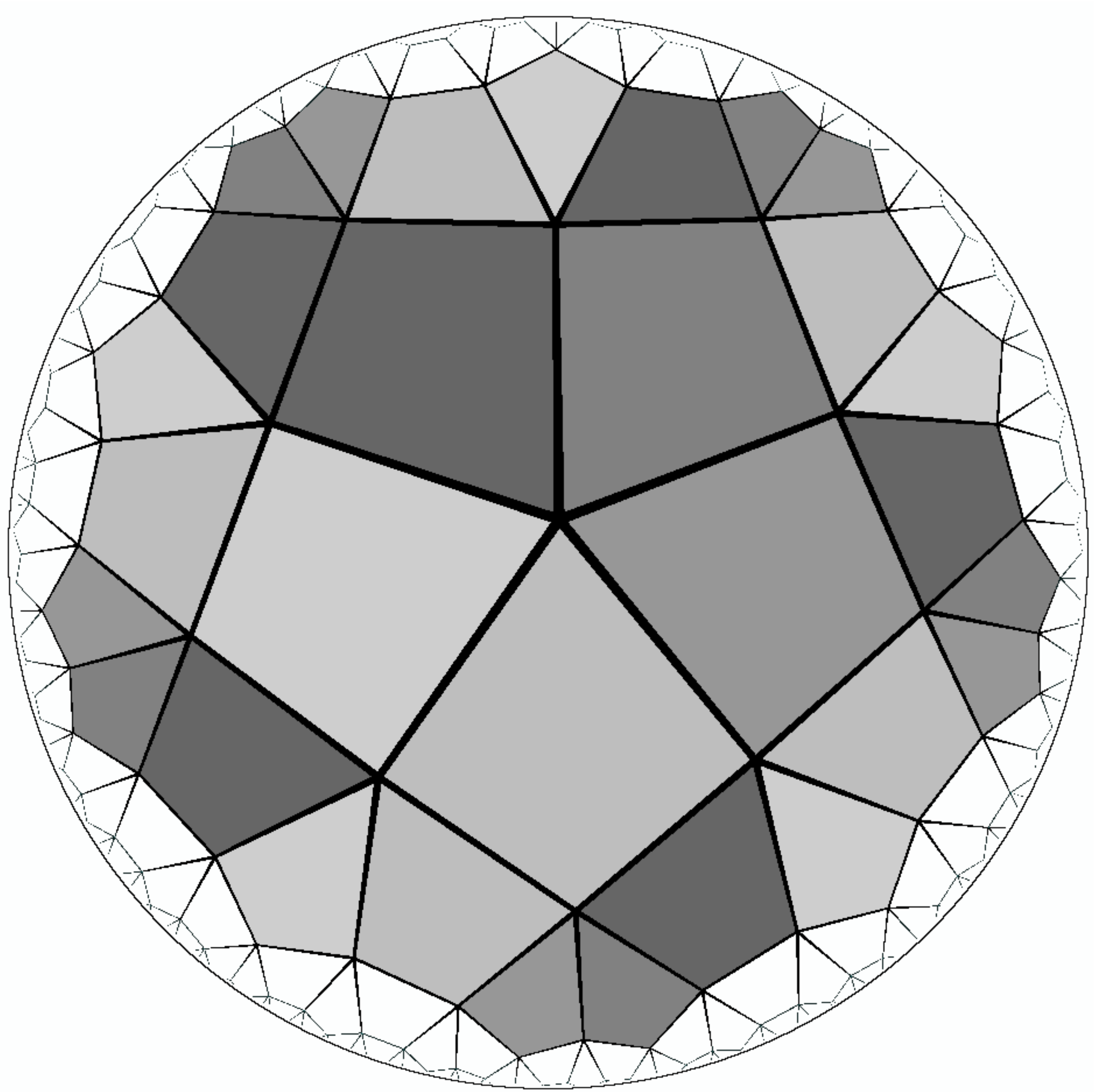}
\end{tabular}
\end{center}
 \caption{The illustration of the three selected lattice geometries ($4,4$), ($5,4$), and
($4,5$). The first two CTMRG iteration steps $k=1$ (upper) and $k=2$ (lower) show the building
process of the lattices by means of the $p$-gonal Boltzmann weight tessellation with the
uniform coordination number $q$. The Boltzmann weights for the given iteration step $k$
are represented by the shaded regular (congruent) $p$-gons. Various shaded intensities
serves as a guide to distinguish the $p$-gons within given $k$.}
  \label{fig:52}
\end{figure}

Figure~\ref{fig:52} depicts the three simplest lattices within the iteration steps $k=1$
and $k=2$ only. The shaded $p$-gons represent the corresponding finite lattice formed
by the Boltzmann weights ${\cal W}_{\rm B}$ at given $k$. The surrounding $p$-gons
shown in white color around the shaded $p$-gons show the entire lattice structure,
where the consequent iteration steps $k$ will expand the lattice to.
The spin variables $\sigma$ are positioned on the vertices of the $p$-gons,
and the sides of the $p$-gons correspond to the constant nearest-neighbor spin
coupling $J=1$. The sizes and the shapes of the polygons are equal for each lattice
geometry ($p,q$). We, therefore, depict each hyperbolic lattice geometry in the 
so-called Poincare disk representation~\cite{Poincare}, where the entire hyperbolic
lattice is projected onto. As the consequence of that projection, the sizes of the
$p$-gons are deformed and they get shrunk from the lattice center towards the
circumference of the circle. Here, the circumference is associated with the lattice
boundary at infinity.

(i) The iterative expansion process is formulated in terms of a generalized corner
transfer tensor notation (for details, see Refs.~\cite{hCTMRG54,hCTMRGp4,hCTMRG3q,
hCTMRG3qn}). The corner transfer tensors ${\cal C}_j$ and the transfer tensors
${\cal T}_j$ undergo an expansion process up to the desired iteration step $k$.
Hence, the expansion process of the three simplest lattices satisfies the following
rules
\begin{equation}
\left.
\begin{split}
{\cal C}_{k+1}&={\cal W}_{\rm B} {\cal T}_{k}^{2} {\cal C}_{k}^{~}\\
{\cal T}_{k+1}&={\cal W}_{\rm B} {\cal T}_{k}^{~}
\end{split}
\ \ \right\} \ \ {\rm for}\ \ (4,4),
\label{rr44}
\end{equation}
\begin{equation}
\left.
\begin{split}
{\cal C}_{k+1}&={\cal W}_{\rm B} {\cal T}_{k}^{3} {\cal C}_{k}^{2}\\
{\cal T}_{k+1}&={\cal W}_{\rm B} {\cal T}_{k}^{2} {\cal C}_{k}^{~}
\end{split}
\ \ \right\} \ \ {\rm for}\ \ (5,4),
\label{rr54}
\end{equation}
\begin{equation}
\left.
\begin{split}
{\cal C}_{k+1}&={\cal W}_{\rm B} {\cal T}_{k}^{2} {\cal C}_{k}^{3}\\
{\cal T}_{k+1}&={\cal W}_{\rm B} {\cal T}_{k}^{~} {\cal C}_{k}^{~}
\end{split}
\ \ \right\} \ \ {\rm for}\ \ (4,5).
\label{rr45}
\end{equation}
The tensors are initialized to the Boltzmann weight
${\cal C}_{1}^{~}={\cal T}_{1}^{~} = {\cal W}_{\rm B}$.

The recurrence relations in Eqs.~\eqref{rr44}--\eqref{rr45} are written in a simplified
form, it means that the indexing of the lattice geometry ($p,q$) they depend on can be
excluded in order to reduce its redundancy, which is actually transformed into the
powers, as we elucidate later. In particular, ${\cal C}_{k}^{(p,q)} \to {\cal C}_{k}^{~}$
and ${\cal T}_{k}^{(p,q)} \to{\cal T}_{k}^{~}$, including the $p$-gonal shape of the
Boltzmann weight is abbreviated to ${\cal W}_{\rm B}$. The partition function
${\cal Z}_{(p,q)}^{[k]}$ at the $k^{\rm th}$ iteration step is given by the
configuration sum (or trace) of the product of the $q$ corner transfer tensors, which
are concentrically connected around the central spin site of the lattice~\cite{hCTMRG54}
\begin{equation}
{\cal Z}_{(p,q)}^{[k]} = {\rm Tr}\left[e^{-{\cal H}_{(p,q)}/T}\right]
={\rm Tr}\ (\underbrace{{\cal C}_{k}{\cal C}_{k}\cdots{\cal C}_{k}}_{q})
\equiv{\rm Tr}\ ({\cal C}_{k})^{q}.
\label{part_fnc_hyp}
\end{equation}
The evaluation of the partition function as the product of the Boltzmann weights of the
$p$-gonal shape can be expressed graphically, which may serve as a visual simplification
of Eq.~\eqref{part_fnc_hyp}. For instance, the size of the square lattice ($4,4$) at the
second iteration step, $k=2$, corresponds to the evaluation of the partition function
${\cal Z}_{(4,4)}^{[k=2]}$ in Fig.~\ref{fig:52}. This is equivalent to the product of
the sixteen Boltzmann weights when applying the recurrence relations~\eqref{rr44}, i.e.,
\begin{equation}
{\cal Z}_{(4,4)}^{[2]}
 = {\rm Tr}\ {\left({\cal C}_{2}\right)}^{4}
 = {\rm Tr}\ {\left({\cal W}_{\rm B}{\cal T}_{1}^{2}{\cal C}_{1}^{~}\right)}^{4}
 = {\rm Tr}\ {\left({\cal W}_{\rm B}\right)}^{16}.
\end{equation}
Thus, the power of ${\cal W}_{\rm B}$ matches the total number of the shaded
squares in Fig.~\ref{fig:52} for given $k$. The number of the square-shaped Boltzmann
weights grows as the power law $4k^2$ on the square ($4,4$) lattice only.

The partition functions of the two hyperbolic lattices ($5,4$) and ($4,5$) have
analogous expressions. For instance, the lattice size at the iteration step $k=2$
(as graphically sketched in Fig.~\ref{fig:52}) is related to taking the configuration
sum over the product of the shaded $p$-gons. For the instructive purpose,
the partition functions satisfy the following
\begin{equation}
{\cal Z}_{(5,4)}^{[2]}
 = {\rm Tr}\ {\left({\cal C}_{2}\right)}^{4}
 = {\rm Tr}\ {\left({\cal W}_{\rm B}{\cal T}_{1}^{3}{\cal C}_{1}^{2}\right)}^{4}
 = {\rm Tr}\ {\left({\cal W}_{\rm B}\right)}^{24}
\end{equation}
and
\begin{equation}
{\cal Z}_{(4,5)}^{[2]}
 = {\rm Tr}\ {\left({\cal C}_{2}\right)}^{5}
 = {\rm Tr}\ {\left({\cal W}_{\rm B}{\cal T}_{1}^{2}{\cal C}_{1}^{3}\right)}^{5}
 = {\rm Tr}\ {\left({\cal W}_{\rm B}\right)}^{30}\, ,
\end{equation}
where the powers of ${\cal W}_{\rm B}$ on the right hand side of the equations again
count the number of the $p$-gonal Boltzmann weights. We recall that the total number
of the Boltzmann weights grows exponentially, as iteration step $k$ increases.
The analytic formula of the exponential dependence of the total number of the spin sites
on $k$ is derived in the next Section, where the free energy is examined in detail.

(ii) In Ref.~\cite{hCTMRGp4}, we have investigated the Ising model on an infinite
sequence of such hyperbolic lattices, for which the coordination number was fixed
to $q=4$, whereas the $p$-gons varied $p=4,5,6,\dots,\infty$. The generalized recurrence
relations satisfying the lattices ($p\geq4,4$) are now summarized into a more compact form
\begin{equation}
\begin{split}
{\cal C}_{k+1}& = {\cal W}_{\rm B} {\cal T}_{k}^{p-2} {\cal C}_{k}^{p-3}\, ,\\
{\cal T}_{k+1}& = {\cal W}_{\rm B} {\cal T}_{k}^{p-3} {\cal C}_{k}^{p-4}\, .
\end{split}
\end{equation}
We have conjectured that the Ising model realized on the sequence of the lattices
$\{(4,4)$, $(5,4)$, $(6,4),\dots,(\infty,4)\}$ converges to the Bethe lattice with the
coordination number $q=4$. The numerical convergence is exponentially fast with
respect to $p$. In other words, the thermodynamic properties of the Bethe lattice
($\infty,4$) are actually numerically indistinguishable with the lattice geometries,
for which $p \geq 15$, even at the phase transition point~\cite{hCTMRGp4}.
After we evaluated the phase transition temperature $T_{\rm pt}^{(\infty,4)}$ of
the Ising model on the Bethe lattice, which had been numerically realized on the
($15,4$) lattice geometry, the CTMRG algorithm resulted in the phase transition
temperature $T_{\rm pt}=2.88539$. Since the Ising model on the Bethe lattice is
an exactly solvable system, the comparison with the exact value $T_{\rm pt}=1/
\ln\sqrt{2}$ exhibits a very high numerical accuracy~\cite{Baxter}.

(iii) If considering an arbitrary ($p,q$) lattice geometry such that $p\geq4$ and
$q\geq4$, the derivation of the recurrence relations is straightforward (after
some algebraic calculations) leads to
\begin{equation}
\begin{split}
{\cal C}_{k+1}& = {\cal W}_{\rm B} {\cal T}_{k}^{p-2} {\cal C}_{k}^{(p-2)(q-3)-1},\\
{\cal T}_{k+1}& = {\cal W}_{\rm B} {\cal T}_{k}^{p-3} {\cal C}_{k}^{(p-3)(q-3)-1}.
\end{split}
\end{equation}
The calculation of the partition function ${\cal Z}$ for any ($p,q$) lattice geometry
at the $k^{\rm th}$ iteration step still persists identical to Eq.~\eqref{part_fnc_hyp}.
An expectation value $\langle O\rangle$ of a local observable $O$ can be evaluated directly.
As an example, the spontaneous magnetization $M=\langle\sigma_c\rangle$ measured in the
center of the lattice ($p,q$), where the spin variable $\sigma_c$ is positioned, has
the following expression in the thermodynamic limit
\begin{equation}
M_{(p,q)}=\langle\sigma_c\rangle
 =\frac{{\rm Tr}\left[\sigma_c\,e^{-{\cal H}_{(p,q)}/T}\right]}
       {{\rm Tr}\left[          e^{-{\cal H}_{(p,q)}/T}\right]}
=\frac{ {\rm Tr}\,\left[ \sigma_c{\left({\cal C}_{\infty}\right)}^{q}\right] }
      {{\cal Z}_{(p,q)}^{[\infty]}}
\label{Mpq}
\end{equation}
for arbitrary ($p,q$).

\subsection{Phase Transition Analysis}\label{PTA_section}

\begin{figure}[tbh]
\begin{center}
\begin{tabular}{ccc}
($4,7$) &  ($7,4$) &  ($7,7$)\\
\includegraphics[width=.28\linewidth]{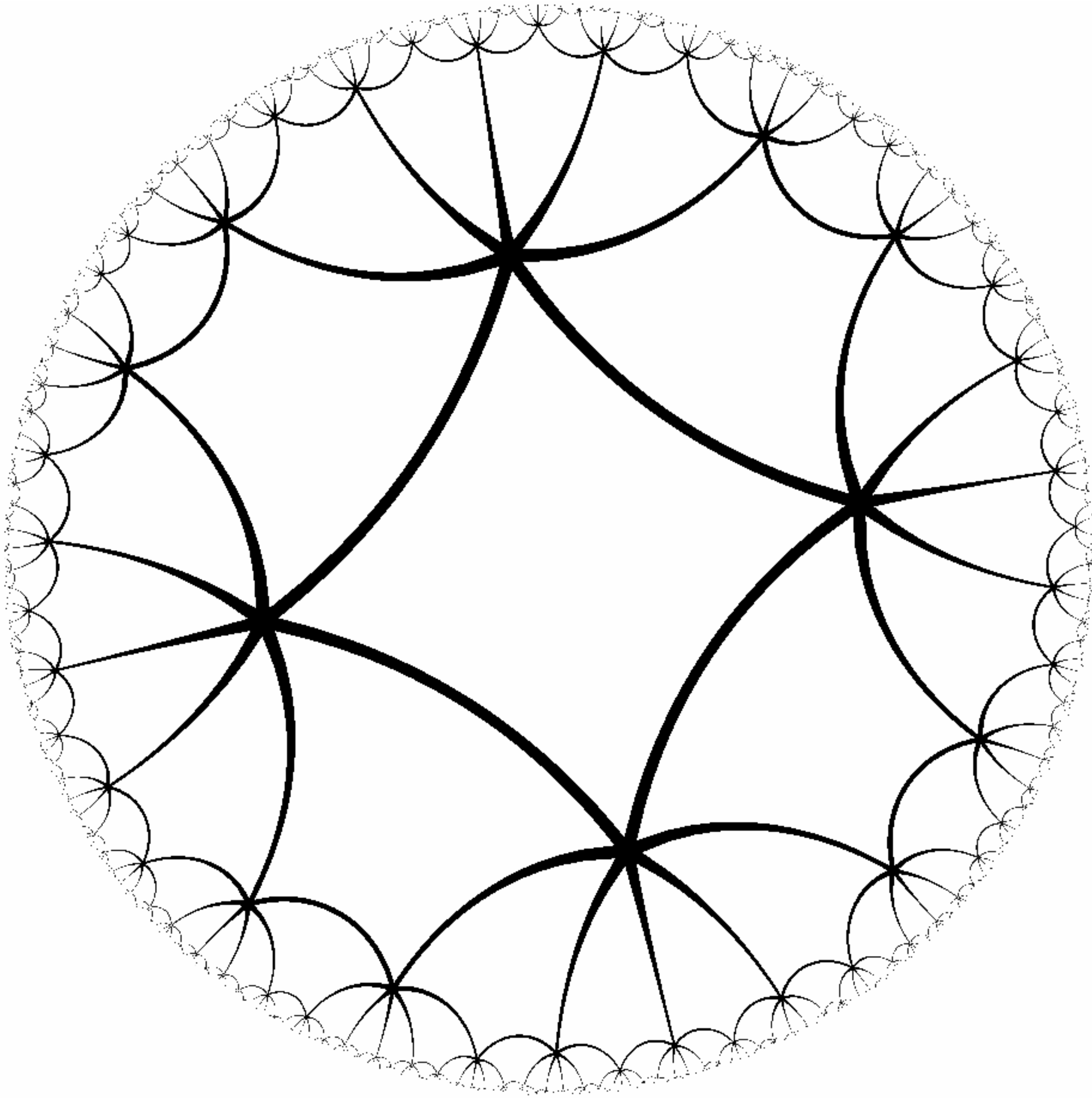} &
\includegraphics[width=.28\linewidth]{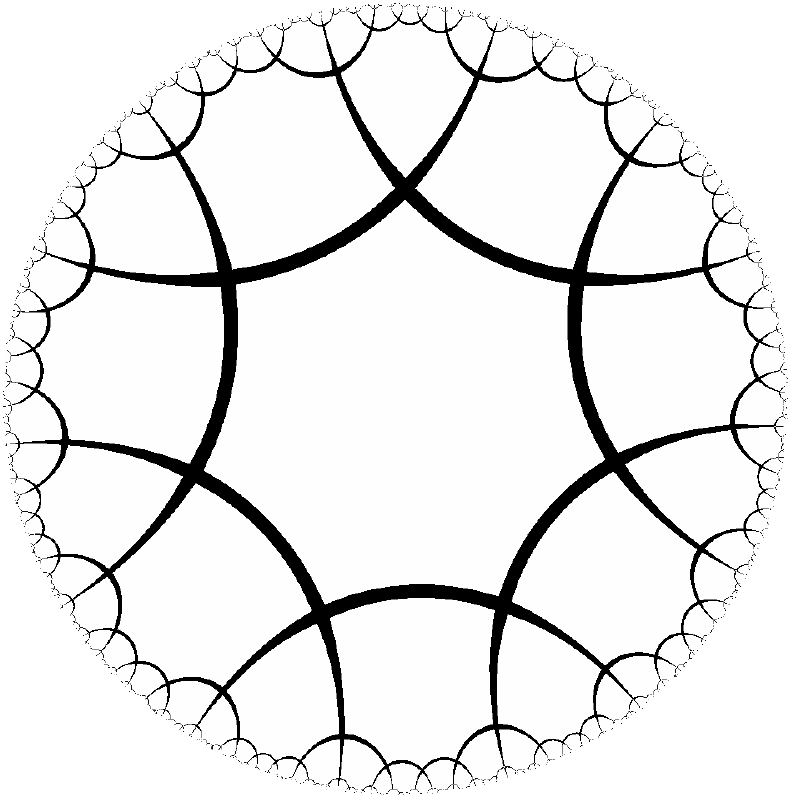} &
\includegraphics[width=.28\linewidth]{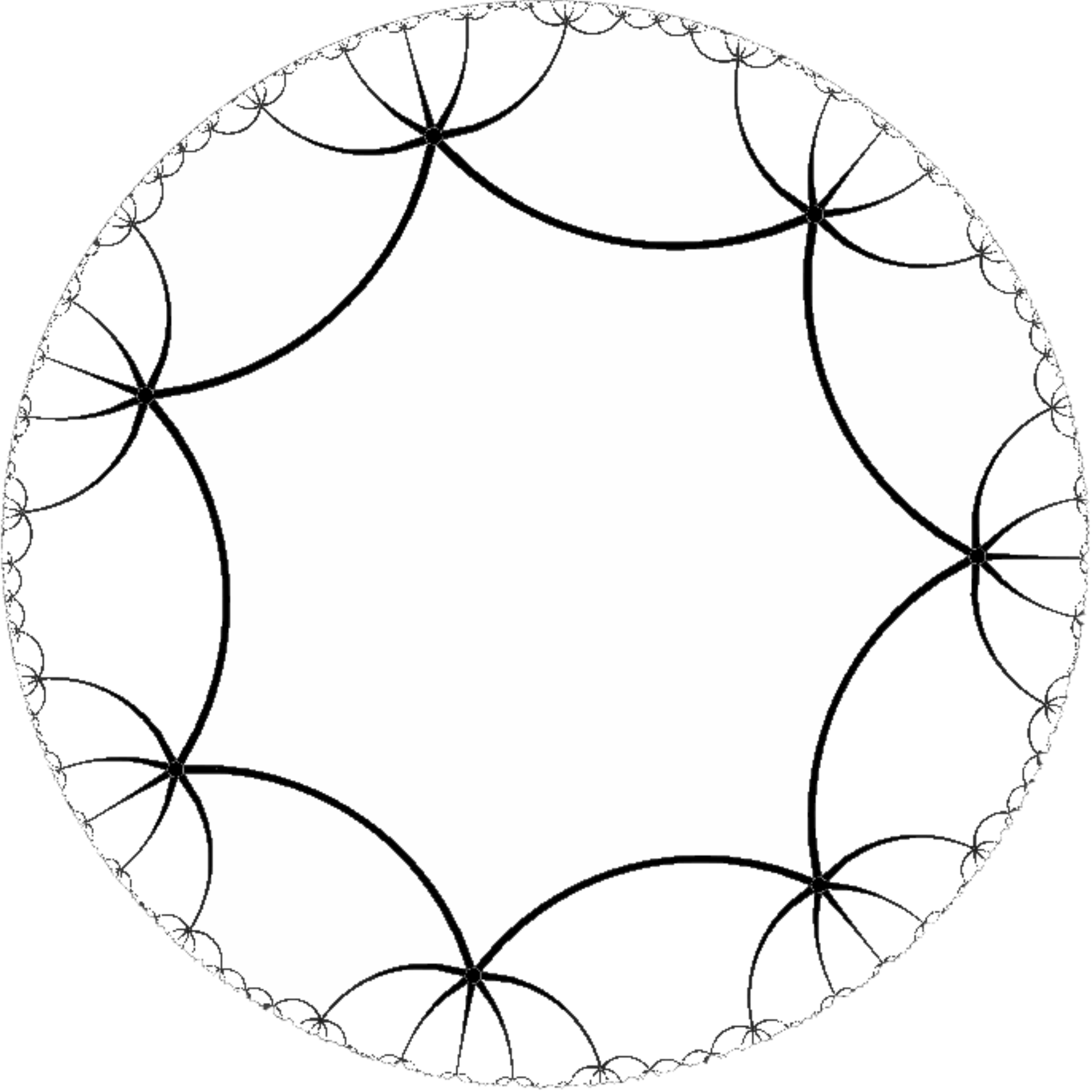}
\end{tabular}
\end{center}
  \caption{The Poincar\'{e} disk representation of the three hyperbolic lattices
chosen for the analysis of the thermodynamic functions of the spin model.}
  \label{fig:53}
\end{figure}
\begin{figure}[!ht]
\begin{center}
\includegraphics[width=3.8in]{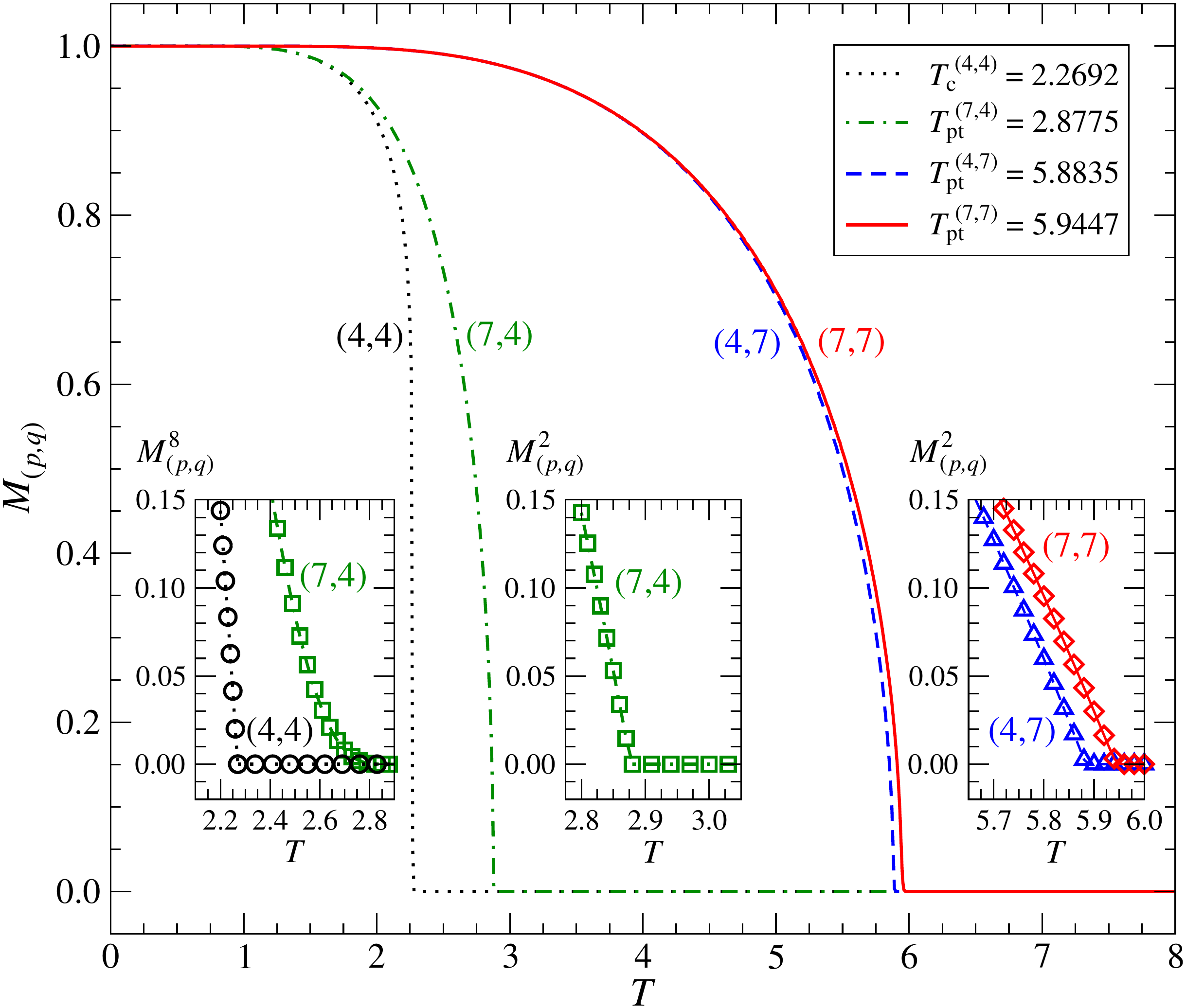}
\end{center}
  \caption{Temperature dependence of the spontaneous magnetization for
the Euclidean square lattice as well as for the three hyperbolic lattices depicted in
Fig.~\ref{fig:53}.}
\label{fig:54}
\end{figure}

For demonstration, we have selected the three non-trivial hyperbolic lattices: ($4,7$),
($7,4$), and ($7,7$), which are depicted in Fig.~\ref{fig:53} in the Poincare representation.
The respective spontaneous magnetizations $M_{(p,q)}$ in the thermodynamic limit
($k\to\infty$) are plotted on the upper graph in Fig.~\ref{fig:54}. For comparison, we have
also included data on the Euclidean ($4,4$) lattice, which can serve as the benchmark of
the exactly solvable system. We have conjectured~\cite{hCTMRGp4,hCTMRG3q,hCTMRG3qn} 
that the Ising model on hyperbolic lattices ($p>4,4$) and ($3,q>6$) belong to the mean-field
universality class. We now expand this conjecture to the ($p,q$) lattices. We
show that the spontaneous magnetization follows the scaling relation $M_{(p,q)}\propto
(T_{\rm pt}^{(p,q)}-T)^{\beta}$ at the phase transition temperature $T_{\rm pt}^{(p,q)}$
yielding the mean-field magnetic exponent $\beta=\frac{1}{2}$ whenever $(p-2)(q-2)>4$.

Here we add an important remark: The mean-field universality is solely the consequence
of the hyperbolic lattice geometry, and has nothing to do with the mean-field approximation.
If the Euclidean geometry ($4,4$) is set, our numerical analysis confirms that
$M_{(4,4)}\propto(T_{\rm pt}^{(4,4)}-T)^{\frac{1}{8}}$ in agreement with theory.
This is clearly manifested by showing the linear dependence of $M^8_{(4,4)}$ on temperature
$T \leq T_{\rm pt}^{(4,4)}$ as depicted on the left-bottom inset in Fig.~\ref{fig:54}.
On the other hand, the mean-field universality class with $\beta=\frac{1}{2}$ can be
read off in the graph if plotting $M^2_{(p,q)}$ for $T \leq T_{\rm pt}^{(p,q)}$.
Obviously, the linearly decreasing dependence of the spontaneous magnetization is present
in the vicinity of the phase transition, i.e., $T \leq T_{\rm pt}^{(p,q)}$,
as shown on the remaining two insets in Fig.~\ref{fig:54}.

The mean-field-like feature of the spin model is always realized on the hyperbolic lattices.
We point out here that such a mean-field-like behavior is not caused by an insufficient
numerical accuracy. The numerical results are fully converged, and any additional
increase of the number of the states kept, $m$, in the renormalization group algorithm
does not improve the thermodynamic functions. The reason for the mean-field-like
feature originates in the exceedance of the critical lattice dimension $d_{c}=4$. It is
so because the Hausdorff dimension is infinite for all the hyperbolic lattices in the
thermodynamic limit. This claim comes from the exact solution of the Ising model on the
Bethe lattice, where the analytically derived mean-field exponents on the Bethe lattice
have nothing to do with the mean-field approximation of the model at all~\cite{Baxter}.
Instead, the {mean-field-like} feature is caused by the hyperbolic lattice geometry,
which is accompanied by the absence of the divergent correlation length at the phase
transition, as we had pointed out in Ref.~\cite{hCTMRG3q}.

\subsubsection{Asymptotic Lattice Geometries}

\begin{figure}[tb]
\begin{center}
\begin{tabular}{ccc}
($\infty,7$) &  ($7,\infty$) & ($\infty,\infty$)\\
\includegraphics[width=.28\linewidth]{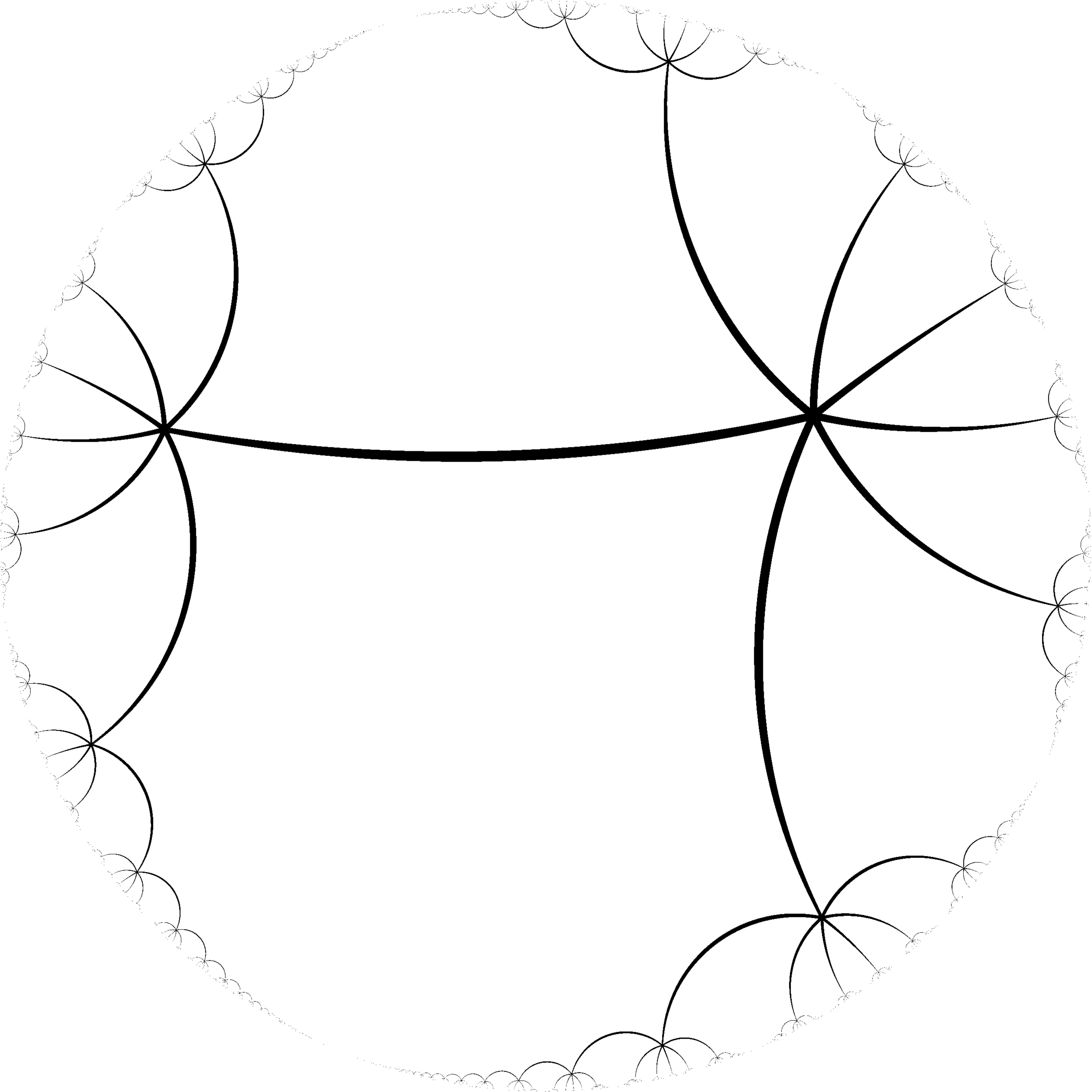} &
\includegraphics[width=.28\linewidth]{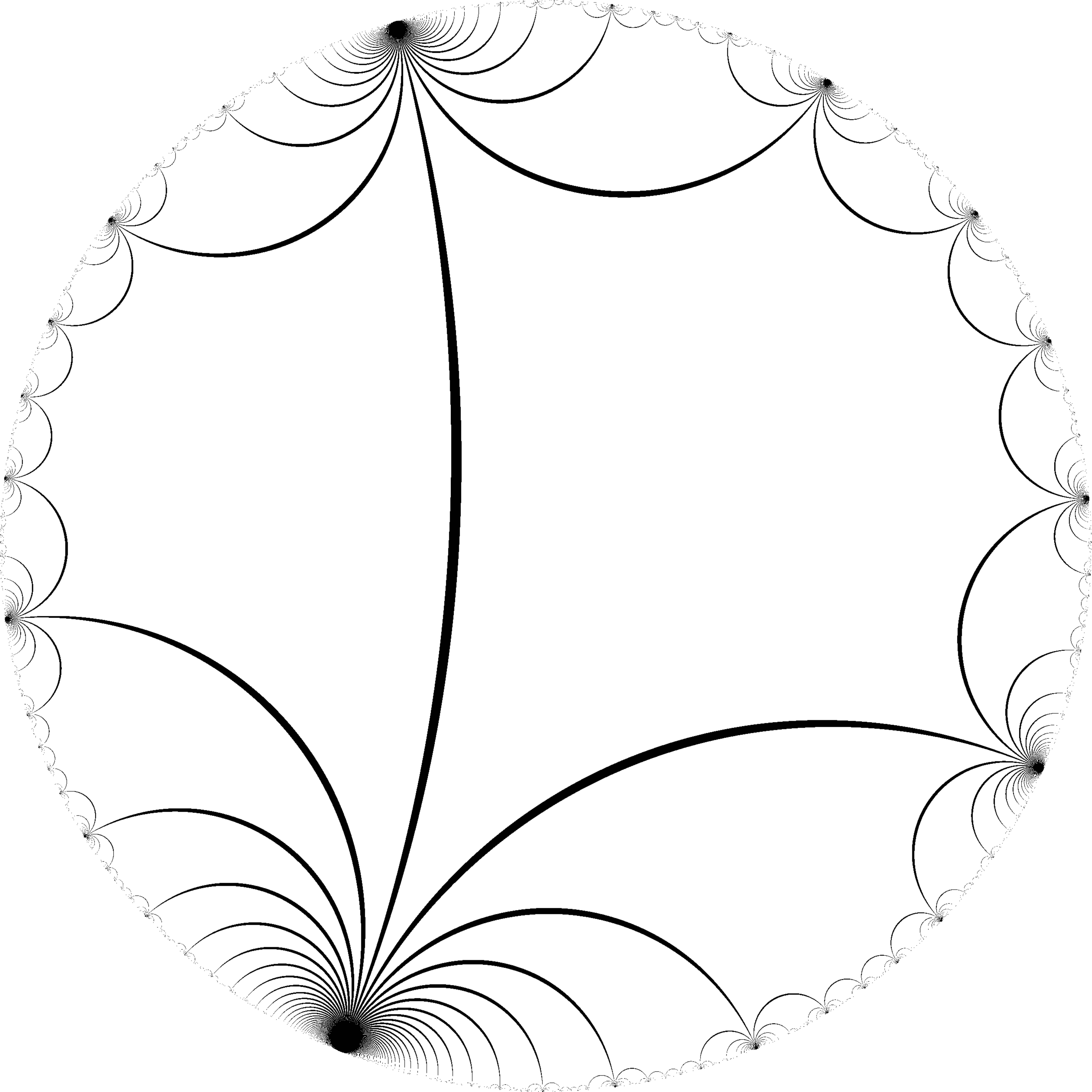} &
\includegraphics[width=.28\linewidth]{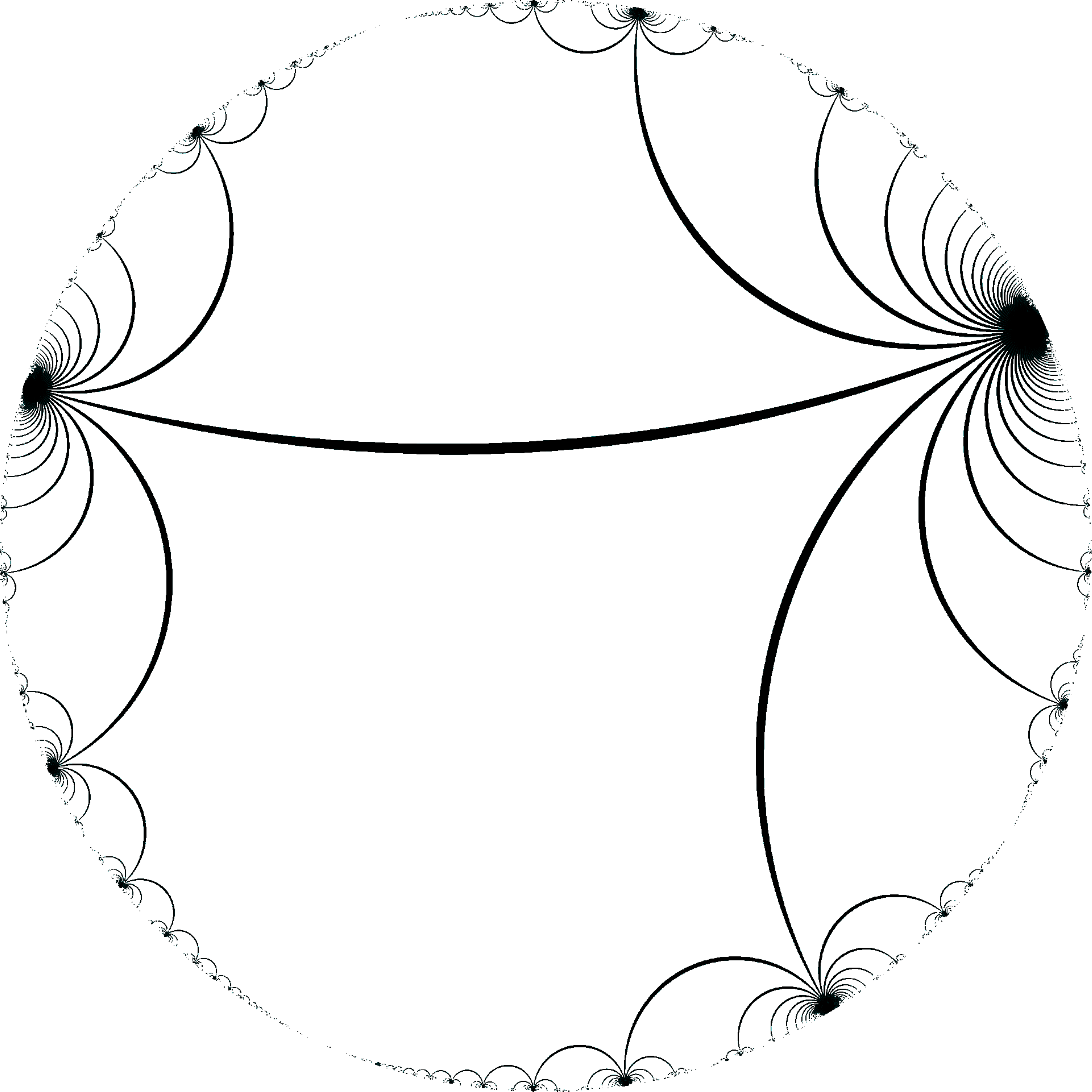}
\end{tabular}
\end{center}
  \caption{The Poincar\'{e} representation of the asymptotic hyperbolic lattices
($\infty,7$) on the left, ($7,\infty$) in the middle, and ($\infty,\infty$) on the right.}
  \label{fig:55}
\end{figure}

Let us investigate the phase transitions of the Ising model on the {\it asymptotic}
lattice geometries, as illustrated in Fig.~\ref{fig:55}. In our earlier studies, two
distinct scenarios have occurred: (1) the coordination number was fixed to $q=4$ while
the $p$-gons gradually expanded $p=4,5,6,\dots,\infty$ and (2) we formed the triangular
tessellation, $p=3$, and the coordination number varied $q=6,7,8,\dots,\infty$. In both
of the cases, a substantially different asymptotic behavior of the phase transition
temperatures was found if either $p$ or $q$ can vary, respectively~\cite{hCTMRGp4,hCTMRG3q}.
In the former case, the phase transition temperature converges to the Bethe lattice
phase transition $T_{\rm pt}^{(\infty,4)}\to\frac{2}{\ln{2}}$.

In the latter case, the triangular tessellation of the lattice types ($3,q\geq6$) leads
to a linear divergence of the phase transition temperature when the coordination
number increases $T_{\rm pt}^{(3,q)}\propto q$. These achievements also remain valid
for arbitrary ($p,q$) lattices when fixing $p$. As another example, we selected the
following infinite set of the hyperbolic lattices ($7,q$) and ($p,7$) with
$p,q=4,5,6,\dots,\infty$, as depicted on the top graph and its inset in Fig.~\ref{fig:55}.

\begin{figure}[tb]
\begin{center}
\includegraphics[width=3.8in]{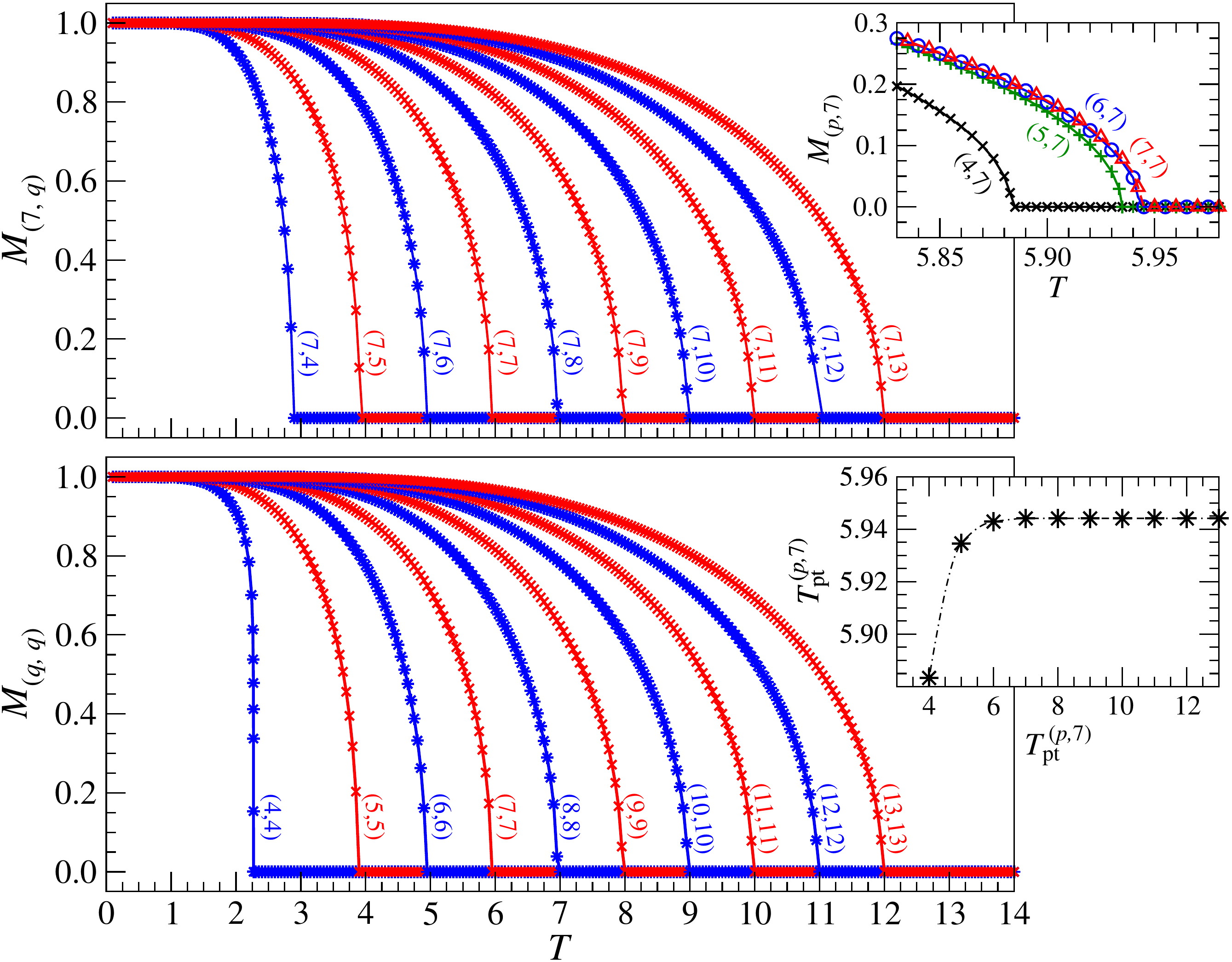}
\end{center}
  \caption{The temperature dependence of the spontaneous magnetization
with respect to $p$ or $q$. The upper and the lower graphs describe the
linear increase of the phase transition temperatures on the lattices ($7,q$)
for $q\geq4$ and ($q,q$) for $q\geq5$, respectively. The two insets show the
fast convergence of the magnetization if fixing $q$ and varying $p$. Notice
that the Ising model on the lattice ($7,7$) exhibits almost identical behavior
of the magnetization on the scale shown if compared with the consequent set of
the lattices from ($8,7$) towards ($\infty,7$).}
\label{fig:56}
\end{figure}

The top graph in Fig.~\ref{fig:56} shows the spontaneous magnetizations $M_{(7,q)}$
on the septagonal lattices for the coordination numbers $q=4,5,6,\dots,\infty$.
The phase transition temperatures, $T_{\rm pt}^{(7,q)}$ tends to grow linearly
with $q$. If generalized, we have always observed (not shown) the linear asymptotic
divergence of the phase transition on the hyperbolic lattices
\begin{equation}
T_{\rm pt}^{(p,q\gg4)}\propto q
\label{Tpt_q}
\end{equation}
irrespective of $p$. When both of the lattice parameters are set to be equivalent,
$p\equiv q$, the effect of the coordination number $q$ prevails over $p$-gonal
feature. The bottom graph in Fig.~\ref{fig:56} depicts the case of the ($q,q$)
lattices for $q=4,5,6,\dots,13$, which, excluding the case $q=4$, also satisfies
the linearity $T_{\rm pt}^{(q,q)}\propto q$ and suppresses the $p$ dependence.

The two insets in Fig.~\ref{fig:56} show the fast convergence of the magnetization
profiles (including the phase-transition temperatures $T_{\rm pt}^{(p,7)}$) toward
the Bethe lattice ($\infty,7$) with the coordination number seven. The `fast'
convergence means that the magnetization profile and the phase-transition
temperature on the ($7,7$) lattice are indistinguishable from those on the
($p>7,7$) lattices (on the scales in the graphs). In particular, we have obtained
the asymptotic phase-transition temperature $T_{\rm pt}^{(p\to\infty,7)}\to5.944002$,
which is in accurate agreement with the Bethe lattice phase-transition
temperature~\cite{Baxter}
\begin{equation}
\lim\limits_{p\to\infty}T_{\rm pt}^{(p,q)}=\frac{1}{\ln\sqrt{\left[{q/(q-2)}\right]}}.
\label{Bethe_q}
\end{equation}

\begin{figure}[tb]
\begin{center}
\centering\includegraphics[width=3.8in]{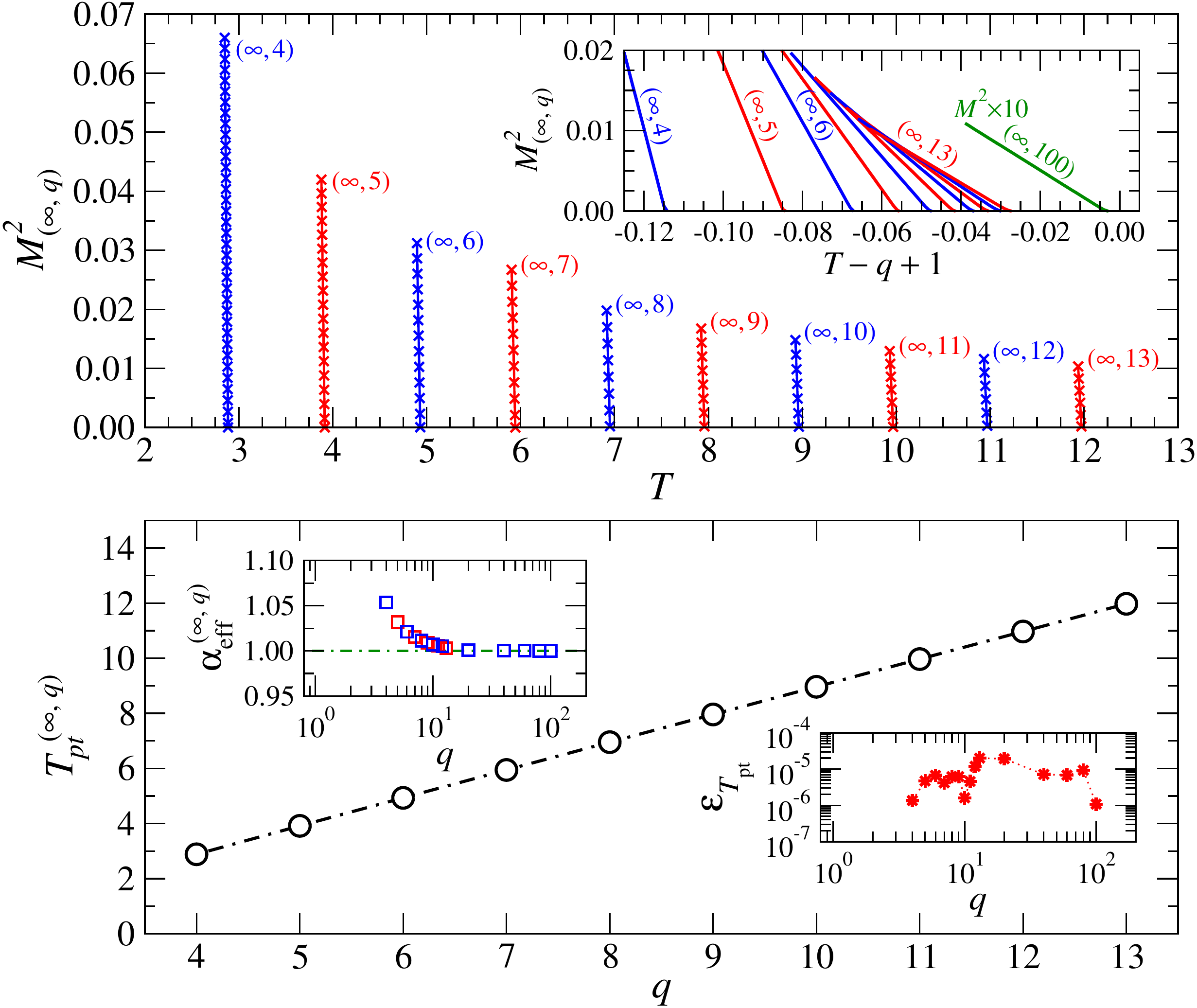}
\end{center}
  \caption{The upper graph shows the temperature dependence of a few nonzero
values of the squared magnetization approaching the phase transition from the ordered phase
$T\leq T_{\rm pt}^{(\infty,q)}$ for the lattices $p=20$ and $4\leq q\leq13$ obtained by CTMRG
which accurately reproduce the Bethe lattice. The inset shows the same data in details
rescaled to temperatures $T-q+1$. In the lower graph, the linearity of $T_{\rm pt}^{
(\infty,q)}$ on the Bethe lattice is satisfied with increasing $q$. The upper left 
inset shows the convergence of the effective exponent $\alpha_{\rm eff}^{(\infty,q)}\to1$
in the log-linear scale. The lower right inset displays numerical accuracy by evaluating
the relative error evaluated at the phase transition temperature on the Bethe lattices
($\infty,q$) in the interval $4\leq q \leq 100$.}
\label{fig:57}
\end{figure}

The mean-field universality is induced by the hyperbolic geometry, which can be spanned
in the infinite-dimensional Hausdorff space and in the thermodynamic limit only. In order
to examine the asymptotics of the lattice geometries shown in Fig.~\ref{fig:55}, we
start by considering the Ising model on the Bethe lattices ($\infty,q$). The linear
decrease of the squared order parameters $M^{2}_{(\infty,q)} \propto
\left(T_{\rm pt}^{(\infty,q)}-T\right)$ toward the phase transition points is plotted
on the upper graph in Fig.~\ref{fig:57} for $q=4,5,6,\dots,13$ and confirms the mean-field
nature. We keep only a few data points right below the phase transition temperature (and
neglect the magnetization data at low temperatures and those in the disordered phase
with $M=0$ for better visibility). The linearity of magnetization thus confirms the
mean-field exponent $\beta=\frac{1}{2}$. 

Now, we focus on the linear divergence of $T_{\rm pt}^{(\infty,q)}\propto q$ in detail.
In the asymptotic regime for the Ising model on the Bethe lattice (for $q\gg4$), it
is evident that
\begin{equation}
T_{\rm pt}^{(\infty,q\gg4)} \to q-1 \, ,
\label{Bethe_asympt}
\end{equation}
where we have made use of
\begin{equation}
T_{\rm pt}^{(\infty,q)} = \frac{1}{\ln\sqrt{\frac{q}{q-2}}}
\equiv \frac{1}{{\rm arctanh}\left(\frac{1}{q-1}\right)} \approx q-1.
\end{equation}
The inset of the upper graph in Fig.~\ref{fig:57} enhances the asymptotic behavior
of $M^{2}_{(\infty,q)}$ with respect to the rescaled horizontal axis $T-(q-1)$. The
data in the inset satisfy the limit in Eq.~\eqref{Bethe_asympt}. We also included
the data of the Ising model on the lattice geometry ($\infty,100$) to see the
tendency of reaching the asymptotic geometry ($\infty,\infty$), as shown in the
inset.

The numerical data at the phase transition can be verified by specifying the
linear dependence of the transition temperatures $T_{\rm pt}^{(\infty,q)}$ on $q$.
In particular, let us assume a $q$-dependent effective exponent
$\alpha_{\rm eff}^{(\infty,q)}$
\begin{equation}
T_{\rm pt}^{(\infty,q)} \propto q^{\alpha_{\rm eff}^{(\infty,q)}} - 1
\label{alpha}
\end{equation}
shown on the bottom graph in Fig.~\ref{fig:57}. The dependence of
$\alpha_{\rm eff}^{(\infty,q)}\to 1$ on $q$ is depicted in the left inset, where
we included additional data with the coordination numbers $q=20,40,60,80,$ and $100$.
It is obvious that the effective exponent converges to $1$, as $q$ grows.
The phase-transition temperatures of the Ising model on the Bethe lattices also
reach a sufficiently high numerical accuracy, see Eq.~\eqref{Bethe_q}.
This accuracy can be visualize by calculating the relative error being as small as
$\varepsilon^{~}_{T_{\rm pt}}\approx10^{-5}$ if calculated at the phase transition
temperature $T_{\rm pt}^{(\infty,q)}$. The relative error is depicted on the right
inset of the bottom Fig.~\ref{fig:57}. (The inset actually demonstrates the worst
(i.e. the lowest) numerical accuracy, which always occurs at the phase transitions.)

Up to this point we have numerically verified the correctness of the recurrence
relations by evaluating and comparing the phase transition temperatures with the
exact solutions on the Bethe lattices. We now proceed with the derivation of the
free energy per spin site as a function of the lattice geometry ($p,q$).

\subsection{Free energy calculation}\label{free_energy_calc_section}

We define the free energy normalized per spin site in order to avoid its divergence
in the thermodynamic limit. The free energy per site as a function of the
iteration step $k$ has the standard expression (recalling that $k_{\rm B}=1$)
\begin{equation}
{\cal F}_{(p,q)}^{[k]} = -\frac{T}{{\cal N}_{(p,q)}^{[k]}}
\ln{\cal Z}_{(p,q)}^{[k]}
\equiv -\frac{T\ln\left[{\rm Tr}\ {\left({\cal C}_k\right)}^q\right]}
   {{\cal N}_{(p,q)}^{[k]}}\, ,
\label{fe}
\end{equation}
The normalization of the free energy per spin site requires to calculate an integer
function ${\cal N}_{(p,q)}^{[k]}$, which counts the total number of the spin sites
at give iteration step $k$ for a particular lattice geometry ($p,q$). The free energy
per site plays a crucial role in deriving all the thermodynamic functions in order
to determine phase transitions accurately. The free energy also involves the effects
originating from the lattice boundaries, which are can be suppressed if the bulk
properties are measured from the mean values $\langle A \rangle = {\rm Tr}(A\rho)$.
The typical example is the spontaneous magnetization (spin polarization) at $A=\sigma$.

If the calculation of the free energy per site in Eq.~\eqref{fe} is carried out directly,
an extremely fast divergence of the partition function ${\cal F}_{(p,q)}^{[k]}$ and the
total number of sites ${\cal N}_{(p,q)}^{[k]}$ is observed on hyperbolic lattices. This
usually happens already at $k\gtrsim 10$. Therefore, numerical operations on the tensors
${\cal C}_k$ and ${\cal T}_k$ require their normalization in each iteration step $k$.
We apply the normalization we have introduced in Eq.~\eqref{norm_CT}, where the
norm is defined as the absolute value of the largest tensor element.

The way of how the normalization constants enter the free-energy derivation on the
Euclidean square lattice ($4,4$) has been given in Section~\ref{CTMRG_Section}, cf.
Eqs.~\eqref{c3_decomp_eqn}--\eqref{fe44}. In the following, we extend this procedure
of the free-energy analysis and first consider hyperbolic lattice ($5,4$) only.
Afterwards, we generalize this procedure for an arbitrary ($p,q$) lattice geometry.

\subsubsection{Free energy on (5,4) lattice}

\begin{figure}[tb]
\begin{center}
\includegraphics[width=.65\linewidth]{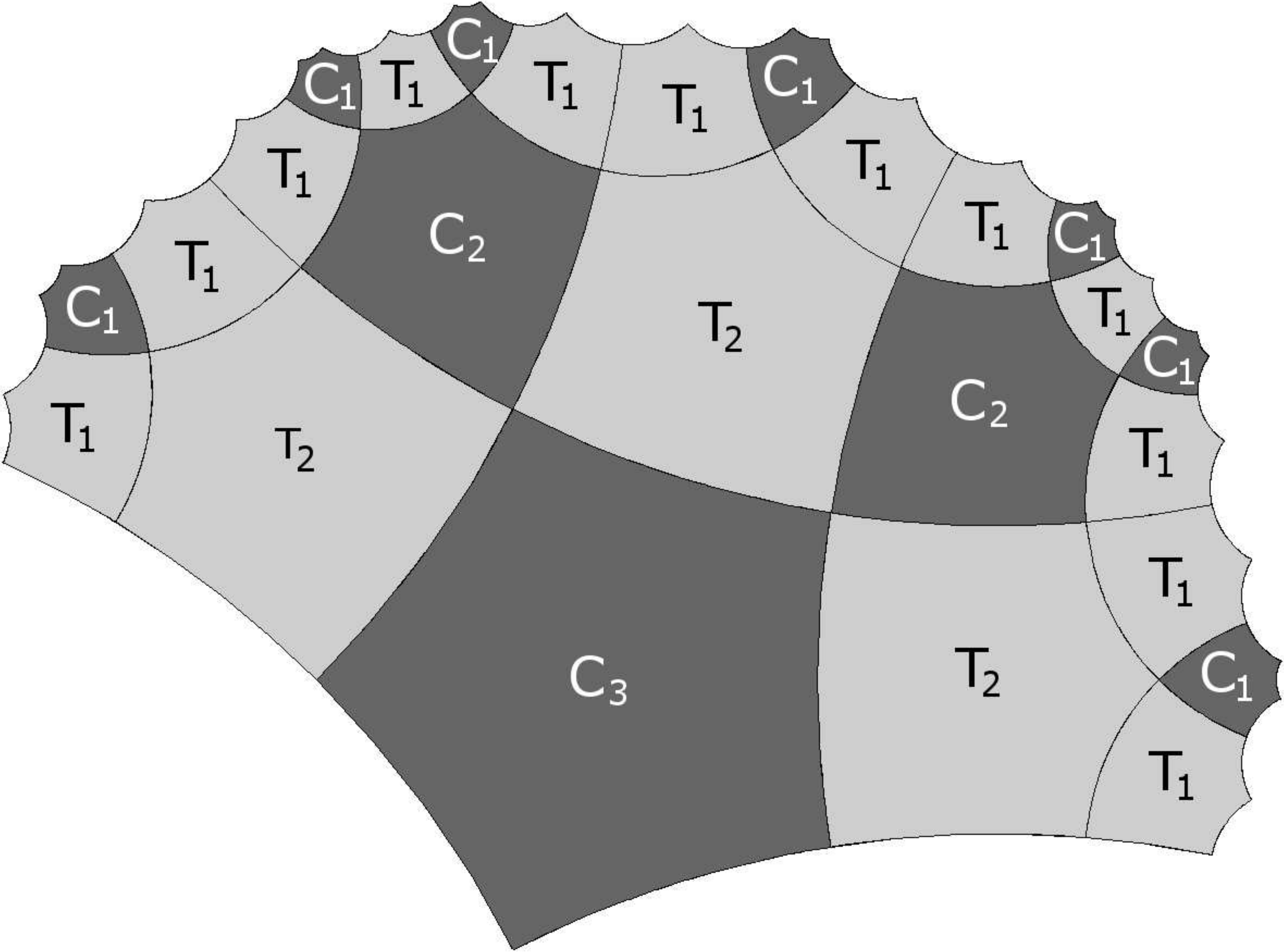}
\end{center}
  \caption{The graphical representation of the corner transfer tensor ${\cal C}_3$ 
  for the hyperbolic lattice ($5,4$) on the in accord with \eqref{rr54} for $k=3$. We
use the dark-gray and bright-gray $p$-gons to distinguish amongst ${\cal C}_k$ and
${\cal T}_k$, respectively.}
  \label{fig:58}
\end{figure}

As we have shown the case of the square lattice ($4,4$), the formalism for the
recurrence relations on the lattice geometry ($5,4$) can be also graphically
visualized. It is instructive to express the structure of the normalized corner
tensor ${\widetilde{\cal C}}_3$ according to structure in Fig.~\ref{fig:58},
which is in agreement with the recurrence relations in Eqs.~\eqref{rr54}. The
normalized tensor ${\widetilde{\cal C}}_3$ can decomposed into the product of
the Boltzmann weights and the related normalization constants $c_j$ and $t_j$
(for $j=1,2,3$) as follows
\begin{equation}
\begin{split}
{\widetilde{\cal C}}_3^{~} & = \frac{{\cal C}_3^{~}}{c_3^{~}}
 = \frac{{\cal W}_{\rm B}^{~}{\widetilde{\cal T}}_2^{3}{\widetilde{\cal C}}_2^{2}}
          {c_3^{~}}
 =\frac{{\cal W}_{\rm B}^{~}{\cal T}_2^{3}{\cal C}_2^{2}}
             {t_2^3 c_3^{~}c_2^{2}} \\
& = \frac{{\cal W}_{\rm B}^{~}{\left({\cal W}_{\rm B}^{~}{\widetilde{\cal T}}_1^{2}
{\widetilde{\cal C}}_1^{~}\right)}^3
{\left({\cal W}_{\rm B}^{~}{\widetilde{\cal T}}_1^{3}{\widetilde{\cal C}}_1^{2}\right)}^2}
{t_2^3 c_2^{2} c_3^{~}}\\
& = \frac{{\cal W}_{\rm B}^6{\cal T}_1^{12}{\cal C}_1^{7}}
{t_1^{12} t_2^{3} c_1^{7}c_2^{2}c_3^{~}}
 = \frac{{\cal W}_{\rm B}^{25}}
{t_1^{12} t_2^{3} t_3^0 c_1^{7}c_2^{2}c_3^{1}}\, .
\end{split}
\label{C_norm_54}
\end{equation}
Evidently, there are $25$ pentagonal Boltzmann weights ${\cal W}_{\rm B}$ on the
lattice ($5,4$), as depicted in Fig.~\ref{fig:58}. The powers associated with the
normalization factors $c_j$ and $t_j$ coincide with the the tensors ${\cal C}_j$
and ${\cal T}_j$.

Since the powers associated with the normalization factors $c_j$ and $t_j$ are
important in the analysis, we denote them by the integers $n_{k-j+1}$ and $m_{k-j+1}$,
respectively. The integer powers are indexed in the reverse ordering for later
convenience (because we intend to rewrite these expressions in a simplified form).
It means the integer powers, appearing in the denominator of Eq.~\eqref{C_norm_54},
have to satisfy the following reverse index ordering $t_1^{m_3} t_2^{m_2} t_3^{m_1}
c_1^{n_3}c_2^{n_2}c_3^{n_1}$. We use them for computing the total number of the
spins ${\cal N}_{(5,4)}^{[k]}$. After some algebraic calculations, one can derive
a general formula for the pentagonal lattice ($5,4$) at arbitrary step $k$
\begin{equation}
{\cal N}_{(5,4)}^{[k]} = 1 + 4 \sum\limits_{j=1}^{k} 3 n_j + 2 m_j \, .
\label{N54}
\end{equation}
In addition, the integer exponents $n_j$ and $m_j$ also satisfy the recurrence relations
\begin{equation}
\begin{split}
n_{j+1}&=2n_{j}+m_{j}\, , \\
m_{j+1}&=3n_{j}+2m_{j}\, , \\
n_{1}&=1\, , \\
m_{1}&=0\, .
\end{split}
\end{equation}
The entire lattice ($5,4$) is made by tiling the four corner tensors ${\cal C}_k$ around
the central spin (cf. Fig.~\ref{fig:52}). Hence, the number $1$ and the prefactor
$4$ (in front of the summation) in Eq.~\eqref{N54}, respectively, correspond to the
central spin and the four joining tensors ($q=4$). The other two prefactors $3$
and $2$ under the summation in Eq.~\eqref{N54} count those spins in each corner
${\cal C}_k$ (composed recursively from ${\cal C}_{k-1}$ and ${\cal T}_{k-1}$, etc.),
which are not shared. A bit deeper algebraic analysis of the ($5,4$) lattice is inevitable
to understand all the details.

\begin{figure}[tb]
\begin{center}
\includegraphics[width=3.8in]{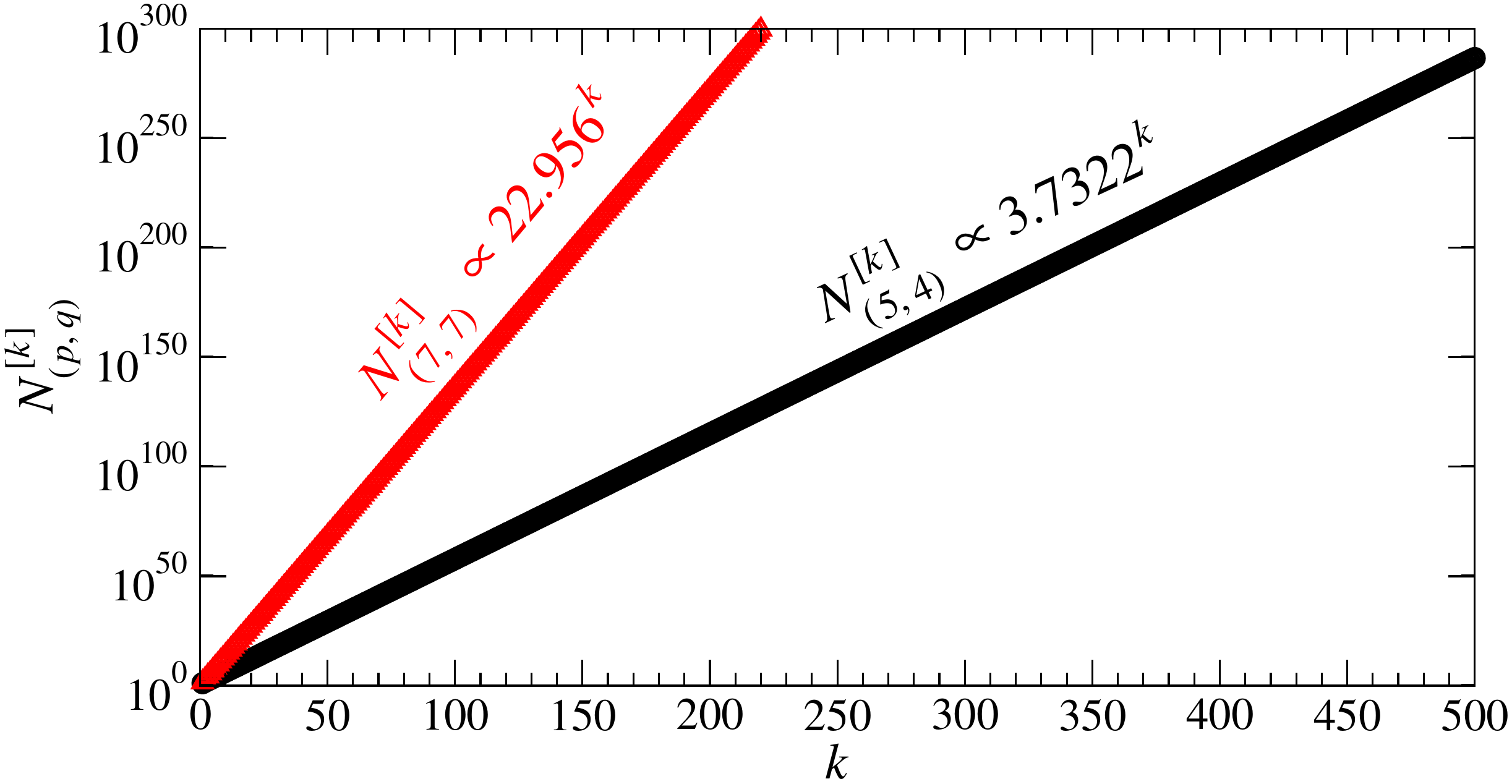}
\end{center}
  \caption{The exponential dependence of the total number of the spins
on the iteration number $k$ (the least-square fitting in the log-log plot) for the two
lattices ($5,4$) and ($7,7$).}
\label{fig:59}
\end{figure}

Finally, the free energy per site at given iteration step $k$ has the following
expression on the hyperbolic lattice ($5,4$)
\begin{equation}
{\cal F}_{(5,4)}^{[k]} =- \frac{4T\ln{\rm Tr}\ \widetilde{C}_{k}^{~}}
      {{\cal N}_{(5,4)}^{[k]}}
-\frac{4T}{{\cal N}_{(5,4)}^{[k]}}
   \sum\limits_{j=0}^{k-1} \left( \ln c_{k-j}^{n_{j+1}} + \ln t_{k-j}^{m_{j+1}} \right) \, .
\label{FE54}
\end{equation}
The first term converges to zero as $k$ increases
\begin{equation}
\lim\limits_{k\to\infty} \frac{4T\ln{\rm Tr}\ \widetilde{C}_{k}^{~}}
                              {{\cal N}_{(5,4)}^{[k]}}  =  0 \, ,
\end{equation}
because the normalized partition function $\widetilde{\cal Z}_{(p,q)}^{[\infty]}
\equiv {\rm Tr}\ \widetilde{C}_{\infty}^{~}$ in the numerator of the first term
is always bounded from both sides and ${\cal N}_{(5,4)}^{[k]}$ increases. At any
temperature $T$ and in the thermodynamic limit $k\to\infty$, we have
\begin{equation}
1 \leq \widetilde{\cal Z}_{(p,q)}^{[k\to\infty]} \leq M \, ,
\label{FElim}
\end{equation}
where $Q$ denotes the $Q$-state spin system. In particular, the lower and the upper
bounds, respectively, correspond to the limits
\begin{equation}
\lim_{T\to 0}\widetilde{\cal Z}_{(p,q)}^{[k\to\infty]} = 1
\end{equation}
and
\begin{equation}
\lim_{T\to\infty}\widetilde{\cal Z}_{(p,q)}^{[k\to\infty]} = M.
\end{equation}
The number of spin sites in the denominator of the first term grows exponentially.
If calculating this term numerically, we obtain ${\cal N}_{(5,4)}^{[k]} \propto 3.7^k$
by least-square fitting, as plotted in Fig.~\ref{fig:59}.

\subsubsection{Free energy on ({$p,q$}) lattices}

The generalization of the free-energy calculation for the $Q$-state spin models
on arbitrary lattice geometries ($p,q$) requires a careful analysis via graphical
representation at finite $k$. This analysis is beyond the scope of this work for
its extensive considerations. The $k$-dependence of the free energy per spin is
\begin{equation}
\begin{split}
{\cal F}_{(p,q)}^{[k]} & = -\frac{qT\ln{\rm Tr}\ \widetilde{C}_{k}^{~}}
      {{\cal N}_{(p,q)}^{[k]}}
-\frac{qT\sum\limits_{j=0}^{k-1} \left( \ln c_{k-j}^{n_{j+1}} + \ln t_{k-j}^{m_{j+1}} \right) }
      {{\cal N}_{(p,q)}^{[k]}} \\
& \overset{k\gg1}{=} -\frac{qT}{{\cal N}_{(p,q)}^{[k]}} 
  \sum\limits_{j=0}^{k-1} \left( n_{j+1} \ln c_{k-j}^{~}+m_{j+1}\ln t_{k-j}^{~} \right) \, ,
\end{split}
\label{FEpq}
\end{equation}
where the total number of the spins is expressed as
\begin{equation}
{\cal N}_{(p,q)}^{[k]} = 1 + q \sum\limits_{j=1}^{k} (p-2) n_j + (p-3) m_j\, ,
\label{Npq}
\end{equation}
and the integer variables $n_j$ and $m_j$ satisfy these recurrence relations
\begin{equation}
\begin{split}
n_{j+1} = &\, [(p-2)(q-3)-1]n_{j} + [(p-3)(q-3)-1]m_{j}\, ,\\
m_{j+1} = &\, (p-2)n_{j}+(p-3)m_{j}\, ,\\
 n_{1} = & \, 1\, ,\\
 m_{1} = & \ 0.
\end{split}
\label{nmpq}
\end{equation}
The evaluation of Eq.~\eqref{Npq} is carried out numerically with strong exponential
dependence on $k$ which is proportional to the increase of $p$ and $q$. As an example,
Fig.~\ref{fig:59} displays the log-log plot of this exponential dependence of the total
number of the spins showing that ${\cal N}_{(7,7)}^{[k]}>{\cal N}_{(5,4)}^{[k]}$.

The final expressions of the free energy in Eqs.~\eqref{FEpq}--\eqref{nmpq} also include
the case of the Euclidean lattice ($4,4$). The complete equivalence with Eq.~\eqref{FE44}
is easily verifiable if assuming that $n_j=n_{j-1}=\cdots=n_1\equiv1$ and $m_j=2n_{j-1}
+m_{j-1}=2(j-1)+m_1\equiv 2j-2$, which reduces the exponential dependence of the total
number of the spins towards the power-law in Eq.~\eqref{N44}
\begin{equation}
{\cal N}_{(4,4)}^{[k]} = 1 + 4 \sum\limits_{j=1}^{k} 2 n_j + m_j = {(2k+1)}^2_{~}\, .
\end{equation}

\subsection{Results}\label{num_reslts_section}

Let us analyze the phase transition of spin models in the thermodynamic limit on the
following four representative lattices: ($4,4$), ($4,7$), ($7,4$), and ($7,7$) we have
used earlier. We have shown that the phase transition temperatures $T_{\rm pt}$,
calculated by the spontaneous magnetization $M_{(p,q)}$ at the lattice center,
correctly reflects the bulk properties since the boundary effects are eliminated.
In other words, if various types of the boundary conditions (such as free and fixed
ones) are imposed, the phase transition of spin models is not affected, provided
that we evaluated the expectation value $\langle \sigma_c\rangle$ by Eq.~\eqref{Mpq}.
Its correctness with high numerical accuracy has been compared with the exactly
solvable Ising model on Bethe lattice~\cite{Baxter}.

\begin{figure}[tb]
\begin{center}
\includegraphics[width=3.8in]{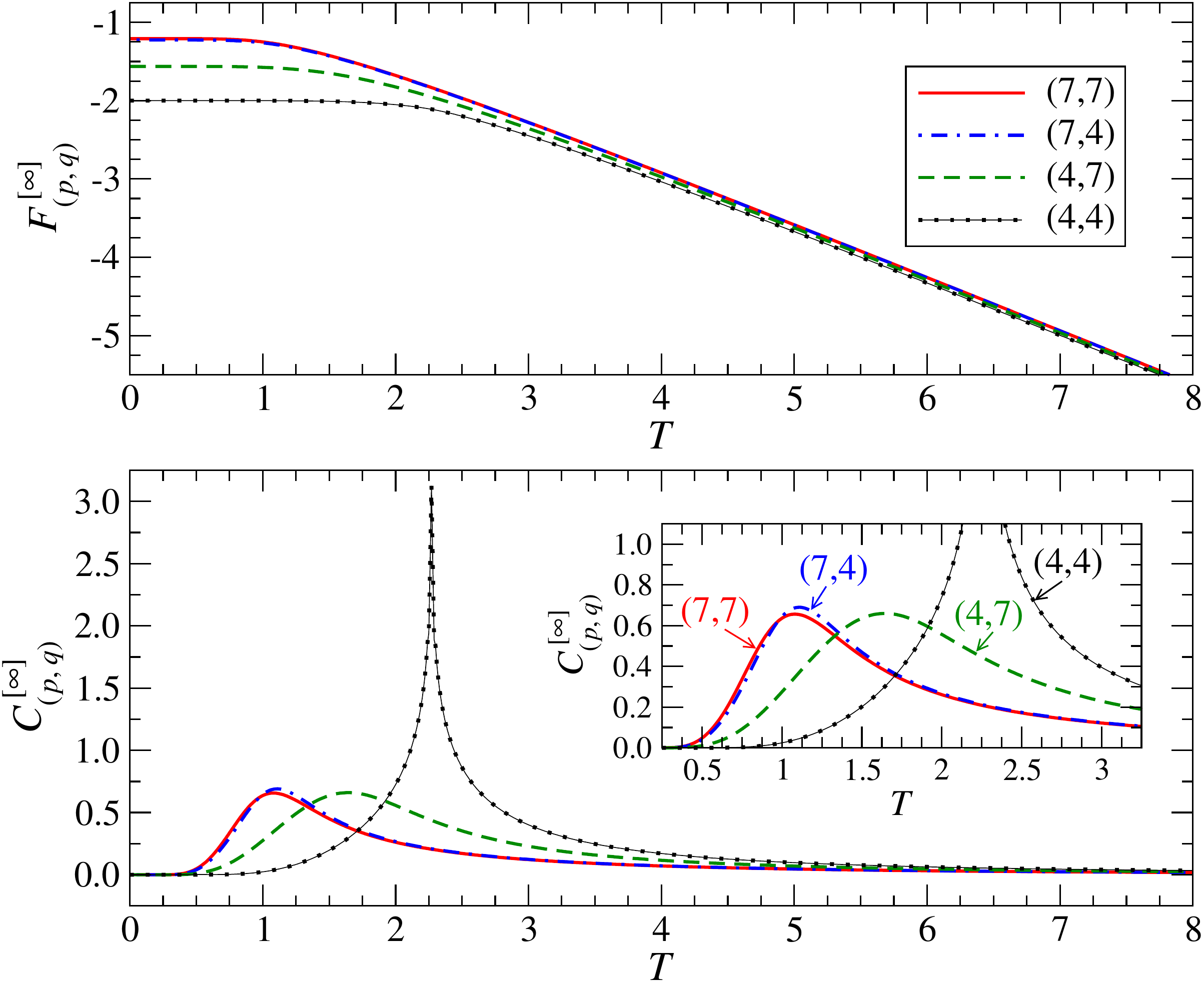}
\end{center}
  \caption{The free energy (top) obtained from Eq.~\eqref{FEpq} and the specific heat
(bottom) from Eq.~\eqref{Cpq} vs temperature on the selected lattices ($4,4$), ($4,7$),
($7,4$), and ($7,7$). The inset shows the details in the vicinity of the broadened
specific heat maxima.}
\label{fig:60}
\end{figure}

\subsubsection{Suppression of phase transitions}

The upper graph of Fig.~\ref{fig:60} shows the free energies on four representative
lattices, which is evaluated by Eq.~\eqref{FEpq}. The phase transition is associated with
a singular (non-analytic) behavior of the specific heat being the second derivative of
the free energy with respect to temperature
\begin{equation}
C_{(p,q)}^{[\infty]} = -T\frac{\partial^2}{\partial T^2}
{\cal F}_{(p,q)}^{[\infty]}\, .
\label{Cpq}
\end{equation}
The temperature dependence of the specific heat of the Ising model on the respective
four lattices is plotted on the lower graph in Fig.~\ref{fig:60}. Evidently, we find
the non-analytic behavior on the square lattice ($4,4$) with the diverging peak
at the temperature, which corresponds to the exact critical temperature of the
Ising model, in which $T_{\rm c} = 2/\ln(1+\sqrt{2})$.

However, none of the three hyperbolic lattice geometries results in an analogous
sharp (non-analytic) peak at the phase transition temperatures $T_{\rm pt}^{(p,q)}$
we had calculated from the spontaneous magnetization plotted in Fig.~\ref{fig:54}.
Instead, rather broader maxima are present for the particular lattices. The maxima
do not correspond to the correct phase transition temperatures we had detected from
the bulk properties by the spontaneous magnetization.

The strong boundary effects on the hyperbolic lattices also prevent Monte Carlo (MC)
simulations from accurate determination of the phase transition on the hyperbolic
lattices~\cite{MC1,MC2,MC3,MC4}. Necessity to provide a correction by means of
`subtraction' of a couple of boundary spin layers can be helpful in order to detect
the correct bulk properties~\cite{boundary}.
The non-negligible contribution from the boundaries is obvious if defining a ratio
between the number of the boundary sites and the number of the remaining (inner) sites.
The ratio converges to zero in the Euclidean case only, whereas it gets a nonzero value
on the hyperbolic lattices (including the thermodynamic limit).

\subsubsection{Bulk Free Energy}

If we intend to specify the phase transition correctly on the hyperbolic lattices,
the free energy needs to be redefined by eliminating contributions from the boundary
layers coming from the total free energy. We have shown how the iteration steps $k$
in the CTMRG algorithm expands the lattice size. As $k$ increases, the boundary sites
are pushed farther from the lattice center. This expansion process can be regarded as
adding a spin layer (or a shell) at a step $k+1$. The boundary spins are always composed
of the initial tensors ${\cal C}_{1}$ and ${\cal T}_{1}$. The tensors thus multiply
themselves exponentially, while the new $q$ tensors ${\cal C}_{k+1}$ are included
in the center of the lattice, cf. Fig.~\ref{fig:58}. Such a hyperbolic lattice
can be thought of as a system of concentric shells indexed by $j$ so that the
$j^{\rm th}$ shell contains the spin sites, which separate the other spin sites in the
tensors ${\cal C}_j$ and ${\cal T}_j$ from the outer spin sites in ${\cal C}_{j-1}$
and ${\cal T}_{j-1}$ within a given ($p,q$) geometry (cf. Fig.~\ref{fig:58}).
This structure of the concentric shells at the $k^{\rm th}$ iteration step enables
us to enumerate the outermost shell (being $j=1$) toward the innermost shell ($j=k$)
while leaving the central spin site apart. This way of the enumeration is related to
the counting of the total number of the spins by Eq.~\eqref{Npq}.

Let the integer $\ell$ denote number of the outermost shells $j=1,2,\dots,\ell<k$.
We introduce a new quantity, the {\it bulk\,} free energy ${\cal B}_{(p,q)}^{\{k,\ell\}}$,
which defines the free energy on the $k-\ell$ inner shells. This is given by the
subtraction of the free energy contributing from the $\ell$ outermost shells from the
total free energy of the entire system. In particular, the bulk free energy in the
$k^{\rm th}$ iteration step after subtracting $\ell$ outer shells is
\begin{equation}
{\cal B}_{(p,q)}^{\{k,\ell\}} = {\cal F}_{(p,q)}^{[k]} - {\cal F}_{(p,q)}^{\ast\,\{k,\ell\}}\, ,
\end{equation}
where the asterisk in the second term denotes the free energy of the $\ell$ outermost
shells so that
\begin{equation}
	{\cal F}_{(p,q)}^{\ast\,\{k,\ell\}} = 
	-\frac{qT}{{\cal N}_{(p,q)}^{\ast\,\{k,\ell\}}}
\sum\limits_{j=k-\ell}^{k-1} \left[ n_{j+1} \ln c_{k-j}^{~}+m_{j+1}\ln t_{k-j}^{~} \right]
\label{FEpq_b}
\end{equation}
and
\begin{equation}
	{\cal N}_{(p,q)}^{\ast\,\{k,\ell\}} = q \sum\limits_{j=k-\ell+1}^{k}
  \left[ (p-2) n_j + (p-3) m_j \right] \, .
\label{Npq_b}
\end{equation}
For the tutorial purpose, it is reasonable to consider $\ell=\frac{k}{2}$, and to
study effects if taking the thermodynamic limit $k\to\infty$. (The dependence of
the bulk free energy on $\ell$ has been thoroughly studied in Ref.~\cite{Yoju}.)
Following the remarks given below Eq.~\eqref{FEpq} (without loss of generality),
we omit the first term in Eq.~\eqref{FEpq_b}, because this term converges to zero
after a few iterations.

\begin{figure}[tb]
\begin{center}
\includegraphics[width=3.8in]{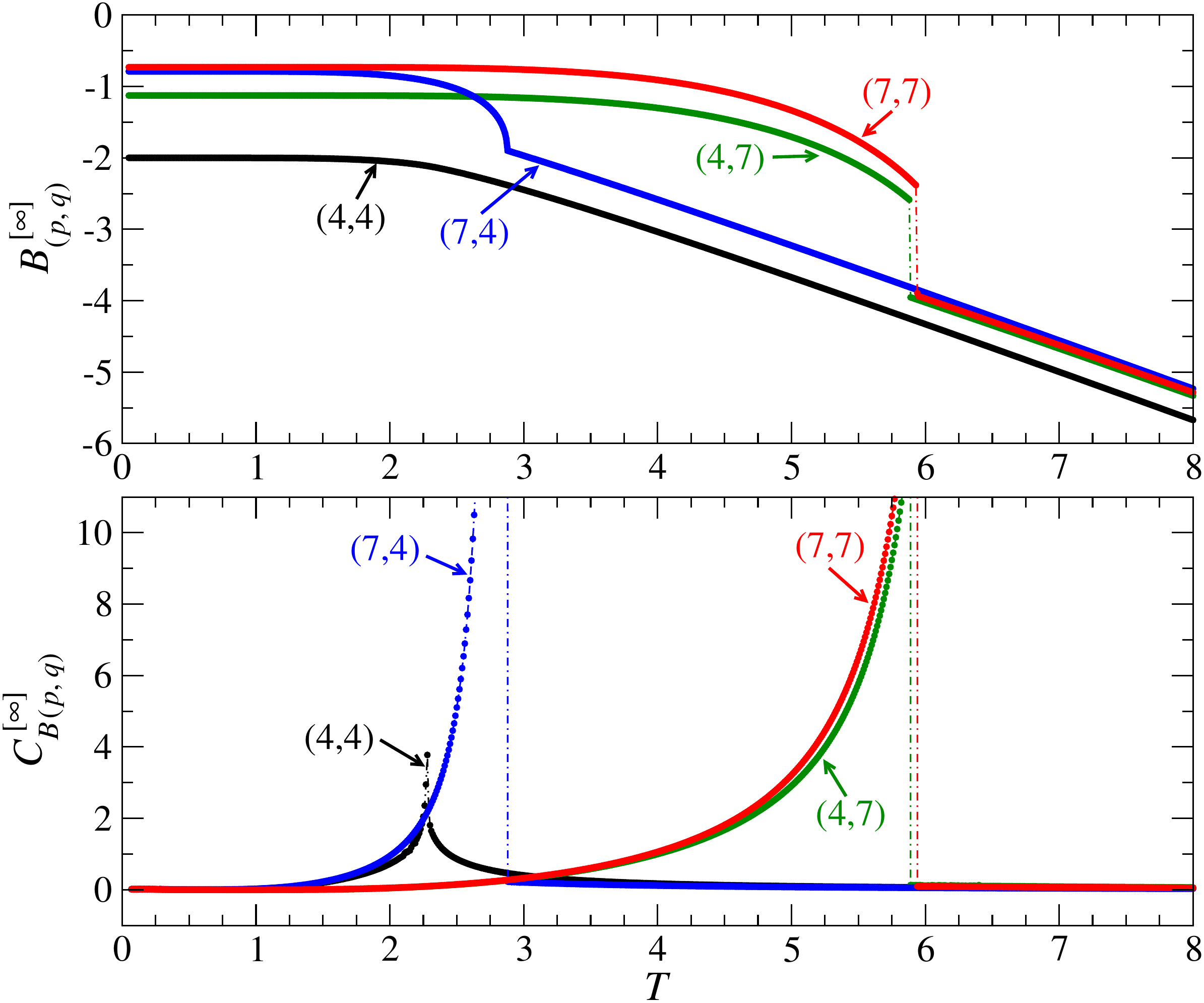}
\end{center}
  \caption{The bulk free energy (the upper graph) and the associated bulk specific
heat (the lower graph) versus temperature for the Ising model on the lattices ($4,4$),
($4,7$), ($7,4$), and ($7,7$). The vertical dot-dashed lines point out the non-analytic
behavior, where the phase transition occurs in the full agreement with the phase transition
we have obtained by the spontaneous magnetization, cf. Fig.~\ref{fig:54}.}
\label{fig:61}
\end{figure}

Figure~\ref{fig:61} (the upper graph) shows the bulk free energy for the Ising model
on the four representative lattices in the thermodynamic limit, i.e.
\begin{equation}
{\cal B}_{(p,q)}^{[\infty]} \equiv
	\lim\limits_{k\to\infty} {\cal B}_{(p,q)}^{\{k,k/2\}}\,.
\end{equation}
In case of the Euclidean lattice, it is not surprising that ${\cal B}_{(4,4)}^{[\infty]}
\equiv {\cal F}_{(4,4)}^{[\infty]}$ since the bulk properties are not affected by the
boundaries. However, the bulk free energy calculated on the hyperbolic lattices exhibits
a remarkable singularity occurring exactly at the phase transition temperature. The
typical non-analytic structure of the free energy does not change irrespective of the
type of the boundary conditions applied (free and fixed ones)~\cite{Yoju}. The maxima
of the bulk specific heat
\begin{equation}
C_{B\,(p,q)}^{[\infty]} = -T\frac{\partial^2}{\partial T^2}
{\cal B}_{(p,q)}^{[\infty]}\, .
\end{equation}
plotted in the lower graph of Fig.~\ref{fig:61} accurately correspond to the phase
transition temperatures $T_{\rm pt}^{(p,q)}$ we have determined in Sec.~II. The
discontinuous jump of the bulk specific heat is associated with the typical mean-field
universality behavior~\cite{hCTMRG3q,hCTMRG3qn}. The vertical dot-dashed lines serve
as guides for the eyes to locate the phase transition temperature. We stress here
that the identical determination of the phase transition temperatures has been
obtained independently by the spontaneous magnetization in Fig.~\ref{fig:54}. 

Our definition of the bulk free energy contains lots of interesting features.
For instance, the $\ell$-dependence enables us to explain the manner of how the
lattice boundary affects the inner bulk part of the lattice. Moreover, the phase
transition can be affected if an additional magnetic field is imposed on the
boundary spins only~\cite{Yoju}. Such a feature of the magnetic field never affects
the thermodynamic properties or the phase transition on Euclidean lattices in the
thermodynamic limit.

We do not follow the Baxter's proposal of calculating the free energy by means of
the numerical integration of the spontaneous magnetization with respect to the
magnetic field for the Bethe lattices~\cite{Baxter}. Although this Baxter's approach
is feasible in our CTMRG analysis, our definition the bulk free energy completely 
reproduces the features of the Bethe lattices as well. Moreover, we can also
study how the magnetic field affects th physics if the field is imposed on the
boundary spins only, which can be studied solely via the bulk free energy.

\subsubsection{Free energy versus lattice geometry}

We have been primarily motivated by the correspondence between the anti-de Sitter spaces
(the hyperbolic geometry) and the conformal field theory within the quantum gravity,
which resulted in a question: {\it "Given an arbitrary spin system on an infinite set
of ($p,q$) geometries, which lattice geometry minimizes the free (ground-state) energy?"}.
This is certainly a non-trivial task to be explained thoroughly. Nevertheless, we
intend to reply the question in the following. It helps us find an insight into
the role of the space geometry with respect to the microscopic description of the
spin interacting system. Although we are currently considering the free energy of the
{\it classical} spin lattice systems, our recent studies of the ground-state
energy of the {\it quantum} spin systems on the lattices ($p,4$) also exhibit
qualitatively similar features, as study in this work~\cite{TPVF54,TPVFp4}.
The free energy for classical spin systems and the ground-state energy of quantum
spin systems are mutually related.

The free energy per site ${\cal F}_{(p,q)}^{[\infty]}$ converges to a negative
value ${\cal F}_{(p,q)}^{[\infty]}<0$ at finite temperatures $T<\infty$ in the
thermodynamic limit. Scanning the entire set of the ($p\geq4,q\geq4$) geometries,
we are going to show in the following that the free energy per site reaches its
minimum on the square lattice ($4,4$) only
\begin{equation}
{\cal F}_{(4,4)}^{[\infty]} = \min_{(p\geq4,q\geq4)}
\left\{{\cal F}_{(p,q)}^{[\infty]}\right\}
\end{equation}
at arbitrary temperature $T$. We, therefore, plot the {\it shifted} free energy per site,
${\cal F}_{(p,q)}^{[\infty]}-{\cal F}_{(4,4)}^{[\infty]}\geq0$ to visualize the figures
better.

\begin{figure}[tb]
\begin{center}
\includegraphics[width=3.8in]{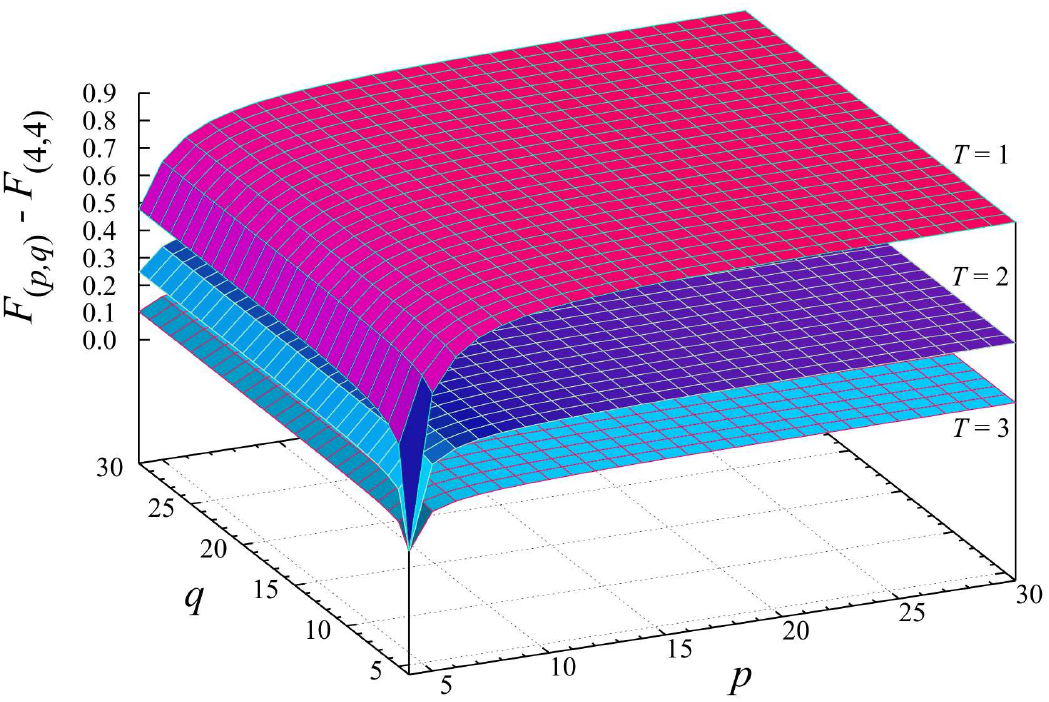}
\end{center}
  \caption{The free energy per site as a function of the lattice geometry
($p,q$) at the selected lower temperatures $T=1,2$, and $3$.}
\label{fig:62}
\end{figure}
\begin{figure}[!tbh]
\begin{center}
\includegraphics[width=3.8in]{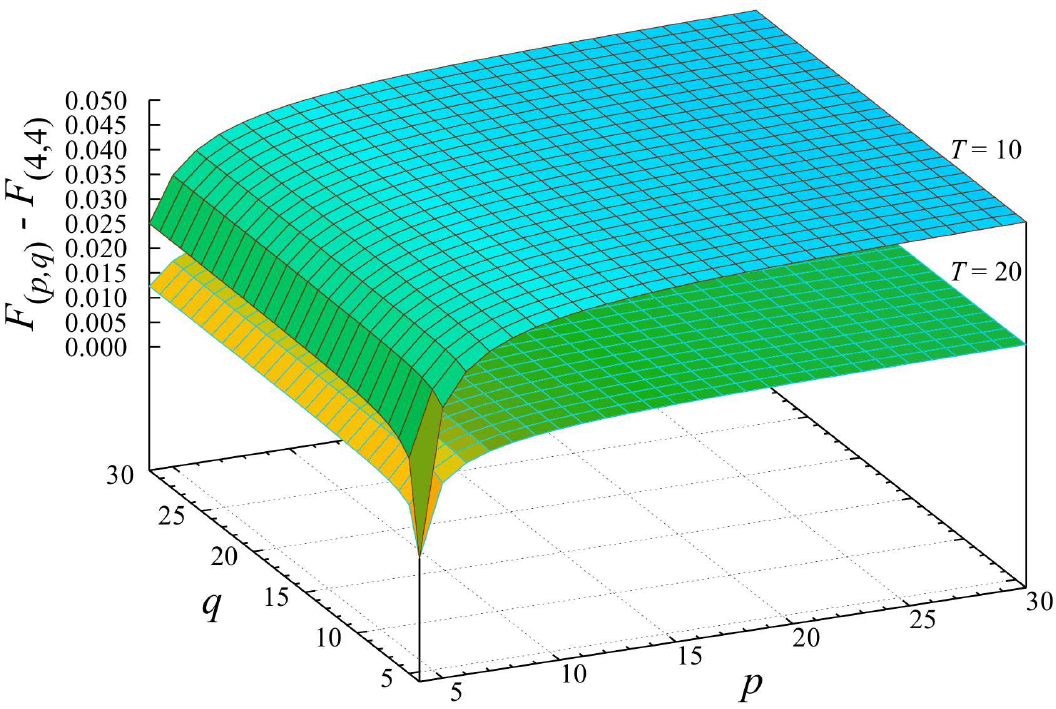}
\end{center}
  \caption{The same as in Fig.~\ref{fig:62} for $T=10$ and $20$.}
\label{fig:63}
\end{figure}

Figures~\ref{fig:62} and \ref{fig:63}, respectively, show the shifted free energy for
the Ising model ($Q=2$) at lower and higher temperatures.
These numerical calculations unambiguously identify the square
lattice geometry as the one, which minimizes the free energy per spin site. The free
energy per site becomes less sensitive for higher values of $p$ and $q$. We observe
a weak increasing tendency in the free energy if $p$ increases while fixing $q$. The
free energy gets saturated to a constant in the opposite case when $q$ increases at
fixed $p$. It is worth to mention that the presence of the phase transition does not
affect the free energy minimum observed. Moreover, as the temperature grows, the
difference between the free energies on the square and the hyperbolic lattices weakens
(Fig.~\ref{fig:63}).

The free energy has specific features, which have to be satisfied for the higher-spin
models, not for the two-state Ising models only. We, therefore, employ the $Q$-state
clock and $Q$-state Potts models we have defined in Eqs.~\eqref{clock}, \eqref{Potts}.
As a simple test, whether the numerical results for the free energy are correct, we
show the high-temperature asymptotics of the free energy for the multi-state spin models.

\begin{figure}[tb]
\begin{center}
\includegraphics[width=3.8in]{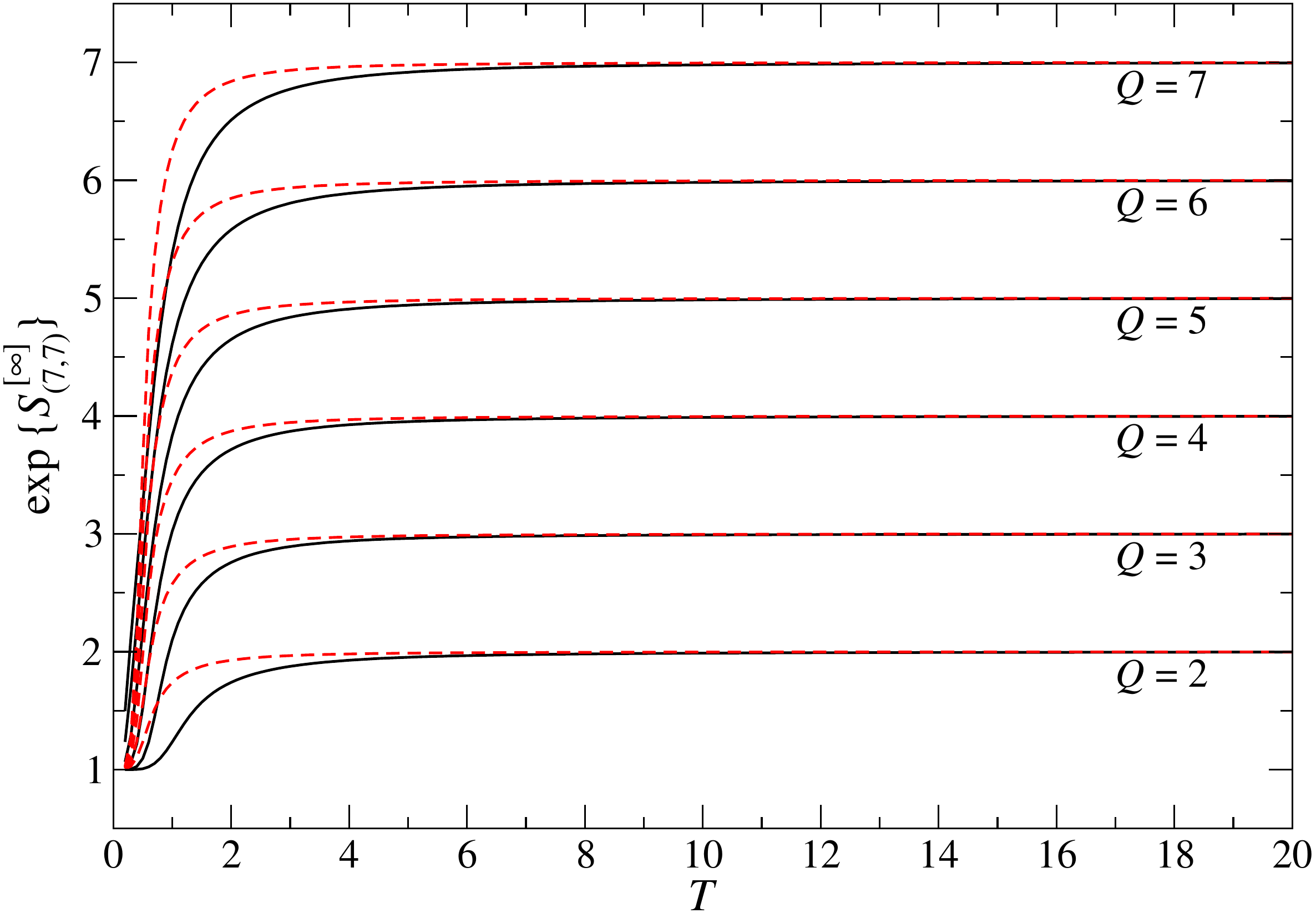}
\end{center}
  \caption{The high-temperature asymptotics of the entropy if applied to the lattice ($7,7$).
The full and the dashed lines correspond to the $Q$-state clock and the $Q$-state
Potts models, respectively, where $Q=2,3,\dots,7$.}
  \label{fig:64}
\end{figure}

The free energy manifests asymptotic behavior at sufficiently high temperatures for any
lattice geometry ($p,q$), where the tensors ${\cal C}_{\infty}$ and ${\cal T}_{\infty}$
prefer more symmetries (on the contrary, less symmetries are present in the ordered
phase when the spontaneous symmetry-braking occurs). The symmetries in the disordered
phase then cause that the normalization factors behave as $c_{k\to\infty} \to Q^{p-2}$
and $t_{k\to\infty} \to Q^{p-3}$ at $T \gg T_{\rm pt}$ (remark here that the exponents
$p-2$ and $p-3$ are associated with the number of the summed up $Q$-state spins in the
tensors, cf. Eq.~\eqref{Npq}). If substituting $c_{k} = Q^{p-2}$ and $t_{k} = Q^{p-3}$
into Eq.~\eqref{FEpq}, one obtains the expression in the high-temperature region
\begin{equation}
\lim\limits_{{k\to\infty}\atop{T\gtrsim 2q}} {\cal F}_{(p,q)}^{[k]} \propto -T \ln Q\, .
\label{F_inf_T}
\end{equation}
This asymptotics of the free energy is remarkable if examined by the thermodynamic entropy
\begin{equation}
{\cal S}_{(p,q)}^{[\infty]} = -\frac{\partial {\cal F}_{(p,q)}^{[\infty]}}{\partial T}\, ,
\label{Spq}
\end{equation}
which reaches a constant ${\cal S}_{(p,q)}^{[\infty]} \to \ln Q$ at $T\gtrsim 2q$.

As a non-trivial reference, we select the hyperbolic lattice ($7,7$), on which we
plotted the asymptotics of the $Q$-state clock and Potts models in Fig.~\ref{fig:64}.
The high-temperature asymptotic behavior of the entropy is most striking when showing
the asymptotics of $\exp\left\{{\cal S}_{(p,q)}^{[\infty]}\right\}=Q$ at high temperatures.

\begin{figure}[tbh]
\begin{center}
\includegraphics[width=3.8in]{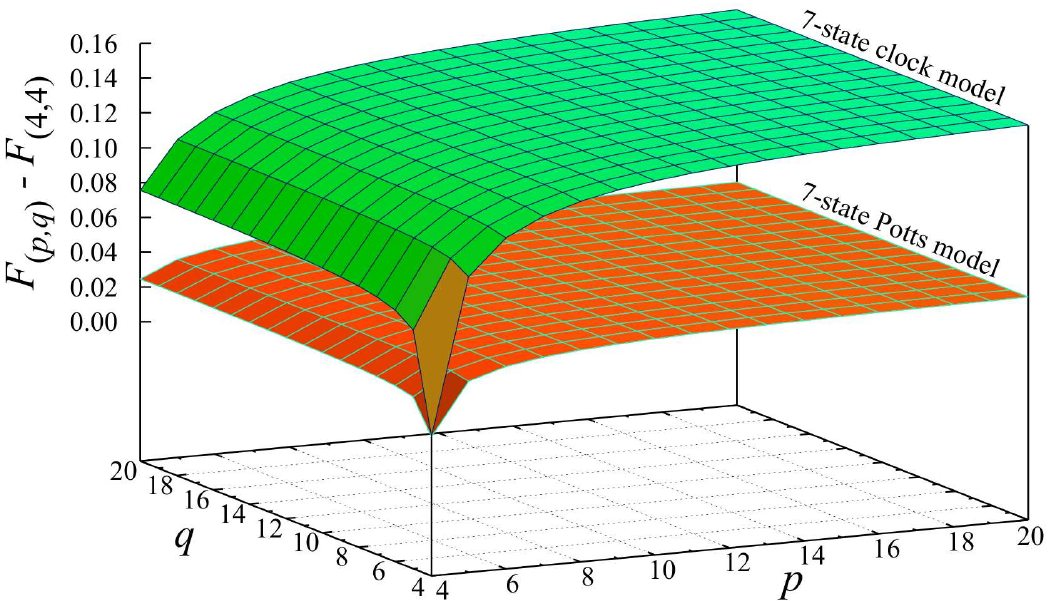}
\end{center}
  \caption{The free energy for the $7$-state clock and Potts models at
$T=5$, which do not differ from the Ising model ($Q=2$) in Figs.~\ref{fig:62} and
\ref{fig:63} qualitatively.}
  \label{fig:65}
\end{figure}

Figure~\ref{fig:65} shows the dependence of the free energy per site on ($p,q$) for
the $7$-state clock and the $7$-state Potts models at $T=5$. Clearly, the free energy
reaches its minimum on the square lattice for the both spin models. Having scanned
the multi-state spin variables $Q=2,3,\dots,7$ (not shown) for various $T$, the free
energy remains minimal exclusively on the square lattice ($4,4$).

\subsubsection{Relation between energy and curvature}

The ($p,q$) lattice geometries can be exactly characterized by the radius of Gaussian
curvature~\cite{Mosseri}, which has the analytical expression
\begin{equation}
 {\cal R}_{(p,q)}^{-1} = -2\,{\rm arccosh}
            \left[ 
                  \frac{\cos \left( \frac{\pi}{p} \right)}
                       {\sin \left( \frac{\pi}{q} \right)}
            \right] .
\label{Rpq}
\end{equation}
For later convenience we include the negative sign in ${\cal R}_{(p,q)}$,
to point out the negative (hyperbolic or Lobachevski) geometry. The radius of
curvature for the square lattice geometry ($4,4$) diverges, $R_{(4,4)}\to-\infty$,
while the remaining hyperbolic lattice geometries ($p,q$) are finite and negative.
The analytical description in Eq.~\eqref{Rpq} results in a constant and position
independent curvature at any lattices ($p,q$). It is a consequence of the constant
distance between the lattice vertices for all geometries ($p,q$), which is equivalent
to keeping the spin-spin coupling $J=1$ in all the numerical analysis of the spin
systems.

\begin{figure}[tb]
\begin{center}
\includegraphics[width=3.8in]{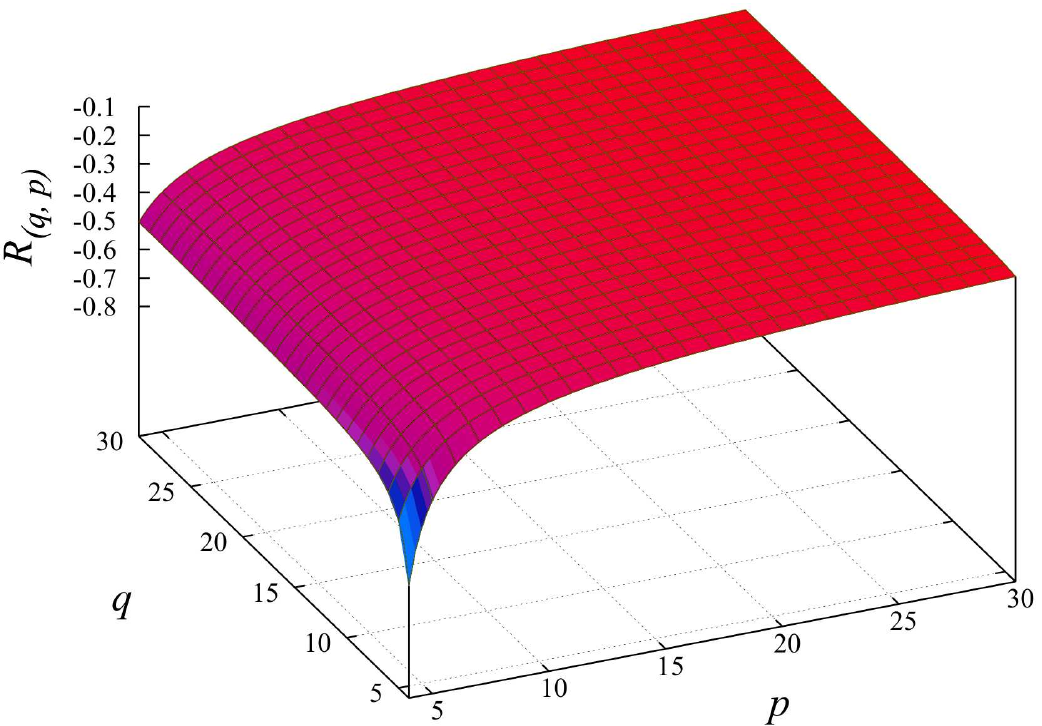}
\end{center}
  \caption{The functional dependence of the Gaussian radius of
curvature ${\cal R}_{(q,p)}$ plotted in the dual lattice geometry ($q,p$).}
\label{fig:66}
\end{figure}

In Fig.~\ref{fig:66} we plot the radius of curvature in the dual geometry ($q,p$),
i.e., $p$ and $q$ are swapped in Eq.~\eqref{Rpq}, i.e., $p \leftrightarrow q$.
The surface shape of ${\cal R}_{(q,p)}$ evidently exhibits a qualitative similarity
if compared to the free energy per site ${\cal F}_{(p,q)}^{[\infty]}$ we have
depicted in Figs.~\ref{fig:62}, \ref{fig:63}, and \ref{fig:65}.

This surprising observation is demanding for a relation between the free energy in
equilibrium and the space (lattice) geometry. In other words, it is equivalent to the
relation between the ground-state energy of quantum systems and the underlying geometry.
We focus our attention on the low-temperature regime, $0<T<1$, where this similarity is
most striking. We demand that the numerical computations are reliable and accurate to
avoid possible under/overflows in the tensors. For this reason, the numerical calculations
require to set a higher quad-precision, i.e., the 32-significant-decimal-digit precision is used.

\begin{figure}[tb]
\begin{center}
\includegraphics[width=3.8in]{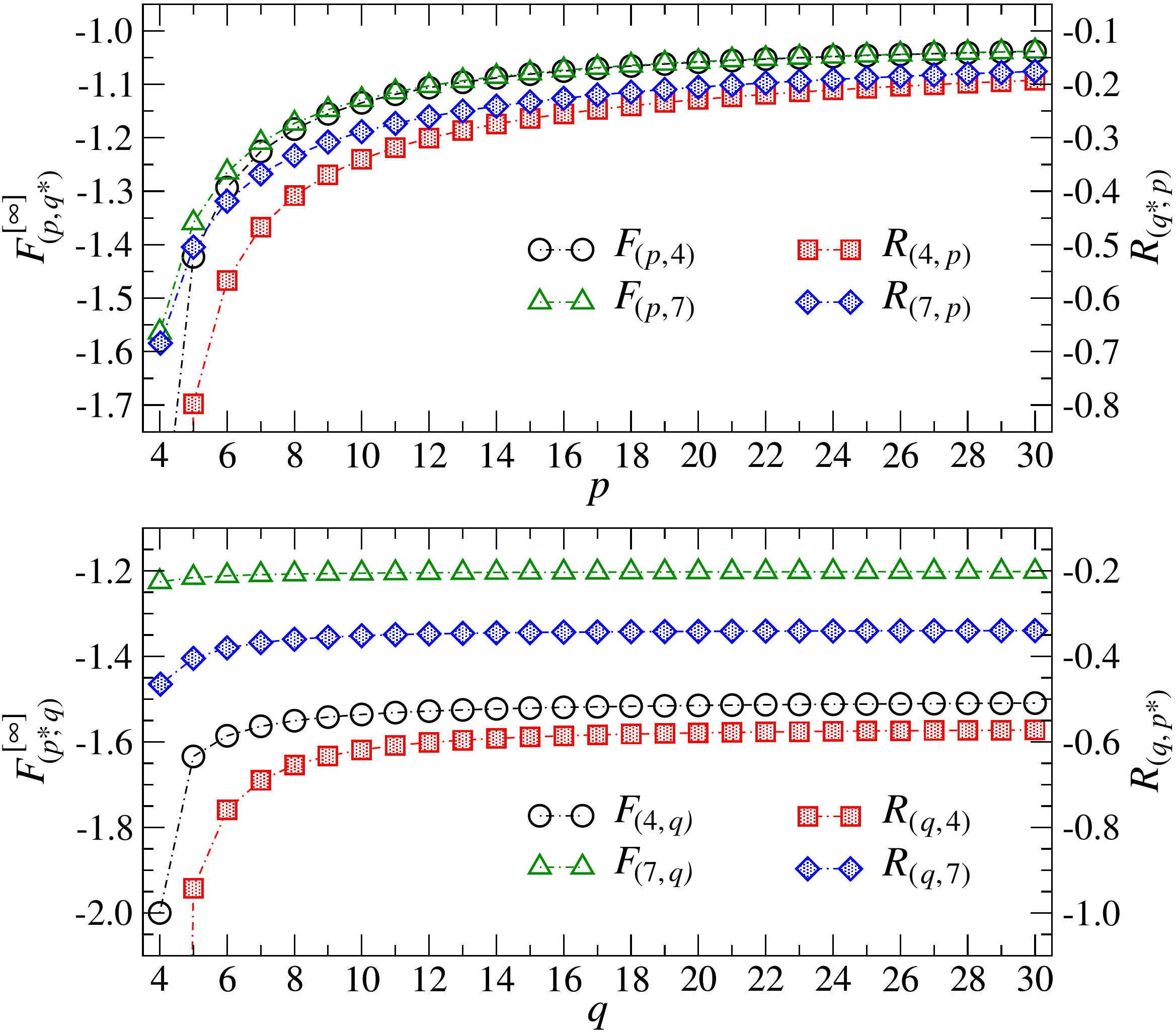}
\end{center}
  \caption{Comparison of the free energy per site for the Ising model at $T=0.5$ with
the Gaussian radius of curvature on the dual geometry. The upper graph shows the case
when the coordination numbers $q^{\ast}=4$ and $q^{\ast}=7$ are fixed, whereas the lower
graph depicts the opposite case when fixing the $p$-gons to be $p^{\ast}=4$ and
$p^{\ast}=7$.}
\label{fig:67}
\end{figure}

In Fig.~\ref{fig:67} we plot the free energy per site ${\cal F}_{(p,q)}^{[\infty]}$
at $T=0.5$ and the radius of curvature ${\cal R}_{(q,p)}$ on the dual lattice with
respect to $p$ (the top graph) and $q$ (the bottom graph). We use notation $\ast$
in $q^{\ast}$ to stress the fact that the parameter $q$ is fixed while $p$ can vary
freely (and vice versa). The top graph shows the free energy and the radius of curvature at
$q^{\ast}=4$ and $q^{\ast}=7$ while varying the $p$-gons within interval $4\leq p \leq 30$.
The bottom graph displays the complementary case, i.e. ${\cal F}_{(p^{\ast},q)}$ and
${\cal R}_{(q,p^{\ast})}$ at fixed $p$-gons $p^{\ast}=4$ and $p^{\ast}=7$ at
varying $4\leq q \leq 30$. In the former case, both of the functions increase with
$p$. In the latter case, the functions saturate at constant values at larger $q$'s.

Let us first inspect the asymptotic behavior of ${\cal R}_{(q,p)}$. If $q$ is fixed
to an arbitrary $q^{\ast}\geq4$, the logarithmic dependence on $p$ is present and
\begin{equation}
{\cal R}_{(q^{\ast},p\gg4)}^{-1}\to -2\ln\left[\frac{2p}{\pi}
    \cos\left(\frac{\pi}{q^{\ast}}\right)\right].
\end{equation}
On the contrary, if fixing $p$ to $p^{\ast}\geq4$ the radius of curvature converges
to a constant and for a sufficiently large $p^{\ast}$, this constant does not depend
on $q$ and
\begin{equation}
{\cal R}_{(q\gg4,p^{\ast})}^{-1} \to -2\ln\left[\frac{2}{\sin(\pi/p^{\ast})}
-\frac{\sin(\pi/p^{\ast})}{2}\right] \approx -2\ln\left(\frac{2p^{\ast}}{\pi}\right).
\end{equation}
It is straightforward to conjecture that the asymptotics of ${\cal R}_{(q,p)}$ is solely
governed by the parameter $p$, i.e.,
\begin{equation}
{\cal R}_{(q\gg4,p\gg4)}^{-1} \to -2\ln\left(\frac{2p}{\pi}\right).
\end{equation}

\begin{figure}[tb]
\begin{center}
\includegraphics[width=3.8in]{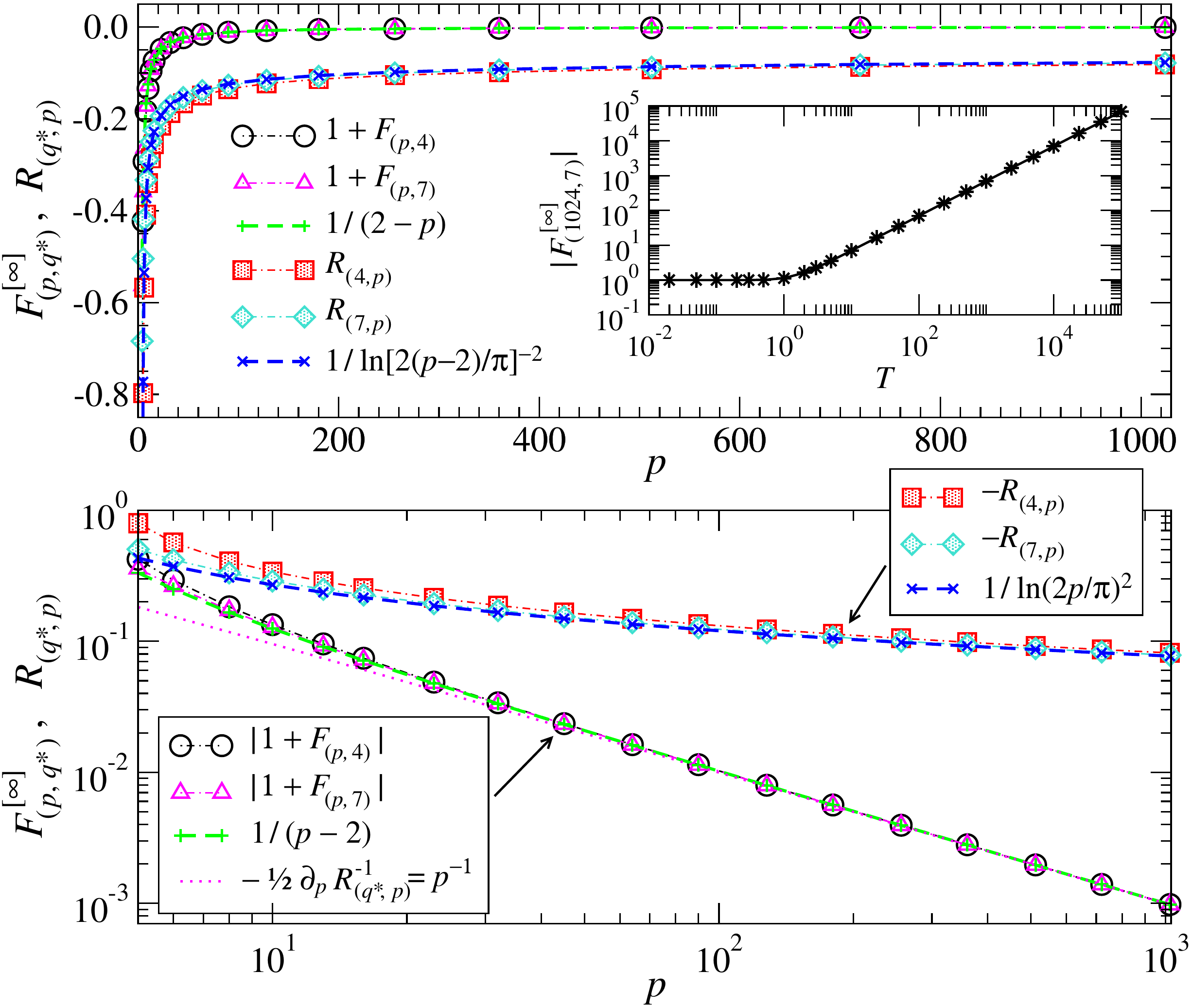}
\end{center}
  \caption{The asymptotic behavior of the free energy per site ${\cal F}_{(p,q^{\ast})}
^{[\infty]}-{\cal F}^{[\infty]}_{(\infty,q^{\ast})}$ for the Ising model at $T=0.1$ and
the Gaussian radius of curvature ${\cal R}_{(q^{\ast},p)}$ at $q^{\ast}=4$ and $q^{\ast}=7$.
The free-energy asymptotics behave as $-\frac{1}{p}$ (the ``$+$'' symbols follow the
least-square fitting of the numerical data). The asymptotics $\propto 1/\ln(2p/\pi)^{-2}$
of the radius of curvature is depicted by the ``$\times$'' symbols. The inset of the top
graph shows the low- and hight-temperature dependence of $\left\vert{{\cal F}^{[\infty]}
_{(p=1024,q^{\ast}=7)}}\right\vert$. The log-log plot on the bottom graph enhances the
asymptotics of $p$.}
\label{fig:68}
\end{figure}

The seemingly similar asymptotic $p$-dependence between the free energy and the Gaussian
radius of curvature, as plotted on the top graph in Fig.~\ref{fig:67}, requires extension
of the current numerical data for the free energy for much larger $p$. This necessity is
due to the slow logarithmic asymptotics of both ${\cal F}$ and ${\cal R}$. Again, we
fix $q^{\ast}=4$ and $q^{\ast}=7$ to study the logarithmic asymptotics of the free energy.
Figure~\ref{fig:68} shows this asymptotic behavior of ${\cal F}_{(p,q^{\ast})}^{[\infty]}$
and ${\cal R}_{(q^{\ast},p)}$ for $4\leq p \leq 1024$ at low temperature $T=0.1$. The top
and bottom graphs display both the free energy per site and the radius of curvature in the
linear scale and the log-log plot, respectively.

Applying the least-square fitting to the free energy data results in the function
$\frac{1}{2-p}+{\cal F}^{[\infty]}_{(\infty,q^{\ast})}$, which reproduces the asymptotics
of the free energy per site for both $q^{\ast}$ correctly. In contrast to the radius of
curvature, which logarithmically converges to zero as $p\to\infty$, the free energy per
site converges to ${\cal F}^{[\infty]}_{(\infty,q^{\ast})}=-1$ for $T\ll1$ and exhibits
linear dependence on high-temperature region, i.e., ${\cal F}^{[\infty]}_{(\infty,q^{\ast}
\lesssim T/2)}=-T\ln Q$ for $T\gg1$ in accord with Eq.~\eqref{F_inf_T}; notice that
the condition $q^{\ast} \lesssim T/2$ has to be fulfilled.

Particularly, we estimate the case of $p\to\infty$ by the polygons $p=1024$ so that
${\cal F}^{[\infty]}_{(p=1024,q^{\ast})}=-1.00098$ for both $q^{\ast}=4$ and
$q^{\ast}=7$ at $T=0.1$. The inset of the top graph in Fig.~\ref{fig:68} displays the
temperature dependence of ${\cal F}^{[\infty]}_{(p=1024,q^{\ast}=7)}$, which is still
numerically feasible for the polygonal size $p=1024$. This graph agrees with the
above-mentioned asymptotic dependence of ${\cal F}^{[\infty]}_{(\infty,q^{\ast})}$
at low and high temperatures.

The double-logarithmic plot on the bottom graph demonstrates the difference in the
asymptotics ($p\gg4$) between the polynomial behavior
\begin{equation}
{\cal F}_{(p,q^{\ast})}^{[\infty]}-{\cal F}^{[\infty]}_{(\infty,q^{\ast})}
= - \frac{1}{p}
\end{equation}
and the logarithmic one
\begin{equation}
{\cal R}_{(q^{\ast},p)}^{-1} = -\ln(2p/\pi)^2.
\end{equation}
The thin dotted line on the bottom graph corresponds to the derivative of
$-\frac{1}{2}{\cal R}_{(q^{\ast},p)}^{-1}$ when $p$ varies. It confirms
that the free energy per site is proportional to $1/p$ in the asymptotic
regime.

We, therefore, propose the following asymptotic dependence between the free
energy per site and the radius of curvature on the dual lattice geometry
\begin{equation}
{\cal F}^{[\infty]}_{(p,q)} - {\cal F}^{[\infty]}_{(\infty,q)}
\propto \frac{\partial}{\partial p}{\cal R}^{-1}_{(q,p)}
\approx -\frac{\pi}{2}\exp\left[ \frac{1}{2}{\cal R}_{(q,p)}^{-1} \right]\, ,
\label{scale_dR-F}
\end{equation}
which remains valid for any $q\geq4$ and $p\gg4$ (typically $p\gtrsim10^2$) at low
temperatures.

We have shown that the free energy of various spin models (or the ground-state energy for
quantum spin systems) on the non-Euclidean lattice geometries can reproduce the properties
of the spatial geometry as we have demonstrated on the Gaussian radius of curvature.
Necessity of other studies is inevitable at this point to support our achievements.
The consequences of the current work are expected to elicit further research, which
may bridge the ground-state properties spin systems with the conformal field theory.

\begin{figure}[tb]
\begin{center}
\includegraphics[width=3.8in]{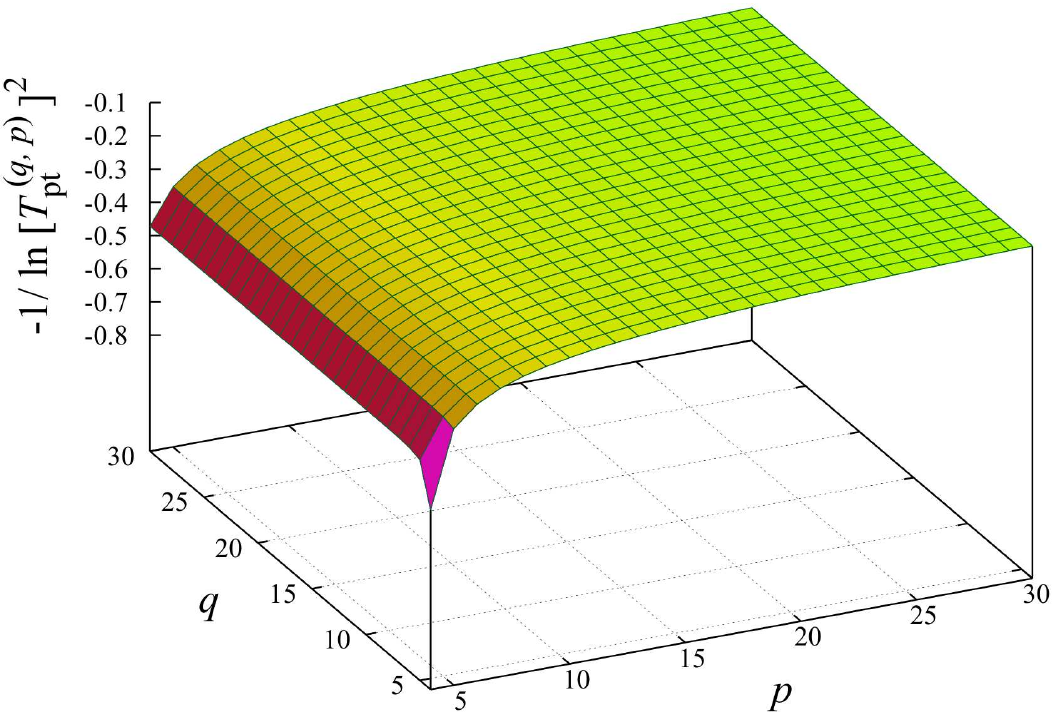}
\end{center}
  \caption{The rescaled phase-transition temperatures with respect to $p$
and $q$ are shown in the dual geometry ($q,p$) to emphasize the similarity with the radius
of Gaussian curvature in Fig.~\ref{fig:66}.}
\label{fig:69}
\end{figure}

Finally, if analyzing the functional dependence of the phase-transition temperatures
$T_{\rm pt}^{(p,q)}$ of the Ising model on the lattice geometries ($p,q$), we point
out another interesting behavior, which connects the phase-transition temperatures
with the radius of Gaussian curvature
\begin{equation}
-1 / \ln { \left[ T_{\rm pt}^{(p,q)} \right] }^2 \propto {\cal R}_{(p,q)}
\label{scale_dR-T}
\end{equation}
as depicted in Fig.~\ref{fig:69}. For better visualization with Fig.~\ref{fig:64},
we plot $-1/[2\ln T_{\rm pt}^{(p,q)}]$ in the dual lattice geometry, where $p$ and
$q$ are swapped. Recall that the higher values of the coordination number $q$ (at
fixed $p$) cause that $T_{\rm pt}^{(p,q)} \propto q$, in accord with Eq.~\eqref{Tpt_q},
whereas if $p$ increases (at fixed $q$), a fast convergence to a constant in
Eq.~\eqref{Bethe_q} is recovered, in accord with the theory.

Hence, the evident mutual similarity of the functional $p,q$-dependence among the free
energy per site (in Figs.~\ref{fig:62}, \ref{fig:63}, \ref{fig:65}), the radius of the
Gaussian curvature (Fig.~\ref{fig:66}), and the phase-transition temperature 
(Fig.~\ref{fig:69}) leads us to conjecture that there must be a theoretical reasoning
which connects these findings together. Or, in other words, our results call for the
necessity to propose a theoretical explanation.

\newpage\setcounter{equation}{0} \setcounter{figure}{0} \setcounter{table}{0}
\section{Models of social behavior} \label{chap4}


The last couple of years has witnessed an increasing interest in the study of the collective behavior of social systems. Social systems are composed of people (individuals) who interact with one another. These interactions influence the people engaged in them, and after many interactions, global properties emerge, i.e., a macro-level behavior of groups or whole societies. An essential question is how these microscopic (local) interactions lead to the global macroscopic (global) properties of the system under investigation. This question can be also answered by tools of the statistical physics. The approach of the statistical physics is used extensively in the study of collective phenomena. 

One may then ask, ``{\it Statistical physics is used for study of interacting particles, aren't humans astonishingly much more complicated?}" This is a reasonable objection. Inevitably, every modeling of social agents may imply severe simplifications of the reality. The social agents are modeled by a rather limited set of variables which represent their properties or states. The microscopic interaction among the agents should be thus
defined sensibly and realistically to capture the dynamics of the social behavior, which is not a trivial task. However, the concept known from the statistical physics as {\it universality} says that the large-scale phenomena do not depend on the microscopic details of the process. They depend only on higher-level
features, like symmetries, dimensions or conservation laws. Keeping this fact in mind, one can learn something about a system just by reproducing the most important properties of the elements and their interactions. 

Studying social dynamics is a rapidly evolving area of scientific studies. It has its applications in disciplines as diverse as sociology, economics, political science, and anthropology. It is a hard task to present the state-of-art in this subject. However, we have found a very useful work in this area~\cite{Castellano_over}. Other resources will be referred to later on. 

Major regularities can emerge spontaneously in social systems, as it is frequently observable in the real world. This can be understood as an order-disorder transition. Among examples of such transitions, we count the spontaneous formation of common language/culture or the emergence of consensus on an issue, collective motion, a hierarchy. The drive toward the ordering is the tendency of agents to become alike as they interact. The term for this mechanism is \textit{social influence}. It is analogous to ferromagnetic interaction in magnetic materials. The Ising model, briefly introduced in Subsection~\ref{Ising_model_subs}, is a model of such a ferromagnetic material. It can, however, be considered as a very simple model of opinion dynamics, wherein the agents are influenced by the state of the majority of their nearest neighbors. 

One way of considering the unknown influence, or some hidden details of the social dynamics, is to introduce the {\it noise}, which corresponds to a variability in the states of the agents. A natural question arises, ``{\it Does the presence of a noise hinder the ordered state?}"
The role of noise in one particular model (the Axelrod model) is to be discussed. 

Traditionally, the statistical physics deals with regular systems such as lattices, where the elements are located on the sites and usually interact with the nearest neighbors only or involve all-to-all interactions (the mean-field approximation). However, social interactions exhibit the interaction pattern which is denoted as complex networks. The role of topology is a highly studied topic in this context; nonetheless, our work is not concerned about it. 

The simplest model of the opinion formation is the Voter model. Each agent $s_i$ in this model is endowed with a binary variable $s_i = \pm 1$. This variable represents, e.g., the answer to the yes-no question. The interaction is defined by the following algorithm: First one randomly selects an agent $i$, chooses one of the neighbors $j$ at random, and then changes the opinion of agent $s_i$ to be equal to the opinion of a selected neighbor $s_j$. This process mimics the homogenization of opinions; however, the convergence to a uniform state cannot be always guaranteed, because the interactions are random and only between two agents at each step. For $D$-dimensional lattice, the described mechanism leads to slow coarse-graining processes, where spatially ordered regions grow, i.e., large regions tend to expand and ``consume" the small ones. 

The evolution of the system can be described by the density of active interfaces between ordered regions $n_a$. The following scaling of evolution of Voter model was found in~\cite{Krapivsky},
\begin{equation} 
n_a(t) \sim \begin{cases}
    t^{-(2-d)/2},  & D<2\text{,}\\
    1/\ln(t),          & D=2\text{,}\\
    a - bt^{-d/2}, & D>2\text{.} 
  \end{cases}
\end{equation}
Amongst some other models of the opinion formation belong: A majority rule model, models of social impact, Sznajd model, and bounded confidence models. For more details concerning the Voter model as well as the other models, see Refs.~\cite{Castellano_over,BBV}.

Besides the opinion dynamics, much interest has been focused on the related field of cultural dynamics. There is no sharp distinction between the two, however, in cultural dynamics, each agent is described by a vector of variables instead of a scalar variable as we have seen in the opinion dynamics. A paradigmatic model of the cultural dynamics is the Axelrod model which we focus on.

As it should be clearer from the text, our approach is to view the agents as being adaptive instead of being rational, in particular, with a focus on communication rather than strategy. The latter approach has, however, been studied extensively. Let us mention at least prisoner's dilemma, where the agents (players) interact (play) pairwise and have two strategies which they can choose, i.e., they can decide whether to cooperate or not. For more details, see the references in~\cite{BBV}. 

We summarize this brief introduction with a note about co-evolution of opinions and topology. So far, the topology of the network was considered to be fixed and served as a playground for a dynamical process. On the other hand, many real networks are of dynamical nature, i.e., the topology of such networks changes with time. The dynamical process taking place on the network can be coupled with the evolution of the topology. This is particularly relevant for social networks. One example is a link removal (or re-wiring) between two agents with dissimilar opinions and creation of new links between other two random agents, or with a preference between similar ones. There is a huge number of works concerning this topic, for survey, see Refs.~\cite{BBV,Castellano_over}.

\subsection{The Axelrod model}\label{AXM}

Axelrod proposed a simple, yet ambitious, model of cultural assimilation and diversity, which is based on two mechanisms: \textit{social influence} and \textit{homophily}~\cite{Axelrod}. The former means, that after interacting, people become more similar than before. In other words, communication reduces differences among the people. The latter is that the probability of social interaction depends on the similarity between two agents. It is based on the idea, that the more similar two people are, the easier their communication becomes. It might seem that this mechanism leads to the homogenization of society. However, it can generate a global polarization, where different cultures are allowed to coexist. The mechanism of this model might be relevant for such topics as state formation, succession conflicts, trans-national integration, or domestic cleavages.

Each agent has $f$ different cultural {\it features}, which we denote as $(\sigma_1, \sigma_2, \dots , \sigma_f)$, and each of the features can assume $q$ different values ({\it traits}), $\sigma_{\alpha} = 0, 1, ..., q-1$. It means that each agent can be in one of $q^f$ possible states. Each feature represents one of the cultural dimension such as, e.g., language, religion, technology, style of dress, and the like. Agents are placed on sites of a regular two-dimensional square lattice $L \times L$ and can interact only with the nearest neighbors (the longer ranged interactions are to be considered later). The dynamics runs in two steps as follows:
\begin{itemize}
\item[1]{An agent at site $i$ and one of his neighbors $j$ are selected randomly.}
\item[2]{The selected agents interact with the probability being proportional to the number of the features for which they share the same value,
\begin{equation}
\omega_{ij} = \frac{1}{f}\sum_{\alpha=1}^{f}\delta\left(\sigma_{\alpha}^{(i)}, \sigma_{\alpha}^{(j)}\right),
\end{equation}
where $\delta(\ast,\ast)$ is the Kronecker's delta.
The interaction consists of random selection of the feature the two agents differ, $\sigma_{\alpha}^{(i)} \neq \sigma_{\alpha}^{(j)}$ and consequently setting this feature of the neighbor to be equal to $\sigma_{\alpha}^{(i)}$. }
\end{itemize}
These two steps are repeated as long as needed. 

\begin{figure}[tb]
\begin{center}
\begin{tabular}{cccc}
\includegraphics[width=.22\linewidth]{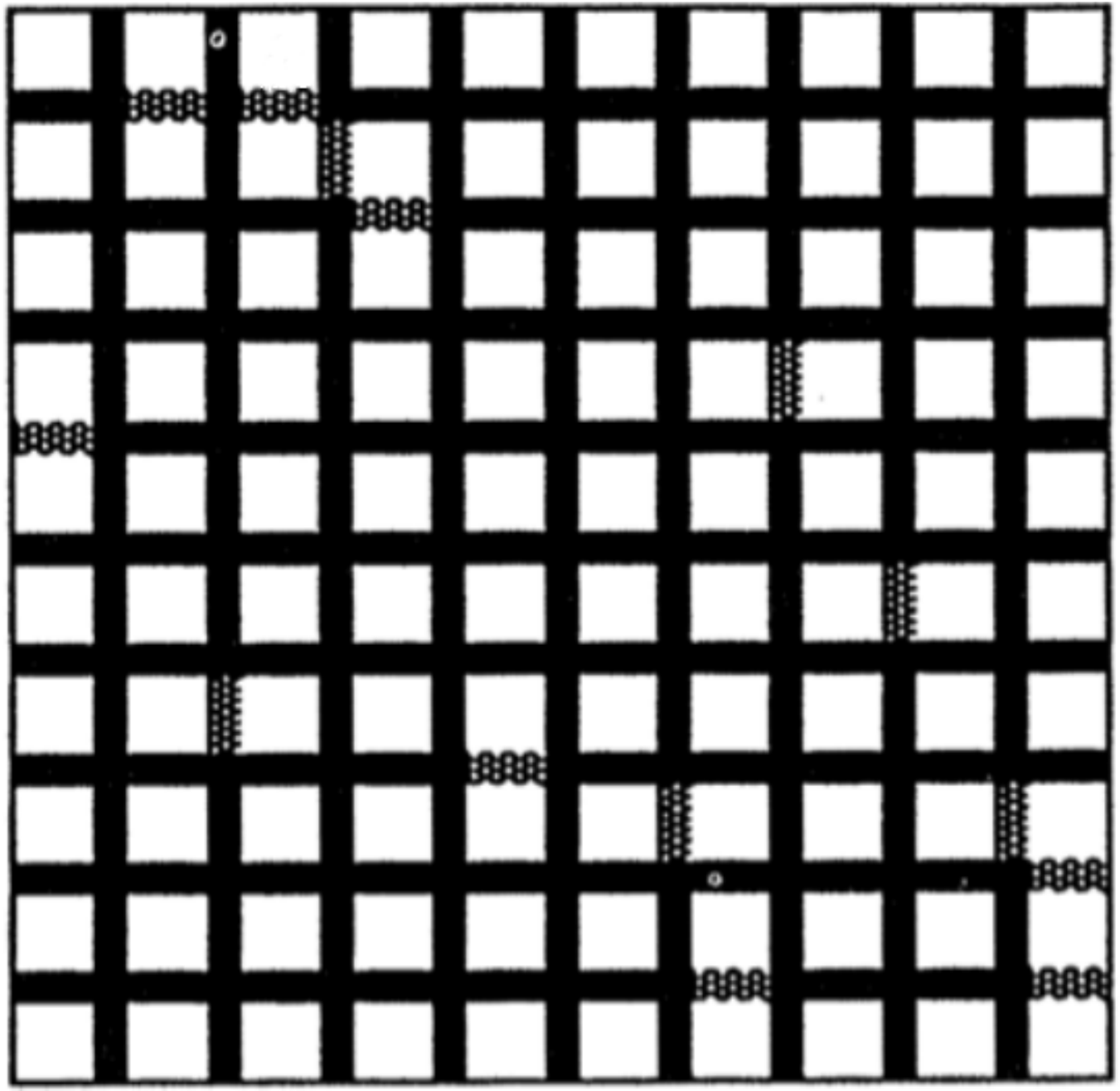} &
\includegraphics[width=.22\linewidth]{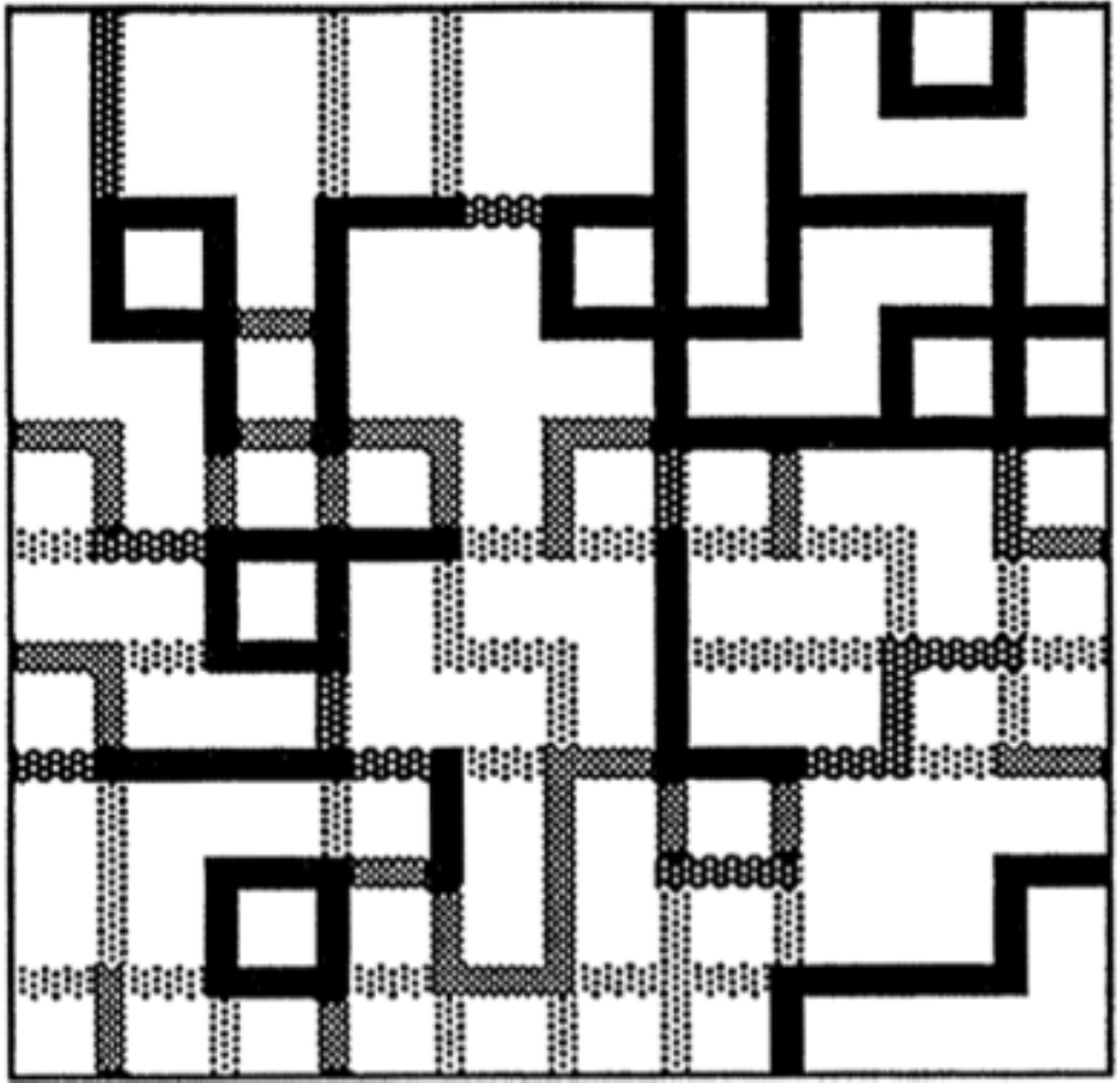} &
\includegraphics[width=.22\linewidth]{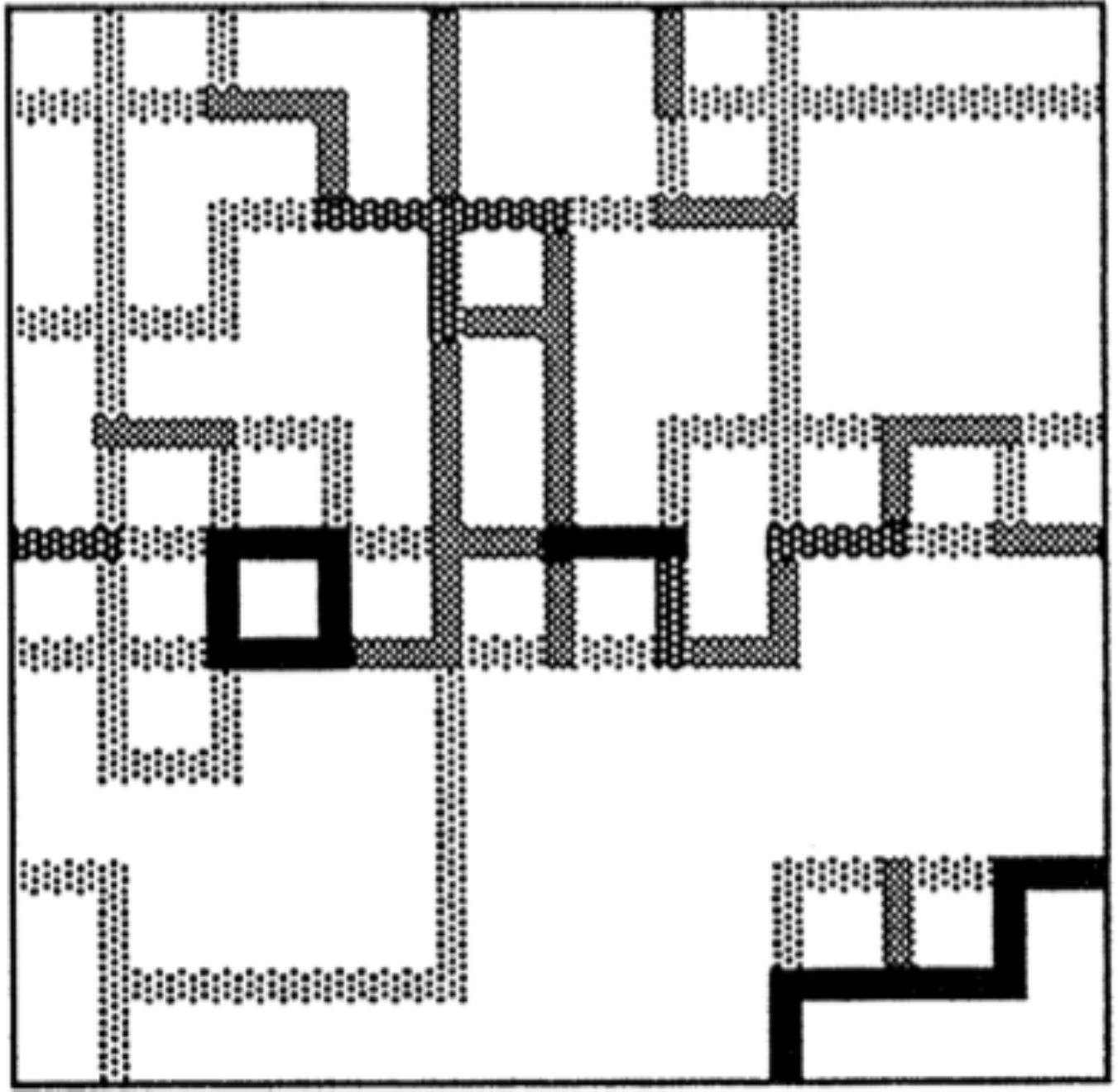} &
\includegraphics[width=.22\linewidth]{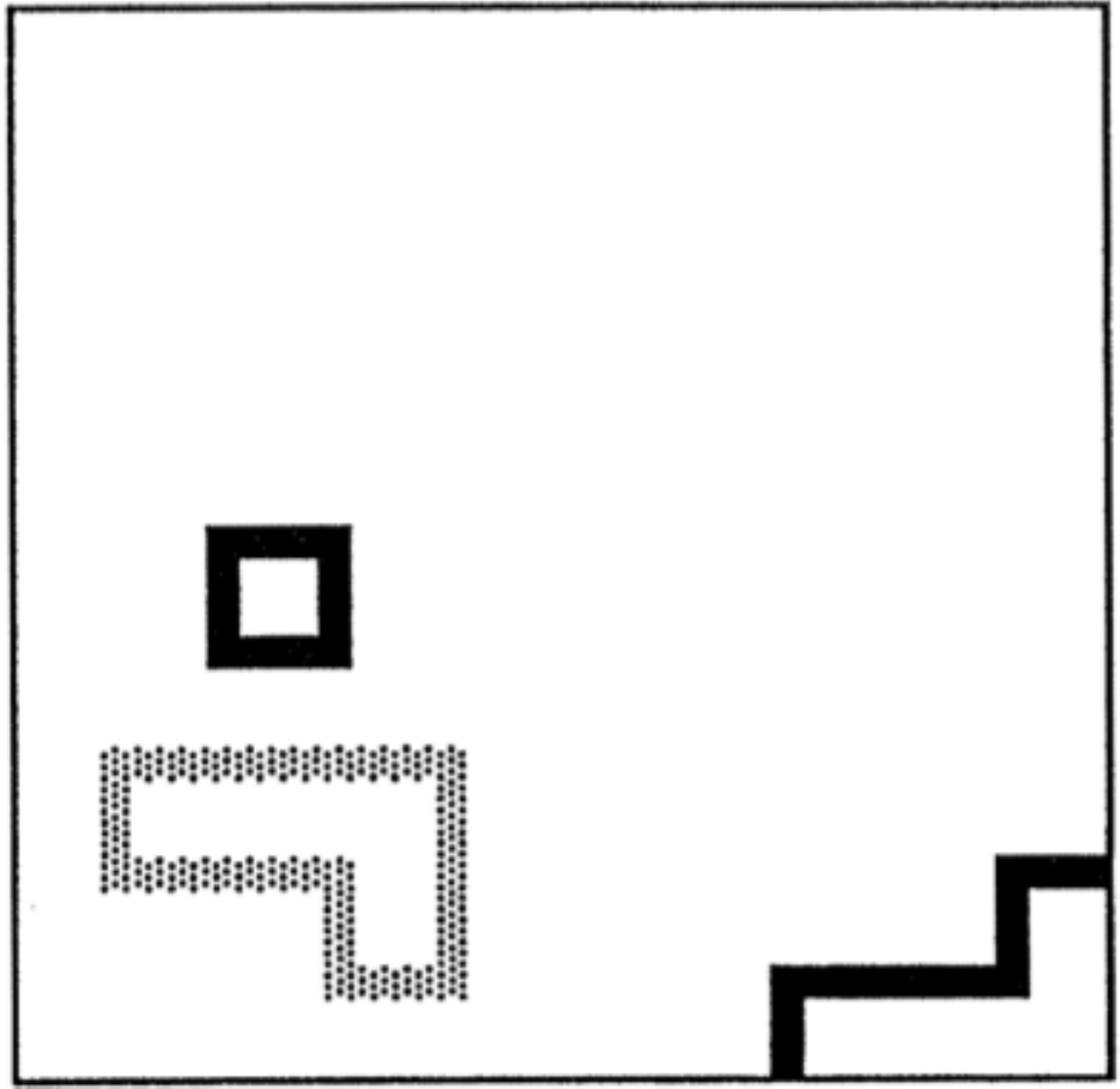} \\
$t=0$ &  $t=2\times10^4$ & $t=4\times10^4$ & $t=8\times10^4$
\end{tabular}
\end{center}
  \caption{Example of the four maps of the cultural similarities. Simulation was conducted for lattice size $10 \times 10$ only, the number of the features is $f = 5$, and the number of possible values of each feature is $q=10$. The cultural similarity (number of the shared features) between adjacent agents is coded as black for $\omega_{ij} \leq 20\%$, dark grey for $\omega_{ij}=40\%$, gray for $\omega_{ij}=60\%$, light grey for $\omega_{ij}=80\%$, and white for $\omega_{ij}=100\%$. The time evolution denotes the number of events $t$ (after~\cite{Axelrod}).}
  \label{map}
\end{figure}

The described process can lead either to a global homogenization or to a fragmented state with coexistence of different homogeneous regions, as shown in Fig.~\ref{map}. As the two agents do interact, they share more specific cultural features. More features tend to be shared over a larger area and a cultural region (with all features being exactly the same) can be created. Eventually, the system may end in a state where no other change is possible. Those features of all neighbors are either identical or there is no match, as they do not interact at all. Several {\it stable regions} can be created. A question of interest is how many.  

As $f$ increases (with $q$ being fixed) the probability of sharing at least one of the features is higher. In that case, the average number of the stable regions (with respect to different runs) decreases. On the contrary, as $q$ increases (with $f$ being fixed) the probability of sharing at least one of the features is smaller and thus the average number of stable regions increases. 

The longer-ranged interactions have also been considered, i.e., the interactions between second nearest neighbors, third nearest neighbors, etc. It is intuitively clear that if the agents could interact over larger distances, the process of cultural convergence would have been made easier. The longer the interactions are allowed, the smaller is the average number of stable regions, as expected.

Another studied question was the dependence of the average number of stable regions on the size of the lattice (size of a territory). Axelrod provided some preliminary results which suggested that this number increases with size of the lattice for small lattices and after certain size it decreases. Small lattices do not have enough space to contain many stable regions, thus it is understood this number would increase as the size of the lattice increases. However, it is surprising that the large lattices would have had fewer stable regions than moderately-sized lattices. 
A detailed phase diagram of the model was studied in Ref.~\cite{Castellano1}. The order parameter was defined as the average size of the largest stable region $\left<S_{\text{max}}\right>$. For fixed $f$, there is a critical value $q_c$ such that $\left<S_{\text{max}}\right>$ increases with the linear system size $L$ and tends to $L^2$ for $q<q_c$, whereas $\left<S_{\text{max}}\right>/L^2 \rightarrow 0$ for $q>q_c$. For the two-dimensional lattices, nature of these phase transitions depends on the value of $f$. For $f=2$ the transition is continuous, whereas for $f>2$ it becomes discontinuous, see Fig.~1 in Ref.~\cite{Castellano1}. 
For the one-dimensional systems, the transition remains continuous for all values of $f$~\cite{Klemmd}.

\textit{Cultural drift} can be modeled as a spontaneous change of a feature at certain rate $r$. The order-disorder transition induced by a noise was demonstrated in Ref.~\cite{Klemmc}. 
The transition depends on the value of $q$, which becomes weaker for larger values of $q$,
as plotted in Fig.~3 in Ref.~\cite{Klemmc}. At small noises, the system tends to homogeneity, because the disordered configurations are unstable with respect to the perturbations enhanced by the noise. On the contrary, when the noise rate gets larger, the system becomes heterogeneous, as domains disappear, which is compensated by creating new ones. 

The effect of the mass media can be modeled by interactions among the agents with a global (uniform) {\it field}. In Ref.~\cite{Gonzalez05}, the global field was introduced to represent the mass-media cultural message as a set of $f$ parameters, $M=(\mu_1, \mu_2, ..., \mu_f)$. A selected agent interacts with the field, as if it was an agent with probability $B$, and with one of his actual neighbors with probability $1-B$. Surprisingly, the global field favors the multi-cultural phase. In other words, $q_c$ is smaller with addition of the global field and depends on $B$. For $B$ larger than a certain threshold, such that $q_c(B)=0$, only the disordered phase is present. A global coupling and a local non-uniform coupling were considered in Ref.~\cite{Gonzalez}.

The social networks have a complex topology as we have mentioned above. With respect to this fact, the Axelrod model was considered on networks with complex topological characteristics. For Watts-Strogatz two-dimensional networks, the previously defined $q_c$, starts growing as the disorder parameter increases~\cite{Klemmb}. For random scale-free networks, $q_c$ diverges with the size of the system as $L^\beta$ (here, $\beta \sim 0.4$ for Barab\'{a}si-Albert type of networks), where $L$ is the size of the network~\cite{Klemmb}.

\subsection{Thermodynamic version of Axelrod model}\label{tAXM}

In Ref.~\cite{Gandica}, a thermodynamic version of the Axelrod model has been proposed. Let us remark that the original Axelrod model is th model out of equilibrium, i.e, it does not satisfy the equilibrial conditions, as detailed balance\footnote{
We assume that each node (agent) $i$ can be in one of the possible states, i.e., $\sigma_i = 0,1, 2, ..., q-1$. The knowledge of the state of all the nodes defines a microscopic state (microstate) $\sigma(t) = (\sigma_1(t), \sigma_2(t), \dots, \sigma_L(t))$, where $L$ is size of the network.
Dynamical description of the system is given by the master equation. The master equation is an evolution equation for probability $P(\sigma, t)$ of finding the system at time $t$ in a microstate $\sigma$. In continuous time approximation, the master equation is
\begin{equation*}
\partial_t P(\sigma, t) = \sum_{\sigma'} \left[P(\sigma', t)W(\sigma' \rightarrow \sigma) - P(\sigma, t) W(\sigma \rightarrow \sigma')\right],
\end{equation*}
where the terms $W(\sigma' \rightarrow \sigma)$ represent the transition rates from one microstate (configuration) to another. 
As we can see, the solution of the master equation provides only statistical information about the system evolution.

One of the interests is to find a stationary state of the system (if exists)
$\lim_{t \to +\infty} P(\sigma, t) = P_{\infty}(\sigma)$.
Stationary distribution for equilibrial physical systems is given by the well-known Boltzmann-Gibbs distribution
$P_{\infty}(\sigma) = P_{\text{eq}}(\sigma) = \exp[-{\cal H}(\sigma)/k_{\text{B}}T]/{\cal Z}$,
where ${\cal H}(\sigma)$ is a Hamiltonian of the system. The partition function ${\cal Z}$ provides the proper normalization 
$ {\cal Z} = \sum_{\sigma} \exp(-{\cal H}(\sigma)/k_{\text{B}}T) $ of the distribution.
The stationary distribution for the equilibrium system may be obtained by the system Hamiltonian and there is no need to solve the master equation. 

At equilibrium, each elementary process should be equilibrated by its reverse process. This is what the so-called {\it detailed balance condition} states
\begin{equation*}
P_{\text{eq}}(\sigma) W(\sigma \rightarrow \sigma') = P_{\text{eq}}(\sigma') W(\sigma' \rightarrow \sigma).
\end{equation*}
This condition gives zero contribution for each pair on the right-hand side of the master equation. 
The detailed balance is not necessarily fulfilled, i.e. $\partial_t P(t) =0$, and the non-equilibrium systems can also reach the stationary state, however, with more complicated cancellation relations in the master equation. Most of the real systems are actually non-equilibrial and we thus cannot use the equilibrium thermodynamic formulations.
} is violated. The Hamiltonian proposed by the authors is
\begin{equation} \label{Hamiltonian_A}
{\cal H} = - \sum_{\alpha=1}^{f} \left[ \sum_{\left<ij\right>}  J_{ij\,} \delta\left(\sigma_{i \alpha}^{~},\sigma_{j \alpha}^{~}\right) + h \sum_{i} \delta\left(\sigma_{i\alpha}^{~},h_{\alpha}^{~}\right) \right]
\end{equation}
with the interaction factor
\begin{equation}
J_{ij} = \sum_{\alpha'=1}^{f}J\delta\left(\sigma_{i\alpha'},\sigma_{j\alpha'}\right). 
\end{equation}
The index $\alpha$ denotes the features, the indices $i$ and $j$ label the lattice sites, where the agents are localized, and the cultural feature $\alpha$ of an agent $i$ can take $q$ values, $\sigma_{i\alpha} = 0, 1, 2, \dots, q-1$. Here, $h_{\alpha}$ is the magnetic field with $q$ values ($h_{\alpha} = 0, 1, 2, \dots, q-1$), $h$ is the magnitude of the field uniformly imposed. The meaning of the interaction factor $J_{ij}$ is such that it defines the number of shared features between two agents $i$ and $j$ multiplied by a constant $J$. 

The Hamiltonian (\ref{Hamiltonian_A}) is inspired by the Potts model; however, the interaction factor $J_{ij}$ depends on the global state of the $f$ variables $(\sigma_{i1}, \sigma_{i2}, \dots , \sigma_{if})$. The thermodynamic and the critical properties were calculated analytically in the one-dimensional case, where the order-disorder phase transition occurred at $T=0$ regardless of $f$ and $q$, as it is in agreement with the one-dimensional Potts (Ising) model~\cite{Baxter}. 

As we have mentioned earlier, the Axelrod model was also studied with the noise (which represents the cultural drift). It has been shown that for the noise rate $r<r_c$ the system converges to the mono-cultural state, whereas for $r>r_c$ the system converges to the multi-cultural state, where $r_c$ is a certain critical value of the noise rate and depends on the system size, not on $q$ \cite{Klemm2005, Klemmc, Toral}. In the one-dimensional case, $r_c$ scales as $r_c\sim1/L^2$ with the size of the system, and in the thermodynamic limit ($L\rightarrow\infty$) there is no phase transition. For any positive $r$ the system converges to the multi-cultural state. In this sense, the noise rate $r$ corresponds to temperature $T$.

\newpage\setcounter{equation}{0} \setcounter{figure}{0} \setcounter{table}{0}

\section{Thermodynamic model of social influence} \label{our model}

In this Section, we consider a classical multi-state spin model of
a social system treated from the point of view of the statistical
mechanics. Our attention is focused on behavior of the model for
a large society in equilibrium. The society is represented by
individuals who can interact via communication channels (e.g.
sharing some interests) between nearest neighbors only.
The society is subject to special rules defined by the model
of the statistical mechanics we introduce for this purpose.
A communication noise plays an important role in this study.
The noise is meant to interfere with the communication channels.
If the noise increases, the communicating individuals are meant
to be less correlated on larger distances. In this way, the
noise can act against the formation of larger clusters of the
individuals with a particular character, i.e., a set of shared features.
Within the cluster, the individuals share a similar social background.
The size of the clusters can be quantified by an appropriate
order parameter or the correlation length, etc., which are commonly
used in the statistical physics. If the model allows a phase transition,
an ordered phase can be unambiguously separated from the disordered one.
The two phases are then identified by the order parameter, which is
either nonzero in the ordered phase or zero in the disordered,
provided that the system is infinite (thermodynamic limit), and the
spontaneous symmetry-breaking mechanism has occurred below that
phase-transition point. The noise can be also regarded as random
perturbations (being a cultural drift) realized as a spontaneous
change in a trait~\cite{Klemmc} and indirectly is associated with
the out-of-equilibrium Axelrod model~\cite{Axelrod}. On the other hand,
the effect of the noise for non-linear dynamical systems can be 
size-dependent~\cite{Toral}.

We have, therefore, proposed a multi-state spin model on the two-dimensional
regular square lattice of the infinite size. Each vertex of the lattice
contains a multi-state spin variable (and represents thus an individual
with a certain cultural setting). We define special nearest-neighbor
interactions among the spins representing the conditional communication
among the individuals. The statistical Gibbs distribution introduces
the thermal fluctuations into our model with the multi-state spin Hamiltonian.
Here, the temperature can be identified with the noise we have introduced above.
If imposing a constant magnetic field on spins, this results in the spins
to be align accordingly, which might have had a similar effect as, for
instance, the mass media or advertisement acting on the society. Having
calculated the effects of the magnetic field, we have observed a typical
response to our model, yielding no phase transition (in accord with the
spin models).

The model is meant to describe the thermodynamic features of the social
influence, which had been studied via the Axelrod model~\cite{Axelrod}.
Gandica {\it et al.}~\cite{Gandica} have recently studied the thermodynamics
features in the coupled Potts models in the one-dimensional lattice, where
the phase transition does not exist since it occurs at zero temperature
(in accord with a thermodynamic one-dimensional interacting multi-state
Potts system as summarized in Subsection~\ref{tAXM}). Our studies go
beyond this thermodynamic Axelrod-model conjectures. We also intend to
study phase transitions on these social systems occurring at nonzero
temperature (noise), and the number of the individuals is considered
infinite. Therefore, the spontaneous symmetry-breaking mechanism
selects a certain preferred cultural character resulting in a large
cluster formation, which is characterized by a nonzero order parameter.
We also identify all of the cultural characters by a special definition
of the order parameters.

This task is certainly nontrivial and our model has not been known to
be exactly solvable. Therefore, we apply a numerical algorithm,
the CTMRG~\cite{ctmrg1}, as the powerful tool in the statistical
mechanics, see Section~\ref{CTMRG_Section}. The CTMRG calculates
all thermodynamic functions to a high accuracy and enables us to analyze
the phase transitions as well as to specify the spontaneous symmetry
breaking. Since we found out that the phase-transition temperature
decreases with the increasing number of the traits $q$, we also
investigate the asymptotic case in this work, i.e., the case when the
number of the traits $q$ of each individual is infinite. We estimate
the phase-transition point in order to answer the question whether the
ordered phase can be permanently present or not. In other words, the
phase-transition point $T_t$ remains nonzero in this asymptotic regime.
Throughout this work we consider the case $f=2$ only.

\subsection{Hamiltonian of the lattice model}

The classical spin lattice model is studied on the regular two-dimensional
square lattice. Only the nearest-neighbor multi-state spins (positioned on
the lattice vertices) interact. Let $\sigma_{i,j} = 0, 1, \dots, n-1$ be the
generalized multi-state spin variable with integer degrees of freedom $n$.
The subscript indices $i$ and $j$ denote the position of each lattice vertex.
The thermodynamic limit means that the square lattice is infinite and the
positions of the spin vertices are $-\infty<i,j<\infty$. We start with the
$n$-state clock (or vector) model~\cite{clock} for this purpose with the
Hamiltonian
\begin{equation} \label{Clock}
{\cal H} = - J \sum\limits_{i=-\infty}^{\infty}
               \sum\limits_{j=-\infty}^{\infty}
               \sum\limits_{k=0}^{1}
                     \cos(\theta_{i,j} - \theta_{i+k,j-k+1})\, .
\end{equation}
The interaction coupling $J$ acts between the nearest-neighbor vector spins
$\theta_{i,j} = \frac{2\pi}{n}\sigma_{i,j}$. The summation over $k$ includes
the horizontal and the vertical directions on the square lattice.

Let us generalize this spin clock model so that the interaction term $J$
contains another special attribute, i.e., more interactions are included.
We introduce extra degrees of the freedom to each vertex. The Hamiltonian
in Eq.~\eqref{Clock} can be further modified into the form
\begin{equation}
{\cal H}=\sum_{ijk} J_{ijk}\cos\,(\theta_{i,j} - \theta_{i+k,j-k+1}).
\end{equation}
The position dependent term $J_{ijk}$ describes the spin interactions $J$ of
the $n$-state clock model controlled by additional $q$-state Potts model
$\delta$-interactions~\cite{FYWu}. The total number of the spin degrees of
the freedom is $nq$ on each vertex $i,j$. We intend to study a simplified
case when $q\equiv n$ starting from the case of $q=2$ up to $q=6$ which is
still computationally feasible. (In more general case when $q\neq n$,
we do not expect substantially different physical consequences).

Hence, our multi-state spin model contains two $q$-state spins placed on
the same vertex, i.e., 
$\sigma_{i,j}^{(1)}=0,1,2,\dots,q-1$ and $\sigma_{i,j}^{(2)}=0,1,2,\dots,q-1$,
which are distinguished by the superscripts ($1$) and ($2$). It is convenient to
introduce a grouped variable with $q^2$ states so that
$\xi_{i,j} = q\sigma_{i,j}^{(1)} + \sigma_{i,j}^{(2)}
= 0,1,\dots,q^2-1$. The Hamiltonian of our model has its final form
\begin{equation}
\label{Final_H}
{\cal H} = \sum\limits_{i,j=-\infty}^{\infty}
            \sum\limits_{k=0}^{1}
               \left\{ J^{(1)}_{ijk}
                  \cos\left[\theta^{\left(2\right)}_{i,j}
                          - \theta^{\left(2\right)}_{i+k,j-k+1}
                      \right]
              +   J^{(2)}_{ijk}
                  \cos\left[\theta^{\left(1\right)}_{i,j}
                          - \theta^{\left(1\right)}_{i+k,j-k+1}
                      \right]
              \right\},
\end{equation}
noticing that $\theta_{i,j}^{(\alpha)} = 2\pi\sigma_{i,j}^{(\alpha)}/q$, where
\begin{equation}
J^{(\alpha)}_{ijk} =
      -J\delta\left(\sigma^{(\alpha)}_{i,j},\,
                    \sigma^{(\alpha)}_{i+k,j-k+1}
              \right)
 \equiv
  \begin{cases}
    -J, & \text{if\ \ \ } \sigma^{(\alpha)}_{i,j} = 
                    \sigma^{(\alpha)}_{i+k,j-k+1},\\
  \phantom{-}0, & \text{otherwise},
  \end{cases}
\end{equation}
and the superscript $(\alpha)$ takes two values only.
The Potts-like interaction $J_{ijk}^{(\alpha)}$ is represented by a diagonal
$q\times q$ matrix with the elements $-J$ on the diagonal.

Thus multi-state spin system describes conditionally communicating (interacting)
individuals of a society. The society is modeled by the individuals ($\xi_{i,j}$)
and each individual has two distinguishable features $\sigma^{(1)}$ and
$\sigma^{(2)}$. Each feature assumes $q$ different values (the traits).
In particular, an individual positioned on $\{i,j\}$ vertex of the square
lattice communicates with a nearest neighbor, say $\{i+1,j\}$, by comparing
the spin values of the first feature $\sigma^{(1)}$. This comparison is
carried out by means of the $q$-state Potts interaction. If the Potts
interaction is nonzero, the individuals continue in communication via the 
$q$-state clock interaction of the other feature with $\alpha=2$.
The cosine enables a broader communication spectrum than the
Potts term.

Since we require symmetry in the {\em Potts-clock} conditional
communication, we include the other term in the Hamiltonian, which
swaps the role of the features ($1$) and ($2$) in our model.
In particular, the Potts-like communication first compares the feature
$J_{ijk}^{(2)}$ followed by the cosine term with the feature $\alpha=1$.
(Enabling the extra interactions between the two features within each individual
and/or the cross-interactions of the two adjacent individuals is to be studied
elsewhere.) The total number of all the individuals is considered infinite
in order to determine and analyze the phase transition when the
spontaneous symmetry breaking is present.

In the framework of the statistical mechanics, we investigate a combined
$q$-state Potts and $q$-state clock model which is abbreviated as the
$q^2$-state spin model. As an example, one can interpret the case of $q=3$
in the following: the feature $\sigma^{(1)}$ can be chosen
to represent {\em leisure-time interests} while the other feature $\sigma^{(2)}$
may include {\em working duties}. In the former case, one could list three
properties such as reading books, listening to music, and hiking, whereas the
latter feature could consist of manual activities, intellectual activities,
and creative activities, as an example. The thermal fluctuations, induced by
the thermodynamic temperature $T$ of the Gibbs distribution, are meant to
describe the noise, which hinders the communication. The higher the noise, the
stronger suppression of the communication is yielded.

\subsection{The reduced density matrix}

We classify the phase transitions of our model by numerical calculation of the
partition function
\begin{equation}
  \label{part_fnc}
  {\cal Z}=\sum\limits_{\{\sigma\}} \exp\left(-\frac{{\cal H}}{k_B T}\right)\, ,
\end{equation}
especially, by its derivatives. The sum has to be taken through all
multi-spin configurations $\{\sigma\}$ on the infinite lattice. The
partition function is evaluated numerically by the CTMRG algorithm.
A typical evaluation of an observable $\langle\hat X\rangle$,
i.e., the averaged value obeys the standard expression
\begin{equation}
  \langle\hat X\rangle = \frac{1}{{\cal Z}} \sum\limits_{\{\sigma\}}
     \hat X\exp\left(-\frac{{\cal H}\{\sigma\}}{k_B T}\right)
     \equiv {\rm Tr}_s \left( \hat X \,\hat \rho_s \right)\, .
\end{equation}
We have introduced a matrix $\hat\rho_s$, which has been commonly called the
{\em reduced density matrix} in quantum mechanics
\begin{equation}
 \label{rdm}
  \hat \rho_s = \frac{1}{{\cal Z}} \sum\limits_{\{\sigma_e\}}
       \exp\left(-\frac{{\cal H}\{\sigma\}}{k_B T}\right)\, .
\end{equation}
This is its classical counterpart. The quantum-classical correspondence between
one-dimensional quantum spin system and the two-dimensional classical spin system
allows as to define a reduced density matrix for a classical subsystem $s$ in
contact with the environment $e$. The reduced density matrix is defined on a line
of the spins $\{\sigma_s\}$ (thus forming the subsystem $s$) between any of the
two adjacent corner transfer matrices, whereas all the remaining spins variables
contribute to the environment $e$.
The configuration sum is taken over all spins of the environment $\{\sigma_e\}$
except those on the subsystem $\{\sigma_s\}$. If the normalization is considered,
we fulfill the condition ${\rm Tr}_s \hat \rho_s = 1$. It is nothing but the
normalized partition function ${\cal Z}=1$ for the classical statistical physics.
For details, compare with the definitions in Section~\ref{CTMRG_Section}.

Another important quantity in quantum system is the {\em entanglement entropy} $S_v$.
The entanglement entropy can be analogously defined for a classical system with
respect to the quantum-classical correspondence (Suzuki-Trotter mapping) so that
\begin{equation}
\label{ee}
S_v = -\text{Tr}_s\left(\hat{\rho}_s \log_2 \hat{\rho}_s \right).
\end{equation}
This quantity (the entanglement entropy) reflects the correlation effects in the
classical systems, which are maximized at the phase transition point.
Our model can be thought of as a system with two non-trivially coupled
sub-lattices, where either sub-lattice is composed of the $q$-state
variables of the given feature $\alpha$.

\subsection{The order parameters}

The order parameter $\langle O\rangle$ can be evaluated
via the reduced density matrix in Eq.~\eqref{rdm}. The order parameter is
either nonzero in an ordered spin phase or zero in the disordered phase.
A continuous dependence of the order parameter leads to the second-order phase
transition, whereas discontinuous behavior of the order parameter signals the
first-order phase transition. However, a detailed analysis of the free energy
and other thermodynamic functions is often necessary to distinguish the
order of the phase transition.

Let us define a {\em sub-site} order parameter for a given feature $\alpha$
\begin{equation} \label{order1}
\langle O_{\alpha}\rangle = {\rm Tr}_s\left(\hat{O}_s^{(\alpha)}\hat{\rho}_s\right)
= {\rm Tr}_s\left[\cos\left(\frac{2\pi\sigma^{(\alpha)}_{i,j}}{q}\right)
\hat{\rho}_s\right]\, ,
\end{equation}
where the sub-site order parameter $\hat O_s^{(\alpha)}$ is measured. For
simplicity, we dropped the subscripts ${i,j}$ from the order parameter
notation. Another useful definition of the order parameter, measuring both
of the spins at the same vertex, is a {\em complete} order parameter
\begin{equation} \label{order2}
\langle O\rangle = {\rm Tr}_s\left(\hat{O}_s\hat{\rho}_s\right)
 = {\rm Tr}_s\left[\cos\left(2\pi\frac{\xi_{i,j}-\phi}{q^2}\right)
\hat{\rho}_s\right]\, .
\end{equation}
We also simplify the expression into $\xi = q\sigma^{(1)} + \sigma^{(2)}$
and extend the definition of the complete order parameter by introducing
a $q^2$-state parameter $\phi$. This parameter $\phi$ specifies the alignment
of $\langle O\rangle$ towards a reference spin level given by $\phi$, where
the multi-state spin projections are measured. Unless stated explicitly in
the text, we consider the parameter $\phi=0$.

\begin{figure}[tb]
\centerline{\includegraphics[width=0.75\textwidth,clip]{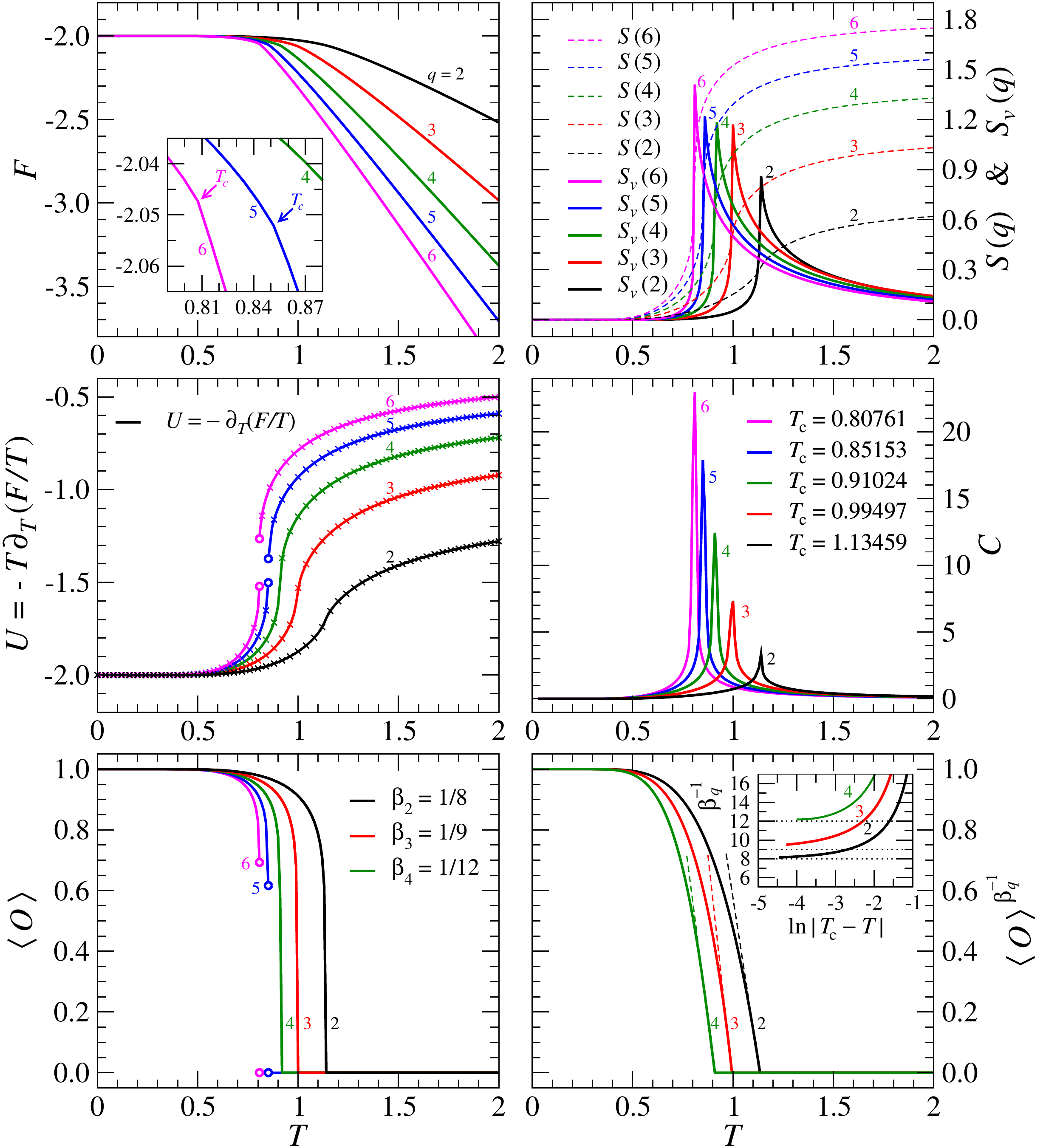}}
\caption{Thermodynamic functions of the $q$-state Potts models calculated
numerically by CTMRG. The discontinuities at the phase-transition temperature
are consequence of the first-order phase transition.}
\label{figAx0}
\end{figure}

Let us first apply CTMRG to the exactly solvable $q$-state Potts model for
$2 \geq q \geq 6$~\cite{FYWu}. Whenever $q>4$, the first-order phase transition is
present. The analytic expression for the phase transition temperature satisfies the
expression $T_{\rm c}=1/\ln(1+\sqrt{q})$. In Fig.~\ref{figAx0} we plotted
the main thermodynamic functions: the free energy $F$, the internal energy
$U=-T\frac{\partial (F/T)}{\partial T}$, the order parameter (spontaneous
magnetization) $\langle O \rangle$, the entropy $S=-\frac{\partial F}{\partial T}$,
the entanglement entropy $S_v$, the specific heat $C=\frac{\partial U}{\partial T}$,
the critical exponent $\beta$ associated with the order parameter, which
is defined only for the second-order (continuous) phase transitions.

\begin{figure}[tb]
\centerline{\includegraphics[width=0.65\textwidth,clip]{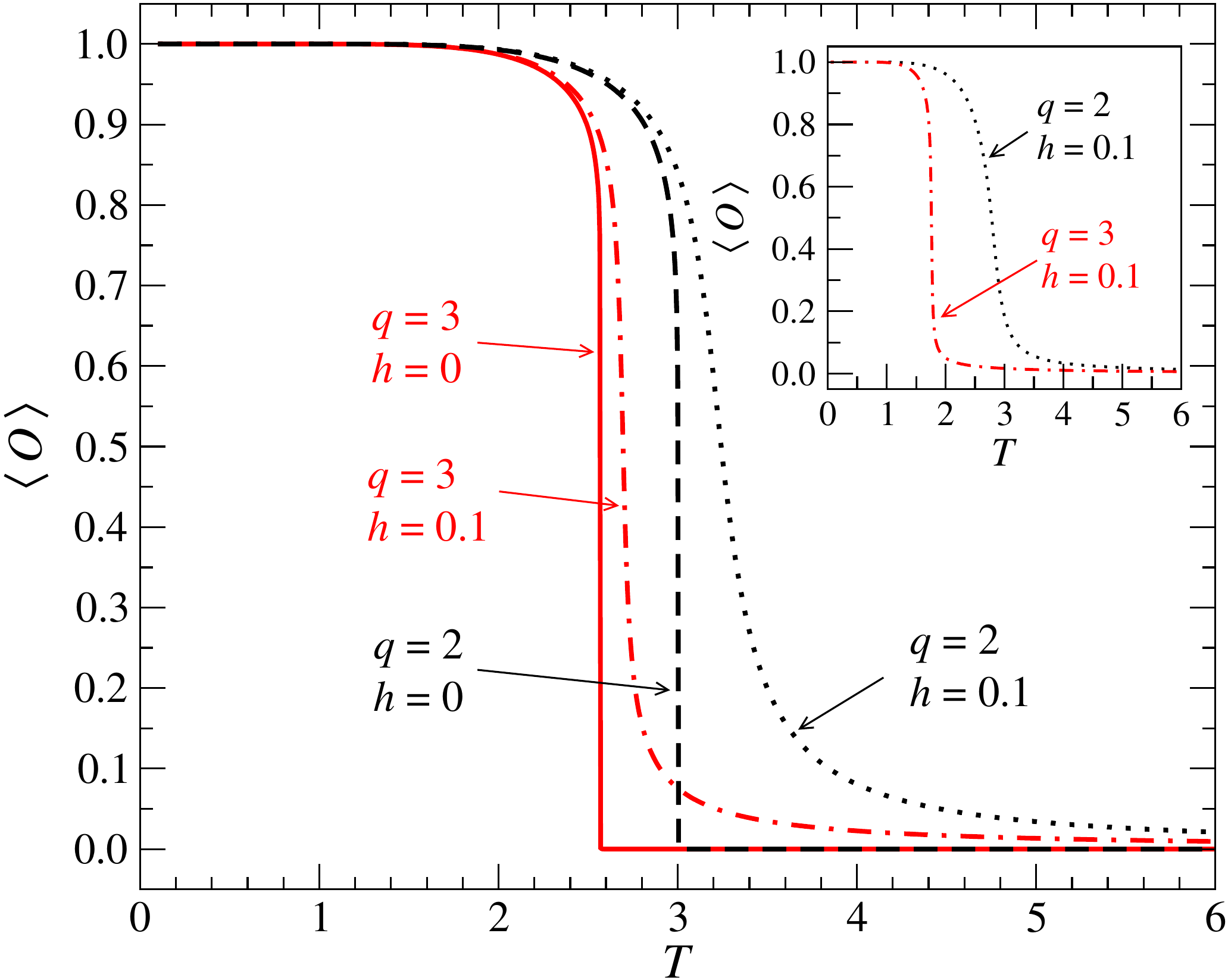}}
\caption{The temperature dependence of the complete order parameter $\langle O\rangle$
on the square lattice of the thermodynamic version of the Axelrod model studied in
Ref.~\cite{Gandica} in the case of $f=2$. A typical response of the model on the
magnetic field $h$ is shown for $q=2$ and $q=3$. (The inset depicts supplemental
information on our model Hamiltonian in Eq.\eqref{Final_H} if the magnetic field
$h=0.1$ is imposed.)}
\label{figAx1}
\end{figure}

If a magnetic field $h$ is nonzero, the thermodynamic functions are always analytic
within all temperature range and no phase transition point is detected.
Figure~\ref{figAx1} shows this case for $h=0$ and $h=0.1$ if CTMRG is applied to
the model Hamiltonian studied in Ref.~\cite{Gandica} on the two-dimensional
the square lattice. It is obvious that the dashed ($q=2$) and the full ($q=3$)
lines for the zero field exhibit the continuous phase transitions with the critical
temperatures and exponents $T_c=3.0012$, $\beta\approx\frac{1}{10}$ and
$T_c=2.5676$, $\beta\approx\frac{1}{20}$, respectively. If imposing the magnetic
field $h=0.1$, no singular behaviors is present, which points out to non-existence
of the phase transition. Since we are interested in the phase transition analysis
in our model, we exclude detailed analysis with nonzero magnetic fields.

\subsection{Numerical results}

\begin{figure}[tb]
\centerline{\includegraphics[width=0.75\textwidth,clip]{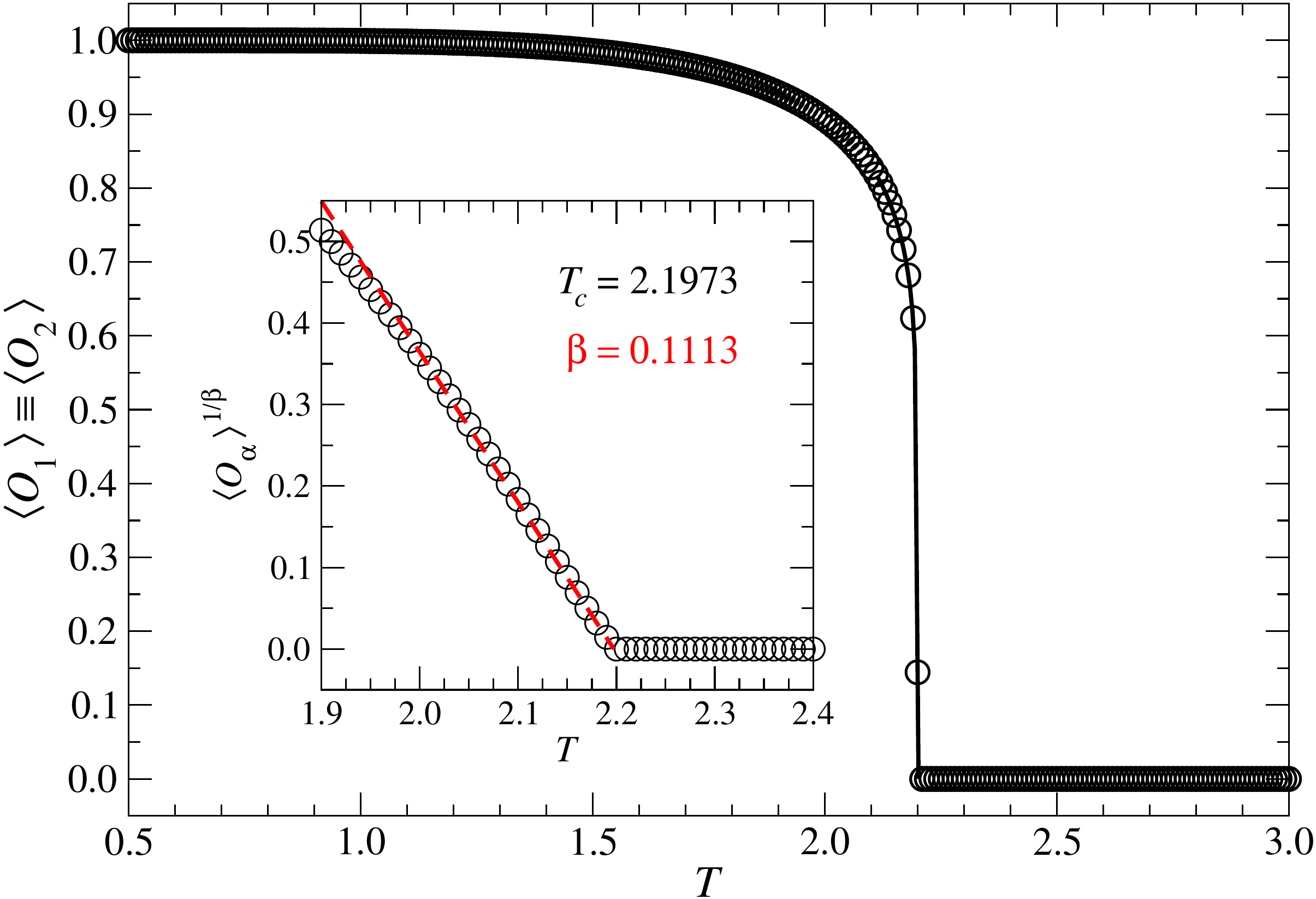}}
\caption{Temperature dependence of the sub-site order parameter
$\langle O_{\alpha}\rangle$ (the circles) for $f=2$ and $q=2$
remains unchanged for both $\alpha = 1, 2$.
The scaling relation with the critical exponent $\beta\approx0.1113$ are
plotted by the dashed line.
The inset depicts the ninth power of $\langle O_{\alpha}\rangle$ with
the expected linearity below $T_c$ coming from the scaling relation.}
\label{figg3}
\end{figure}

The phase transitions in the classical spin systems are governed by thermal
fluctuations, i.e. temperature $T$ in Eq.~\eqref{part_fnc}. We set $J = 1$,
which corresponds to the ferromagnetic spin ordering. The simplest non-trivial
case of $q=2$ is shown in Fig.~\ref{figg3}, where the sub-site order parameter
$\langle O_{\alpha}\rangle$ is plotted as a function of $T$. For both $\alpha=1$
and $\alpha=2$ the sub-site order parameters are identical.
The second order phase transition results in the critical temperature
$T_c = 2.1973$. The associated universality scaling
$\langle O_{\alpha}\rangle\propto\left(T-T_c\right)^{\beta}$ gives the
critical exponent $\beta\approx0.1113$. The inset shows the asymptotic linearity
if plotting $\langle O_{\alpha}\rangle^{1/\beta}$ below the critical point.
The critical exponent of our model at $q=2$ is is almost identical to the
$3$-state Potts model universality class~\cite{FYWu}, where $\beta=\frac{1}{9}$.
This analogy between the two models is non-trivial and requires a clarification.
Notice that the exponent $\beta\approx 0.1113$ differs from the well-known Ising
($2$-state clock) universality, where $\beta=\frac{1}{8}$. It belongs neither
to the $4$-state Potts nor the $4$-state clock model universality classes.

\begin{figure}[tb]
	\centerline{\includegraphics[width=0.75\textwidth,clip]{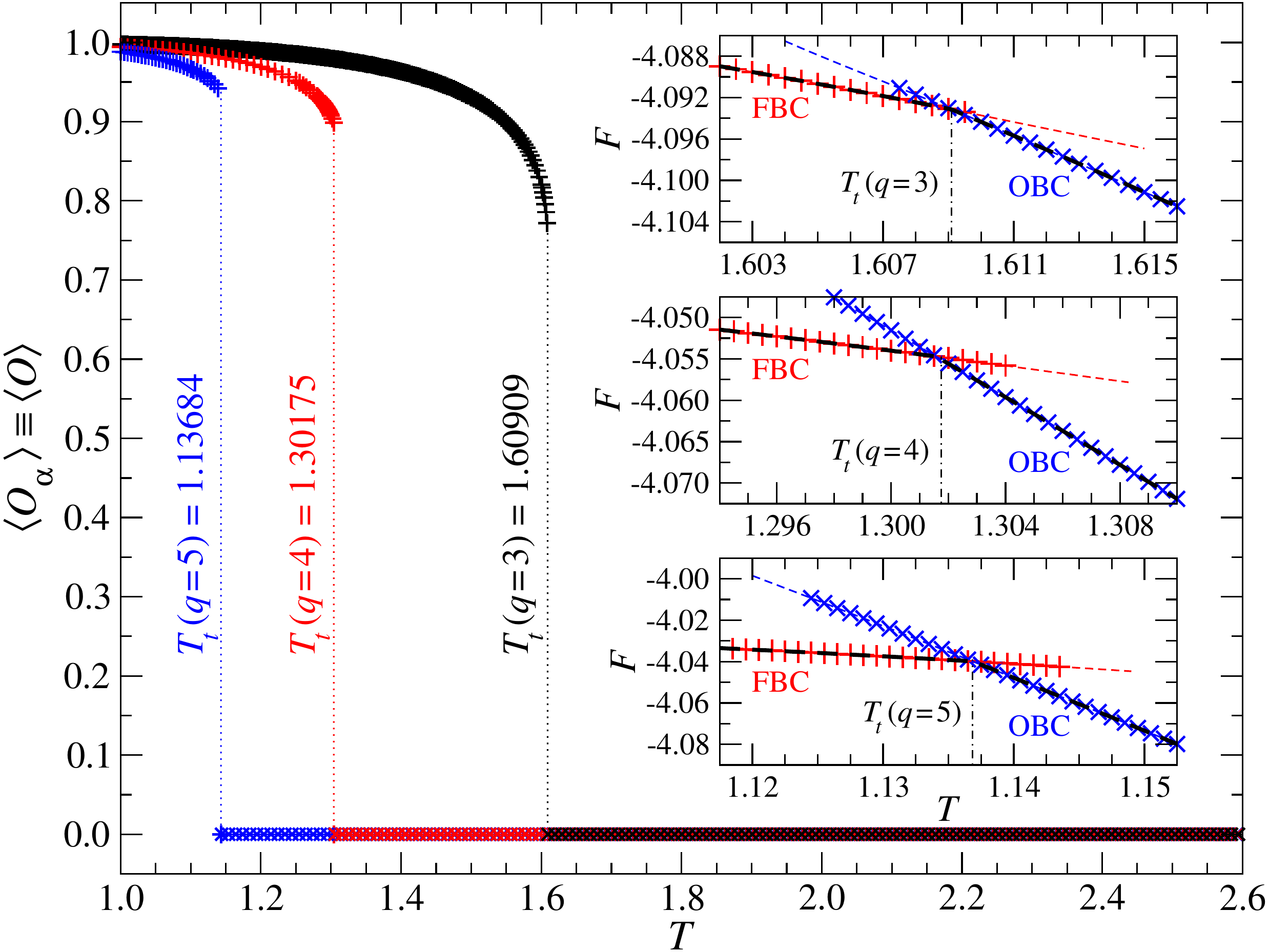}}
\caption{The order parameter $\langle O_{\alpha}\rangle$
is discontinuous for $q\geq3$ ($f=2$) and reflects the first-order phase
transition in all three cases when $3\leq q\leq5$. The free energy $F$ is
depicted in the respective three graphs on the right. We show $F$ for the
fixed BCs (red $+$ symbols) and for the open BCs (blue $\times$ symbols)
around the phase-transition temperature.}
\label{figg4}
\end{figure}

The sub-site order parameter $\langle O_{\alpha}\rangle$ for $q=3$, $4$, and
$5$ is depicted in Fig.~\ref{figg4}. It gradually decreases with increasing
temperature, but at certain temperature it discontinuously jumps to zero.
Such behavior usually signals the first-order phase transition. To confirm
this statement, the normalized Free energy $F=-k_{\rm B}T\ln{\cal Z}$ 
per spin is plotted for fixed boundary conditions (FBC) and open boundary
conditions (OBC). Both of the boundary conditions (BCs) are imposed
at the beginning of the iterative CTMRG scheme in order to enhance or suppress
the spontaneous symmetry breaking resulting in the ordered or disordered
phase in a small vicinity around the phase transition point. In particular,
if the FBC are applied, the spontaneous symmetry-breaking mechanism
selects one of $q^2$ free-energy minima as specified by the FBC.
On the contrary, the OBC prevent the spontaneous symmetry breaking from
falling into a minimum and makes the system be in a metastable state below
the phase transition. The first-order phase transition is known to exhibit
the coexistence of both the phases in a small temperature interval around the phase
transition. Therefore, the two different BCs are inevitable to apply to locate
the phase transition accurately. The insets for the three cases, $q=3,4,5$,
show the normalized free energy around the transition temperature.
The red $+$ and blue $\times$ symbols of the free energy correspond to
FBC and OBC, respectively. The temperature interval, in which two distinguishable
free energy are measured, defines the region, in which the ordered and disordered
phases can coexist. The true phase-transition temperature $T_t(q)$ is located at
the free energy crossover. The equilibrium free energy is shown by the thick dashed
line corresponding to the correct free energy, being the lower one. In this case,
the free energy becomes non-analytical at $T_t(q>2)$ and exhibits a typical kink
for the first-order phase transition (further details on the first-order phase
transition analysis can be found in Ref.~\cite{3dPotts}). Taking the
derivatives of $F$ with respect to $T$, a discontinuity at $T_t(q>2)$
in Eqs.~\eqref{int_eng} and \eqref{spec_heat} is observed. (We remark
here that the free energy is not sensitive to the different BCs if the
second-order phase transition is present, i.e., if $q=2$ in this work.)

The phase-transition temperatures for $q>2$ are calculated within a high
accuracy yielding $T_t(3)=1.60909$, $T_t(4)=1.30175$, $T_t(5)=1.12684$,
and $T_t(6)=1.03234$ ($q=6$ is not plotted) at the crossing point of the free energy.
It is obvious that $T_t(q)$ gradually decreases with increasing $q$. Below
we study the asymptotic case when $q\to\infty$. It is also worth to mention
that the first-order phase transition is not critical in sense of the non-diverging
correlation length at the phase-transition temperature. In contrast to the
second-order phase transition, when the correlation length always diverges.
For this reason, we reserve the term {\em critical} temperature, $T_c(q)$, for
the second-order phase transitions only, which is resulted in our model only
if $q=2$. Otherwise, we use the notation {\em transition} temperature $T_t(q)$.

\begin{figure}[tb]
\centerline{\includegraphics[width=0.75\textwidth,clip]{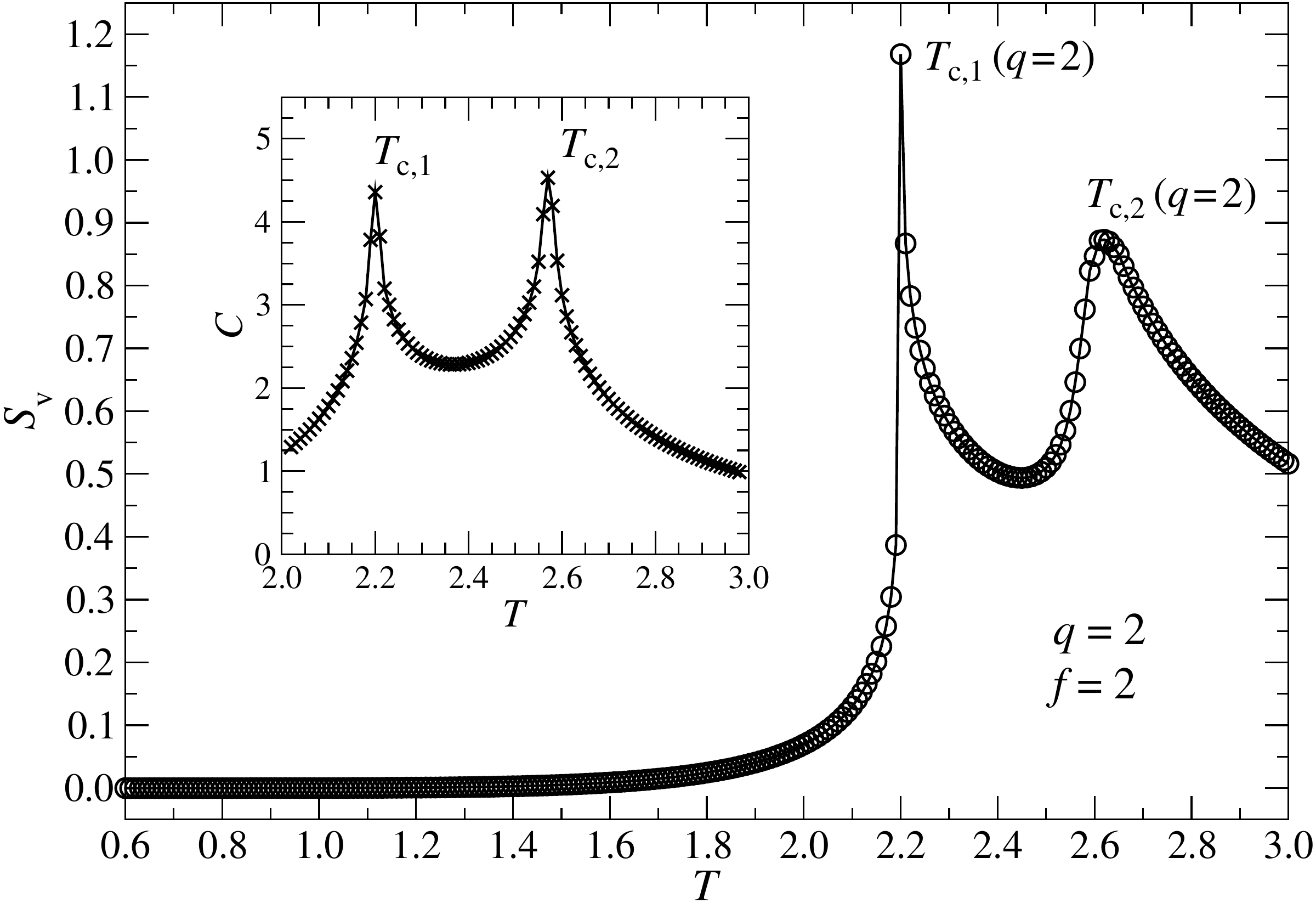}}
\caption{The temperature dependence of the entanglement entropy
$S_v$ for $q=2$ and $f=2$. The first maximum in $S_v$ coincides with the critical
temperature $T_{c,1}(2)$ plotted in Fig.~\ref{figg3}, and the second transition
appears at $T_{c,2}(2) = 2.57$. The specific heat, plotted in the inset,
reveals two maxima corresponding to the phase transition temperatures $T_{c,1}(2)$
and $T_{c,2}(2)$.}
\label{figg5}
\end{figure}

The entanglement (von Neumann) entropy $S_v$ when $q = 2$ is plotted in
Fig.~\ref{figg5}. Here our calculations of $S_v$ evidently result in two maxima,
not only a single maximum as expected for the single phase transition observed
in Fig.~\ref{figg3}. Hence, the entanglement entropy indicates existence
of another phase transition, which could not be detected by the sub-site order
parameter $\langle O_{\alpha}\rangle$. The phase transition at lower temperature,
$T_{c,1}(q=2)=2.1973$, coincides with the one plotted in Fig.~\ref{figg3},
however, a higher-temperature phase transition appears at $T_{c,2}(q=2)=2.57$.
To support this finding obtained by $S_v$, we calculated the specific heat
$C$, as shown in the inset. Again, there are two evident maxima in $C$, which
remain present in our model at the identical critical temperatures $T_{c,1}(2)$
and $T_{c,2}(2)$. Thus, the sub-site order parameter $\langle O_{\alpha}\rangle$
in Fig.~\ref{figg3} cannot reflect the higher-temperature phase transition
at all. We have achieved a new phase transition point, which is likely
pointing to a topological ordering. A second-order transition has been
found in the out-of-equilibrium Axelrod model~\cite{AxRev}. In addition,
the existence of a modulated order parameter with two different phase-transition
temperatures has been reported earlier (often being associated with experimental
measurements of the magnetization in crystal alloys~\cite{Ito,Sakon}).

\begin{figure}[tb]
\centerline{\includegraphics[width=0.75\textwidth,clip]{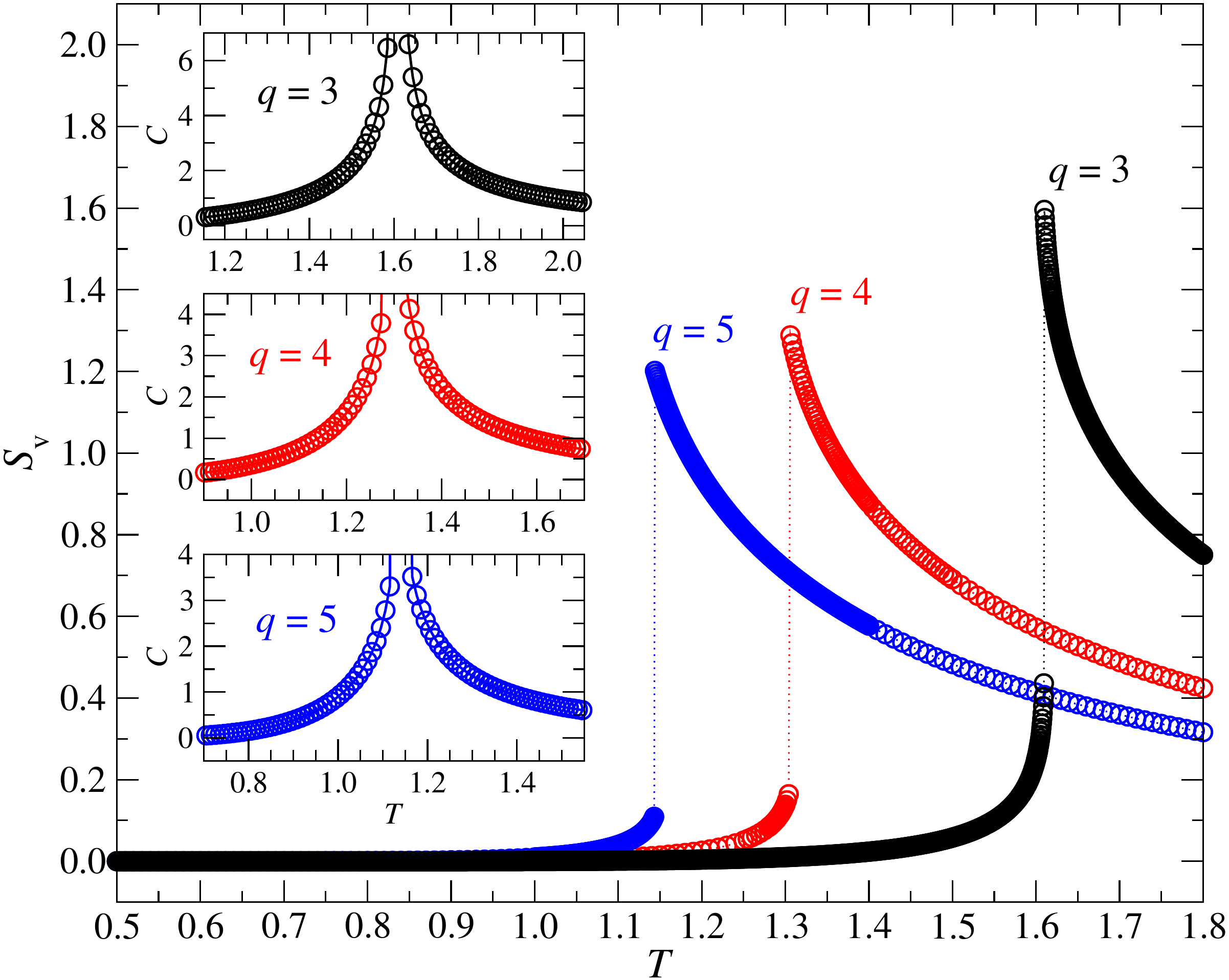}}
\caption{The entanglement entropy for $q=3$, $4$, and $5$ exhibits a single
maximum. The inset shows the specific heat $C$ reflects the single
(first-order) phase-transition temperature.}
\label{figg6}
\end{figure}
\begin{figure}[htb!]
\centerline{\includegraphics[width=0.75\textwidth,clip]{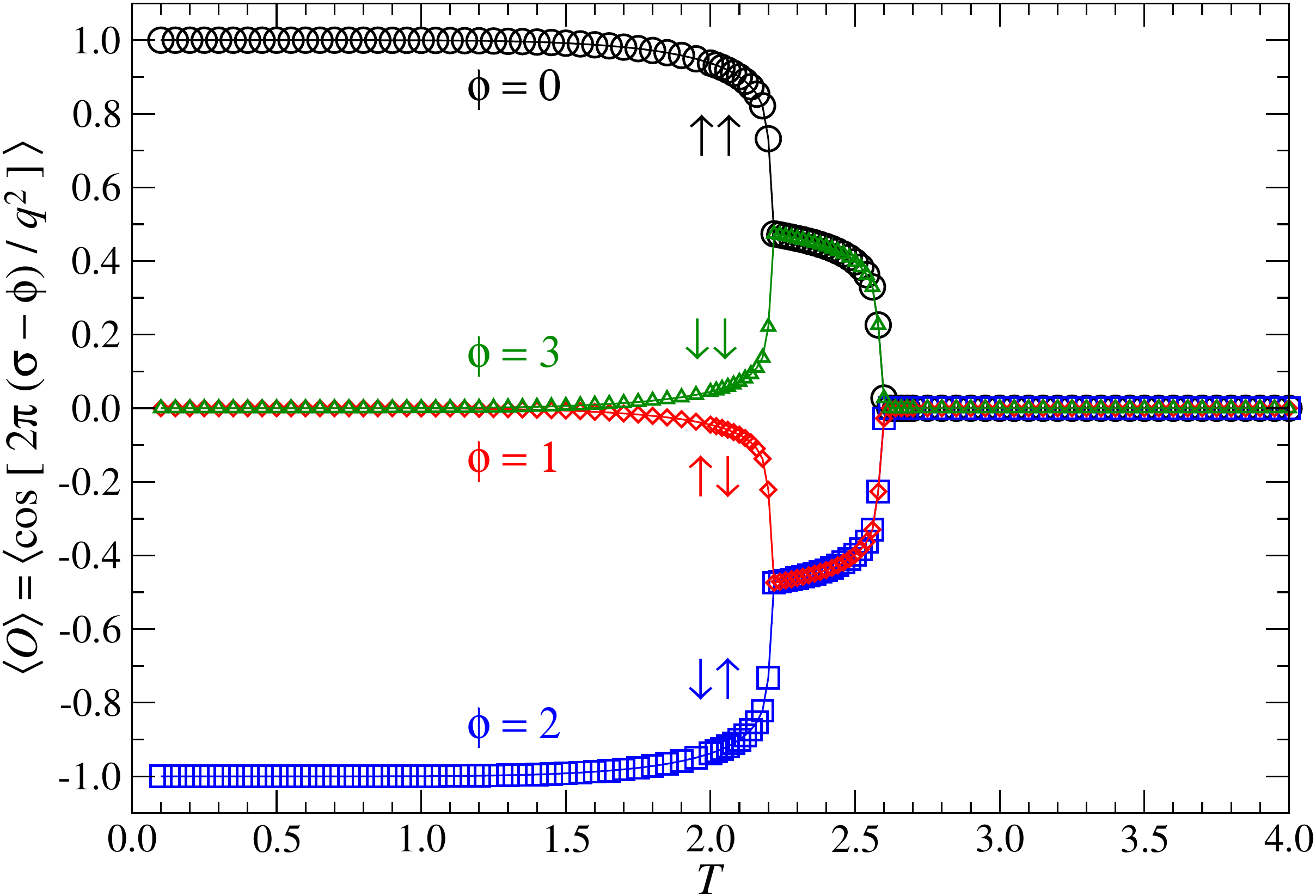}}
\caption{The complete order parameter acting on the $q^2$-state
variable $\xi$ exhibits the presence of the two phase transition temperatures
if $q=2$ and $f=2$. All of the four reference spin levels (labeled by $\phi$)
are displayed after the spontaneous symmetry breaking occurs.}
\label{figg7}
\end{figure}

The entanglement entropy $S_v$ exhibits only a single maximum for any $q>2$ as seen
in Fig.~\ref{figg6}. The discontinuity of $S_v$ at the phase-transition temperature
$T_t(q)$ is characteristic for the first-order phase transition. The three
insets display the specific heat with the single maximum for each $q>2$ at the
transition temperature, which is in full agreement with the observation of the
sub-site order parameter. Therefore, we conclude existence of the single
phase-transition point of the first order whenever $q>2$.

Figure~\ref{figg7} shows the complete order parameter when $q=2$ as defined in
Eq.~\eqref{order2}. Obviously, the non-analytic behavior of $\langle O\rangle$
points to the two well-distinguishable critical temperatures $T_{c,1}(2)$ and
$T_{c,2}(2)$, which coincide with the critical temperatures depicted in
Fig.~\ref{figg5}. The $q^2$-state spin variable $\xi$ has four degrees of
freedom targeted by the parameters $\phi=0,1,2,3$. The complete order
parameter is explicitly evaluated for each $\phi$ separately. It satisfies
a condition, for which the sum of all the four complete order parameters at
temperature $T$ has to be zero. The mechanism of the spontaneous symmetry
breaking at low temperatures thus causes that the free energy becomes
four-fold degenerate at most. This feature is associated with four
equivalent free energy minima with respect to the complete order parameter.
Accessing any of the four free energy minima by CTMRG is numerically feasible
just by targeting the reference spin state $\phi$.

Let us denote the four spin state at the vertex by the notation
$|\sigma^{(1)}\sigma^{(2)}\rangle$.
There are four possible scenarios for the order parameter $\langle O\rangle$
as shown in Fig.~\ref{figg7}. These scenarios are depicted by the black circles
($\phi=0$), the red diamonds ($\phi=1$), the blue squares ($\phi=2$), and the
green triangles ($\phi=3$), which correspond to the following vertex
configurations $|\hspace{-0.12cm}\uparrow\uparrow\rangle$,
$|\hspace{-0.12cm}\uparrow\downarrow\rangle$,
$|\hspace{-0.12cm}\downarrow\uparrow\rangle$,
and $|\hspace{-0.12cm}\downarrow\downarrow\rangle$, respectively.

At zero temperature there are three minima of the free energy leading to
the three different complete order parameters $\langle O\rangle$ being $-1$,
$0$, and $+1$. There are four minima of the free energy at $0<T<T_{c,1}(2)$
so that the order parameter has four different values $\langle O\rangle =
-1+\varepsilon$, $-\varepsilon$, $+\varepsilon$, and $+1-\varepsilon$ with
the condition $0<\varepsilon\leq\frac{1}{2}$. It means the two states share
the identical free-energy minimum when the order parameter is zero at $T=0$ and
$\varepsilon=0$. In the temperature interval $T_{c,1}(2)\leq T< T_{c,2}(2)$,
there are only two free-energy minima present. The pairing of the order
parameter for $\phi=0$ and $\phi=3$ is identical to the pairing for $\phi=1$
and $\phi=2$. Finally, a single free-energy minimum is resulted at $T\geq
T_{c,2}(2)$ when the order parameter is zero, which is typical in the
disordered phase.

Let us stress that at the temperatures in between
$T_{c,1}(2)$ and $T_{c,2}(2)$, the pairing of the site configurations
$|\hspace{-0.12cm}\uparrow\uparrow\rangle$ and
$|\hspace{-0.12cm}\downarrow\downarrow\rangle$
is indistinguishable by the complete order parameter
(i.e. the black and green symbols coincide), and the same topological
uniformity happens for the pairing of the site configurations
$|\hspace{-0.12cm}\uparrow\downarrow\rangle$ and
$|\hspace{-0.12cm}\downarrow\uparrow\rangle$. In other words,
the anti-parallel alignments between the spins $\sigma^{(1)}$ and
$\sigma^{(2)}$ are preferable in the temperature region
$T_{c,1}(2)\leq T< T_{c,2}(2)$.

If calculating the critical exponent $\beta$ of the complete order parameter
at the critical temperatures $T_{c,1}(2)$ and $T_{c,2}(2)$, we obtained
$\beta\approx\frac{1}{18}$ if $T\to T_{c,1}(2)$, whereas the other
critical exponent remains identical to the previous one, as we have discussed
above, in particular, $\beta\approx\frac{1}{9}$ if $T\to T_{c,2}(2)$.

\begin{figure}[tb]
\centerline{\includegraphics[width=0.75\textwidth,clip]{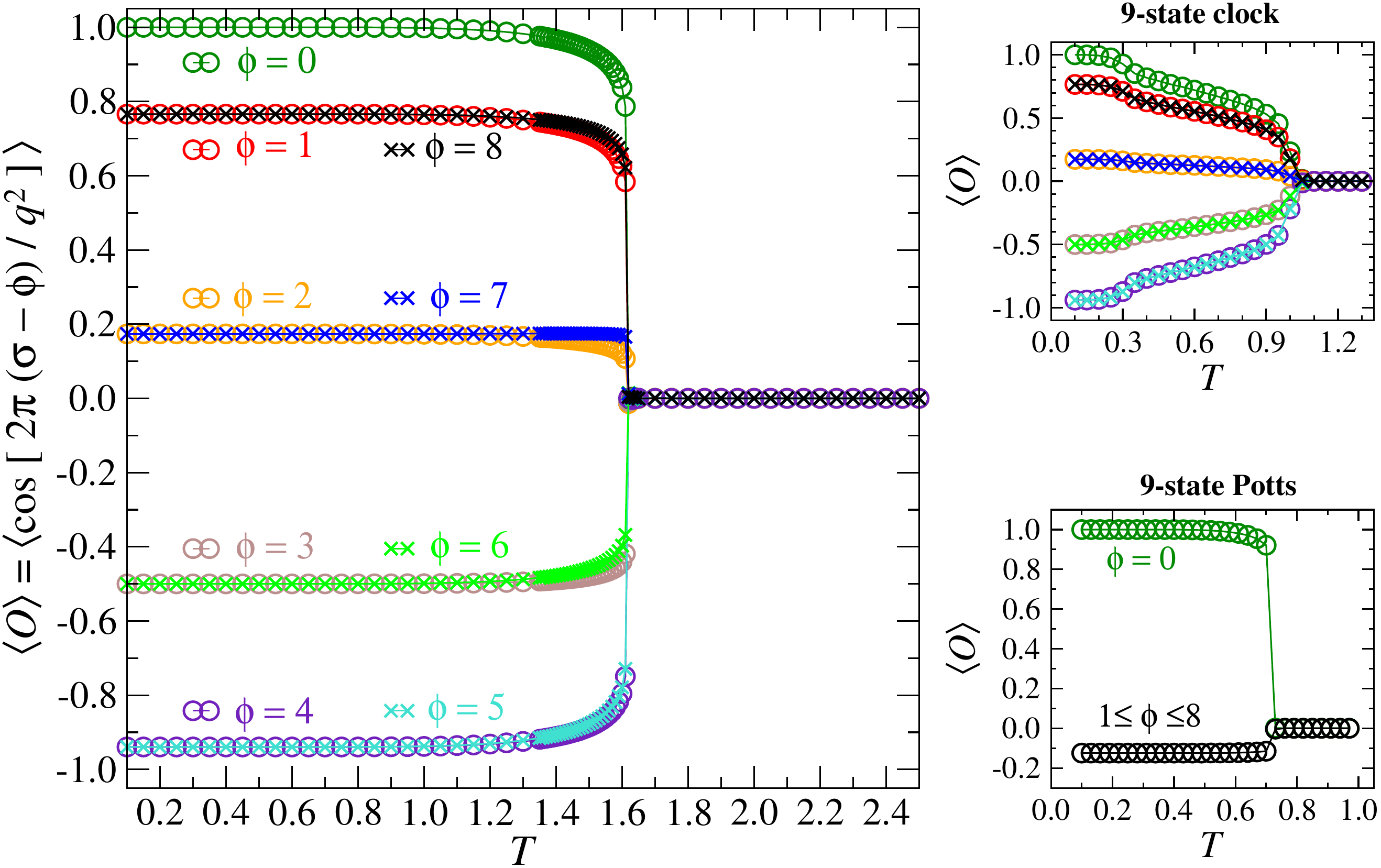}}
\caption{The complete order parameter acting on the $q^2$-state spin model
at $q=3$ plotted for the reference levels $\phi=0,1,...,8$ ($f=2$).
For comparison, the order parameter of the $9$-state clock and Potts models
are shown on the right top and bottom graphs.}
\label{figg8}
\end{figure}
\begin{figure}[htb!]
\centerline{\includegraphics[width=0.7\textwidth,clip]{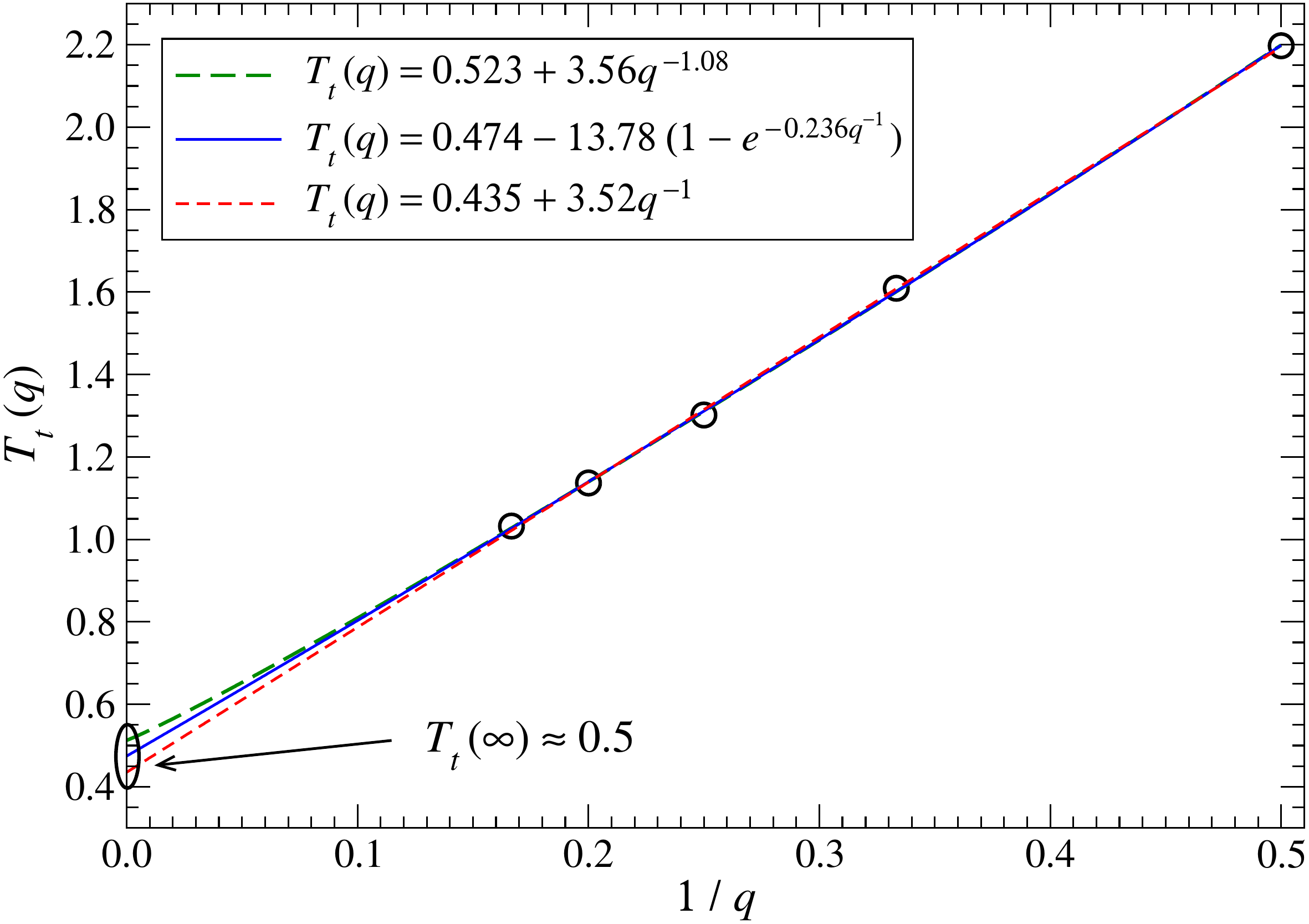}}
\caption{The three variants of extrapolating the transition temperature
$T_t(q\to\infty)$ by the power-law least-square fitting (the green long-dashed
line), the exponential fitting (the blue full line), and the inverse
proportionality (the red short-dashed line).}
\label{figg9}
\end{figure}

In analogy, we plotted the complete order parameter for $q=3$ in
Fig.~\ref{figg8}. The free energy is five-fold degenerated at zero temperature
unless the symmetry breaking mechanism (enhanced by $\phi$) selects one of them.
This mechanism results in the five distinguishable order parameters within
$0\leq\phi\leq8$, which decouple into nine different order parameters when
$0<T<T_t(3)$. A single free-energy minimum is characteristic in the disordered
phase at $T\geq T_t(3)$, which exhibits a uniform $\langle O\rangle=0$.
In order to compare the main differences of the complete order parameter
between our model and the standard $9$-state clock model and the $9$-state
Potts models, we calculated the respective order parameters shown in the insets
of Fig.~\ref{figg8}. In the case of the $9$-state clock model, there are
five distinguishable order parameters originating in the five-fold
degeneracy of the free energy below the phase transition. Thus the five-fold
degeneracy persists within the interval $0<T<T_t(3)$. (We also remark that the
Berezinski-Kosterlitz-Thouless phase transitions~\cite{BKT1,BKT2} occurs in the
$q\geq5$-state clock models~\cite{BKT3}.) In the case of the $9$-state Potts model,
there are only two distinguishable order parameters out of nine below the phase
transition point. (Recall again that the total sum of $\langle O\rangle$ over all
$\phi$ has to be zero.) The discontinuity in the complete order parameter at
$T_t(3)$ in our model and the $9$-state Potts model reflects the first-order
phase transition, as it is common for $q$-state Potts models on the square lattice
with $q>4$~\cite{FYWu}.

If the number of the spin degrees of freedom $q$ increases, numerical calculations
become memory and time demanding. If extrapolating the spin degrees of the freedom
$q\to\infty$, a nonzero phase transition temperature $T_t(\infty)$ is resulted.
We have carried out three independent extrapolations as depicted in Fig.~\ref{figg9}
by means of the least-square fitting. In particular, the power-law
$T_t(q)=T_t(\infty)+a_0\,q^{-a_1}$, the exponential $T_t(q)=T_t(\infty)+a_0
(1-e^{-a_1/q})$, and the inverse proportional $T_t(q)=T_t(\infty)+a_0\, q^{-1}$
fitting functions have been used to estimate the asymptotic values of $T_t(\infty)$,
$a_0$, and $a_1$. All of them yielded a nonzero transition temperature being
$T_t(\infty)\approx0.5$. Based on these extrapolations, we conjecture existence
of the ordered phase regardless of $q$ in our model, i.e., the nonzero
phase-transition temperature $T_t(q)$ persists for any $q\geq2$.

\newpage\setcounter{equation}{0} \setcounter{figure}{0} \setcounter{table}{0}

\section{Fractal geometries} \label{chap5}

\begin{figure}[b!]
\begin{center}
\includegraphics[width=2.8in]{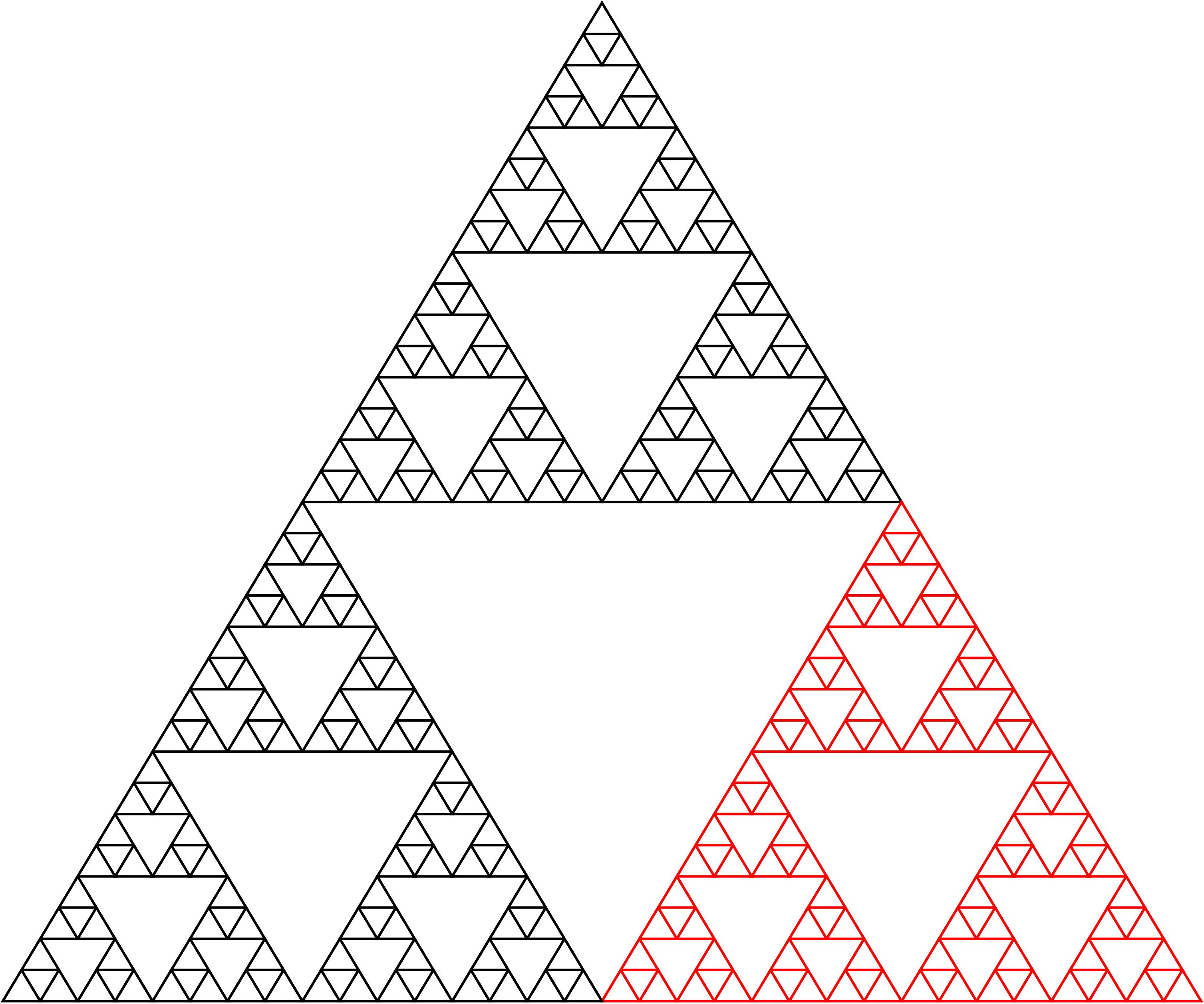}
\end{center}
\caption{
The Sierpinski gasket. 
}
\label{sierpinski}
\end{figure}

A fractal is a geometric structure which exhibits the property of \textit{self-similarity} at every scale, i.e., 
as we zoom in (or zoom out), the same (self-similar) pattern is repeated. Let us demonstrate this concept 
on the Sierpinski gasket, see Fig.~\ref{sierpinski}. Each triangle can be decomposed into the three smaller 
triangles, which are the exact replicas of the original. For instance, zooming to the lower-left triangle (red), 
we obtain the triangle we have started with (for simplicity, only the six levels of the Sierpinski gasket are depicted). 

Another important property of the fractal is its fractional dimension. 
The Hausdorff dimension $d^{ ( {\rm H} ) }_{~}$ can be understood in terms of the relation
\begin{equation} \label{hausdorff_dim}
N = L^{d^{ ( {\rm H} ) }_{~}} \, , 
\end{equation}
where $L$ is the linear dimension (i.e. the magnification factor) and $N$ is the number of copies (i.e. the self-similar pieces). 
For example, doubling a line segment gives two copies of the original line (one dimensional case, $2=2^1$), 
doubling both the length and the width of a square gives four copies of the original square (two dimensional case, $4=2^2$),
or doubling the all three linear dimensions of a cube gives eight copies of the original cube (three dimensional case, $8=2^3$),
all in accord with Eq.~\eqref{hausdorff_dim}. In the case of the Sierpinski gasket, doubling the linear dimension gives three copies of the 
original triangle, therefore $d^{ ( {\rm H} ) }_{~} = \ln 3 / \ln 2 \approx 1.585$. 

The dimension can be also introduced in an alternative way, which may be even more relevant in the context of this work. 
It is reasonable to ask about the scaling of the size of the boundary $M$ with respect to the linear dimension $L$
\begin{equation} \label{ML_relation}
M = L^{d-1}_{~} \, . 
\end{equation}
Again, for example, doubling a line segment does not change the size of the boundary (which always consists of two points, thus $d=1$), 
doubling the linear dimensions of a square increases the size of the boundary by a factor of two ($d=2$), 
and doubling the linear dimensions of a cube increases the area of the cube by a factor of four ($d=3$). 
Considering the lattices, the size of the boundary $M$ naturally represents the number of outgoing bonds. 
Doubling the linear dimension of the Sierpinski gasket does not change the number of outgoing bonds, thus $d=1$. 

In this Chapter, the phase transition of the Ising model is investigated on a planar lattice that has a fractal 
structure. On the lattice, the number of bonds that cross the border of a finite area is 
doubled when the linear size of the area is extended by a factor of four. The free energy 
and the spontaneous magnetization of the system are obtained by means of the HOTRG method. 
Our modification of the HOTRG method used in the study is 
explained in Section~\ref{frac_meet_hotrg}. As shown in Section~\ref{frac_num_results}, 
the system exhibits an order-disorder phase transition,
where the critical indices are different from those of the square-lattice Ising model. An
exponential decay is observed in the density matrix spectrum even at the critical point.
It is possible to interpret the system as being less entangled because of the fractal geometry.

\subsection{Introduction}

The phase transitions and critical phenomena have been one of the central issues in statistical 
analyses of the condensed matter physics~\cite{Domb_Green}. When the second-order phase 
transition is observed, thermodynamic functions, such as the free energy, the internal energy, 
and the magnetization, show non-trivial behavior around the transition temperature 
$T_{\rm c}^{~}$~\cite{Fisher, Stanley}, see also Section~\ref{PT_Section}. 
This critical singularity reflects the absence of any 
scale length at $T_{\rm c}^{~}$, and the power-law behavior of the thermodynamic functions 
around the transition can be explained by the concept of the renormalization 
group~\cite{Kadanoff, Kadanoff2, Wilson-Kogut,Domb_Green}.

An analytic investigation of the renormalization group flow in $\varphi^4_{~}$-model shows 
that the Ising model exhibits a phase transition when the lattice dimension is larger than 
one, which is the lower critical dimension~\cite{Wilson-Kogut, Justin}. In a certain sense, 
the one-dimensional Ising model shows rescaled critical phenomena around 
$T_{\rm c}^{~} = 0$. When the lattice dimension is larger than four, which is the upper 
critical dimension, and provided that the system is uniform, then the classical Ising model on regular 
lattices exhibits mean-field-like critical behavior.

Compared with the critical phenomena on regular lattices, much less is known on fractal 
lattices. Renormalization flow is investigated by Gefen et al.,~\cite{Gefen1, Gefen2,
Gefen3, Gefen4} 
where correspondence between lattice structure and the values of critical indices is
not fully understood in a quantitative manner. For example, the Ising model on the Sierpinski 
gasket does not exhibit any phase transition at any finite temperature, although the Hausdorff 
dimension of the lattice, $d^{ ( {\rm H} ) }_{~} = \ln 3 / \ln 2 \approx 1.585$, is larger than one~\cite{Gefen5, Luscombe}.
The absence of the phase transition could be explained by the fact that the number of 
interfaces, i.e., the outgoing bonds from a finite area, does not increase when the size 
of the area is doubled on the gasket. A non-trivial feature of this system is that there is
a logarithmic scaling behavior in the internal energy toward zero temperature~\cite{log}. 
The effect of anisotropy has been considered recently~\cite{Ran}.
In case of the Ising model on the Sierpinski 
carpet, the presence of the phase transition is proved~\cite{Vezzani}, and its critical indices 
were roughly estimated by Monte Carlo simulations~\cite{Carmona}. It should be noted 
that it is not easy to collect sufficient number of data points for finite-size
scaling~\cite{FSS} 
on such fractal lattices by means of Monte Carlo simulations, because of the exponential 
blow-up of the number of sites in a unit of fractal. 

\begin{figure}[tb]
\begin{center}
\includegraphics[width=0.8\textwidth]{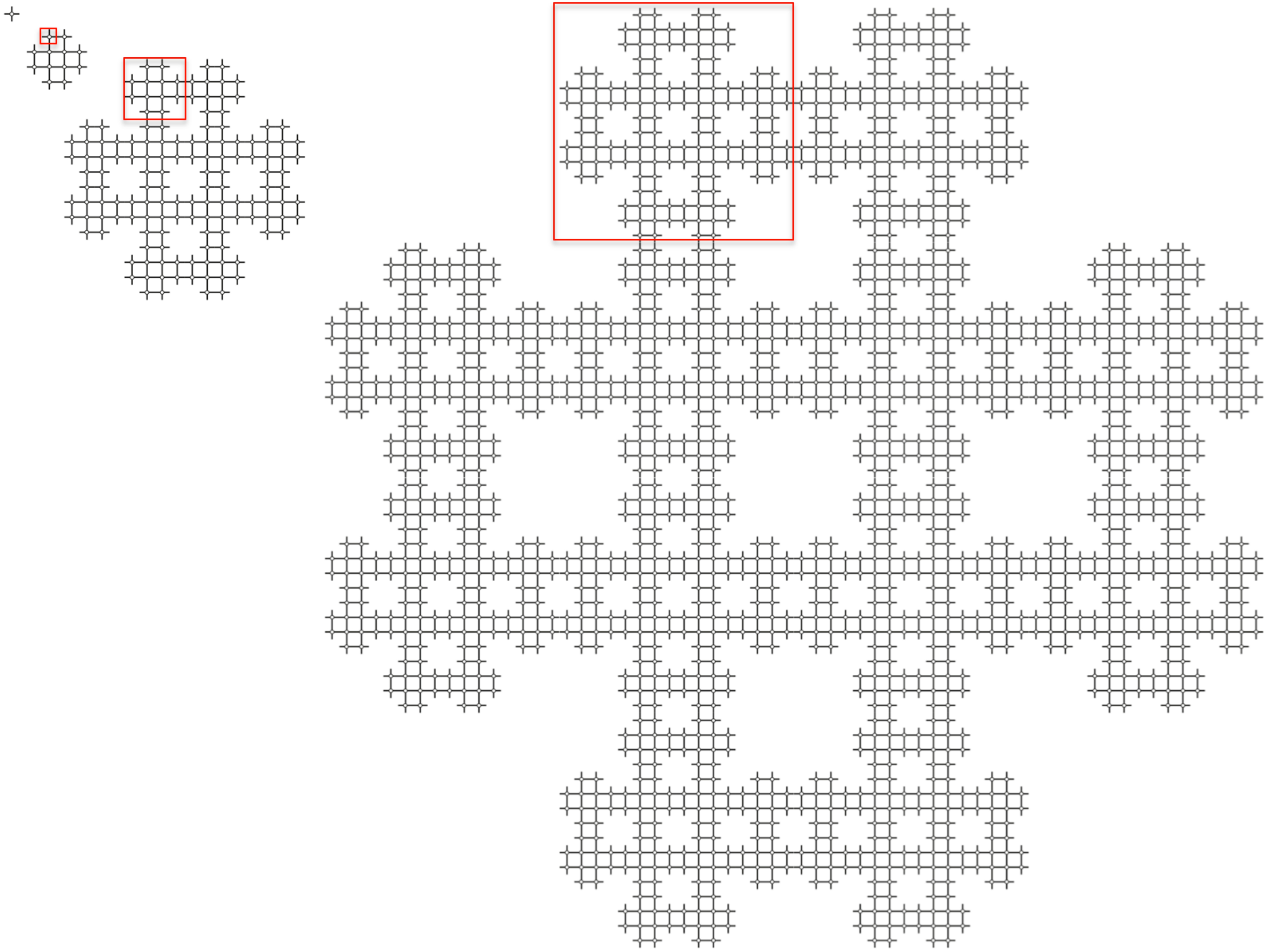}
\end{center}
\caption{
Composition of the fractal lattice. Upper left: a local vertex around an Ising spin
shown by the empty dot in the zeroth iteration step $n=0$. Middle: the basic cluster which contains $N_{n=1}^{~} = 12$ 
vertices in the first iteration step. Lower right: the extended cluster which contains $N_{n=2}^{~} = 12^2_{~}$ 
vertices at the second iteration step. In each step $n$ of the system extension, the linear size of the system increases 
by the factor of 4, where only 12 units are linked, and where 4 units at the corners 
are missing, if it is compared with a 4 by 4 square cluster. 
}
\label{fig:Fig_1}
\end{figure}

In this study, we investigate the Ising model on a planar fractal lattice, shown in 
Fig.~\ref{fig:Fig_1}. The lattice consists of vertices around the lattice points, which are 
denoted by the empty dots in the figure, where the Ising spins are positioned. The whole lattice 
is constructed by recursive extension processes, where the linear size of the system 
increases by the factor of four in each step. If the lattice is a regular square one, 
$4 \times 4 = 16$ units are connected in the extension process, whereas only 12 units 
are connected on this fractal lattice; 4 units are missing in the corners. As a result, 
the number of sites contained in a cluster after $n$ extensions is $N_n^{~} = 12^n_{~}$, 
and the Hausdorff dimension of this lattice is $d^{ ( {\rm H} ) }_{~} = \ln 12 / \ln 4 \approx 1.792$. 
The number of outgoing bonds from a cluster is only doubled in each extension process 
since the sites and the bonds at each corner are missing. If we evaluate the lattice dimension 
from the second relation Eq.~\eqref{ML_relation}
between the linear dimension $L$ and the number of outgoing bonds $M$, we get $d = 1.5$, 
since $M$ is proportional to $\sqrt{L}$ on the fractal. Remark that the value is different from 
$d^{ ( {\rm H} ) }_{~} \approx 1.792$

\subsection{Fractal meets HOTRG}\label{frac_meet_hotrg}

\begin{figure}[tb]
 \centering
 \includegraphics[width=2.0in]{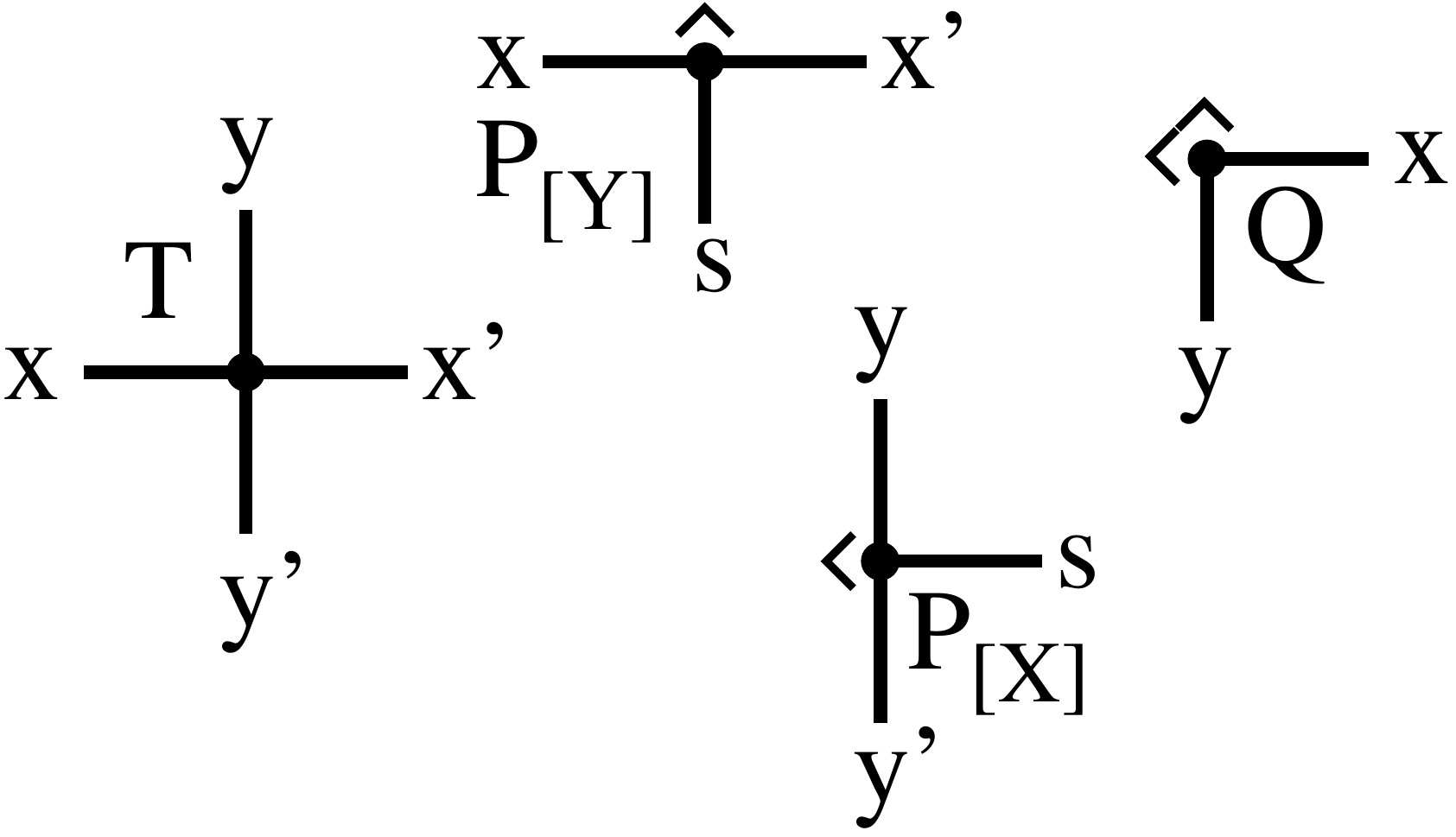}
 \caption{Four types of local tensors. A tensor network state can be decomposed into the local tensors $T$, $P$ ($P_{\text{[Y]}}$ and $P_{\text{[X]}}$), and $Q$. The missing legs are indicated by the carets ``$\wedge$'' and ``${\small <}$''. } \label{Initial}
 \end{figure}

The partition function of the Ising model defined on the fractal lattice can be represented as a tensor network state
 with three (four) types of the local tensors $T$, $P$ ($P_{\text{[Y]}}$ and $ P_{\text{[X]}}$), and $Q$ (see Fig.~\ref{Initial}),

\begin{equation}
T_{x_i^{~} x'_i y_i^{~} y'_i} = \sum_{\sigma} W_{\sigma x_i^{~}} W_{\sigma x'_i} W_{\sigma y_i^{~}} W_{\sigma y'_i} \, ,
\end{equation}

\begin{equation}
P_{\text{[Y]} x_i^{~} x'_i s_i^{~}}  = \sum_{\sigma} W_{\sigma x_i^{~}} W_{\sigma x'_i} W_{\sigma s_i^{~}} \, ,
\end{equation}

\begin{equation}
P_{\text{[X]} y_i^{~} y'_i s_i^{~}}  = \sum_{\sigma} W_{\sigma y_i^{~}} W_{\sigma y'_i} W_{\sigma s_i^{~}} \, ,
\end{equation}

\begin{equation}
Q_{x_i^{~} y_i^{~}} = \sum_{\sigma} W_{\sigma x_i^{~}} W_{\sigma y_i^{~}} \, , 
\end{equation}
where $W$ is a $2\times2$ matrix determined by the bond weight factorization\footnote{See the explanation of the tensor-network representation in Subsection~\ref{TN_representation}.}. For instance, let's choose the asymmetric factorization

\[ W =  \left(\begin{array}{cc} 
\sqrt{\cosh(1/T)}, & \sqrt{\sinh(1/T)}   \\
\sqrt{\cosh(1/T)} ,& -\sqrt{\sinh(1/T)} \end{array} \right) \, . \]  
The absent legs in the tensors $P$ and $Q$ are graphically indicated by the carets ``$\wedge$'' and ``${\small <}$'', this notation becomes clear from the coarse-graining procedure explained in this Section.

In order to calculate the partition function, we adapted the coarse-graining renormalization procedure from \cite{HOTRG}, which is explained in Section~\ref{HOTRG_Section}. However, the construction of the fractal lattice is slightly more intricate, so we explain our adapted procedure in more depth in the following. 

\begin{figure}[tb]
 \centering
 \includegraphics[width=3.1in]{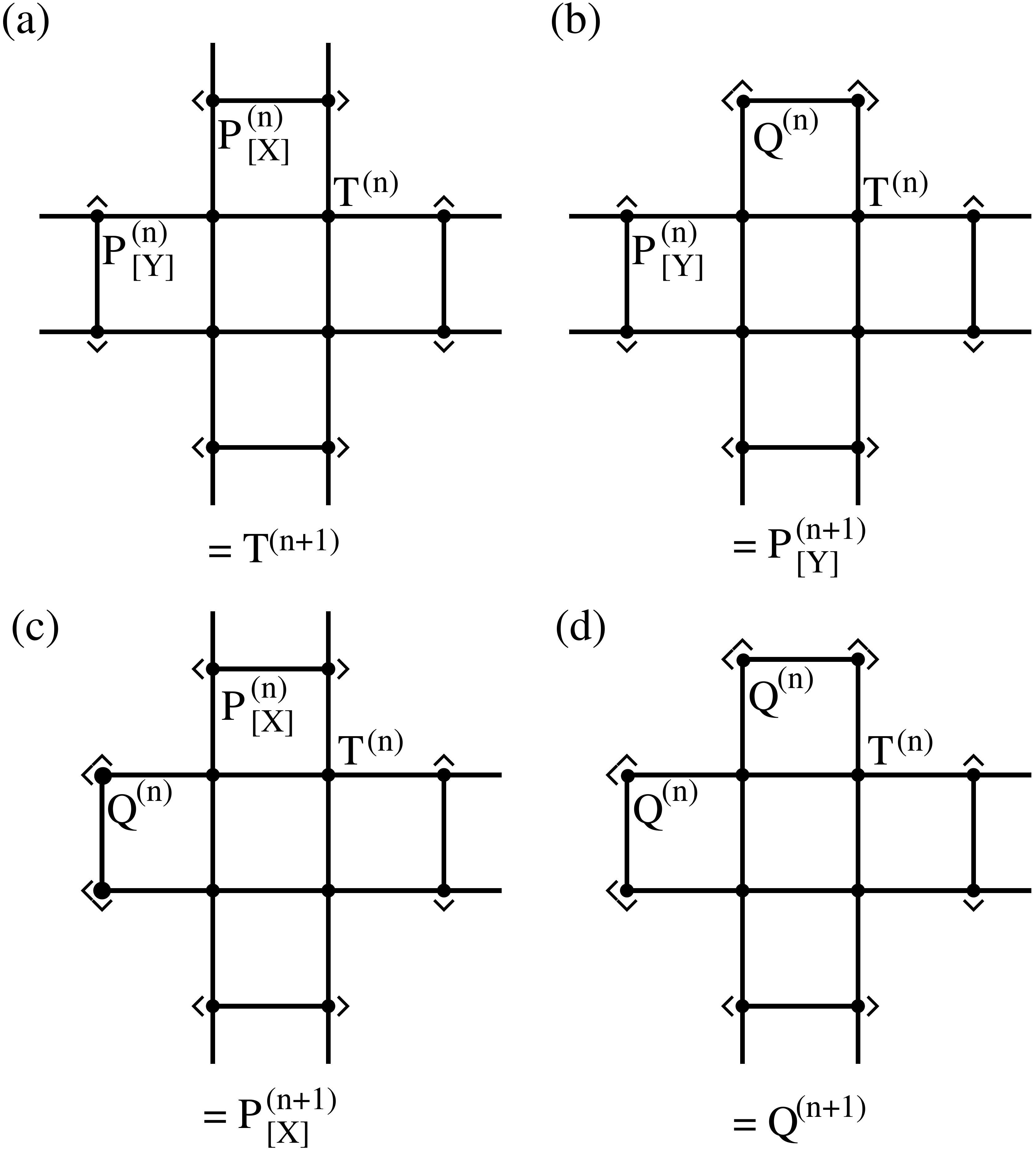}
 \caption{Composition of new tensors: (a) A new tensor $T^{(n+1)}$ is created from four tensors $T^{(n)}$, four tensors $P^{(n)}_{\text{[Y]}}$, and four tensors $P^{(n)}_{\text{[X]}}$. (b) A new tensor $P_{\text{[Y]}}^{(n+1)}$ is created from four tensors $T^{(n)}$, four tensors $P^{(n)}_{\text{[Y]}}$, two $P^{(n)}_{\text{[X]}}$, and two $Q^{(n)}$. (c) A new tensor $P_{\text{[X]}}^{(n+1)}$ is created from four $T^{(n)}$,  two $Q^{(n)}$, two $P^{(n)}_{\text{[Y]}}$, and four $P^{(n)}_{\text{[X]}}$. (d) A new tensor $Q^{(n+1)}$ is created from four $T^{(n)}$, four $Q^{(n)}$, two $P^{(n)}_{\text{[Y]}}$, and two $P^{(n)}_{\text{[X]}}$. } \label{TPQ}
 \end{figure}

\begin{figure}[tb]
 \centering
 \includegraphics[width=3.8in]{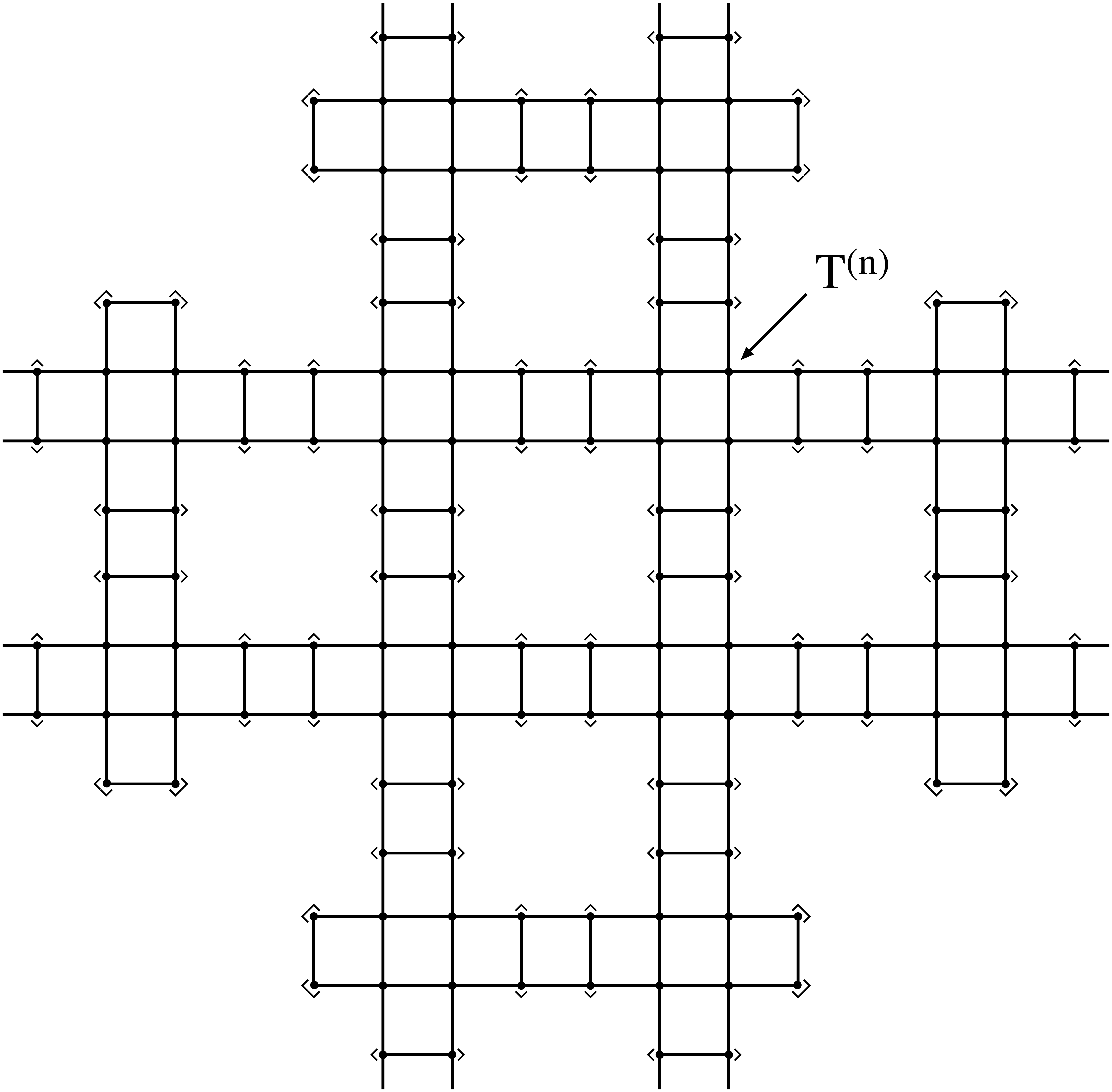}
 \caption{Structure of a tensor $T^{(n+2)}$. For clarity, one tensor $T^{(n)}$ is denoted in the picture. } \label{Large}
 \end{figure}

At each iterative step $n$, the new tensors $T^{(n+1)}$, $P^{(n+1)}$, and $Q^{(n+1)}$ are created from the preceding tensors $T^{(n)}$, $P^{(n)}$, and $Q^{(n)}$ (see Fig.~\ref{TPQ}). Practically, this is achieved in several steps. Firstly, two tensors $T^{(n)}$ are contracted and renormalized along the $y$ axis. Subsequently, the resulted tensor is contracted and renormalized along the $x$ axis. At this stage,  the central tensor $S^{(n)}$ and the unitary matrices $U_{\text{[Y]}}^{(n)}$ and $U_{\text{[X]}}^{(n)}$ have been created. The unitary matrices $U_{\text{[Y]}}^{(n)}$ and $U_{\text{[X]}}^{(n)}$ are calculated in the process of HOSVD of the tensors contracted along the $y$ axis and $x$ axis, respectively. Notice that the central tensor $S^{(n)}$ is composed of four tensors $T^{(n)}$ and can be found in the center of the new tensors $T^{(n+1)}$, $P^{(n+1)}$, and $Q^{(n+1)}$. Depending on what type of tensor is constructed, different legs ($L_{\text{[Y]}}$ or $L_{\text{[X]}}$) or carets ($C_{\text{[Y]}}$ or ($C_{\text{[X]}}$) are attached to the central tensor (see Fig.~\ref{Update}). By repeating this procedure, one can construct a lattice structure as large as required (e.g. the next iterative step yields to a tensor $T^{(n+2)}$ as depicted in Fig.~\ref{Large}).

\paragraph{Central tensor construction}

\begin{figure}[tb]
 \centering
 \includegraphics[width=2.5in]{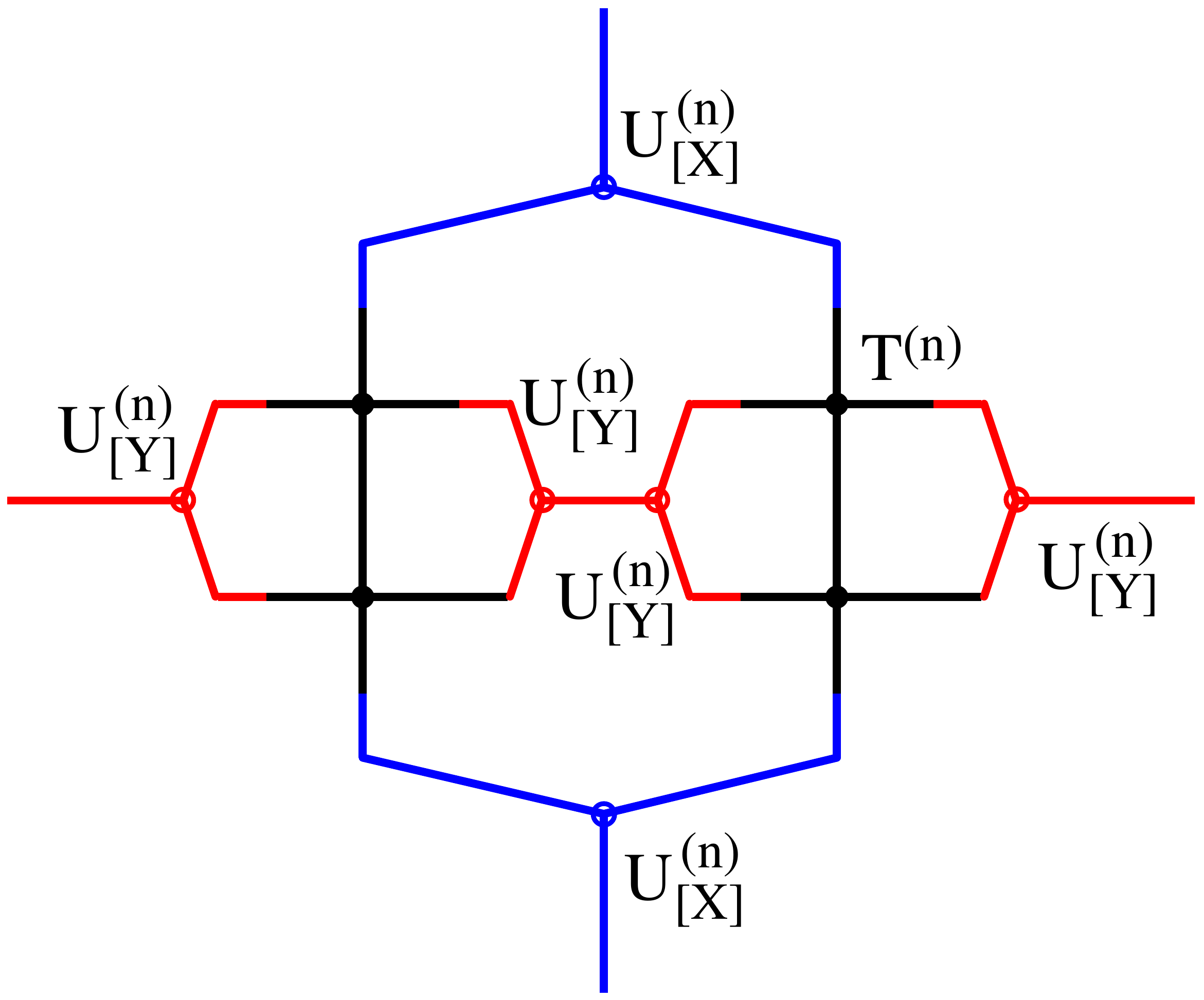}
 \caption{Structure of a central tensor $S^{(n)}$. Tensors are contracted and renormalized along the $y$ axis by the unitary matrix $U_{\text{[Y]}}^{(n)}$ (red color), and along the $x$ axis by $U_{\text{[X]}}^{(n)}$ (blue color). } \label{Central_S}
 \end{figure}

The central tensor $S^{(n)}$ is constructed in two steps: contraction and renormalization along the $y$ axis followed by the same procedure along the $x$ axis on the resulted tensor (see Fig.~\ref{Central_S}). 

First, by contraction of two tensors $T^{(n)}$ along the $y$ axis, we define

\begin{equation}
M^{(n)}_{\text{[Y]} xx'yy'} = \sum_i T^{(n)}_{x_1 x'_1 y i}  T^{(n)}_{x_2 x'_2 i y} \, , 
\end{equation}
where $x = x_1 \otimes x_2$ and $x' =x'_1 \otimes x'_2$. To truncate the tensor $M^{(n)}_{\text{[Y]}}$ by HOSVD, two matrix unfoldings are prepared

\begin{equation}
M_{\text{[Y]} (1) x, x'yy'} = M_{\text{[Y]} xx'yy'} \, , 
\end{equation}
and

\begin{equation}
M_{\text{[Y]} (2) x', yy'x} = M_{\text{[Y]} xx'yy'} \, .
\end{equation}
Then, SVD for these two matrices is performed

\begin{equation}
M_{\text{[Y]} (1)} = U_{\text{[Y]} (1)} \Sigma_{\text{[Y]} (1)}^{} V_{\text{[Y]} (1)}^{\dag} \, ,
\end{equation}

\begin{equation}
M_{\text{[Y]} (2)} = U_{\text{[Y]} (2)} \Sigma_{\text{[Y]} (2)}^{} V_{\text{[Y]} (2)}^{\dag} \, ,
\end{equation}
where $U_{\text{[Y]} (1)}$, $V_{\text{[Y]} (1)}$, $U_{\text{[Y]} (2)}$, and $V_{\text{[Y]} (2)}$ are unitary matrices of respective index dimensions, and $\Sigma_{\text{[Y]} (1)}$ and $\Sigma_{\text{[Y]} (2)}$ are matrices with singular values as its diagonal entries
\begin{equation}
\Sigma_{\text{[Y]} (.)} = \text{diag}(\sigma_{(.) 1}, \sigma_{(.) 2}, \text{\ldots}) \, .
\end{equation}
The singular values are ordered in decreasing order by convention. To obtain the best approximation of the tensor $M^{(n)}_{\text{[Y]}}$, the two errors

\begin{equation}
\varepsilon_1 = \sum_{i>D} \sigma_{(1) i}^2 
\end{equation}
and

\begin{equation}
\varepsilon_2 = \sum_{i>D} \sigma_{(2) i}^2
\end{equation}
are calculated and compared. If $\varepsilon_1 < \varepsilon_2$, we truncate the second index dimension of $U_{\text{[Y]} (1)}$ down to $D$ and set $U_{\text{[Y]}} = U_{\text{[Y]} (1)}$. Otherwise, the second index dimension of  $U_{\text{[Y]} (2)}$ is truncated and $U_{\text{[Y]}} = U_{\text{[Y]} (2)}$. 

After the truncation, we can create a new tensor

\begin{equation}
T^{(n)}_{\text{[Y]} x x' y y'} = \sum_{ij} U^{(n)}_{\text{[Y]} i x} M^{(n)}_{\text{[Y]} i j y y'} U^{(n)}_{\text{[Y]} j x'} \, .
\end{equation}

The contraction and the renormalization along the $x$ axis is performed identically. By the contraction of two tensors $T^{(n)}_{\text{[Y]}}$ along the $x$ axis, we define

\begin{equation}
M^{(n)}_{\text{[X]} xx'yy'} = \sum_i T^{(n)}_{\text{[Y]} x i y_1 y'_1}  T^{(n)}_{\text{[Y]} i x' y_2 y'_2} \, ,
\end{equation}
where $y = y_1 \otimes y_2$ and $y' = y'_1 \otimes y'_2$. Analogously, the matrix unfoldings are prepared

\begin{equation}
M_{\text{[X]} (3) y,y'xx'} = M_{\text{[X]} xx'yy'} \, ,
\end{equation}

\begin{equation}
M_{\text{[X]} (4) y',xx'y} = M_{\text{[X]} xx'yy'} \, ,
\end{equation}
on which SVD is performed again. As before, the errors $\varepsilon_3$ and $\varepsilon_4$ are compared and the chosen unitary matrix (associated with the smaller error $\varepsilon$) is truncated and set to $U_{\text{[X]}}$. Finally, we can define the central tensor as

\begin{equation}
S^{(n)}_{x x' y y'} = \sum_{kl} U^{(n)}_{\text{[X]} k y} M^{(n)}_{\text{[X]} x x' k l} U^{(n)}_{\text{[X]} l y'} \, .
\end{equation}

\begin{figure}[tb]
 \centering
 \includegraphics[width=2.3in]{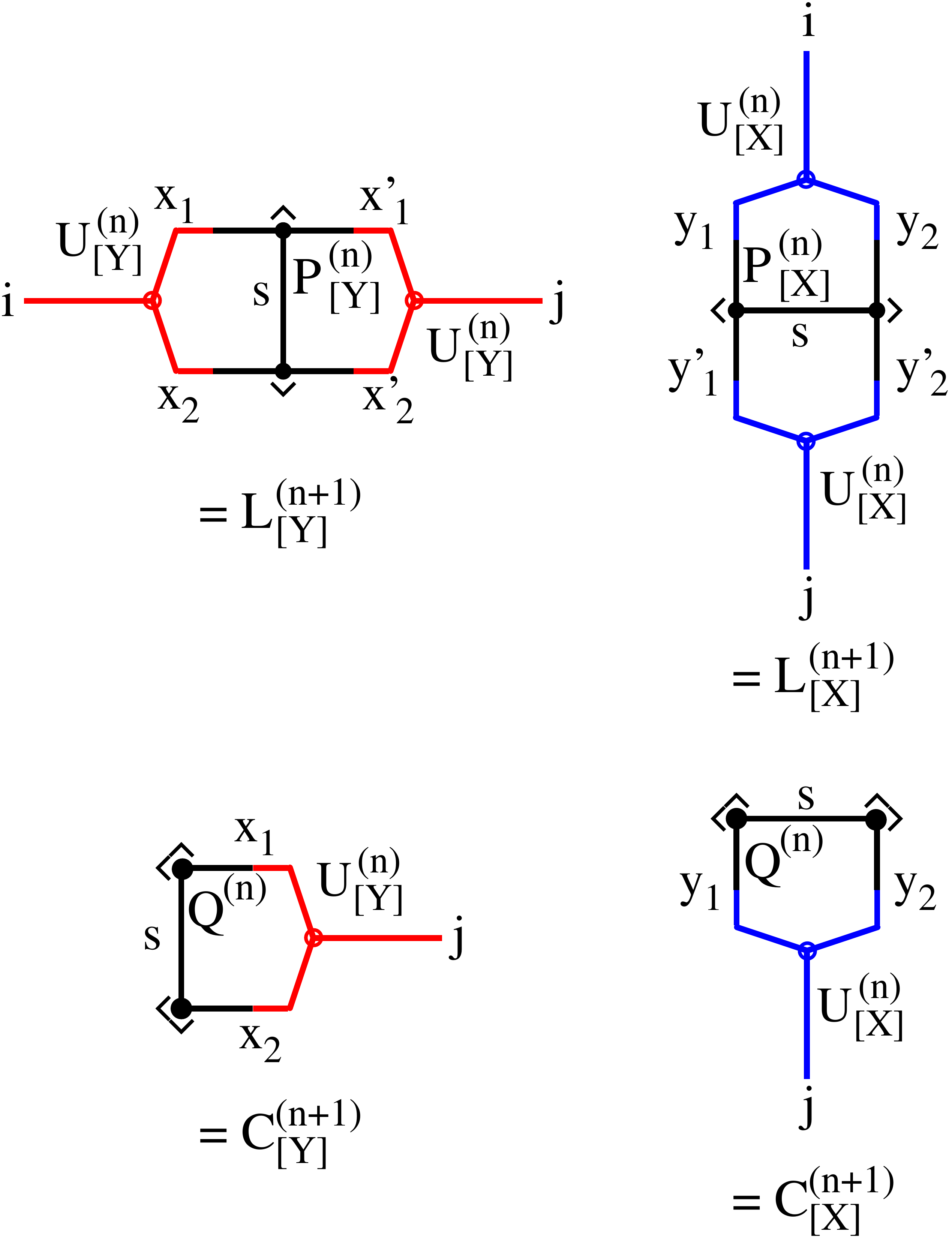}
 \caption{Composition of the legs and the carets (auxiliary objects).} \label{Legs_Carets}
 \end{figure}

\paragraph{Legs and carets construction} The legs and the carets are auxiliary objects used in updating the local tensors. These objects are composed of the tensors $P$ or $Q$ contracted with the unitary matrices $U_{\text{[Y]}}$ or $U_{\text{[X]}}$ (see Fig.~\ref{Legs_Carets}). 

Let us begin with the preparation of the leg $L^{(n)}_{\text{[Y]}}$

\begin{equation}
L^{(n)}_{\text{[Y]} ij} = \sum_{s x_1 x_2 x'_1 x'_2} U^{(n)}_{\text{[Y]} x_1\otimes x_2 i} P^{(n)}_{{\text{[Y]}} x_1 x'_1 s} P^{(n)}_{{\text{[Y]}} x'_2 x_2 s} U^{(n)}_{\text{[Y]}  x'_1 \otimes x'_2 j} \, .
\end{equation}
Note that the $P_{\text{[Y]}}$ tensors are symmetric, i. e., $P^{(n)}_{{\text{[Y]}} x x' s} = P^{(n)}_{{\text{[Y]}} x' x s}$. Hence, in this way, the calculation does not depend on the order of the first two indices of $P_{\text{[Y]}}$ (similar remark holds for $P_{\text{[X]}}$). 

The $L^{(n)}_{\text{[X]}}$ leg is constructed as

\begin{equation}
L^{(n)}_{\text{[X]} ij} = \sum_{s y_1 y_2 y'_1 y'_2} U^{(n)}_{\text{[X]} y_1\otimes y_2 i} P^{(n)}_{\text{[X]} y_1 y'_1 s } P^{(n)}_{\text{[X]} y'_2 y_2 s} U^{(n)}_{\text{[X]} y'_1\otimes y'_2 j} \, .
\end{equation}

Let us now proceed with the creation of the carets. The carets $C^{(n)}_{\text{[Y]}}$ and $C^{(n)}_{\text{[X]}}$ are defined as 

\begin{equation}
C^{(n)}_{\text{[Y]} j} = \sum_{s x_1 x_2} Q^{(n)}_{x_1 s} Q^{(n)}_{x_2 s} U^{(n)}_{\text{[Y]}x_1\otimes x_2 j} \, ,
\end{equation}

\begin{equation}
C^{(n)}_{\text{[X]} j} = \sum_{s y_1 y_2} Q^{(n)}_{s y_1} Q^{(n)}_{s y_2} U^{(n)}_{\text{[X]} y_1 \otimes y_2 j} \, ,
\end{equation}
respectively. Note that carets are just vectors (they have only single index $j$). 

\paragraph{Update of local tensors}

With all auxiliary objects prepared (central tensor, legs and carets), we are ready to create the new tensors $T^{(n+1)}$, $P^{(n+1)}_{\text{[Y]}}$,  $P^{(n+1)}_{\text{[X]}}$,  $Q^{(n+1)}$ for the next iteration step $n+1$. The local tensors are updated as follows (see Fig.~\ref{Update}):

 \begin{figure}[tb]
 \centering
 \includegraphics[width=4.5in]{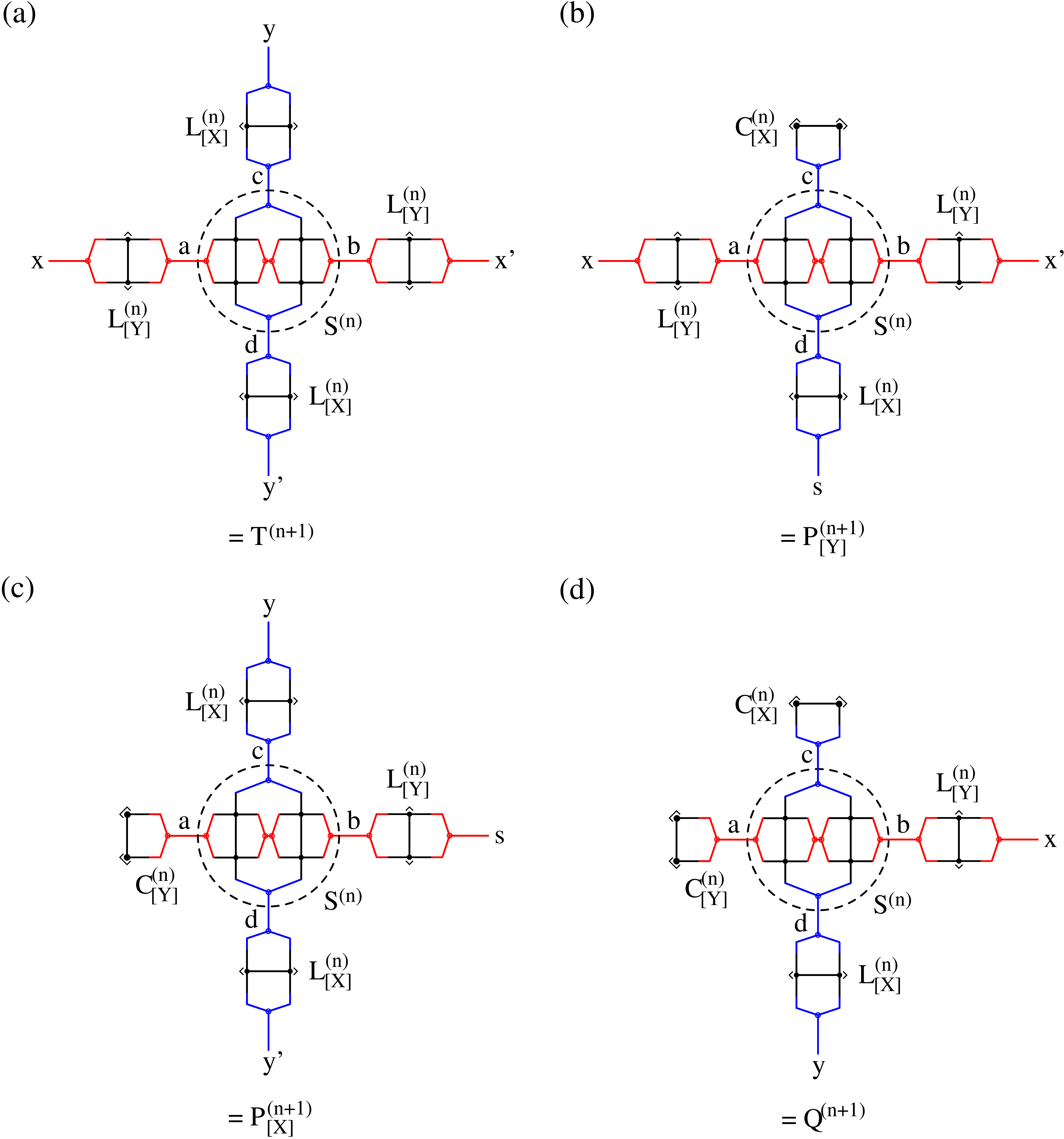}
 \caption{Update of local tensors.} \label{Update}
 \end{figure}

\begin{itemize}
\item creation of $T^{(n+1)}$ 

\begin{equation}
T^{(n+1)}_{xx'yy'} = \sum_{abcd} S^{(n)}_{abcd} L^{(n)}_{\text{[Y]} xa} L^{(n)}_{\text{[Y]} bx'} L^{(n)}_{\text{[X]} y c} L^{(n)}_{\text{[X]} d y'}
\end{equation}

\item creation of $P^{(n+1)}_{\text{[Y]}}$

\begin{equation}
P^{(n+1)}_{\text{[Y]} xx's} = \sum_{abcd} S^{(n)}_{abcd} L^{(n)}_{\text{[Y]} xa} L^{(n)}_{\text{[Y]} bx'} C^{(n)}_{\text{[X]} c} L^{(n)}_{\text{[X]} ds}
\end{equation}

\item creation of $P^{(n+1)}_{\text{[X]}}$ 

\begin{equation}
P^{(n+1)}_{\text{[X]} y y' s} = \sum_{abcd} S^{(n)}_{abcd} C^{(n)}_{\text{[Y]} a} L^{(n)}_{\text{[Y]} b s} L^{(n)}_{\text{[X]} y c} L^{(n)}_{\text{[X]} d y'}
\end{equation}

\item creation of $Q^{(n+1)}$

\begin{equation}
Q^{(n+1)}_{xy} = \sum_{abcd} S^{(n)}_{abcd} C^{(n)}_{\text{[Y]} a} L^{(n)}_{\text{[Y]} bx} C^{(n)}_{\text{[X]} c} L^{(n)}_{\text{[X]} dy}.
\end{equation}

\end{itemize}

\subsection{Numerical Results}\label{frac_num_results}

Throughout the numerical analysis, we keep using the setting $J = k_{\rm B} = 1$ so that
$K = 1 / T$. The numerical calculations by HOTRG have been carried out at $D = 24$, and
when analyzing the critical point, we used the auxiliary variable up to $D=32$. We have
also verified that the choice $D = 24$ is sufficient for obtaining the completely converged
free energy per site.
\begin{equation*}
F_n^{~}( T ) = - \frac{T}{N_n}  \ln {\cal Z}_n^{~}( T )
\end{equation*}
in the entire temperature region~\footnote{Larger values of $D$ are necessary if small density-matrix eigenvalues are required for the 
purpose of accurate analyzing their asymptotic decay.}. 
We treat the free energy per site in the thermodynamic limit
\begin{equation}
f( T ) = \lim_{n \rightarrow \infty}^{~} \, F_n^{~}( T ) \, .
\end{equation}
Our numerical analysis has always reached a complete convergence of the free energy whenever the number of the extensions is $n \lesssim 30$.

\begin{figure}[tb]
\begin{center}
\includegraphics[width=3.8in]{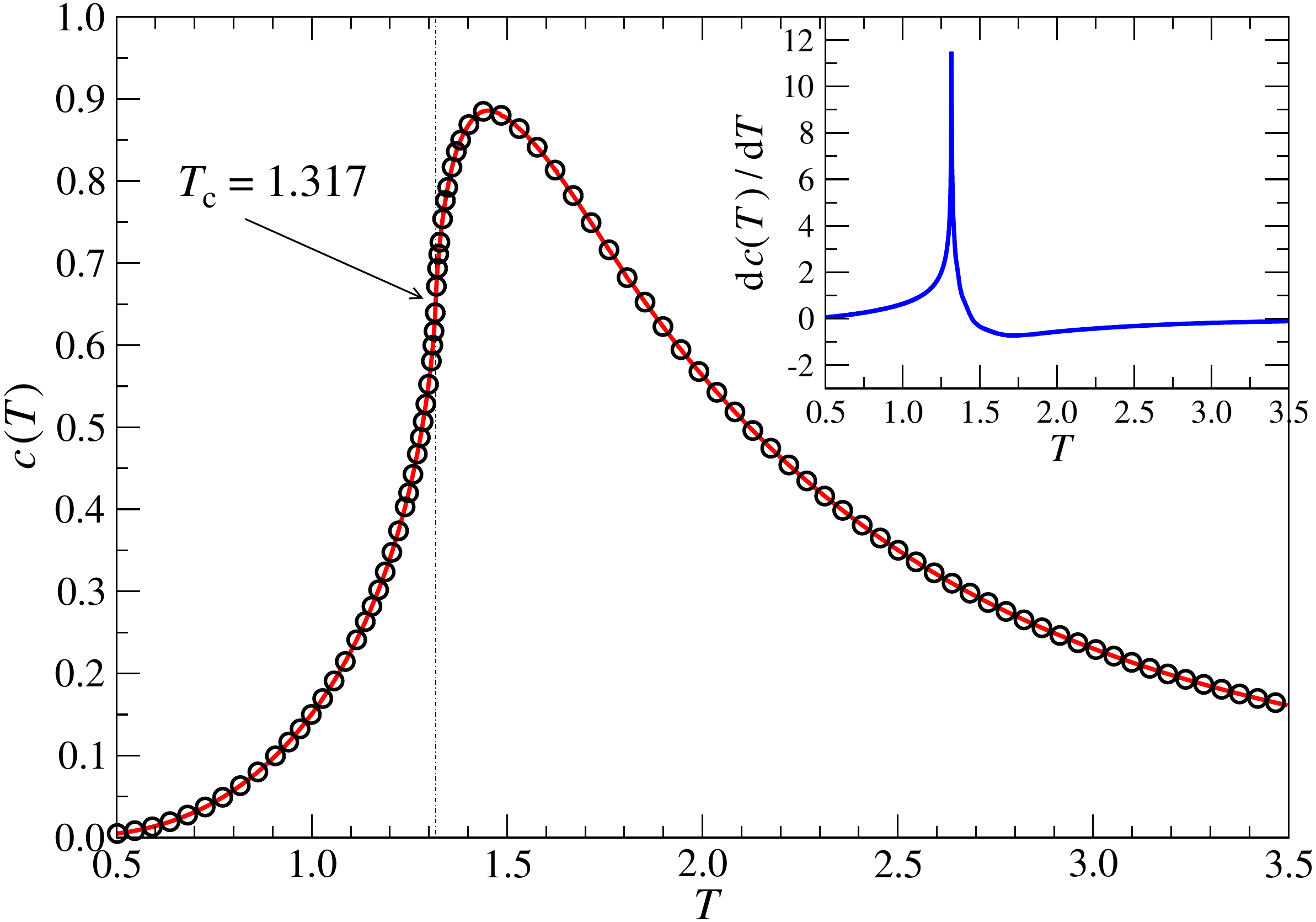}
\end{center}
\caption{Temperature dependence of the specific heat $c( T )$. 
Inset: if taking the numerical derivative of $c( T )$ with respect to temperature, a sharp peak appears exactly at $T_{\rm c}^{~} \approx 1.317$.}
\label{fig:Fig_2fr}
\end{figure}

Figure~\ref{fig:Fig_2fr} shows the temperature dependence of the specific heat per site $c( T )$, which is evaluated numerically as the second derivative of the free energy per site
\begin{equation}
	c( T ) = -T\frac{\partial^2 f(T)}{\partial T^2}
\label{Cv}
\end{equation}
in accord with Eq.~\eqref{spec_heat}. Surprisingly, there is no singularity in $c( T )$ around its maximum located at $T \approx 1.45$. Usually, a divergent peak in $c(T)$ is associated with a phase transition of the second order. However, one finds a weak non-analytic behavior at $T_{\rm c}^{~} \approx 1.317$, which is marked by the vertical dotted line in the figure. To visualize it, we performed a numerical derivative of $c( T )$ with respect to temperature (plotted in the inset), which exhibits a sharp peak at the correct critical temperature $T_{\rm c}^{~}$, which we analyze below in detail.

It is, however, numerically not feasible to determine the critical exponent $\alpha$ from the scaling $c(T) \propto |T-T_c|^{-\alpha}$, because of the weakness at the singularity. As shown in the figure, $c( T )$ around $T_{\rm c}^{~}$ is almost linear in $T$, and the exponent $\alpha$ is expected to be very close to zero.

\begin{figure}[tb]
\begin{center}
\includegraphics[width=3.8in]{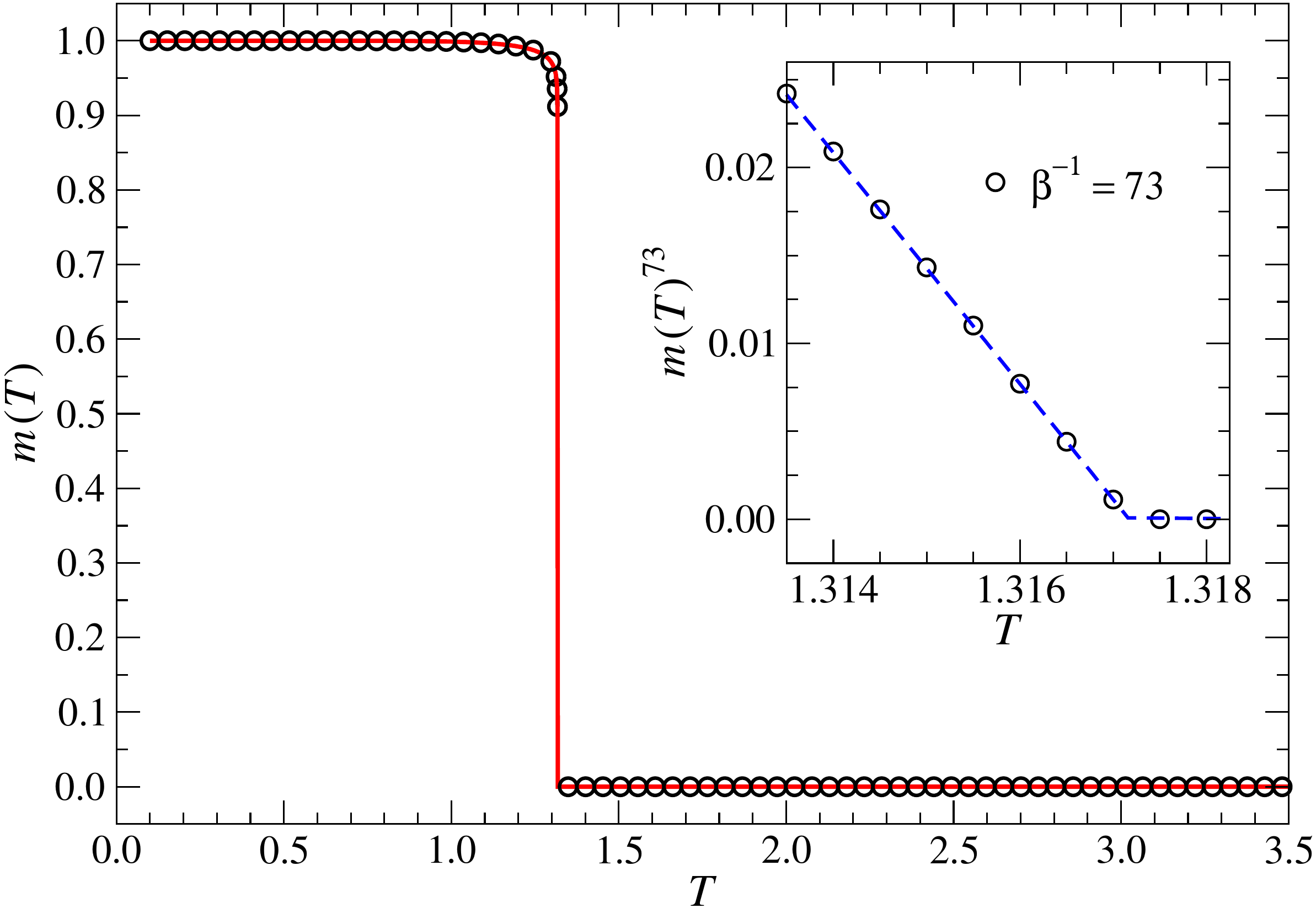}
\end{center}
\caption{The spontaneous magnetization per site $m( T )$. 
Inset: the power-law behavior below $T_{\rm c}^{~}=1.31716$.}
\label{fig:Fig_3fr}
\end{figure}

Figure~\ref{fig:Fig_3fr} shows the temperature dependence of the spontaneous magnetization per site $m( T )$. The calculation of the magnetization is obtained by means of an impurity tensor, i.e., we have inserted a $\sigma$-dependent local tensor into the system, as in Eq.~\eqref{Ising_Imp_init}. Since the fractal lattice is inhomogeneous, the value of the magnetization weakly depends on the location of the observation spin site. However, the critical behavior of the model is not affected by the location. We have chosen a spin site out of the four spin sites located in the middle of the 12-site cluster shown in Fig.~\ref{fig:Fig_1}.

The numerical calculation by HOTRG captures the spontaneous magnetization $m( T )$ below $T_{\rm c}^{~}$ since any tiny round-off error is sufficient for breaking the symmetry inside the low-temperature ordered state. Around the phase-transition temperature, the magnetization satisfies a power-law behavior (see Eq.~\eqref{beta_exponent})
\begin{equation}
m( T ) \propto
| T_{\rm c}^{~} - T |^{\,\beta}_{~} \, ,
\end{equation}
which is typical for the second-order phase transition. At a first glance, the sudden drop of the magnetization to zero just below the phase transition may suggest the presence of the first-order phase transition, where a discontinuous jump at the phase transition is present. It is not the case, since we were able to determine $T_{\rm c}^{~}=1.31716$ and $\beta=0.0137$ at $D=24$. If we have increased the numerical precision up to $D=32$, we got a better precision, where $T_{\rm c}^{~}=1.31717$ and $\beta=0.01388$. The linearity of $m(T)^{1/\beta}$ in temperature $T \leq T_{\rm c}^{~}$, plotted in the inset of Fig.~\ref{fig:Fig_3fr}, also confirms the correctness of the second-order phase transition at $T_{\rm c}^{~}$ and $\beta$.

\begin{figure}[tb]
\begin{center}
\includegraphics[width=3.8in]{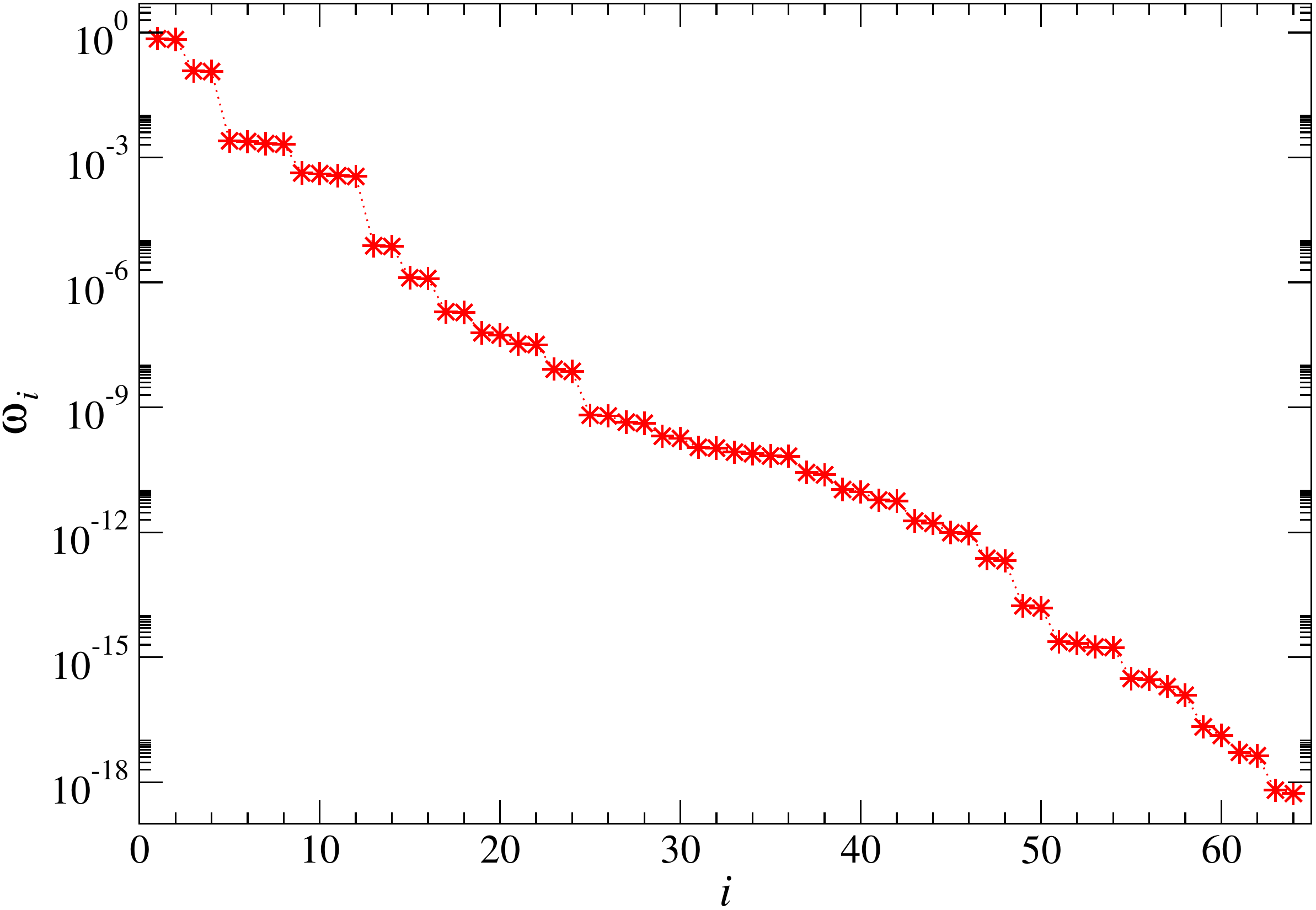}
\end{center}
\caption{Decay of the singular values after $n=8$ extensions.}
\label{fig:Fig_4fr}
\end{figure}

As a byproduct of the numerical HOTRG calculation, we can roughly
observe the entanglement spectrum~\footnote{It is possible to identify the system boundary of a finite area of two-dimensional classical lattice models as ``a wave function'' of a certain one-dimensional quantum system. In this manner, one naturally finds the quantum-classical correspondence, and can introduce the notion of entanglement in classical lattice models.}, which is the distribution of the eigenvalues $\omega$ of the density matrix that is created for the purpose of obtaining the renormalization transformation. Since the effect of  {\it environment} is not considered in our implementation of the HOTRG method, the eigenvalues $\omega_i^{~} = \lambda^2_i$ are the squared singular values $\lambda_i^{~}$ for $1\leq i \leq D^2$ in the higher-order
singular value decomposition applied to the extended tensors.
Figure~\ref{fig:Fig_4fr} shows a typical spectrum $\omega$ at the phase transition $T = T_{\rm c}^{~}$ and $D=8$ ordered decreasingly. The decay is exponential. The additional increase of the block-spin states (e.g. $D=16$) does not significantly improve the numerical precision of the partition function ${\cal Z}_n^{~}$. For comparison, the difference in the calculation of the free energy $f( T_{\rm c}^{~} )$ at $D = 8$ and at $D = 16$ is lower than $10^{-6}_{~}$. It should be noted that the eigenvalues are not distributed equidistantly in logarithmic scale; the corner double line structure is absent~\cite{CDL1, CDL2}.

\subsection{Outlook}\label{fractal_outlook}

This Section proposes several possible paths of our future work. 
Our preliminary data, which are to be further improved, extended, and later published, are presented in the following.

\begin{figure}[tb]
\begin{center}
\includegraphics[width=4.5in]{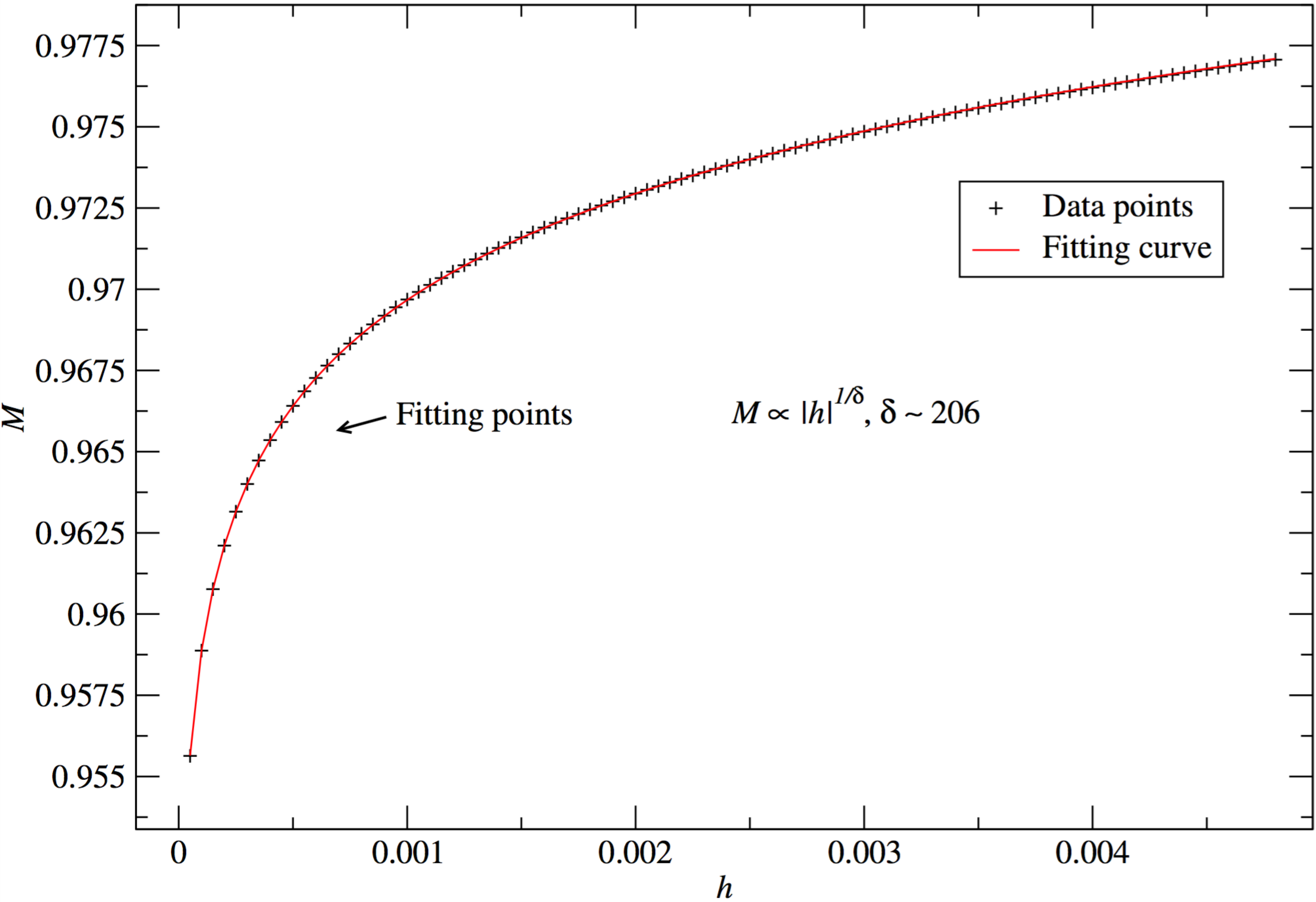}
\end{center}
\caption{The magnetic field response of the spontaneous magnetization at the temperature $T=T_c$ for the fractal-lattice Ising model studied earlier in this Chapter (for $D=12$). }
\label{frac_d12_fit2}
\end{figure}

\paragraph{Hyperscaling hypothesis for fractals}
A first interesting task is to verify the validity of the scaling relations Eq.~\eqref{scaling}--Eq.~\eqref{hyperscaling} numerically in the case of the fractional dimension. 
For this purpose, we obtained $\delta \approx 206$ by analysis of the field response in the fractal-lattice Ising model (for $D=12$)
\footnote{For $D=16$ (data not presented here), we have obtained $\delta \approx 205$, which is a good verification of the achieved accuracy.}, see Fig.~\ref{frac_d12_fit2}. 

Using the previously obtained exponents $\alpha=0$ and $\beta=1/73$ and the dimension of the lattice $d=3/2$, we can derive from the scaling relations Eq.~\eqref{scaling}-Eq.~\eqref{hyperscaling} following output

\begin{align}
\gamma &= 144 / 73 \, , \\
 \delta &= 145 \, , \\
\nu &= 4 / 3  \, .
\end{align}
Or, alternatively, by taking the dimension $d = d^{ ( {\rm H} ) }_{~} = 1.792$ we have $\nu = 1.116$.

Note that our numerical analysis yielded $\delta \approx 206$, not $\delta = 145$ as implied from the scaling hypothesis. 
This discrepancy will be analyzed and explained in our future studies. A possible explanation is that the scaling hypothesis is not applicable to fractal lattices. 
Another source of the discrepancy is the fact that the HOTRG is not accurate enough to be used for the determination of the exponent $\delta$. 
Notice that the calculation of $\delta$ (via the scaling relation) has nothing to do with the lattice dimension $d$. 

A question of high interest is to estimate the exponent $\nu$ numerically, 
which appears in the hyperscaling relation Eq.~\eqref{hyperscaling}.  
It is not clear if the exponent $\nu$ is well-defined in the case of the fractal-lattice Ising model as 
we have observed exponentially decaying spectrum of the singular values at the phase transition temperature; 
it means there is no power-law decay as it is characteristic at the phase transitions on Euclidean lattices. 
The power-law decay is connected to the algebraic decay of the correlation function, 
out of which one can calculate the critical exponent $\nu$. Since we have observed the exponential decay of the singular values 
with respect to tensor entanglement,  
the associated exponential decay of the correlation function at $T_c$ is expected, which cannot be used to obtain $\nu$ (as for mean-field models). 
However, we keep in mind that these are all very preliminary conjectures and we are still at the beginning of the study on the fractal 
geometries. 

\paragraph{Legs extension}

\begin{figure}[tb]
\begin{center}
\includegraphics[width=4.5in]{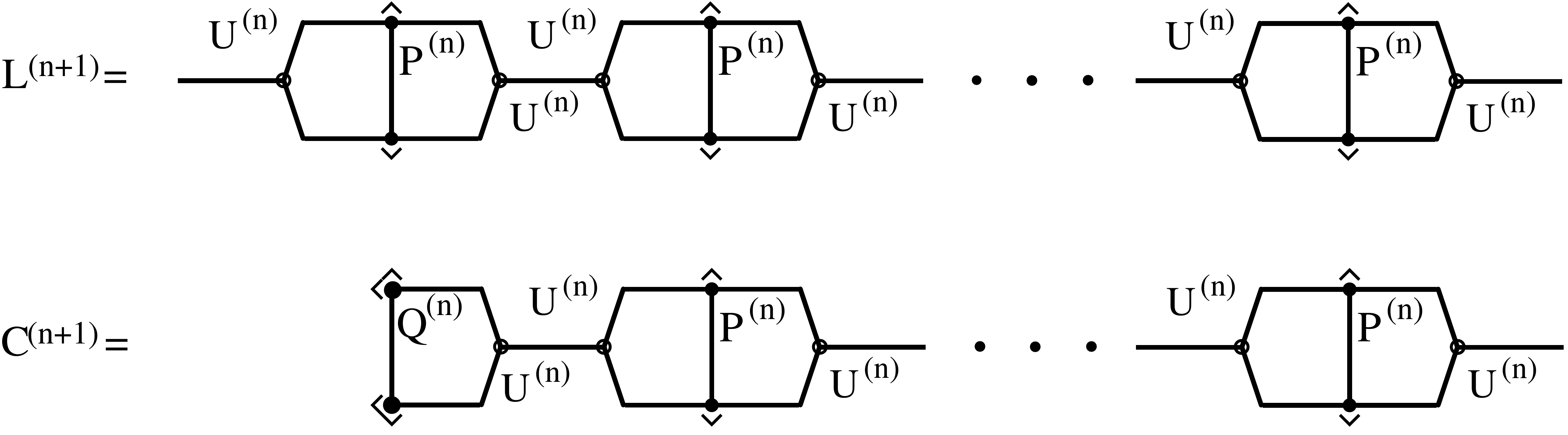}
\end{center}
\caption{Composition of the legs $L^{(n)}$ and the carets $C^{(n)}$. The leg is just $L$ times the leg of the original fractal as depicted in upper half of Fig.~\ref{Legs_Carets}. 
The caret is composed of the original caret (lower half of Fig.~\ref{Legs_Carets}) attached to the $L-1$ copies of the original leg.}
\label{extended_legs_carets}
\end{figure}

To progress in these studies, we intend to generalize the original fractal lattice so that we are able to control the lattice dimension by a tuning parameter $L$.
We propose an infinite series of fractal lattices, whose fractal dimensions $ \left\{ d_L \right\}_{i=0}^{\infty} $, $ d_0 \equiv 2 > d_1 > d_2 > \dots d_{\infty} \equiv 1 $, 
converge to the one-dimensional lattice monotonously. 
A simple way to decrease the dimension is to extend the leg size $L$, see Fig.~\ref{extended_legs_carets}. The resulted extended legs (and carets) 
are connected to the central tensor (``body'') in the same way as explained in the update of local tensors in the case of the original fractal lattice, cf. Fig.~\ref{Update}. 

\begin{figure}[tb]
	\vspace{-0.2cm}
\begin{center}
\includegraphics[width=3.3in]{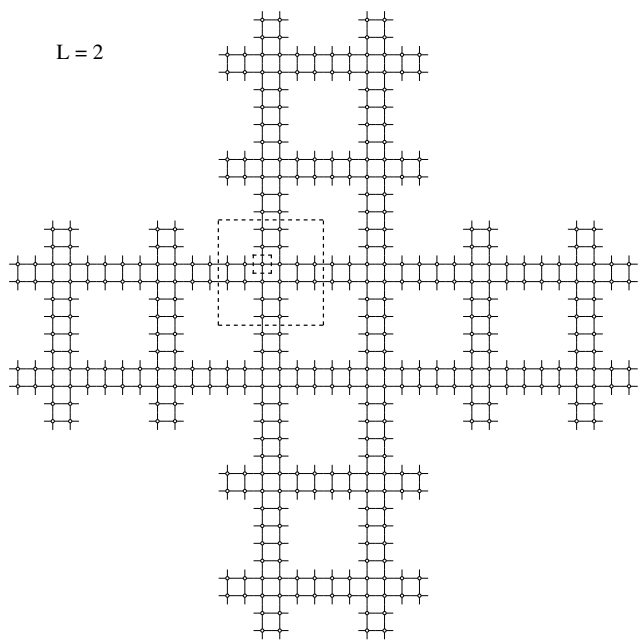}
\end{center}
	\vspace{-0.4cm}
\caption{Composition of the fractal lattice in the case of $L=2$. The smallest dashed square: a local tensor with the single spin site in the zeroth iteration step $n=0$. 
The bigger dashed square: basic cluster with 20 spin sites in the first iteration step $n=1$. The entire picture: extended cluster in the second iteration step $n=2$. 
The number of sites is  $20^n$ in the $n^{\text{th}}$ iteration step. }
\label{Large1}
\end{figure}

For better understanding, the extension process of the lattice when $L=2$ is graphically represented in Fig.~\ref{Large1}.
\begin{figure}[!tbh]
\begin{center}
\includegraphics[width=3.9in]{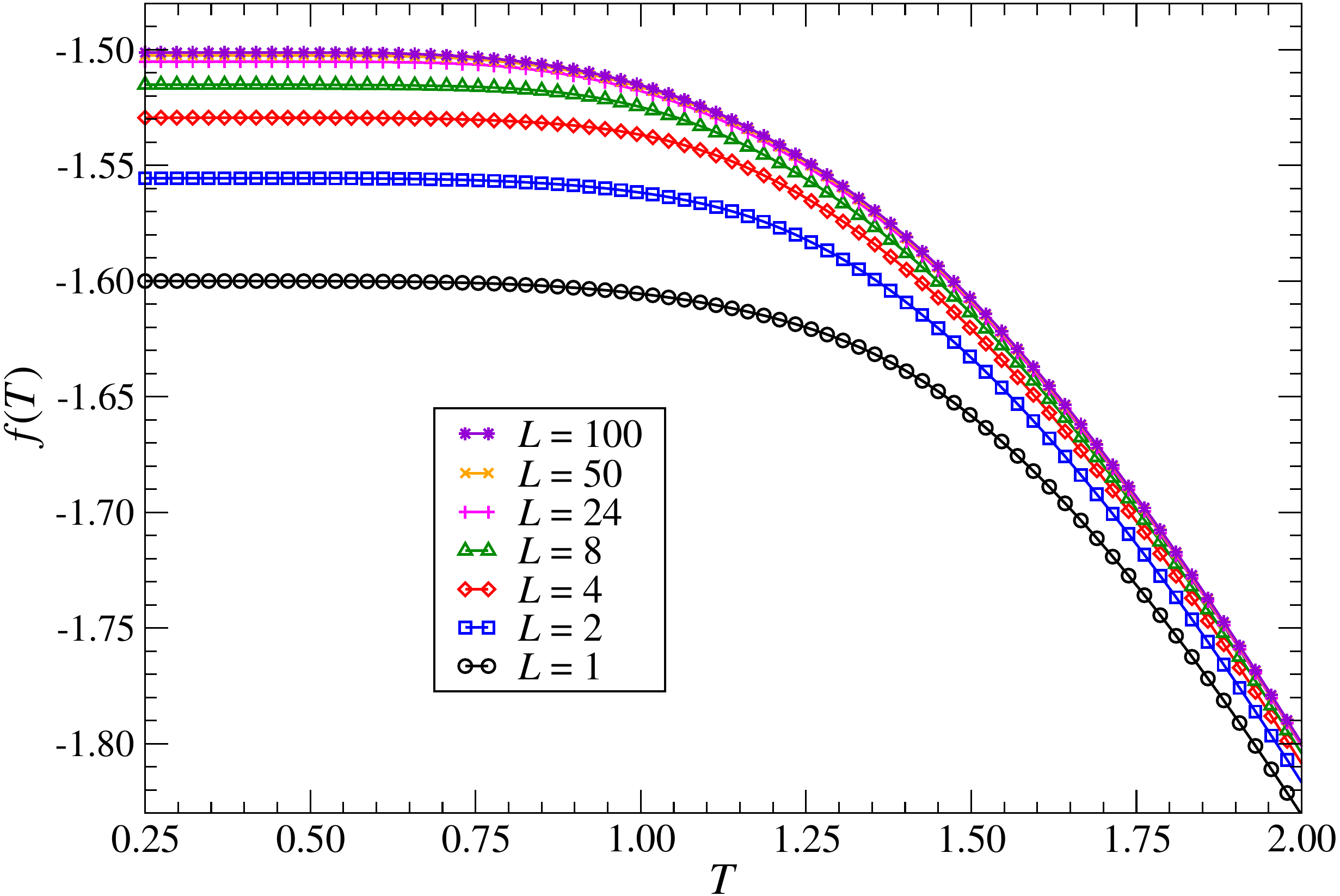}
\end{center}
	\vspace{-0.3cm}
\caption{The free energy for the fractal-lattice Ising model for the different legs $L$. }
\label{frac_F}
\end{figure}
The number of sites grows as $(4 + 8 L)^n$ with the iteration step $n$. 
One can easily find out that the Hausdorff dimension $d^{(H)}_L$ depends on $L$ as 

\begin{equation} \label{haus_dL}
d^{(H)}_L = \frac{\log (4 + 8L)}{\log(2 + 2L)} \, , \qquad L = 0, 1, 2, \dots, \infty \, , 
\end{equation}
whereas the other dimension $d_L$ is

\begin{equation} \label{tomo_dL}
d_L = 1 + \frac{\log(2)}{\log(2+2L)} \, . 
\end{equation}
Therefore, the infinite series of the fractal lattices allows us to study the thermodynamic properties of the spin models, which depend on their (fractional) dimension. 
So far, we have calculated the free energy and the spontaneous magnetization for different values of $L$, as shown in Figs.~\ref{frac_F} and~\ref{frac_L_D10}. 
\begin{figure}[tb]
\begin{center}
\includegraphics[width=3.9in]{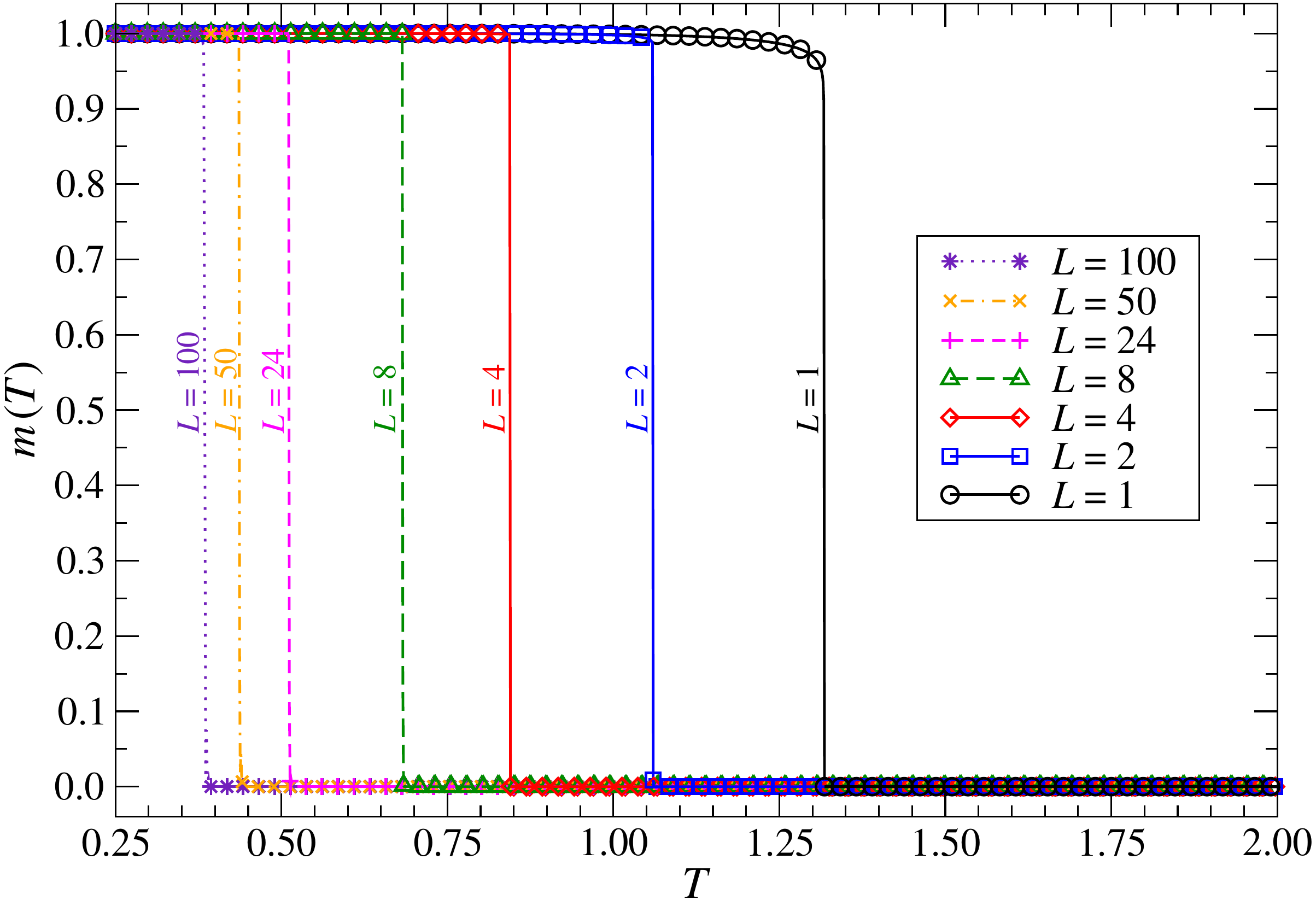}
\end{center}
	\vspace{-0.3cm}
\caption{The magnetization for the fractal-lattice Ising model for different values of the lengths of legs $L$. }
\label{frac_L_D10}
\end{figure}
\begin{figure}[tb]
	\vspace{-0.3cm}
\begin{center}
\includegraphics[width=1.0in]{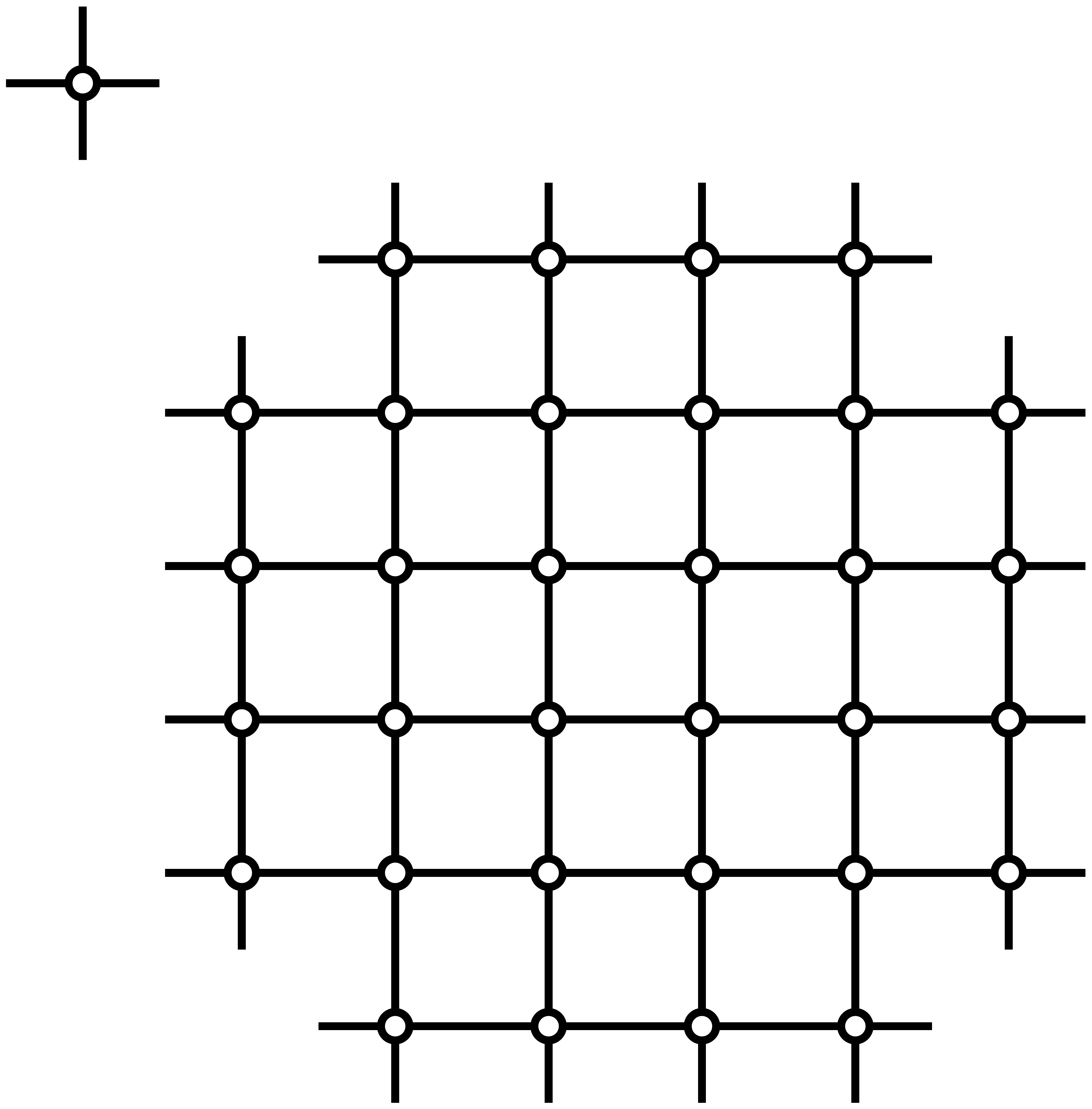}
\end{center}
	\vspace{-0.3cm}
\caption{The $6 \times 6$ fractal lattice. }
\label{fractal_6by6}
\end{figure}
It is immediately evident from the Fig.~\ref{frac_L_D10} that the expected critical temperature decreases as $L$ increases. 
Note that according the Eq.~\eqref{haus_dL} and Eq.~\eqref{tomo_dL}, both dimensions $d_L^{(H)}$ and $d_L$ converge to $1$ as $L$ goes to infinity 
(i.e. $\lim_{L \rightarrow \infty} d_L^{(H)} = \lim_{L \rightarrow \infty} d_L^{} = 1$). 
Further details and rigorous results will be published elsewhere. 

\paragraph{Body extension: 6 by 6 fractal}

Next step in generalization of the fractal lattice is to increase the size of the fractal ``body''. 
This approach is meant to propose a complementary infinite series of fractal lattices of 
fractal dimensions $\left\{d_K\right\}_{K=1}^{\infty}$ such that $d_1 < d_2 < \dots d_{\infty} \equiv 2$. 
In other words, the fractal dimensions $d_K$ converge monotonously to the two-dimensional square lattice. 
Let us return to the original fractal lattice and recall that 
it is composed of 4 by 4 spin blocks with the four corners removed.
By repeating the coarse-graining procedure in the process of creation of the central part, 
it is possible to construct the generalized series of the fractal lattices of the size $(2^K+2)$ by $(2^K+2)$, 
where $2^K$ is the linear dimension of the square-shaped body part of the lattice (i.e., without the legs). 
For instance, if $K=1$, we reproduce the original fractal lattice and for $K=2$ we define the following 6 by 6 lattice (see Fig.~\ref{fractal_6by6}), 
with the four missing corner spin sites.

\begin{figure}[tb]
	\vspace{-0.2cm}
\begin{center}
\includegraphics[width=3.8in]{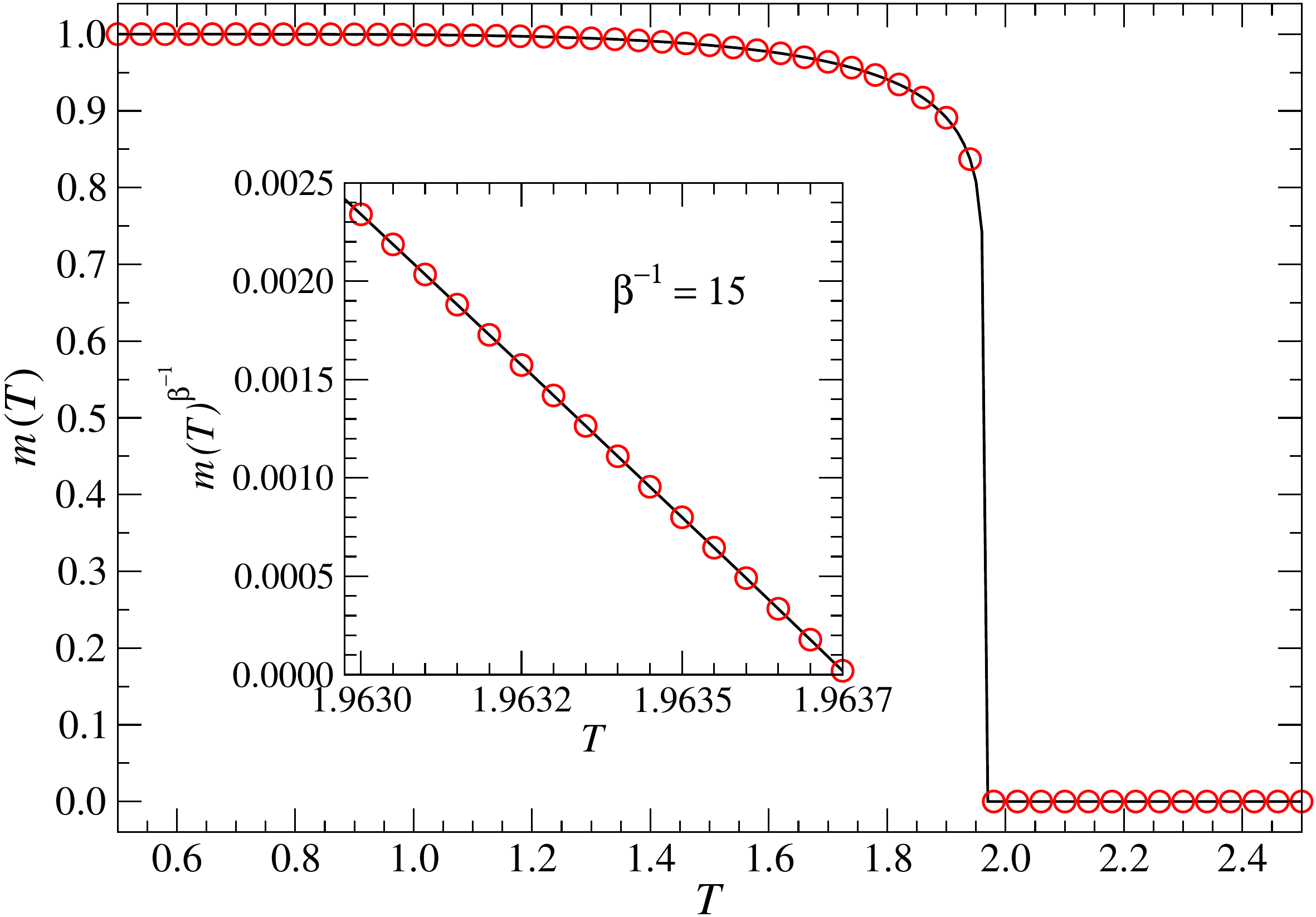}
\end{center}
	\vspace{-0.3cm}
\caption{The spontaneous magnetization $m(T)$ for the 6 by 6 fractal lattice (for $D=16$). Inset: the linear behavior of $\left[m(T)\right]^{1/\beta}$ below $T_c \approx 1.96376$. }
\label{magnetization_6x6_D16}
\end{figure}
\begin{figure}[!tbh]
	\vspace{-0.2cm}
\begin{center}
\includegraphics[width=3.8in]{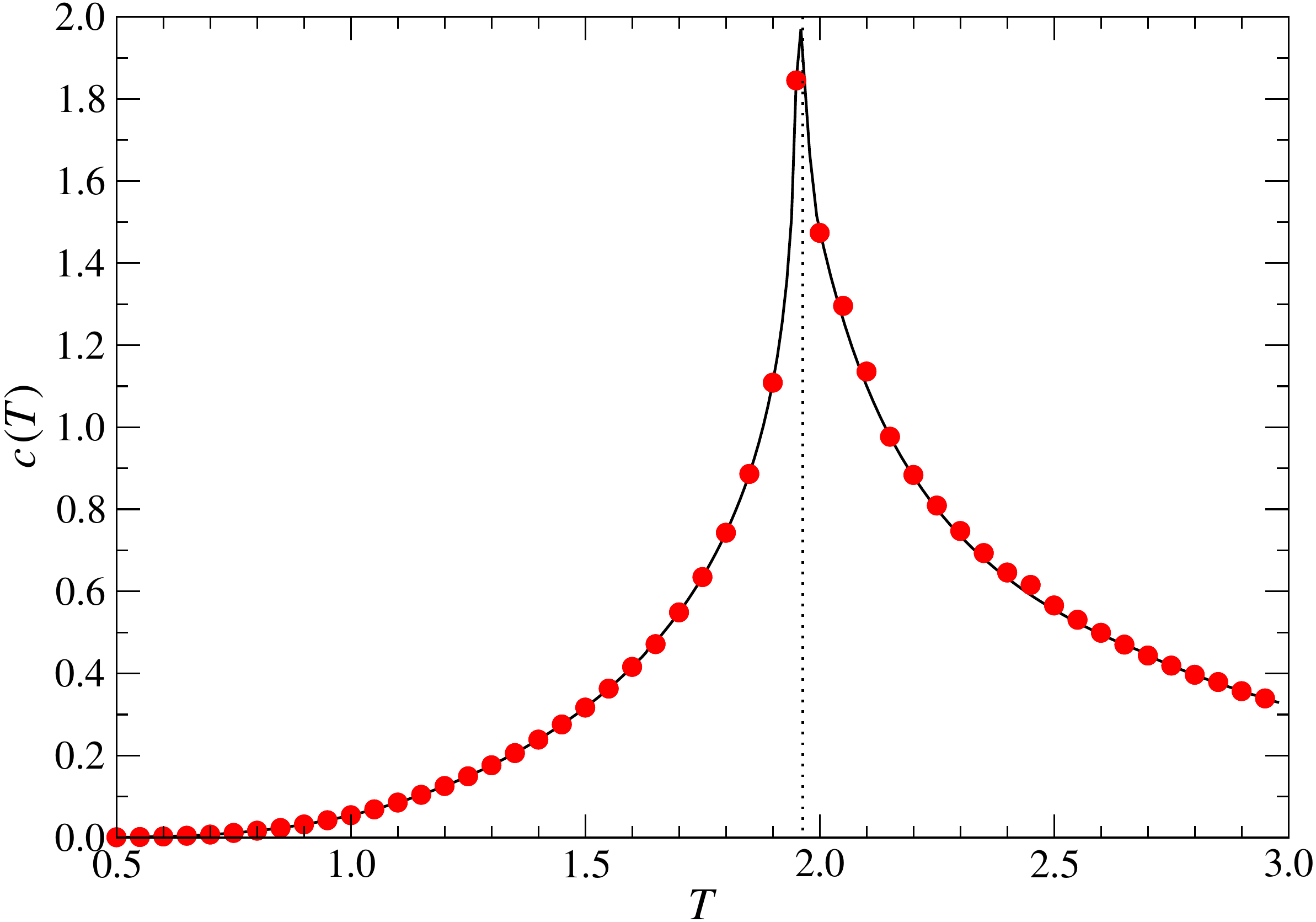}
\end{center}
	\vspace{-0.3cm}
\caption{The specific heat $c(T)$. The sharp singularity at $T_c \approx 1.96376$ corresponds to the second order phase transition 
of the Ising model on the fractal lattice with $K=2$ (for $D=16$).}
\label{specific_heat_6x6_D16}
\end{figure}

This generalization leads to the following dimensions

\begin{equation}
d^{(H)}_K = \frac{\log \left[ \left(2^K + 2\right)^2 - 4 \right] }{\log \left(2^K + 2 \right)} \, , \qquad d_K = 1 + \frac{\log \left( 2^K \right)  }{\log \left(2^K + 2 \right)} \, .
\end{equation}
Thus, for the 6 by 6 fractal lattice we get $d^{({\rm H})}_{K=2} \approx 1.934$ and $d_{K=2} \approx 1.774$. We estimated the critical temperature from the magnetization 
in this case to be $T \approx 1.96376$ and the critical exponent $\beta \approx 0.0658652 \approx 1/15$ (for $D=16$), see Fig.~\ref{magnetization_6x6_D16}. 
However, the accuracy in determining the critical temperature as well as in the critical exponent $\beta$ can be further improved by taking larger values of $D$. 
We have observed a sharp peak in the specific heat at $T_c \approx 1.96376$, see Fig.~\ref{specific_heat_6x6_D16}. 

\newpage\setcounter{equation}{0} \setcounter{figure}{0} \setcounter{table}{0}
\section{Conclusions and outlook}

This work aimed to explore the phase transitions of various spin systems on
non-Euclidean lattice geometries. Due to high complexity of this task,
non-standard mathematical concepts were required based on generalizations of
the tensor-product states. For this purpose, we have focused on the two methods,
CTMRG and HOTRG, which were found appropriate for treating the critical phenomena
of the multistate spin systems on the hyperbolic and fractal lattices, respectively.
 
We have been motivated by the two tensor-network-based algorithms in order to
extend their applicability to the multistate spin system on the non-Euclidean
geometries, which are exclusively specified by the topological structure of the
spin interaction in Hamiltonians. Such missing results have been demanding, as
they are considered to be the key for understanding various complex systems,
such as neural networks, social behavior analysis, as well as the general theory
of relativity, where the Euclidean geometry cannot reproduce the real systems.
Let us make another remark: None of these tasks has been known to be exactly
solvable or standard numerical methods, such as Monte Carlo simulations, exact
diagonalization, Density Matrix Renormalization Group, etc., also could not be
applicable straightforwardly.

The concluding remarks of our research are grouped into the three
work-packages (to be discussed separately in the following order):
\begin{itemize}
\item[{\bf{(1)}}] The unique free-energy analysis of the multistate spin systems on the
infinite set of hyperbolic geometries with respect to the radius of Gaussian curvature;
\item[{\bf{(2)}}] The application of the CTMRG method on a multistate model of social
behavior, which originates in the statistical physics;
\item[{\bf{(3)}}] The development of the algorithm (based on HOTRG), which can be used
to classify the phase transitions on fractal geometries.
\end{itemize}

\newpage

\paragraph{(1)}

We have analyzed the free energy per site of the multistate models on variety of
the underlying lattice geometries ($p\geq4,q\geq4$). For this purpose, we derived
an analytic expression for the free energy per site. It was a set of the recurrence
relations, which were required when calculating the free energy by the generalized
CTMRG algorithm for the regular non-Euclidean geometries. The derived free energy
per site can be applied to any spin model. We have chosen to study the $Q$-state clock 
and Potts models for $Q\geq2$. The numerical results yielded high numerical accuracy
with regard to the exact solutions known for the Ising models on the square and Bethe
lattices. If minimizing the free energy with respect to the underlying geometry
($p,q$), the lowest bulk free energy per site has yielded the multistate spin
models on the Euclidean square lattice ($4,4$) irrespective of temperature.

The free energy contains complete information on the spin system and incorporates
the boundary structure of the complex hyperbolic geometry. This is the essential
feature, which is important if describing the AdS (hyperbolic) spaces. We have been
trying to find a unique relation between the solid-state physics and the general
theory of relativity if classifying the regular AdS spaces. In particular, this 
relation lies in a direct calculation of the entanglement entropy by CTMRG of
a subsystem ${\cal A}$ in the quantum Heisenberg model on ($4,q\geq4$) lattice
geometries. We intend to understand the concept of the so-called holographic
entanglement entropy from the different viewpoint~\cite{Takayanagi}. This concept
states that a non-gravitational theory can live on the subsystem boundary
$\partial {\cal A}$ of $(d+1)$-dimensional hyperbolic spaces. Hence, the
entanglement entropy $S_{\cal A}$, which is associated with a reduced density
matrix of the subsystem ${\cal A}$, provides a correct measure of the information
contained in the AdS-CFT correspondence. The entanglement entropy $S_{{\cal A}}$
is related to a surface region $\partial {\cal A}$  (also known as the minimal
area surface) in the AdS space, which is bounded by a geodesic line, which we
can find by proper combining of the corner tensors. Moreover, the entanglement
entropy $S_{\cal A}$ is proportional to the corresponding $d$-dimensional region
${\cal A}$ defined within CFT. Our future aim is to obtain the von Neumann
entanglement entropy of quantum spin systems, which depends on the underlying AdS
lattice geometry.

Our results have revealed another surprising feature: there exists an inherited physical
similarity between the ground-state energy of microscopic multispin models and the
Gaussian curvature. Such an achievement certainly deserves a deeper understanding
supported by theoretical reasoning in the future. Our current numerical findings
cannot unambiguously justify the incomplete conjectures within the scope of this
work. In future, we intend to broaden our analyses to explain how the intrinsic
structure of the space-lattice geometry (being mapped onto microscopic spin-interaction
networks) affects the lowest energy of the entire system. Let us note that the energy
inherits information about the geometry of the entire system. We, therefore, conjecture
that the free-energy analysis of the multistate systems intrinsically contains the
underlying regular hyperbolic structure being proportional to the radius of curvature.
We are now collecting data on the complete set of lattices ($p\geq3,q\geq3$), where all
regular, spherical, and hyperbolic geometries are taken into account.

\newpage
\paragraph{(2)}

Having been motivated by the Axelrod model known for its applicability to mimic
social behavior, we have proposed a multistate thermodynamic spin model. Our
model was defined on the two-dimensional $(4,4)$ lattice in the thermodynamic limit
(whereas the original Axelrod model was considered on a square-shaped lattice
of a small size.) The thermodynamic model was analyzed numerically with the
aim to obtain equilibrial properties of our social model. We have been analyzing
its phase transitions for this non-trivial spin models.

We have considered a simplified case restricted to mimic two ($f=2$) cultural
features only, where each feature can assume $q$ different cultural traits
($2\leq q\leq6$).  Such constraints of our model have significantly affected
the thermodynamic properties resulting in the $q$-dependent phase-transition
point being associated with a critical noise. The value of the critical noise
was found to decrease with the increasing number of traits per feature $q$.
We have thus proposed a thermodynamic analog of the Axelrod model in two
dimensions, in which we do not consider the Potts-like interactions only.
Instead, we allowed a higher variability by incorporating the clock-like
interactions leading to the substantially richer communication structure
(which has its analog with the multistate spin interactions).

Such a multispin model could be again mapped onto mutually communicating
individuals subject to an external noise. The noise prevents the mutual
communication among the individuals. If the noise increases gradually,
the formation of larger clusters of the individuals is suppressed because
they do not share the identical cultural features (e.g., interests) any longer. 
The clusters were quantified by the order parameter in our model. If the noise
increases, the correlations are suppressed at longer distances. The noise has
the analogous character as the thermal fluctuations in the multistate spin model. 

We have identified two phase transitions in our social system for $q=2$.
The language of the social systems can be used to interpret our results
in the following example: let the first feature represent \textit{leisure-time
interests} taking two values: e.g. `reading books' and `listening to music'.
Let the second feature represent \textit{working duties} with the two values:
e.g., `manual activities' and `intellectual activities'. Both of the phase transitions
are continuous separating the three phases, which are classified into (i)
the low-noise regime, (ii) the medium-noise regime, and (iii) the high-noise
regime.

\begin{itemize}
\item[(i)] In the low-noise regime, the individuals tend to form a single dominant
cluster, where the associated complete order parameter can possess four states
(restricted to three states only in the limit of the zero noise), see Fig.~\ref{figg7}. 
The statistical probability of forming dominant clusters is proportional to
the evaluation of the complete order parameter $\langle O\rangle$. By increasing the
noise towards the phase transition between the low- and the medium-noise regimes,
the complete order parameter $\langle O\rangle$ does not drop to zero. Instead, it
becomes $\langle O\rangle=\frac{1}{2}$.
\item[(ii)] Within the medium-noise regime, another interesting topological phase reveals.
Two equally likely traits of the individuals are formed in it. In the social terms,
the pairing of the cultural settings coincides with two cases. The first one: (1a)
the equal mixture of those individuals who `read books' and `do manual activities' and
(1b) the individuals who `listen to music' and `do intellectual activities'. The
second one: (2a) the equal mixture of those who `listen to music' and `do manual
activities' and (2b) those who `read books' and `do intellectual activities'.
\item[(iii)] In the high-noise regime, the clusters of common interests become
less relevant, i.e., the correlation between the individuals weakens with the increasing
noise. As a consequence, the individuals behave in a completely uncorrelated way.
\end{itemize}

The further results, being associated with the only discontinuous phase transition
between the low- and high-noise regimes, are present if the trait number is larger
than two, i.e., for $q>2$. The larger clusters of individuals possessing $q^2$
cultural setting are formed inside the low-noise regime. Again, the order parameter
$\langle O\rangle$ measures the proportionality with the selected cultural setting
of the dominant cluster sizes. Therefore, this region corresponds to the ordered
multistate spin phase right below the phase transition noise $T_t(q)$. The high-noise
regime characterizes the fully uncorrelated individuals (the disordered phase) above
the phase transition noise. The low-noise regime is separated from the high-noise
regime by the discontinuity of the cluster size, in particular, the complete order
parameter exhibits a jump in agreement with the phase transition of the first order.

It is worth mentioning that the phase transition noise is found to be nonzero in the
asymptotic limit of the trait number $q\rightarrow\infty$, in particular,
$T_t(\infty)\approx\frac{1}{2}$. We, therefore, conjecture the permanent existence
of the correlated clusters of individuals below the nonzero phase transition noise
$T_t(\infty)$.

Recently, we have been investigating thermodynamic properties of the extended
social influence on the non-Euclidean and fractal communication geometries for
any $f\geq 2$ and $q\geq2$. The properties of the hyperbolic geometries with the
infinite dimensionality $d$ resemble the so-called \textit{small-world} effect,
which is the basic property of the many real-world networks, including the social
systems~\cite{BBV}. At the opposite spectrum, the fractal structure allows us to
study the social model in a range of fractional dimensions $1 \leq d \leq 2$,
which might provide additional insight into the character of the robustness 
of our model. We have completed interesting features which are to be published
elsewhere.

\newpage
\paragraph{(3)}

Finally, we have investigated the simple Ising model on the fractal lattice (as depicted
in Fig.~\ref{fig:Fig_1}) by means of the HOTRG algorithm. Although there was no evident
singularity in the specific heat, deeper analyzing suggests that the model exhibits
the second-order phase transition. Qualitatively, such an atypical existence of the weak
singularity in the specific heat is in agreement with the $\varepsilon$-expansion,
which exhibits increasing nature of the critical exponent in the specific heat with
respect to the space dimension $d~$~\cite{Wilson-Kogut}. At the same time, the spontaneous
magnetization reveals features of the second-order phase transition, for which
we have obtained the critical exponent $\beta_{\rm fractal}^{~} \approx 0.0137$.
Notice that the exponent $\beta$ is smaller by an order of the magnitude than the
Ising model critical exponent $\beta_{\rm square}^{~} = 1/8 = 0.125$ on the
two-dimensional square lattice.

The fractal structure of the lattice caused that the spectrum of the entanglement
entropy also differs from the square lattice, as explained by the corner double
line picture~\cite{CDL1,CDL2}. The process of the renormalization-group transformation
results in absorption of the short-range entanglement, which has an origin in the
missing four corners of the basic tensor cell (it forms the fractal structure
of the lattice, cf. Fig.~\ref{fig:Fig_1}). Therefore, a few degrees of the freedom
suffice to manipulate with the renormalized tensors. The situation is similar to the
entanglement structure, as reported in the tensor renormalization~\cite{TNR1, TNR2,
TNR3, TNR4, TNR5, Loop}.

One can also create a variety of fractal lattice geometries by modifying the basic
tensor cell. We consider the following three processes of proposing an infinite series
of the fractal lattices to be investigated in future, as in Sec.~\ref{fractal_outlook}.
\begin{itemize}
\item[(i)] The first process can be carried out by a gradual extension of the length of
the connecting legs. We would have thus specified an infinite series of the fractal
lattices with monotonously decreasing dimensions $ \left\{ d_L \right\}_{L=0}^{\infty}$,
which converge to the one-dimensional chain, i.e., $ d_0 \equiv 2 > d_1 > d_2 > \dots
d_{\infty} \equiv 1$;
\item[(ii)] Instead of considering the leg extensions, we could have expanded the body
size of the basic tensor cells. This might lead to a different infinite series of the
fractal lattices with the fractal dimensions $\left\{d_K\right\}_{K=1}^{\infty}$, which
satisfy another monotonous increasing sequence of the dimensions
$d_1 < d_2 < \dots d_{\infty} \equiv 2$ converging to the two-dimensional square
lattice.
\item[(iii)] Furthermore, an appropriate combination of the two processes is also
available. Such a process would have been useful to construct fractal lattices of
the desired non-integer dimension.
\end{itemize}

The justification for considering such specific processes lies in a long-lasting open
problem of verifying the validity of the scaling hypotheses for the fractional systems. 
Further, numerical analyses of the quantum spin systems on a variety of the fractal
lattices is another challenging extension of the task~\cite{Voigt1, Voigt2}. All of these
studies will help clarify the role of the entanglement in the universality of the
phase transition in both the regular and the fractal lattices. 

\newpage

\addcontentsline{toc}{section}{Acknowledgement}
\begin{ack}

Here, J.G. would like to thank his supervisor Andrej Gendiar. 
Thanks for his humor, enthusiasm, and contagious interest in the research,
it has always been fun to work together. He taught me everything about CTMRG 
and encouraged me to write my code (CTMRG, HOTRG, TEBD) in C++. 
This work would have never been possible without his support.

The fractal lattices have been studied in collaboration with Tomotoshi Nishino, 
who came up with this idea in the first place. This collaboration has proved to
be fruitful, and new results are expected to be published soon~\cite{GGT}. 
J.G. is grateful to colleagues Yoju Lee and Roman Krcmar for the interesting and
helpful discussions. 

This work was supported by Vedeck\'{a} Grantov\'{a} Agent\'{u}ra M\v{S} VVa\v{S} SR
a SAV VEGA-2/0130/15 and Agent\'{u}ra na Podporu V\'{y}skumu a V\'{y}voja
QETWORK APVV-14-0878.
\end{ack}

\newpage

\addcontentsline{toc}{section}{Bibliography}
\bibliographystyle{apalike}
\bibliography{main}

\end{document}